\documentclass[3p]{elsarticle}

\let\today\relax
\makeatletter
\def\ps@pprintTitle{%
    \let\@oddhead\@empty
    \let\@evenhead\@empty
    \let\@evenfoot\@oddfoot
    }
\makeatother
\usepackage{framed,multirow}

\usepackage{amssymb}
\usepackage{latexsym}

\usepackage{graphicx}
\usepackage[T1]{fontenc}
\usepackage{mwe}
\usepackage{array}
\usepackage[usestackEOL]{stackengine}
\usepackage{booktabs}
\usepackage[export]{adjustbox}
\usepackage{textcomp}
    
\usepackage{mathtools}
\usepackage{animate}
\usepackage{float}
\usepackage{bm}
\usepackage{amsmath,amsfonts, amsthm}
\usepackage{caption}
\usepackage{subcaption}
\usepackage{chngcntr}
\usepackage{soul}
\usepackage{color}
\usepackage{dsfont} 

\theoremstyle{plain}
\usepackage[normalem]{ulem}

\usepackage{pgf,tikz}
\usetikzlibrary{arrows, positioning}

\usepackage{lineno}

\usepackage{hyperref}
\usepackage{cleveref}
\hypersetup{
	colorlinks=true,       
	linkcolor=green,        
	citecolor=green,        
	filecolor=magenta,     
	urlcolor=green         
}
\usepackage[toc]{appendix}
\newcommand{\bx}{\boldsymbol{X}}
\newcommand{\by}{\boldsymbol{Y}}
\newcommand{\bz}{Z}
\newcommand{\bq}{\boldsymbol{Q}}
\newcommand{\q}{\boldsymbol{q}}

\newcommand{\x}{\boldsymbol{x}}

\newcommand{\m}{m}
\newcommand{\boldm}{M}

\newcommand{\Ent}{\mathrm{H}}

\newcommand{\n}{w}
\newcommand{\y}{\boldsymbol{y}}
\newcommand{\z}{z}
\newcommand{\zspace}{V^{z}}

\newcommand{\rsample}{\boldsymbol{r}}
\newcommand{\br}{\boldsymbol{R}}

\newcommand{\w}{\boldsymbol{w}}
\newcommand{\s}{\boldsymbol{s}}
\newcommand{\bw}{\boldsymbol{W}}
\newcommand{\floor}[1]{\lfloor #1 \rfloor}

\newcommand{\bstate}{U}
\newcommand{\state}{u}
\newcommand{\R}{\mathbb{R}}
\newcommand{\noise}{\boldsymbol{N}}
\newcommand{\dimx}{x}

\newcommand{\dimy}{y}
\newcommand{\statespace}{V^{\state}}
\newcommand{\dimz}{z}
\newcommand{\dimw}{w}
\newcommand{\identity}{\mathrm{Id}}

\newtheorem{assumption}{Assumption}

\DeclareMathOperator*{\argmax}{arg\,max}
\DeclareMathOperator*{\argmin}{arg\,min}

\newtheorem{observation}{Observation}

\begin{document}


\begin{frontmatter}

\title{Bayesian Model Calibration for Block Copolymer Self-Assembly: Likelihood-Free Inference and Expected Information Gain Computation via Measure Transport}

\author[mit]{Ricardo Baptista\corref{cor1}\fnref{1}}
\ead{rsb@mit.edu}
\author[ut]{Lianghao Cao\corref{cor1}\fnref{1}}
\ead{lianghao@oden.utexas.edu}
\author[ut]{Joshua Chen\fnref{1}}
\ead{joshuawchen@utexas.edu}
\author[ut]{Omar Ghattas}
\ead{omar@oden.utexas.edu}
\author[mit]{Fengyi Li\fnref{1}}
\ead{fengyil@mit.edu}
\author[mit]{Youssef M. Marzouk}
\ead{ymarz@mit.edu}
\author[ut]{J. Tinsley Oden}
\ead{oden@oden.utexas.edu}
\fntext[fn1]{Authors contributed equally to this manuscript.}

\cortext[cor1]{Corresponding authors}

\address[mit]{Massachusetts Institute of Technology, Cambridge, MA 02139, USA}
\address[ut]{The University of Texas at Austin, Austin, TX 78712, USA}

\begin{abstract}
We consider the Bayesian calibration of models describing the phenomenon of block copolymer (BCP) self-assembly using image data produced by microscopy or X-ray scattering techniques. To account for the long-range disorder in BCP equilibrium structures we introduce auxiliary variables to represent this aleatory uncertainty. These variables, however, result in an integrated likelihood for high-dimensional image data that is generally intractable to evaluate. We tackle this challenging Bayesian inference problem using a likelihood-free approach based on measure transport together with the construction of summary statistics for the image data. We also show that expected information gains (EIG) from the observed data about the model parameters can be computed with no significant additional cost. Lastly, we present a numerical case study based on the Ohta--Kawasaki model for diblock copolymer thin film self-assembly and top-down microscopy characterization. For calibration, we introduce several domain-specific energy- and Fourier-based summary statistics and quantify their informativeness using EIG. We demonstrate the power of the proposed approach to study the effect of various data corruptions and experimental designs on the calibration results.
\end{abstract}


\end{frontmatter}

\tableofcontents

\section{Introduction}\label{sec:introduction}
\subsection{The self-assembly of block copolymers: Application and modeling}
\begin{figure}
    \centering
    \begin{tabular}{c c}
        \includegraphics[width = 0.4\linewidth]{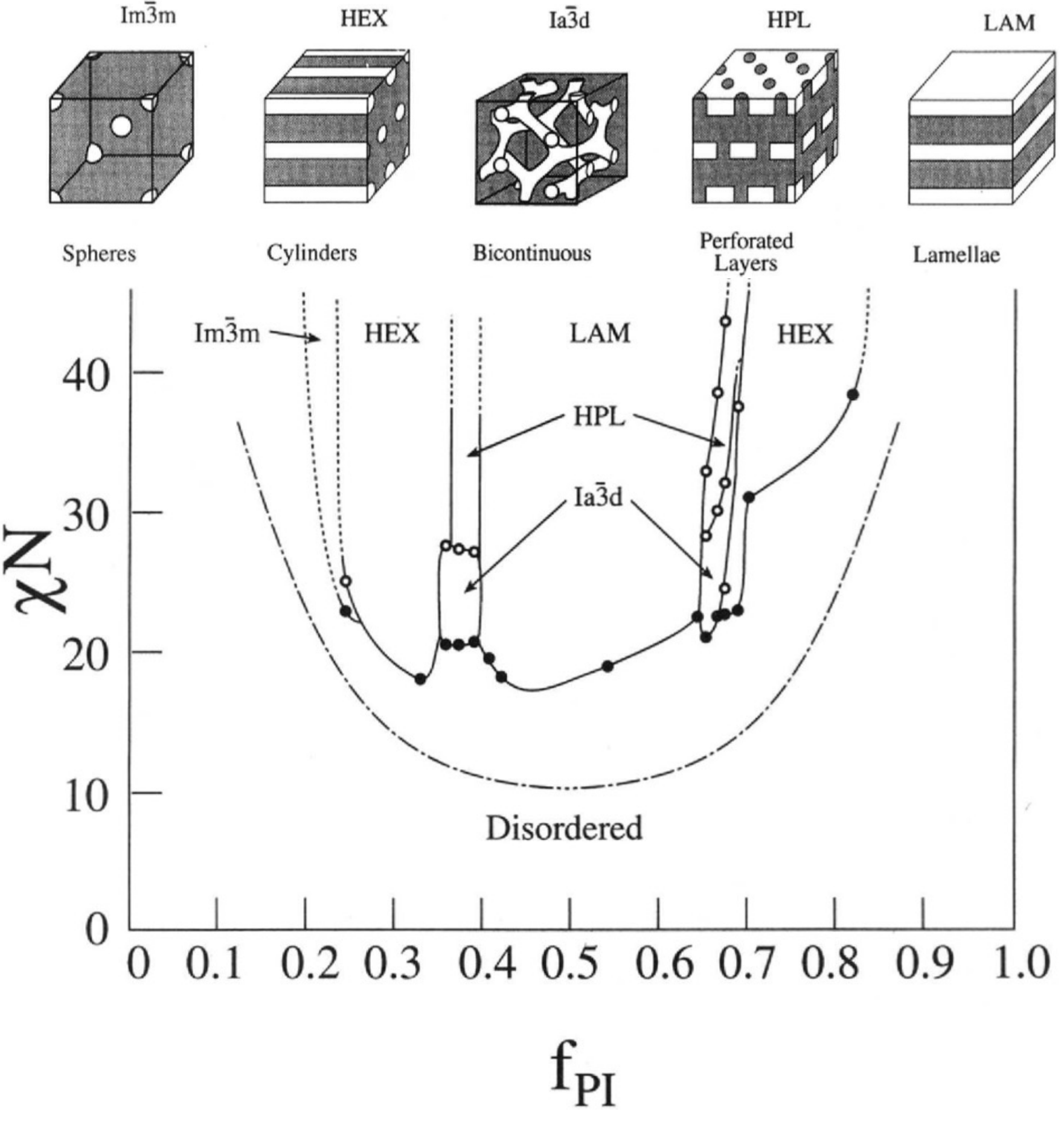} & \includegraphics[width = 0.4\linewidth]{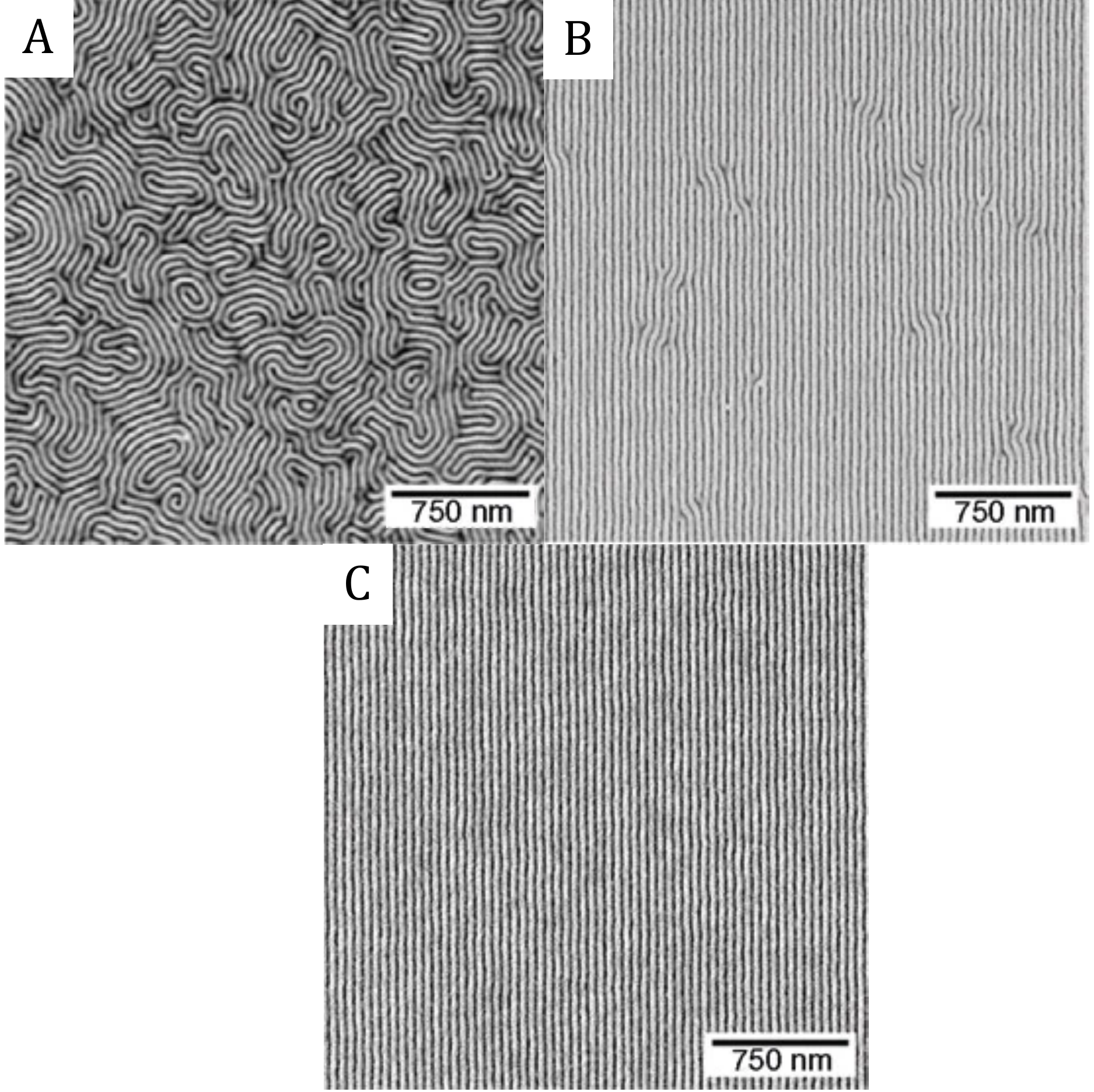} 
    \end{tabular}
    \caption{(\emph{left}) The experimental phase diagram for the self-assembly of Polyisoprene-Polystyrene diblock copolymers. The parameter $f_{PI}$ denotes the volume ratio of the Polyisoprene block, $\chi$ denotes the Flory--Huggins interaction parameter, and $N$ denotes the degree of polymerization. Reprinted (adapted) with permission from \cite{Khandpur1995}. Copyright \textcopyright 1995 American Chemical Society. (\emph{right}) Top-down scanning electron microscope images of lamella phase PS-b-PMMA copolymer films with (A) a chemically neutral surface; (B) a chemically nanopatterned surface with period smaller than the natural period of the copolymer, thus leading to defects; (C) a chemically nanopatterned surface with period close to the natural period of the copolymer. Reprinted (adapted) with permission from~\cite{Kim2003}. Copyright \textcopyright 2003 Springer Nature.}
    \label{fig:experimental_phase_diagram}
\end{figure}

Block copolymer (BCP) melts are large collections of one or more types of BCPs, or polymers consisting of several blocks of distinct monomers joined together by covalent bonds. Originally mixed in high temperature equilibrium, the distinct monomer blocks in a BCP melt spontaneously segregate upon thermal or solvent annealing as a result of their thermodynamic incompatibility. Periodically ordered nanoscale structures emerge as the melt achieves a new equilibrium configuration. This physical phenomenon is known as the \textit{self-assembly} of BCPs~\cite{Hamley1998, Bates1999}. Examples of the equilibrium structures for the self-assembly of diblock copolymers (Di-BCPs), i.e., BCPs with two distinct blocks, are lamellae, spheres, cylinders, etc.; see 
Fig.~\ref{fig:experimental_phase_diagram} (\emph{left}) along with an experimental phase diagram. The massive attention given to BCP self-assembly stems from the flexibility in their achievable equilibrium structures when subject to external guidance. Specifically, carefully engineered techniques, such as substrate-based techniques~\cite{Segalman2001, Kim2003, Bita2008} or techniques using thermal and electric fields~\cite{Thurn-Albrecht2002, Berry2007, Wang2013}, can influence the self-assembly process to form equilibrium structures with desired features. The application of these techniques is known as \emph{directed self-assembly} (DSA) of BCPs. The DSA of BCPs has quickly attracted attention for its wide-ranging commercial applications, the most notable being \emph{patterning} for lithography at the sub-10 nm scale~\cite{Park2003, Stoykovich2007, Bates2014, Ji2016}.

Computer simulation of BCP self-assembly based on mathematical models has played an important role in the study of self-assembly. These simulations have proven to be valuable in determining the effects of material properties, film thickness, and polymer-substrate interactions on the self-assembly \cite{Ginzburg2015}. The popular models and simulation techniques for self-assembly are predominantly derived from statistical polymer theory, such as Monte Carlo simulations \cite{Wang1999, Binder2000}, theoretically-informed coarse-grained (TICG) simulation \cite{Detcheverry2009}, self-consistent field theory (SCFT) \cite{Muller2005, Fredrickson2006}, and density functional theory (DFT) \cite{Fraaije1993, Uneyama2005}\textemdash listed in decreasing order by granularity\footnotemark. While model-based studies of the self-assembly mainly serve as aides to interpreting empirical findings, there is an increased interest in using models to make potentially consequential predictions. Most noticeably, recent efforts have been made on the \emph{inverse design}\footnotemark of DSA, where model simulations and iterative optimization schemes suggest designs of external guidance and BCP material properties that are best suited for achieving a target equilibrium pattern \cite{Hannon2013, Qin2013, Khaira2014, Hannon2014, Khadilkar2017}. Model-based inverse design is often considered the critical pathway towards transitioning DSA technology from the laboratory into commercial manufacturing of integrated circuit devices \cite{itrs2013}.

\footnotetext{We define the granularity of a model as the finest length scale at which it can faithfully resolve polymer conformation, monomer interaction, and thermal fluctuation.}
\subsection{Calibration of block copolymer self-assembly models: Uncertainties and the Bayesian approach}\label{subsubsec:Bayes_justification}
\begin{figure}
    \centering
    \begin{tabular}{c c}
        \includegraphics[width =0.3\linewidth]{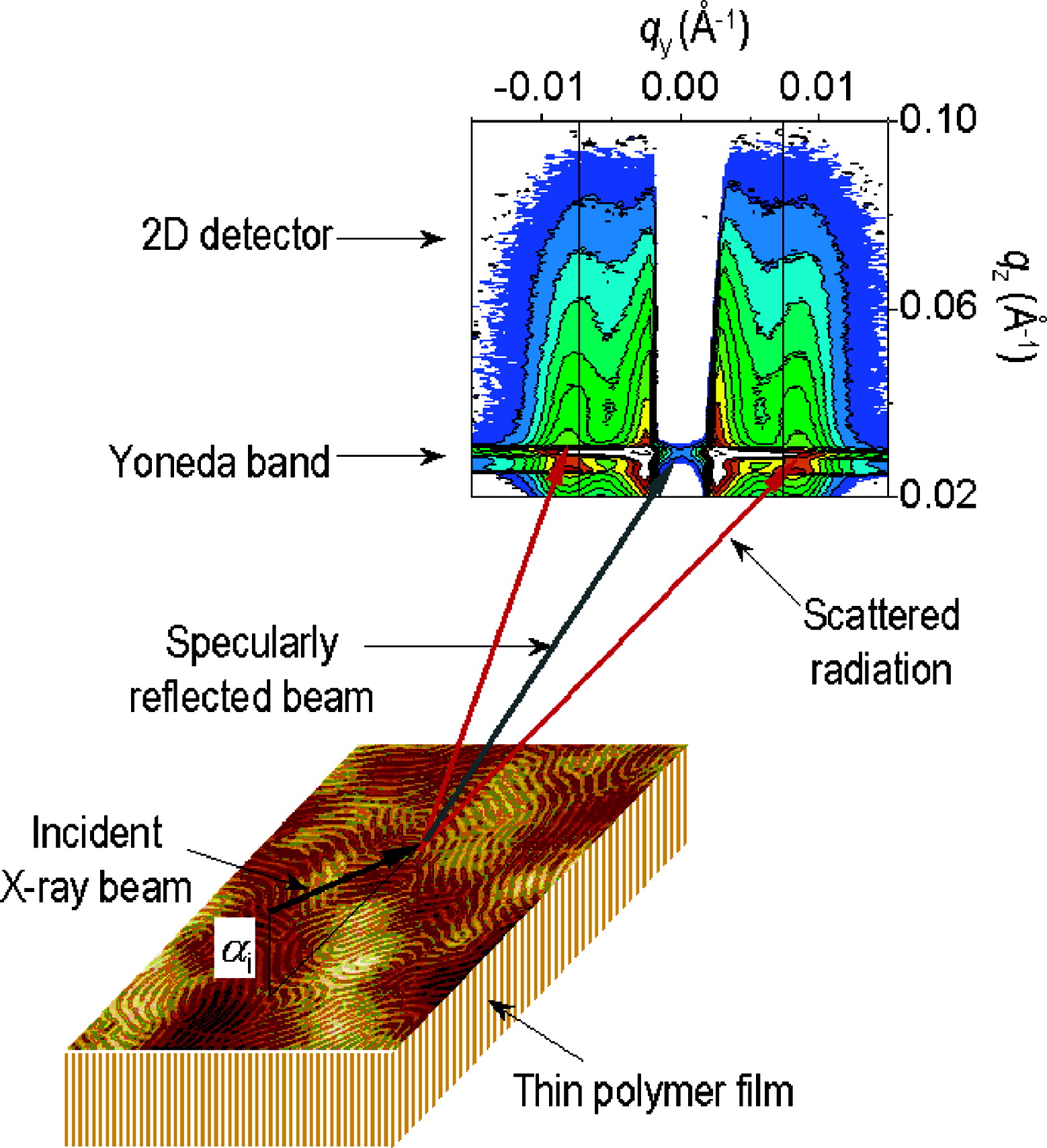}     &  \includegraphics[width = 0.59\linewidth]{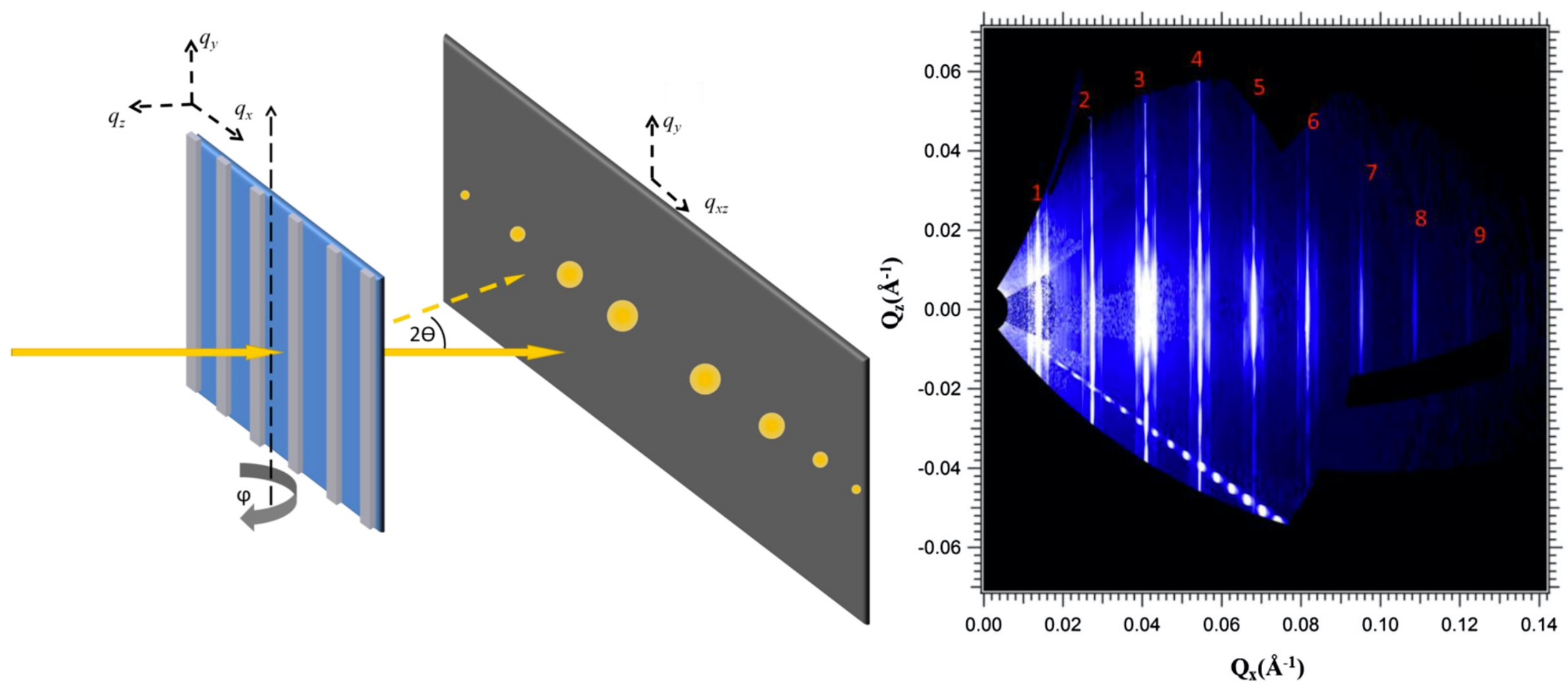}
    \end{tabular}
    \caption{(\emph{left}) A schematic drawing of the GISAXS experimental setup for a Di-BCP film with perpendicular lamellae. Reprinted (adapted) with permission from \cite{Busch2007}. Copyright \textcopyright 2007 American Chemical Society. (\emph{right}) A schematic drawing of res-CDSAXS experimental setup and a reconstructed scattering image of PS--b--PMMA film strongly guided by a chemically nanopatterned substrate \cite{Sunday2013}. 
    }
    \label{fig:x-ray_scattering}
\end{figure}

With the increasing usage and growing impact of model-based predictions for BCP self-assembly, it is increasingly important to ensure the reliability of models. Model parameter estimation, or model calibration, is necessary in order to produce reliable model-based predictions~\cite{Oden2017} and is the focus of this work. In particular, we formulate and propose a suitable methodology for calibrating models of BCP self-assembly. Models are calibrated according to observational data obtained by experimental characterizations \cite[Vol. 2]{Matyjaszewski2012}, via microscopy or X-ray scattering techniques (see, for example, Fig. \ref{fig:x-ray_scattering}), of BCP melts, and various intrinsic uncertainties are also explicitly accounted for.

One source of intrinsic uncertainty stems from corruptions introduced in the characterization process. This is often modeled as stochastic sensor noise. We also address an \emph{aleatoric uncertainty}, related to the well-known phenomenon that unguided or weakly guided self-assembly of BCPs results in a range of possible equilibrium morphologies with long-range disorder. A characteristic example for such long-range disorder is the `finger-print' patterns for the lamellae phase of Di-BCPs thin films under unguided self-assembly. For DSA, the aleatoric uncertainty results in \emph{defects} in the equilibrium patterns. Examples of a `finger-print' pattern and DSA pattern defects are shown in Fig.~\ref{fig:experimental_phase_diagram} (\emph{right}). Pattern defects have significant implications on product yield and quality in the nanolithography application of BCP self-assembly \cite{Bates2014, Li2015}.

Statistical polymer theory \cite{Grosberg1994} accounts for the impact of aleatoric uncertainty by describing the variety of achievable equilibrium morphologies for a particular polymer melt system via a \emph{statistical ensemble}. The statistical ensemble assigns a probability of occurrence for each achievable equilibrium morphology \cite{Nagpal2012}. In this work, we consider all BCP self-assembly models as \emph{statistical models}, through which the computer-generated solutions are samples from a model-specific statistical ensemble. We assume the stochastic component of the models is accounted for by \emph{auxiliary random variables} which are possibly infinite-dimensional and/or parameter-dependent.

To perform model calibration that accounts for these uncertainties, we advocate for a Bayesian probabilistic approach that uses probability distributions to describe the uncertainty in model parameters. Under such an approach, the output of the model calibration is a posterior probability distribution of the model parameters conditioned on the observed data. In addition to performing parameter estimation, we are also interested in studying the effects of instrument-specific parameters that describe the strength of various kinds of corruptions introduced in the characterization process. These effects can be quantified by the amount of information content, i.e., the Shannon entropy \cite{Shannon1948}, preserved by the characterization process. Doing so 
informs decision-making for appropriate specifications of characterization instruments, so that any experimental data are as effective as possible for model calibration. For the numerical case study presented in Sections~\ref{sec:model_definitions} and~\ref{sec:numerical_results}, we focus on top-down microscopy characterization of Di-BCP thin films, and study the effect of (i) various degrees of image blurring and/or noise intensities, and (ii) various levels of magnification on the calibration of the Ohta--Kawasaki (OK) model \cite{Ohta1986, Uneyama2005, Choksi2003}, a DFT model, for Di-BCP self-assembly.
\subsection{Likelihood-free inference and expected information gain computation via measure transport }\label{subsec:methodology}

To solve the proposed Bayesian model calibration problem, we must first tackle key challenges that may seem insurmountable, especially if addressed via conventional likelihood-based methods for Bayesian inference. Specifically, analytical evaluations of \textit{integrated likelihood functions} \cite{berger1999integrated} are deemed intractable due to (i) the high-dimensional auxiliary random variables, and (ii) the potentially complex forms of sensor noise present in the characterization data. To address these challenges, we adopt the \textit{measure transport} approach to likelihood-free inference (LFI) that was considered in~\cite{Papamakarios2016, Marzouk2017}. This approach addresses the challenges of intractable likelihoods, given that it only requires samples from the joint distribution of the parameters and model-generated data. Moreover, the generation of these samples can be performed offline in a scalable manner.

Performing LFI with high-dimensional image data can be challenging, however, as the regions of the parameter space with large likelihood values are typically highly concentrated. Therefore, we advocate constructing low-dimensional summary statistics of the images to capture information in the data that is relevant to the model parameters of interest. This approach is commonly used for the analysis of the image data for BCP self-assembly~\cite{Lee2005,Sunday2013, Murphy2015}. Furthermore, using summary statistics is natural for the model calibration of strongly guided Di-BCP thin film self-assembly using X-ray scattering image data \cite{Khaira2014, Hannon2018}, as these scattering images are inherently low-dimensional. In our numerical case study in Section~\ref{sec:model_definitions}, we propose several energy-based and Fourier-based summary statistics of top-down microscope images of Di-BCP thin film self-assembly, and their effectiveness for model calibration is quantitatively studied.

The ability to produce posterior densities for Bayesian calibration of BCP models is a prerequisite for the more computationally difficult task of computing the expected information gain (EIG) from summary statistics of image data under various forms of corruptions introduced by the characterization process. To do so, we present a novel likelihood-free estimator for EIG based on measure transport estimators for the posterior distribution. While several EIG estimators based on variational approximations have been recently proposed~\cite{Long2013, Foster2019}, the estimators we use in this work possess nice theoretical properties, such as having the potential of being consistent in the infinite sample setting. Furthermore, we show that the estimation of the EIG can be computed using the proposed LFI approach with no significant additional cost.
\subsection{Related work}\label{subsec:related_work}
Parameter estimation of models for self-assembly based on experimental characterization of the equilibrium patterns has been previously studied. Khaira et.\thinspace al~\cite{Khaira2017} determined material and geometric parameters in the TICG model for chemical substrate guided self-assembly of lamellae phase Di-BCP thin films. The parameter estimation is based on data obtained from resonant critical-dimension small-angle X-ray scattering (res-CDSAXS) and grazing-incidence small-angle X-ray scattering (GISAXS) characterization data. Hannon et.\thinspace al \cite{Hannon2018} accomplished similar calibration tasks for the SCFT model with res-CDSAXS data.

The parameter estimation in previous works generally follow a \emph{deterministic} approach. Since deterministic approaches do not quantify the impact of various intrinsic uncertainties on calibration confidence, the reliability of calibrated models may be poor. In particular, by calibrating only to characterization data of Di-BCP equilibrium states strongly influenced by DSA external guidance, the aleatoric uncertainty can be reasonably omitted, since it has a negligible effect on the data. However, in the scenario where the self-assembly is unguided or weakly guided, the aleatoric uncertainty cannot be neglected, as model predictions and experimental characterizations can differ in nontrivial ways due to the long-range disorder or abundance of defected states.

We present related works in LFI and estimation of EIG in Sections~\ref{subsec:LFI_methods} and~\ref{subsec:related_work_eig}, respectively.

\subsection{Paper layout}
The paper is organized as follows. In Section~\ref{sec:prob_data_model}, the relevant components of probabilistic data models for BCP self-assembly are defined. These components include the forward operator of computer simulations and the statistical model for experimental characterization. In Section~\ref{sec:bayesian_inference_and_uq}, we introduce the Bayesian framework for the calibration of BCP models and provide a short review of relevant concepts from information theory. In Section~\ref{sec:addressing_the_challenges}, we outline the challenges in the Bayesian calibration of BCP models and discuss how these challenges can be addressed with the LFI approach and constructing summary statistics of image data. In Section~\ref{sec:transportLFI}, we present a measure transport approach to LFI for estimating the posterior distribution. In Section~\ref{sec:eig}, we introduce a methodology for estimating EIG via triangular transport maps. In Section~\ref{sec:model_definitions}, we describe the data models for the numerical case study: Bayesian calibration of the OK model for the self-assembly of Di-BCP thin films with top-down microscopy image data. Several energy- and Fourier-based summary statistics of the images are proposed. The prior distribution is defined via a careful analysis of an admissible parameter regime for the OK model. Lastly, in Section~\ref{sec:numerical_results} we present numerical results for the case study obtained through the proposed LFI and EIG estimation methods.
\subsection{Contributions} \label{subsec:Contributions}
\noindent The contributions of this work are summarized, in order of appearance, as follows.
\begin{enumerate}
    \item We introduce a Bayesian approach for calibration of a 
    BCP self-assembly model that accounts for both measurement and aleatoric uncertainties. 
    \item We overview a quantitative approach for understanding the informativeness of experimental characterizations about parameters of BCP self-assembly models.
    \item We highlight the computational
    challenges of integrated likelihood evaluations for calibrating a BCP self-assembly model and demonstrate that
    a LFI approach is suitable for overcoming such challenges.
    \item We overview one methodology for LFI, based on measure transport, that is well-suited for inference problems with generative models, given that it only requires a limited number of offline realizations of parameters and model evaluations.
    \item We introduce the notion of extracting summary statistics from BCP characterization image data and propose an efficient approach to measure the utility of selected summary statistics in the LFI setting via a noval EIG estimator that is, in principle, consistent. 
    \item We present a numerical case study on the Bayesian calibration and EIG computation of the OK model for Di-BCP thin film self-assembly with top-down microscopy characterization.
\end{enumerate}
\section{A probabilistic data model of block copolymer self-assembly}\label{sec:prob_data_model}

In this section, we conceptualize and formulate generic probabilistic data models of BCP self-assembly that consist of (i) statistical models of experimental characterization, introduced in Section \ref{subsec:characterization_model}, and (ii) statistical models of BCP self-assembly, introduced in Section \ref{subsec:statistical_models}. The former account for measurement uncertainty in the characterization data, while the latter account for the aleatoric uncertainty. A particular example of the data model that is used for a numerical case study is provided in Section~\ref{sec:model_definitions}.

\subsection{Statistical models of experimental characterization}\label{subsec:characterization_model}
We consider the process that generates an image, $\y\in\R^y$, that characterizes a latent BCP melt $u\in V^u(\mathcal{D})$ in a physical domain $\mathcal{D}\subset\R^3$.  The material state $u$ may be described by discrete spatial positions of monomers or coarse-grained monomer segments. On the meso- or continuum scale, the material state can also be modeled as the local fractions or normalized densities fields of distinct monomers. We denote the set of all possible Di-BCP material states as $V^u(\mathcal{D})$.

Let the characterization process refer to the map from $u \mapsto \y$. We model this process as the composition of a \textit{state-to-observable map} $\mathcal{J}$ and a \textit{statistical sensor operator} $\boldsymbol{\mathcal{S}}$, respectively representing signal formation and image formation in the characterization process. That is,
\begin{equation}\label{eq:instrument_model_deterministic}
    \y = \boldsymbol{\mathcal{S}}(\mathcal{J}(\state),\boldsymbol{n}),
\end{equation}
where $\boldsymbol{n}$ represents the stochastic noise introduced during the characterization process.

The state-to-observable map can be further decomposed into a restriction operator $\mathcal{R}$ and an imaging operator $\mathcal{I}$, i.e., $\mathcal{J} \coloneqq\mathcal{I}\circ \mathcal{R}$. The restriction operator\footnote{In many Bayesian inference problems associated with real-space observations of a physical domain, this restriction operator is a point-wise observation operator. See \cite{Bui-Thanh2013}.}  $\mathcal{R}\colon \statespace(\mathcal{D}) \to \statespace\big(R(\mathcal{D}))$ extracts the observed portion of the material state in the subdomain $R(\mathcal{D}) \subset \mathcal{D}$. For microscopy characterization, $R(\mathcal{D})$ may represent regions of top or cross-section surfaces in the field of view of the microscope. For X-ray scattering characterization, it may represent the subdomain around the incident X-ray beam that affects its trajectory.

The operator $\mathcal{I}\colon \statespace\big(R(\mathcal{D})\big) \to  V^o$ models the imaging mechanism, which depends on the experimental setup and certain properties of the latent BCP material, such as the refractive indices of the distinct monomers that are relevant for X-ray scattering imaging~\cite{Sunday2015}. Most microscopy imaging methods are real-space methods, where the imaging operator is often constructed using identity maps. One may also include effects related to image resolution, contrast, brightness, depth of field, etc., in the microscopy imaging operator~\cite{Sawyer2008}. For X-ray scattering characterization, one may additionally incorporate transformations to the reciprocal space (e.g.\thinspace via Fourier transform)~\cite{Lazzari2002, Lee2005, Muller-Buschbaum2009, Boldon2015} in the imaging operator for reproducing scattering patterns from given material states.

The statistical sensor operator $\boldsymbol{\mathcal{S}}\colon V^o\times \mathbb{R}^{n} \to \mathbb{R}^{y}$ models the discretization and noise corruptions introduced by image-forming sensors. The sensor noise $\boldsymbol{n}\in \mathbb{R}^n$ is a realization of the random variable $\noise$, often attributed to infinitesimal fluctuations in sensor voltage. A simple additive model with independent noise is often considered:
\begin{equation}\label{eq:additive_noise}
    \boldsymbol{\mathcal{S}}\left(\mathcal{J}(u), \boldsymbol{n}\right) = \boldsymbol{\mathcal{D}}\left(\mathcal{J}(u)\right) + \boldsymbol{n},
\end{equation}
where the components of the stochastic noise are independent and identically distributed (i.i.d.), and $\boldsymbol{\mathcal{D}}\colon V^o\to \mathbb{R}^{y}$ represents the local averaging of signals introduced by image-forming sensors. Let us note that realistic sensor noise models also account for more complex spatial and signal intensity dependence \cite{Kirian2011,Roels2014, Haider2016}.
\subsection{The aleatoric uncertainty and statistical models of block copolymer self-assembly} \label{subsec:statistical_models}
While the self-assembly of block copolymers leads to the formation of ordered periodic structures, such as those depicted in the experimental phase diagram in Fig.~\ref{fig:experimental_phase_diagram} (\emph{left}), they are typically not realizable in experiments without strong external guidance, even though they are often considered to be the most stable configurations. Instead, a rich variety of states that exhibit long-range disorder, often referred to as \textit{defected states} or \textit{metasable states} are observed; see Fig.~\ref{fig:experimental_phase_diagram} (\emph{right}).

To account for the impact of aleatoric uncertainty on the experimentally observed equilibrium states of BCP self-assembly, we assume that the various states attainable under the same experimental scenario $S_E$ to be realizations of a $V^u$-valued random variable, $\bstate(S_E)$. We refer to $\bstate(S_E)$ as the \textit{state random variable}. The state random variable gives rise to a probability distribution of the achievable states, whose structure depends on the material, geometrical, thermal, and various other controllable macroscopic factors specified by $S_E$. When the self-assembly is strongly guided, the distribution of the state random variable is close to a Dirac mass, meaning that only a handful of BCP equilibrium states are attainable. When the self-assembly is unguided or weakly guided, the distribution can be quite complex, yet it typically concentrates on states in the same phase or mixture of phases that have similar structures, such as similar radius and periodicity length of spheres in the sphere phase.

A \textit{statistical model} of BCP self-assembly can be generalized as a mathematical model derived from statistical polymer theory for specifying the probability distribution, i.e., the \textit{statistical ensemble}, of $\bstate(S_E)$ for a range of experimental scenarios. It typically contains a model of polymer chains, models of the inter-polymer and polymer--environment interactions, and possibly a model of thermal fluctuations and/or dynamics of polymers during the self-assembly. A comprehensive review of these models can be found in~\cite[Vol. 1]{Matyjaszewski2012}, \cite{Grosberg1994, Gompper2005, Fredrickson2006}, and references therein.

Computer simulations of BCP self-assembly can be treated as stochastic simulators for generating independent samples from model-specific statistical ensembles of equilibrium states. Let us denote the forward operator of a statistical model of BCP self-assembly as $\mathcal{F}\colon \R^x\times V^z\to V^u$. This model assumes the equality in distribution
\begin{equation}\label{eq:state_variable}
    \bstate(S_E) \stackrel{d}{=} \mathcal{F}(\boldsymbol{x}, Z),
\end{equation}
where $\boldsymbol{x}\in \R^x$ are parameters\footnote{This work considers the model parameters $\boldsymbol{x}$ to be finite dimensional and belonging to a Euclidean space. See~\cite{Stuart2010} for a general treatment in which $\boldsymbol{x}$ belongs to an infinite-dimensional function space.} that represents the material, geometrical, thermal, and/or other macroscopic factors of the experimental scenario $S_E$ that are considered in the model. The $V^z$-valued random variable $Z$ represents the stochastic component of the forward operator. 

The origin of the stochastic component $Z$ may be purely numerical. For example, $Z$ can partly represent random initial configurations of the polymeric system and the sampling mechanism for accepting perturbed configurations in Monte Carlo simulations of equilibrium BCP melts \cite{Binder2012}. It can also be high-dimensional or even infinite-dimensional. For example, $Z$ can be specified by the stochastic forcing term that resides in infinite-dimensional function spaces for dynamical DFT models \cite{Fraaije1993}. We refer to $Z$ as an \textit{auxiliary random variable}, or \textit{nuisance variable}, as it is either non-physical or represents uncontrollable microscopic factors in experiments that are difficult to observe. For the numerical case study in Section~\ref{sec:model_definitions} and \ref{sec:numerical_results}, we use the OK model, in which the stochastic component is specified by random initial states for free energy minimization.
\subsection{A generic data model of block copolymer self-assembly} \label{subsec:data_model}

We call the combination of the model of BCP self-assembly and model of experimental characterizations of equilibrium states as the \emph{data model}. For notational convenience, we first define the \textit{parameter-to-observable map} $\mathcal{O}\coloneqq \mathcal{J}\circ\mathcal{F}$, where $\mathcal{O}(\bx, Z)$ is the \textit{observable random variable}. For example, for X-ray scattering characterization, $\mathcal{O}(\boldsymbol{x}, Z)$ may be the (noise-free) model-predicted latent scattering pattern where variability comes from the model-predicted random BCP equilibrium structures. We define the image data as a $\R^y$-valued random variable arising from both the measurement uncertainty and the aleatoric uncertainty. The \textit{data model} $\boldsymbol{\mathcal{G}}:\R^x\times V^z\times\R^n\to\R^y$ summarizes the relationship between model parameters and data as
\begin{equation} \label{eq:data_model}
    \by = \boldsymbol{\mathcal{G}}(\boldsymbol{x}, Z, \noise)\coloneqq \boldsymbol{\mathcal{S}}\big(\mathcal{O}(\boldsymbol{x}, Z), \noise\big).
\end{equation}
\section{Bayesian inference and quantifying uncertainty}\label{sec:bayesian_inference_and_uq}

In this section, we present the Bayesian inference framework for calibrating models for BCP self-assembly. In Section~\ref{subsec:bayes_problem} we introduce the posterior distributions 
that arise from the likelihood and the, so called, conditional likelihood. To summarize the uncertainty in calibrated parameters, we introduce some information-theoretic quantities in Section~\ref{subsec:entropy_and_ig}. Lastly, in Section~\ref{subsec:eig_intro}, we present the expected information gain as a natural quantity to consider in a Bayesian inference context for measuring the information contained in the data about the unknown model parameters.

\subsection{The posterior and joint posterior}\label{subsec:bayes_problem}

For a Bayesian formulation to model calibration, we introduce a random variable for the parameters $\bx\in \R^{\dimx}$ and seek to characterize its distribution given an  observation of $\by$. Initially, the parameter $\bx$ is distributed according to the \emph{prior} measure $\nu_{\bx}\in \mathcal{P}(\R^{\dimx})$ with density $\pi_{\bx}$. The prior measure represents our \emph{epistemic uncertainty} in the model parameters before collecting observations. Similarly, the (possibly infinite-dimensional) auxiliary random variable is distributed according to the conditional prior measure $\nu_{Z|\bx}$ that may depend on the model parameter. In our numerical case study, a prior distribution for the OK model parameters is constructed in Section~\ref{subsec:prior_transform} and the distribution for the auxiliary variable is described in \ref{app:our_z_model}. 

The data model in Section~\ref{subsec:data_model} defines the conditional distribution  
for the observations. Let $\by = \boldsymbol{\mathcal{G}}(\bx,\bz, \noise)$ be a \textit{data random variable} and assume the sensor noise random variable $\noise \sim \pi_{\noise}$ is independent of ($\bx$, $\bz$). This noise yields a conditional density for the data given the inputs $\x$ and $\z$ that we denote by $\pi_{\by|\bx,\bz}(\cdot | \x,\z)$. For example, if the sensor operator in the data model~\eqref{eq:data_model} is given by the additive form in~\eqref{eq:additive_noise}, then this conditional density is given by $\pi_{\noise}(\cdot - \boldsymbol{\mathcal{D}}(\mathcal{O}(\x,\z)))$. 

For the BCP model calibration problem, however, we are not interested in inferring the auxiliary variable $\bz$. Thus, we seek to describe the conditional distribution of the data given only the model parameters. The uncertainties in the sensor noise and the auxiliary variable define a marginal conditional distribution for $\by$ given $\bx$, whose density we denote by $\pi_{\by|\bx}(\cdot | \x)$. For these conditional distributions of the data, we define both the \emph{conditional likelihood} and the \emph{likelihood} for an observation $\y^*$ as the functions which arise when fixing the data to $\y^*$ in the conditional densities. That is, $\pi_{\y^*|\bx,\bz}(\x,\z) \coloneqq \pi_{\by|\bx,\bz}(\y^*|\x,\z)$ is the conditional likelihood for both the model parameters and the auxiliary variable, while $\pi_{\y^*|\bx}(\x) \coloneqq \pi_{\by|\bx}(\y^*|\x)$ is the likelihood 
for only the model parameters. The two likelihood functions represent, respectively, the likelihood of observing an image $\y^*$ for (i) a particular model-predicted equilibrium state generated at $(\x, z)$, and (ii) a variety of random model-predicted equilibrium states generated at $\x$. We will discuss how both are defined in depth in Section~\ref{sec:addressing_the_challenges}.

Given a prior and likelihood function for an observation $\y^*$, Bayes' theorem provides a rule for producing the conditional distributions of the parameters. The \emph{joint posterior} density for the parameters and the auxiliary variable is given by\footnotemark
\footnotetext{Similar to the notation of the likelihood function, we use $\pi_{\bx, \bz|\y^*}(\x, \z)$ to emphasize that the posterior is conditioned on a particular observation. For the more general case of arbitrary observations $\y\in\R^y$, we use the notation $\pi_{\bx, \bz|\by}(\x, \z|\y)$ instead. Note that in the setting where the auxiliary variable $Z$ is infinite dimensional, Bayes' rule in \eqref{eq:joint_Bayes_rule} is no longer mathematically valid and it should be expressed using the Radon–Nikodym derivative between the posterior and the prior~\cite{Stuart2010}.}
\begin{align}\label{eq:joint_Bayes_rule}
    \pi_{\bx, \bz|\y^*}(\x, \z) = \frac{\pi_{\y^*\vert\bx, \bz}(\x,\z)\pi_{\bx, \bz}(\x, \z)}{C_{\y^*}},
\end{align}
where $C_{\y^*}=\pi_{\by}(\y^*) = \int_{\mathbb{R}^{x}}\int_{\zspace}\pi_{\y^*\vert\bx,\bz}(\x,\z) \textrm{d}\nu_{\bz,\bx}(\z | \x)\pi_{\bx}(\x)\textrm{d}\x >0$ is the normalization constant for the posterior density 
and is often referred to as the \emph{model evidence} or the \emph{marginal likelihood}. To calibrate the model parameters, our interest is in the \emph{marginal posterior} density of the parameters alone given by 
\begin{align}\label{eq:marginal_Bayes_rule}
    \pi_{\bx|\y^*}(\x) = \frac{\pi_{\y^*\vert\bx}(\x)\pi_{\bx}(\x)}{C_{\y^*}}.
\end{align}

Many computational tasks in Bayesian inference rely on evaluating these densities to draw samples from the posterior distributions. While evaluations of~\eqref{eq:joint_Bayes_rule} depend on evaluations of the conditional likelihood $\pi_{\y^*\vert\bx, \bz}$, evaluations of~\eqref{eq:marginal_Bayes_rule} inherit difficulties from evaluating the likelihood $\pi_{\y^*\vert\bx}$ that stem from the influence of the nuisance variable $\bz$ on the observations. We discuss these computational difficulties in Section~\ref{sec:addressing_the_challenges}. For a thorough treatment on these challenges in Bayesian inference, in general, see~\cite{robert2004monte, gelman2014bayesian}.
\subsection{Entropy and information gain}\label{subsec:entropy_and_ig}

To understand and quantify the effectiveness of Bayesian model calibration, we turn to information theory that formally quantifies the information content in random variables. One information-theoretic measure for the uncertainty of a parameter $\bx$ is the \emph{Shannon entropy} $\Ent\colon \mathcal{P}(\mathbb{R}^{x})\to \mathbb{R}$ given by
\begin{equation}\label{eq:entropy}
    \Ent(\bx) \coloneqq \mathbb{E}_{\pi_{\bx}} [-\log \pi_{\bx}].
\end{equation} 
The entropy is larger for a parameter that is less certain. For example, if a parameter is known, i.e., its distribution is a Dirac mass, then its entropy is zero. After conditioning on the data $\y^*$, we can measure the entropy of the parameters by computing the conditional entropy $\Ent(\bx|\by=\y^*)$. If the entropy decreases after conditioning, i.e., $\Ent(\bx|\by=\y^*)< \Ent(\bx)$, we say that observing the data $\y^*$ has \emph{decreased our uncertainty} or \emph{increased our confidence} in the values of model parameters. The expectation of $\Ent(\bx|\by=\y)$ over the distribution of the data random variable defines the conditional entropy of the parameters given the data $\Ent(\bx|\by)$ and it always satisfies $\Ent(\bx|\by) \leq \Ent(\bx)$. If the conditional entropy is \emph{less} than the prior's entropy, we say that the data is, on average, informative of the parameters.

To quantify the information gained from performing Bayesian inference, we measure the change in the entropy of the parameters $\bx$ before and after conditioning on data. This change quantifies the value of acquiring certain noisy, corrupted, and/or partial observations of the system to learn about the unknown model parameters. In the context of Bayesian inference, this change compares the prior density $\pi_{\bx}$ and the posterior density $\pi_{\bx|\y^*}$ for the parameters. An information-theoretic quantity for measuring information gain based on the change in entropy is the \textit{Kullback--Leibler (KL) divergence} $\mathcal{D}_{\textrm{KL}}$. The KL divergence from the prior to the posterior is given by 
\begin{equation} \label{eq:KL_prior_posterior}
\mathcal{D}_{\textrm{KL}}(\pi_{\bx|\y^*}||\pi_{\bx}) = \mathbb{E}_{\pi_{\bx|\y^*}} \left[\log \frac{\pi_{\bx|\y^*}}{\pi_{\bx}} \right].
\end{equation}
The KL divergence is always non-negative and is zero if and only if the posterior is equal to the prior almost everywhere. Furthermore, a large value for the KL divergence implies that the posterior has lower entropy relative to the prior and its density is typically more concentrated than the prior density. A large departure of the posterior from the prior indicates that we gained more information from the observations about the parameters.
\subsection{Expected information gain}\label{subsec:eig_intro}

Before observing any particular data, we can measure the information gained about the parameter in a Bayesian inference problem by considering the \textit{expected information gain} (EIG) $I\colon\mathcal{P}(\mathbb{R}^x) \times \mathcal{P}(\mathbb{R}^y) \to \mathbb{R}_{+}$ between the random variables $\bx$ and $\by$. We say that $I(\bx; \by)$ represents the the EIG for learning about $\bx$ from $\by$. The EIG can be defined by the integration of the KL divergence in~\eqref{eq:KL_prior_posterior} with respect the data $\y$. That is,
\begin{equation} \label{eq:EIG}
  I(\bx;\by) = \int_{\mathbb{R}^{y}} \mathcal{D}_{\textrm{KL}}(\pi_{\bx|\y}||\pi_{\bx}) \pi_{\by}(\y)\textrm{d}\y. 
\end{equation}
Within the field of Bayesian optimal experimental design (BOED),  the EIG is commonly used to assess the utility of an experimental measurement for reducing uncertainty in the model parameters~\cite{Lindley1956}. In this context, the EIG is typically a function of the experimental design variables, and is maximized to find the optimal design. An example of BOED is the optimal design of a discrete observation operator, e.g., choosing sensor locations that maximize the EIG on certain quantities of interest~\cite{Krause2008}. In the context of microscopy characterization of BCP melts, for example, we can measure $I(\bx;\by)$ for different restriction operators $\mathcal{R}$, introduced in Section \ref{subsec:characterization_model}, to determine the optimal level of magnification for inferring a particular subset of the model parameters. In our numerical case study in Section~\ref{sec:model_definitions} and~\ref{sec:numerical_results}, we will demonstrate results on the computation of EIG that enables such a type of BOED study.

We note that the expression for the EIG $I(\bx; \by)$
in~\eqref{eq:EIG} is equivalent to the mutual information (MI) between random variables $\bx$ and $\by$. The MI is always a non-negative random variable. While a zero value implies that the two random variables are independent, a larger value indicates that the data is more informative on average for inferring the parameter. Using the form of the KL divergence in~\eqref{eq:KL_prior_posterior} and Bayes' rule in~\eqref{eq:marginal_Bayes_rule}, the EIG can also be written in the two equivalent ways
\begin{subequations}\label{eq:EIG_p_and_l}
\begin{align}
    I(\bx;\by) 
    &= \mathbb{E}_{\nu_{\bx,\by}}\left [\log \frac{\pi_{\bx|\by}}{\pi_{\bx}}\right] \label{eq:EIG_posterior} \\
    &= \mathbb{E}_{\nu_{\bx,\by}} \left[\log \frac{\pi_{\by|\bx}}{\pi_{\by}}\right]. \label{eq:EIG_likelihood}
\end{align}
\end{subequations}
While the first expression uses the ratio of the posterior and prior densities, the second expression uses the ratio of the conditional likelihood and the marginal density for the observations. We note that as compared to standard Bayesian inference problems that characterize the posterior for a single realization of the data, computing the EIG requires evaluating the posterior densities for many possible realizations of the data. In Section~\ref{sec:eig}, we propose an efficient estimator for the EIG in~\eqref{eq:EIG_posterior} that is based on a variational approximation for all of the posterior densities.
\section{Addressing the challenges of auxiliary variables and high-dimensional image data via likelihood-free inference}\label{sec:addressing_the_challenges}

Performing Bayesian calibration and computing EIG requires \emph{characterizing posteriors}, i.e., producing representative samples from the the posterior distribution 
or providing an exact or approximate expression of the density $\pi_{\bx | \y^*}$. In this section, we outline distinctive challenges for the Bayesian calibration of BCP self-assembly models. In Section~\ref{subsec:integrated_likelihoods} we describe intractable integrated likelihood functions that arise when accounting for the aleatoric uncertainty in the data model. Sections~\ref{subsec:computation_wo_likelihoods} and~\ref{subsec:LFI_methods} discuss computational methods for dealing with integrated likelihoods, including LFI methods. Lastly, in Section~\ref{subsec:summary_statistics} we address inference with high-dimensional image data by presenting an approach that constructs low-dimensional summary statistics of the data to capture relevant information about the model parameters of interest. 

\subsection{Evaluating integrated likelihoods} \label{subsec:integrated_likelihoods}

To calibrate the model parameters $\x$ given image data $\y^*$, it is necessary to prescribe the likelihood function $\pi_{\y^*|\bx}$ in~\eqref{eq:marginal_Bayes_rule}. This likelihood is derived from the stochastic procedure for generating the image data using the data model for BCP self-assembly in~\eqref{eq:data_model}. The procedure depends on, not only the sensor noise $\noise$, but also the auxiliary variable $\bz$ entering the parameter-to-observable map. In particular, the likelihood of an observed image $\y^*\in \mathbb{R}^{\dimy}$ given model parameters $\x$ accounts for \emph{all} realizations of the auxiliary variable and stochastic noise that could possibly generate the particular image. Mathematically, this is achieved by summing the contributions of the conditional likelihoods $\pi_{\y^* | \bx,\bz}(\x, \z)$ weighted by the probability of generating each auxiliary variable $\z$. Hence, the likelihood function for the parameters is defined as the following \emph{integrated likelihood}~\cite{berger1999integrated} that marginalizes out the auxiliary variables
\begin{equation} \label{eq:marginal_likelihood} 
\pi_{\y^* | \bx}(\x) = \int_{\zspace} \pi_{\y^* | \bx, \bz}(\x, \z) \textrm{d}\nu_{\bz|\bx}(\z|\x).
\end{equation}
Using this likelihood leads to the posterior density for the parameters $\x$  in~\eqref{eq:marginal_Bayes_rule} that is a low-dimensional marginal of the joint posterior distribution for $(\bx,\bz)$\footnotemark.
\footnotetext{Characterizing the joint posterior involves a tractable likelihood $\pi_{\y^*|\bx,\bz}$ that does not require marginalization, but requires performing inference on a potentially much higher-dimensional distribution.}

The integrated likelihood in~\eqref{eq:marginal_likelihood} has a closed-form expression only in very special situations. One setting is when both the conditional distribution for the auxiliary variables $\nu_{\bz|\bx}(\cdot | \x)$ and the conditional density for the observations $\pi_{\by|\bx,\bz}(\cdot | \x,\z)$ are Gaussian, and the parameter-to-observable map is affine. In general, however, the integral we consider over the infinite-dimensional auxiliary variables $z$ is unavailable in closed form\footnotemark. In these cases, we say that the integrated likelihood is \emph{unavailable} and consider it \emph{intractable} to evaluate.
\footnotetext{The conditional likelihood is an example of a \emph{semi-parametric} model that contains both finite-dimensional and infinite-dimensional inputs; our work seeks to build on inference for semi-parametric models, where the infinite-dimensional auxiliary variables can include stochastic forcing terms, stochastic initial conditions or random fields, which are not relevant in inference. We refer the reader to~\cite{kosorok2007introduction} for more details on these models.}

Despite their intractability, integrated likelihoods are frequently encountered in many fields. Other interesting inference problems that involve integrated likelihoods include time, phase, and distance marginalization in gravitational wave inference~\cite{Thrane2019}, boundary condition marginalization for linear parabolic PDEs~\cite{Ruggeri2017}, and hyper-parameter discovery for hierarchical models in image deblurring and X\textendash ray computational tomography~\cite{Zhou2018}. Therefore, developing efficient computational inference methods that work with integrated likelihoods is an important ongoing area of research. 
\subsection{Inference methods for intractable likelihoods} \label{subsec:computation_wo_likelihoods}

In settings with intractable likelihoods, it is not possible to compute posteriors using likelihood-based sampling algorithms, such Markov chain Monte Carlo (MCMC), or other variational methods which require evaluating the likelihood at multiple 
values of the parameters. To satisfy this constraint, three classes of inference methods have been developed that do not require explicit likelihood evaluations. These classes include surrogate likelihood functions, pseudo-marginal Monte Carlo methods~\cite{Andrieu2009, alenlov2021pseudo}, which assume tractable conditional likelihood function evaluations, and likelihood-free inference, often referred to as simulation-based inference~\cite{cranmer2020frontier}.

The \emph{profile likelihood} method~\cite{Murphy2000} is one popular choice of surrogate that replaces integration in~\eqref{eq:marginal_likelihood} with a maximization over $\z$. Specifically, the profile likelihood $\widehat{\pi}_{\y^*|\bx}$ is defined as the conditional likelihood evaluated at the maximum likelihood estimate (MLE) for $\z$ as a function of $\x$, i.e., $z_{\textrm{MLE}}(\x)= \argmax_{\z \in V^{\dimz}} \pi_{\y^*|\bx,\bz}(\x,\z)$. Thus, if a well-defined unique $\z_{\textrm{MLE}}$ exists, the profile likelihood can be tractably evaluated as $\widehat{\pi}_{\y^*|\bx}(\x) \coloneqq \pi_{\y^*|\bx,\bz}(\x,\z_{\textrm{MLE}}(\x))$. This approach, however, replaces $\z$ by a point estimate and ignores the impact of its uncertainty on the likelihood. As a result, the profile likelihood is inconsistent and often requires adding factors to correctly characterize the posterior uncertainty~\cite{mccullagh1990simple}.

If conditional likelihood evaluations are available, pseudo-marginal methods can perform inference using Monte Carlo estimates of the integrated likelihood function in~\eqref{eq:marginal_likelihood}. These approaches generate $N$ samples of the auxiliary variable $\{\z^i\}_{i=1}^N \sim \nu_{\bz|\bx}(\cdot | \x)$ and estimate the integrated likelihood as $\widehat{\pi}_{\y^*|\bx}(\x) \approx \frac{1}{N} \sum_{i=1}^{N} \pi_{\y^* | \bx, \bz}(\x,\z^i)$. If the variance $\mathbb{V}\textrm{ar}(\pi_{\y^* | \bx, \bz}(\x,\cdot))$ is finite, the standard deviation of this estimator decreases with more samples at the rate $\mathcal{O}(1/\sqrt{N})$. Thus, we can reduce the sample variance of $\widehat\pi_{\y^*|\bx}$ by increasing the number of samples until sufficient accuracy of the integration has been reached. While unbiased Monte Carlo estimators of the integrated likelihood have been shown to yield consistent posterior sampling~\cite{Andrieu2009}, these methods may necessitate a large number of samples to achieve a reasonable computational efficiency in practice~\cite{Pitt2012}. On the other hand, the evaluation of the conditional likelihood function at each sample of the auxiliary variable requires evaluating the forward operator, which can be computationally expensive. Furthermore, estimates of the integrated likelihood are not shared for different parameter values and need to be recomputed at each iteration of some sampling methods, such as MCMC.

In addition to the challenge of marginalization, the conditional likelihood $\pi_{\y^*\vert\bx,\bz}$ can also become analytically intractable to evaluate for many non-trivial sensor noise models, $\boldsymbol{\mathcal{S}}(\cdot, \noise)$, which rules out the use of pseudo-marginal methods and the profile likelihood. The analytical expression for the conditional likelihood stems from a change of variables between the noise random variable and the data random variable. For example, the sensor operator with multiplicative noise, i.e., $\boldsymbol{\mathcal{S}}(\cdot, \noise) = \boldsymbol{\mathcal{D}}(\cdot)\noise$, with noise $\noise \sim \pi_{\noise}$ has the conditional likelihood
\begin{align*}
    \pi_{\y^* | \bx,\bz}(\x,\z) = \pi_{\noise}\Big(\y^*/\boldsymbol{\mathcal{D}}\big(\mathcal{O}(\x,\z)\big)\Big)\Big(\textrm{det}\big|\boldsymbol{\mathcal{D}}(\mathcal{O}(\x,\z))\big|\Big)^{-1}\;.
\end{align*} 
In general, evaluating the conditional likelihood 
requires computing the inverse and Jacobian inverse of the sensor noise model, which is often unfeasible when it has a complicated form\footnotemark. When the data arises from a ``black-box'' simulator, i.e., when the data model $\boldsymbol{\mathcal{G}}$ relies on legacy scientific code whose mathematical expressions may be unavailable, it is not possible to derive the conditional likelihood analytically. In polymer modeling and characterization, complex mathematical expressions and scientific codes underlay both the forward operators $\mathcal{F}$ and realistic sensor operators $\boldsymbol{\mathcal{S}}$. Therefore, we consider the evaluation of conditional likelihoods to be generally unfeasible.
\footnotetext{We note that, $\pi_{\by | \bx,\bz}(\y|\x,\z) = \sum_{A_k}\mathds{1}_{\mathcal{S}_{\x,\z}(A_k)}(\y) \pi_{\noise}(\mathcal{S}_{\x,\z}^{-1}(\y))\det\big|\nabla\mathcal{S}_{\x,\z}^{-1}(\y)\big|$ where $\mathcal{S}_{\x,\z} \coloneqq \boldsymbol{\mathcal{S}}(\mathcal{O}(\x,\z),\cdot):\mathbb{R}^n \to \mathbb{R}^{y}$ is presumed to be a generally non-monotonic differentiable transformation, and $\pi_{\by | \bx,\bz}(\y|\x,\z)$ is a weighted sum of densities over a partition of the input space $A_k\subset\mathbb{R}^{n}$ into regions for which $\mathcal{S}_{\x,\z}$ is monotonic. Unless this reduces to a simpler closed\textendash form expression with explicit computations, the evaluation of $\pi_{\by | \bx,\bz}(\y|\x,\z)$ can be quite complex for arbitrary $\boldsymbol{\mathcal{S}}$, due to the difficulties of evaluating generalized inverses $\mathcal{S}_{\x,\z}^{-1}$, computing Jacobian determinants (generally $O(n^3)$ computational complexity, where $n\gg 1$ for our sensor noise), and determining the partitions $\{A_k\}$.}

On the other hand, \textit{likelihood-free inference} (LFI) methods enable statistical inference in generative models. They rely \emph{only} on sampling from the joint distribution of model parameters and model-generated data. 
In the context of the integrated likelihood function in~\eqref{eq:marginal_likelihood}, this is accomplished by
\begin{enumerate}
    \item sampling the model parameters, $\x^i\sim \pi_{\bx}$,
    \item sampling the auxiliary variable, $\z^i \sim \nu_{\bz|\bx}(\cdot|\x^i)$,
    \item evaluating the parameter\textendash to\textendash observable map, $o^i = \mathcal{O}(\x^i, \z^i)$,
    \item sampling the sensor noise $\boldsymbol{n}^i\sim \pi_{\noise}$, and
    \item evaluating the statistical sensor model, $\y^i = \boldsymbol{\mathcal{S}}(o^i, \boldsymbol{n}^i)$.
\end{enumerate}
The resulting samples $(\x^i,\y^i)$, with $z^i$ and $\boldsymbol{n}^i$ discarded, are samples from the marginal distribution of $(\bx,\by)$. 
This collection of marginal samples avoids the step of explicitly marginalizing the conditional likelihood $\pi_{\by|\bx,\bz}$ to evaluate the integrated likelihood in~\eqref{eq:marginal_likelihood}.

Given a set of i.i.d.\thinspace joint samples $(\x^i, \y^i)$ that is large enough to be representative of their joint distribution, 
the goal of LFI methods is to approximate the posterior density $\pi_{\bx|\y^*}$ or to sample from it, without evaluating the likelihood function. Thus, LFI is agnostic to the form of the likelihood and applicable for inference with various sources of intractability, including the high-dimensional integration with respect to the auxiliary variable. This makes LFI appealing for the BCP model calibration problem.
\subsection{Likelihood-free inference: Related work and practical desiderata}\label{subsec:LFI_methods}

The family of LFI methods for posterior sampling includes both classical techniques such as approximate Bayesian computation (ABC) and synthetic likelihoods, as well as modern machine learning methods. ABC methods~\cite{Beaumont2002, Marin2012, Sisson2018} define a non-parametric approximation to the likelihood function, and thereby the posterior, using an accept/reject sampling scheme. For each pair of joint samples $(\x^i,\y^i)$, this scheme accepts parameters whose simulated data $\y^i$ is close to the observed data $\y^{*}$ with respect to a discrepancy function $d$ and tolerance level $\epsilon$, i.e., ABC accepts $\x^i$ if $d(\y^i,\y^{*}) < \epsilon$. While ABC can be shown to be a consistent approximation of the posterior~\cite{frazier2015consistency} in the limit as $\epsilon \rightarrow 0$, small values for $\epsilon$ lead to rejection of many candidate parameters, and often an impractical number of simulations is required to accurately 
estimate posterior statistics. Furthermore, the quality of the non-parametric approximation in ABC deteriorates as the data dimension increases~\cite{nott2018high}.

An alternative to ABC that is biased, but typically more efficient, is a parametric approximation to the likelihood function. As an example, the Bayesian synthetic likelihood (BSL) method uses samples $\y^i \sim \nu_{\by|\bx}(\cdot|\x)$ to create a multivariate Gaussian approximation to the likelihood at each parameter value $\x$ given by $\widehat{\pi}_{\by|\bx} = \mathcal{N}(\overline{\boldsymbol{y}}(\x),C(\x))$ where $\overline{\boldsymbol{y}}(\x)$ and $C(\x)$ denote the estimated mean and covariance matrix. The approximate likelihood evaluations $\widehat{\pi}_{\y^*|\x}(\x) = \widehat{\pi}_{\by|\x}(\y^*)$ can then be efficiently used within MCMC algorithms~\cite{Wood2010, Price2018}. While BSL can be more efficient than ABC for high-dimensional data, this approach requires sampling multiple data, i.e., evaluating the data model, for each parameter in order to determine $\overline{\boldsymbol{y}}(\x)$ and $C(\x)$. As a result, both ABC and synthetic likelihood approaches are less suitable for applications that involve computationally expensive forward operators. Lastly, the inference results of both methods are applicable only to the particular set of observed data $\y^*$ and thus require repeating the computation for any new set of data.

To amortize the cost of performing inference across many sets of observed data, recent machine learning methods propose to learn models that are functions of both the parameters and the data~\cite{cranmer2020frontier}. These methods use the collection of joint samples $(\x^i,\y^i)$ to learn variational approximations for either the posterior density $\pi_{\bx|\by}$~\cite{Papamakarios2016, greenberg2019automatic}, the likelihood function $\pi_{\by|\bx}$~\cite{papamakarios2019sequential}, or the likelihood ratio $\pi_{\bx|\by}/\pi_{\bx}$~\cite{Thomas2020}. The resulting models are then used in an inference step to define a tractable approximation to the posterior density or to generate approximate posterior samples for any observed data $\y^*$. This enables the inference results to be used for downstream tasks that require access to the posteriors for all data realizations, such as EIG computation and BOED. To avoid constraining the structural forms for these models, many techniques use flexible neural network architectures that can capture complex posterior distributions and can learn the structure of high-dimensional and domain-specific data (e.g., using architectures designed for images). Many of these models, however, often require a large upfront cost to learn them reliably and adapting these methods to small-sample settings remains challenging. 
To select a LFI method for the calibration of BCP self-assembly models, we have the following desiderata: 
\begin{enumerate}
\item \textbf{Sample efficiency}: Method provides reasonable posterior approximations in settings where it is computationally expensive to generate simulated data.
\item \textbf{Consistency}: Method does not impose arbitrary approximations on the form of the likelihood or the posterior distribution in order to capture non-Gaussianity without biases from the inference procedure. 
\item \textbf{Amortized inference}: Method provides posterior densities $\pi_{\bx|\y^*}$ or samples for almost \emph{all} values of the data without needing to repeat inference for each realization of $\y^*$.
\end{enumerate}

To produce \emph{consistent} posterior approximations, without discarding or rejecting samples, we consider ABC and BSL to be less appealing for our application. To also benefit from amortization similarly to the machine learning methods, we propose to use a measure transport approach for LFI in this work, which relies on a careful parameterization for settings with limited sample sizes. The details of the proposed approach are discussed in Section~\ref{sec:transportLFI}.

\subsection{Informative summary statistics of high-dimensional image data}
\label{subsec:summary_statistics}

When performing LFI with high-dimensional data, it is common to construct and work with low-dimensional summary statistics of the data~\cite{Fearnhead2012}, $\bq=\mathcal{Q}(\by)\in \mathbb{R}^q$. Working with this type of summary statistics has two inherent advantages: (i) it can improve the quality and tractability of LFI, and (ii) it can provide insight into what information contained in the data is relevant for inference. 

In principle, arbitrarily high-dimensional data could be used to perform inference.
This data, however, can lead to concentrated regions of the parameter space with large likelihood values, which hampers the effectiveness of many LFI methods. 
Here we provide two numerical examples to visualize this phenomenon. For both examples, we consider the calibration of the OK model for Di-BCP thin film self-assembly using top-down microscopy image data. We consider the characterization model with additive Gaussian noise $\noise \sim \mathcal{N}(0,C_{\noise})$ as described in \eqref{eq:additive_noise}. That is 
\begin{equation}\label{eq:num_exp_chara_model}
    \by = \boldsymbol{\mathcal{D}}\left(\mathcal{F}(\bx,\bz)\right) + \noise,
\end{equation}
where the state-to-observable map is taken as the identity resulting in $\mathcal{O}\equiv \mathcal{F}$. 
The OK model, its associated forward operator $\mathcal{F}$, and sensor signal discretization $\boldsymbol{\mathcal{D}}$ are defined in Section~\ref{sec:model_definitions}. Here we use the squared noise-covariance--weighted $l^2$ data misfit\footnotemark $\mathcal{M}_f$ defined as 
\footnotetext{The subscript $C_{\noise}^{-1}$ stands for $l^2$ inner product weighted by the matrix $C_{\noise}^{-1}$. Here, $C_{\noise} = 0.04 \identity_{\dimy}$ is a diagonal matrix, corresponding to a signal-to-noise ratio of $5$.}
\begin{equation}
    \mathcal{M}_f(\x, z; \y^*, C_{\noise}) \coloneqq \Big\lVert \y^* - \boldsymbol{\mathcal{D}}\big(\mathcal{O}(\x, z)\big)\Big\rVert_{C_{\noise}^{-1}}^2\,,
\end{equation}
to compare the synthetic data $\y^*$ that is generated from the fixed latent inputs $(\x^*, \z^*)$ via \eqref{eq:num_exp_chara_model}, with model-predicted patterns at other inputs $(\x,\z)$. We note that the data misfit is proportional to the negative log-likelihood generated by the sensor noise in \eqref{eq:num_exp_chara_model}.

\begin{figure}[!ht]
    \centering
    \begin{subfigure}{0.4\linewidth}
    \centering
        \includegraphics[width= 0.9\linewidth]{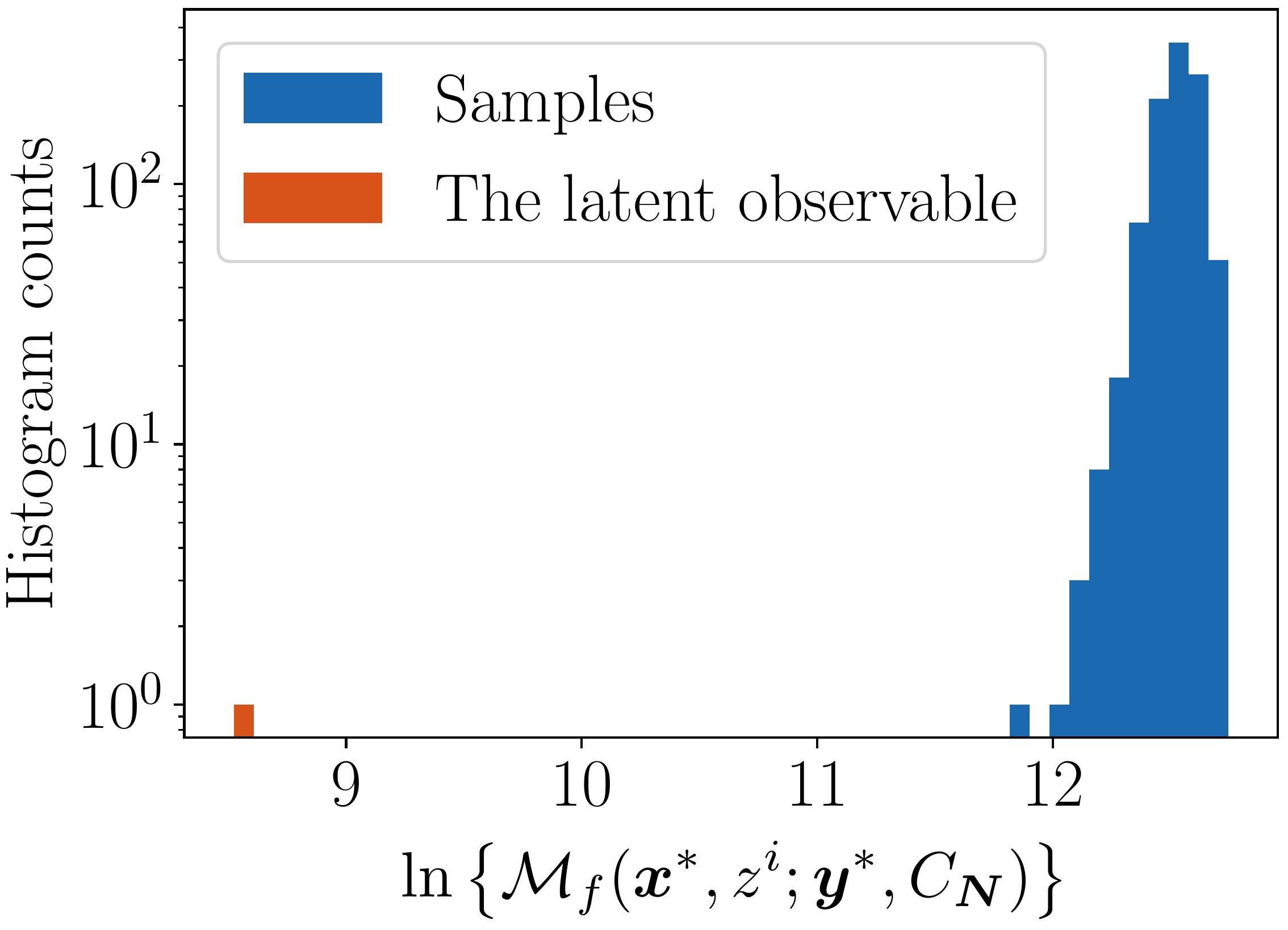}
    \end{subfigure}
    \begin{subfigure}{0.5\linewidth}
    \centering
    \begin{tabular}{|c|}\hline
        {\footnotesize The synthetic data $\y^*$ } \\\hline
        \includegraphics[width = 0.4\linewidth]{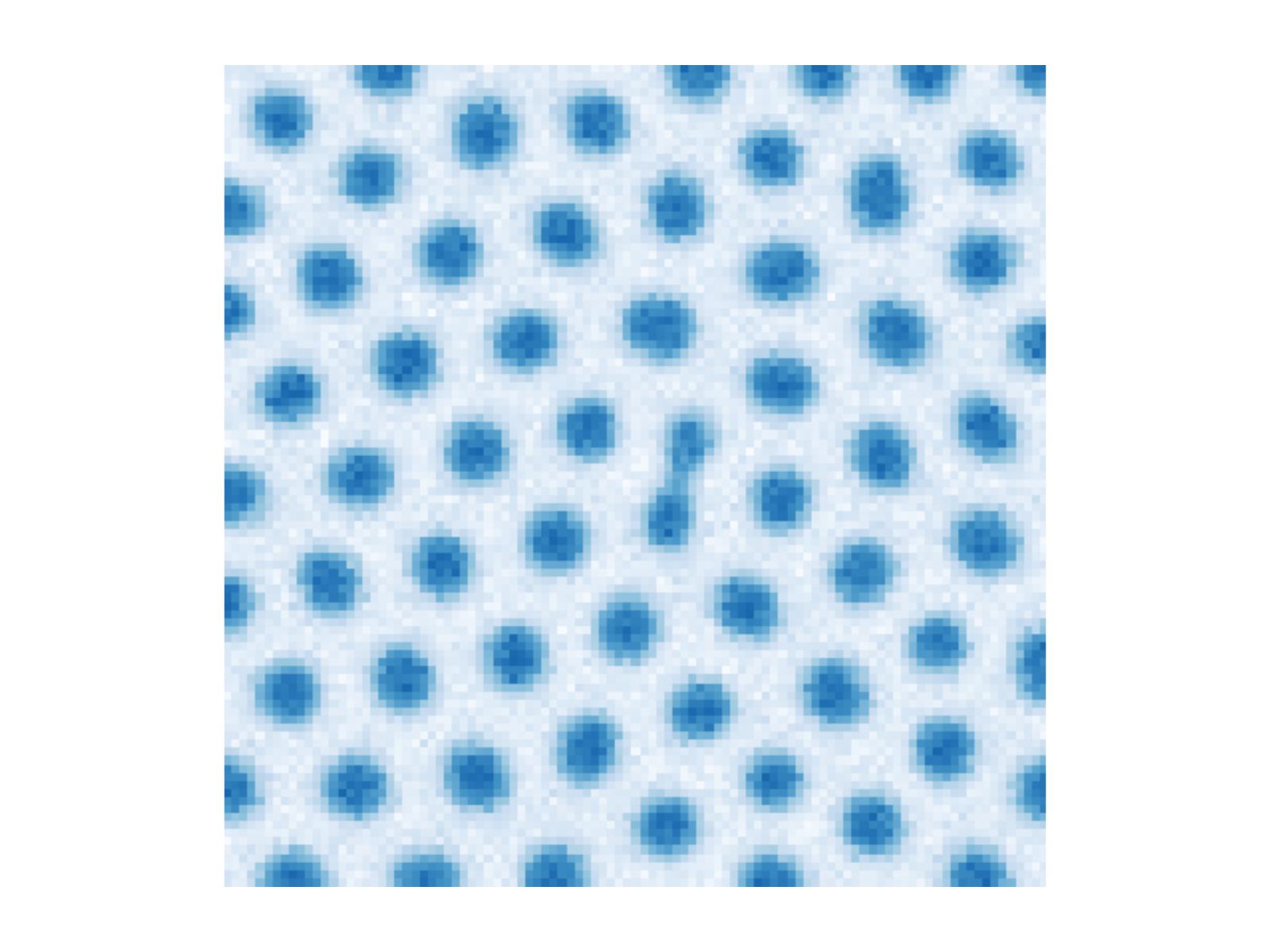}\\\hline
    \end{tabular}
    \begin{tabular}{|c|c|}
    \hline
        {\footnotesize The most likely observable} & {\footnotesize The least likely observable} \\\hline
        \includegraphics[width = 0.4\linewidth]{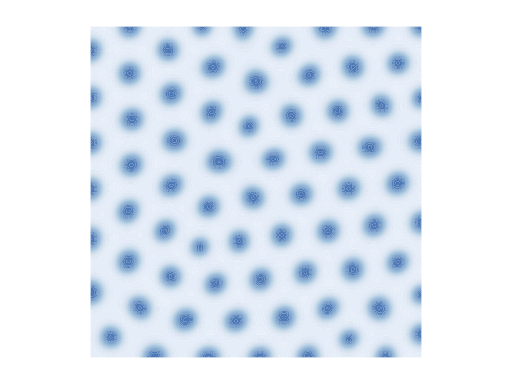} & \includegraphics[width = 0.4\linewidth]{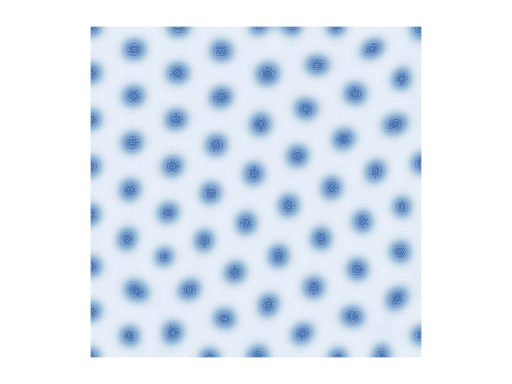}\\\hline
    \end{tabular}
    \end{subfigure}
    \caption{Observables $o^i = \mathcal{O}(\x^*, z^i)$, generated with fixed model parameters $\x^*$ and random samples of the auxiliary variable, $\z^i \sim \nu_{\bz|\bx}(\cdot|\x^*)$, have large misfit values (\emph{left}) with respect to a synthetic top-down microscopy characterization image $\y^*$ (\emph{right top}), generated by the data model at $\x^*$ and $z^*$. However, the most likely and the least likely observables (\emph{right bottom}) are visually similar in terms of phase, spot radius, and spot periodicity.} \label{fig:image_likelihood_fixed_x_parameter}
\end{figure}

\begin{figure}[!ht]
    \centering
        \centering
    \begin{subfigure}{0.4\linewidth}
    \centering
        \includegraphics[width= 0.9\linewidth]{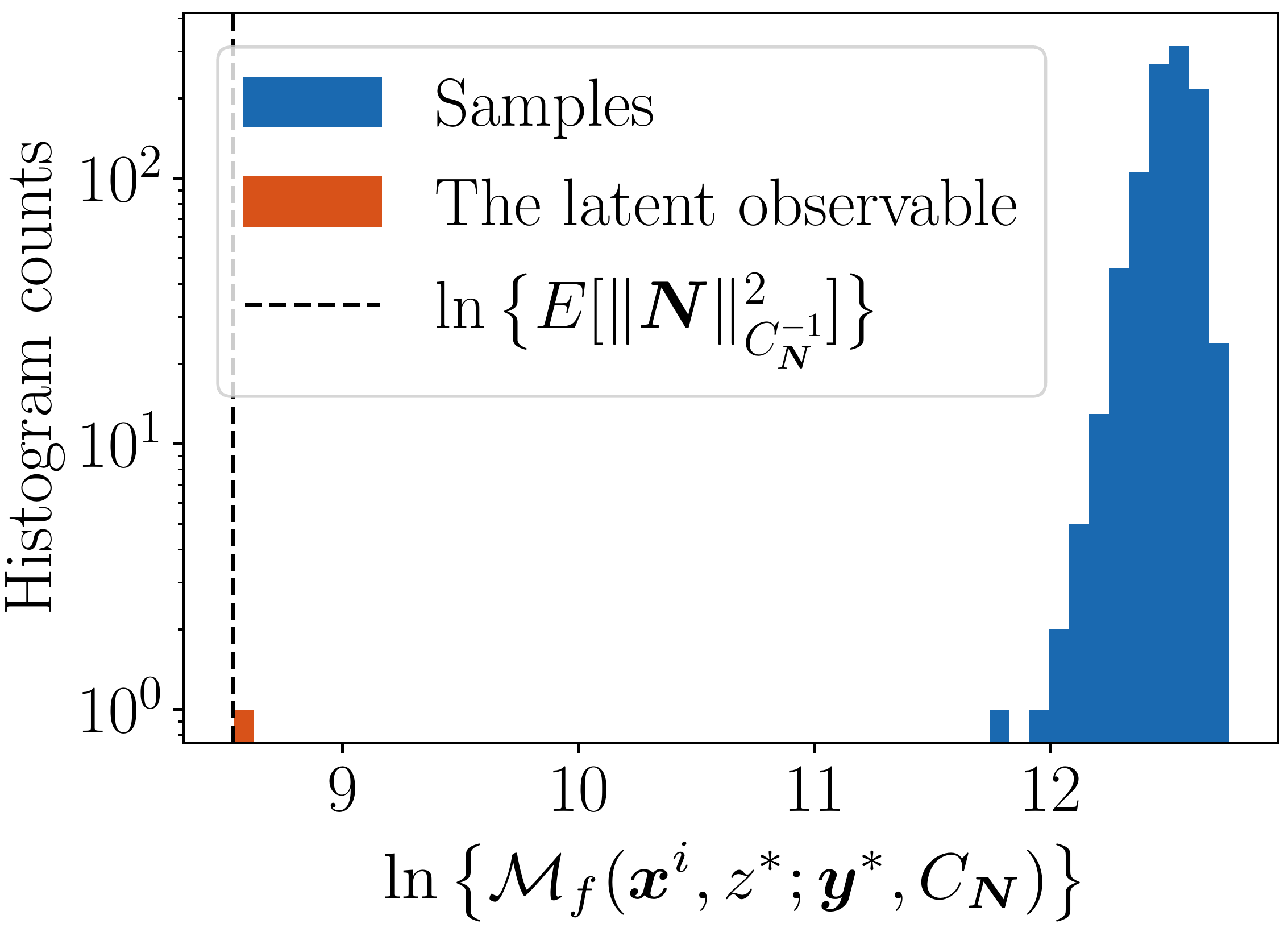}
    \end{subfigure}
    \begin{subfigure}{0.5\linewidth}
    \centering
    \begin{tabular}{|c|}\hline
        {\footnotesize The synthetic data $\y^*$ } \\\hline
        \includegraphics[width = 0.4\linewidth]{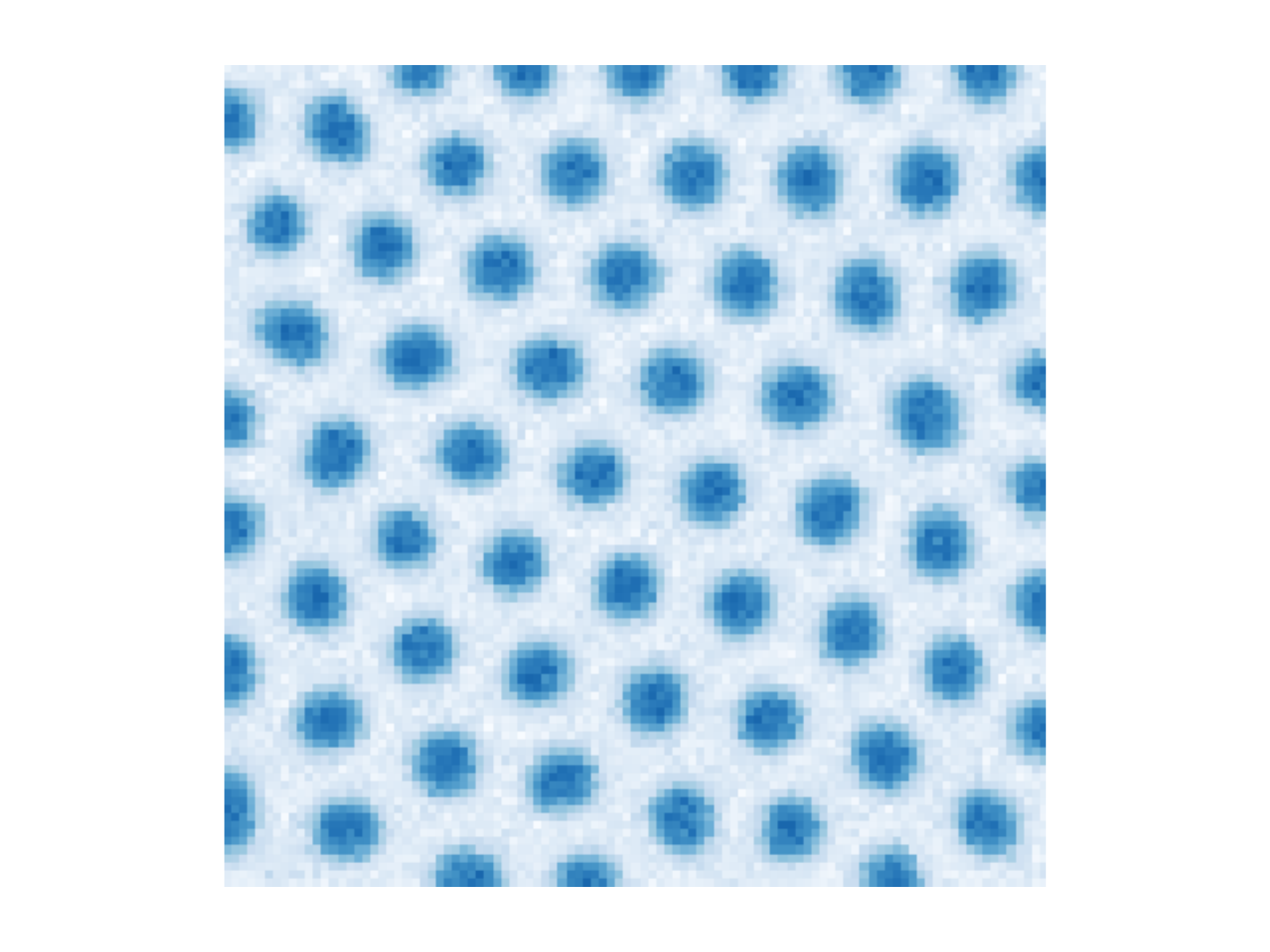}\\\hline
    \end{tabular}
    \begin{tabular}{|c|c|}
    \hline
        {\footnotesize The most likely observable} & {\footnotesize The least likely observable} \\\hline
        \includegraphics[width = 0.4\linewidth]{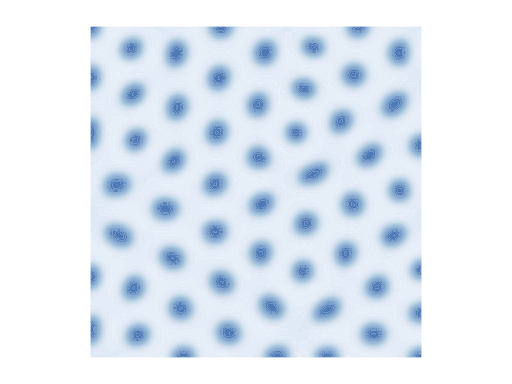} & \includegraphics[width = 0.4\linewidth]{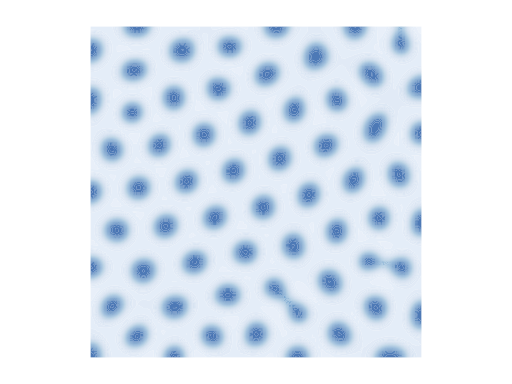}\\\hline
    \end{tabular}
    \end{subfigure}
    \caption{Observables $o^i = \mathcal{O}(\x^i, z^*)$, generated with small perturbations in the model parameters, $\x^i\sim \mathcal{N}(\x^*, \delta^2 \identity_{\dimx})$, and fixed auxiliary variable $\z^*$, have large misfit values (\emph{left}) with respect to a synthetic top-down microscopy characterization image $\y^*$ (\emph{right top}), generated by the data model at $\x^*$ and $z^*$. However, the most likely and the least likely observables (\emph{right bottom}) are visually similar in terms of phase, spot radius, and spot periodicity. The expected misfit for only the sensor noise $\noise$ is included (\emph{left, dashed line}) as a reference to contrast with the large misfit values that arise from perturbing the model parameters.} \label{fig:image_likelihood_fixed_z_parameter}
\end{figure}

In the first numerical example, the data misfit is evaluated at $\x^*$ and $10^3$ i.i.d. samples of the auxiliary variable, $\z^i \sim \nu_{\bz|\bx}(\cdot|\x^*)$. In Fig.~\ref{fig:image_likelihood_fixed_x_parameter}, a histogram of the squared $l^2$ data misfit evaluations is shown along with the synthetic data $\y^*$, the most likely observable corresponding to the lowest $l^2$ misfit, and the least likely observable corresponding to the highest $l^2$ misfit. In the second numerical example, the data misfit is evaluated at $\z^*$ and $10^3$ i.i.d. samples of the model parameter $\x^i\sim \mathcal{N}(\x^*, \delta^2 \identity_{\dimx})$, where $\delta$ is chosen to be small\footnotemark. Similar plots for the second example are shown in Fig.~\ref{fig:image_likelihood_fixed_z_parameter}. \footnotetext{We choose $\delta = 0.05$ for the same computational domain defined in \ref{app:function_space}. The choice of $\delta$ is small comparing to the admissible parameter region defined in Section~\ref{subsec:prior_transform}.}

In both figures, we observe that, while the true parameters reproduce observables identical to the image up to the additive noise, most realizations of the nearby parameters or auxiliary variables produce observables that are drastically different from the image data according to the $l^2$ data misfit. As a result, the misfit values for most of the samples are very large, relative to misfit values for the latent input that generated the synthetic data. This leads to a likelihood concentration phenomenon that  results in LFI methods requiring many joint samples to recover the true parameters.

Furthermore, similar features in characterization images $\y^*$ of BCP self-assembly indicate that certain characteristics of their latent BCP equilibrium structures are similar. This should, in general, imply certain closeness in their underlying model parameters. However, the likelihood function $\pi_{\y^*|\bx}(\x)$ measures the probability of the model parameterized by $\x$ matching a \emph{specific} image. As a result, the observables that are visually similar to the image data, but are mismatched in entry-wise values, will not be weighted heavily by the likelihood function, as demonstrated through the two numerical examples above. This leads to unnecessarily ineffective parameter inference.

Although using entire images for inference ensures that all of the available information about the latent state is preserved, some information may be irrelevant and/or redundant for the purpose of model calibration and prediction. For example, the specific orientation of phase interfaces in a top-down microscope image of unguided lamella phase BCP thin films (i.e. the `figer-print' patterns) are irrelevant (when ignoring boundary effects) if our interest is to predict the natural periodicity length of the BCP material. In other words, compressing irrelevant information in the data should not affect the posterior density for the parameters. 

The notion of data reduction is very common for image data. While the $l^2$ distance can be easily computed, it does not capture the visual similarity between images. For instance, the distance can be arbitrarily large when the two images are identical, up to a rotation or reflection. Therefore, it is often recommended to measure the discrepancy between images in a more natural way. Different metrics that mitigate the issues for high-dimensional images have been considered by many authors~\cite{Fienup1997, Morel2009, DiGesu1999} and some image processing techniques have been applied to image data for BCP self-assembly in the past~\cite{Lee2005, Murphy2015, Khadilkar2017}. These metrics  
and the corresponding likelihood functions are designed to be relatively insensitive to certain transformations that do \emph{not} change the fundamental characteristics of the image related to the model parameters of interest.

In this work we advocate for constructing low-dimensional summary statistics of BCP characterization images that quantitatively describe image features 
that are related to characteristics of their latent BCP equilibrium structures. Ideally, the summary statistics $\bq$ are chosen to be \textit{sufficient statistics}. This means that they do not introduce any bias in the inference procedure, i.e., they satisfy $\pi_{\bx|\by}(\x|\y) = \pi_{\bx|\mathbf{Q}}(\x|\mathcal{Q}(\y))$ for all $\x,\y$. Furthermore, we can also look for minimal sufficient statistics, i.e., the summary statistics that most reduce the dimension of the data without discarding any relevant information. Unfortunately, sufficient statistics are not readily available outside of exponential family models, and consequently many methods have been proposed to select summary (but not necessarily sufficient) statistics for LFI in practice~\cite{blum2013comparative}. These summary statistics may discard some information during the inference, as a trade-off for suffering less from the  concentration phenomenon of the likelihood for high-dimensional images~\cite{csillery2010approximate}. 

It is important to note that working with summary statistics of the data remains feasible with LFI methods. While the conditional distributions for the summary statistics $\pi_{\bq|\bx}(\cdot | \x)$ may not be analytically available or feasible to evaluate, LFI methods only require joint samples of the parameters $\x^i$ and summary statistics $\q^i$. These samples can be acquired by an additional step of computing the summary statistics $\q^i = \mathcal{Q}(\y^i)$ for each data sample $\y^i$.
\section{Likelihood-free inference via triangular transport maps}\label{sec:transportLFI}

In this section, we present a measure transport approach for characterizing the family of conditional densities for the parameters given the observations $\pi_{\bx \vert \bq}$ using only samples from the joint distribution $\nu_{\bx,\bq}$. In Section~\ref{subsec:transport} and~\ref{subsec:triangular_maps} we introduce the notion of measure transport and triangular transport maps, respectively. In Section~\ref{subsec:estimating_maps} we show how to estimate the maps from samples. Then, in Section~\ref{subsec:cde_maps}, we demonstrate how these maps can be used for likelihood\textendash free inference. A numerical implementation of this approach is available online in a \textsc{MATLAB} package for transport map estimation\footnote{\url{https://github.com/baptistar/ATM}}. 

\subsection{Measure transport} \label{subsec:transport}
\begin{figure}
\centering
\scalebox{0.55}{\begin{tikzpicture}
	\node (D1) {\includegraphics[width=5.5cm]{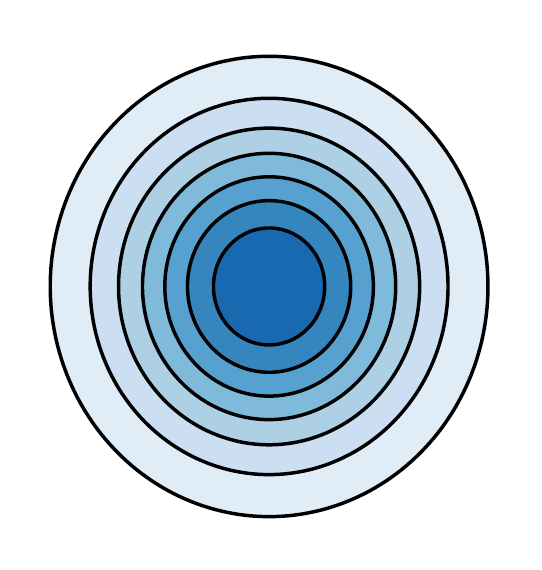}};
	\node[below=1mm of D1] (D2) {\includegraphics[width=7cm]{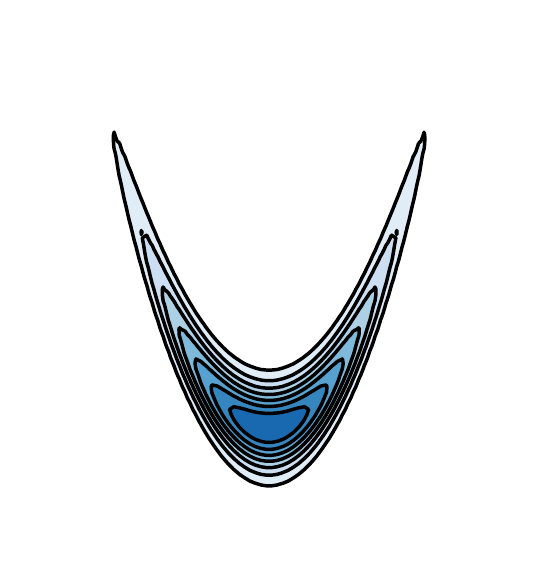}};

	\node[right=25mm of D1] (M1) {\includegraphics[width=5.5cm]{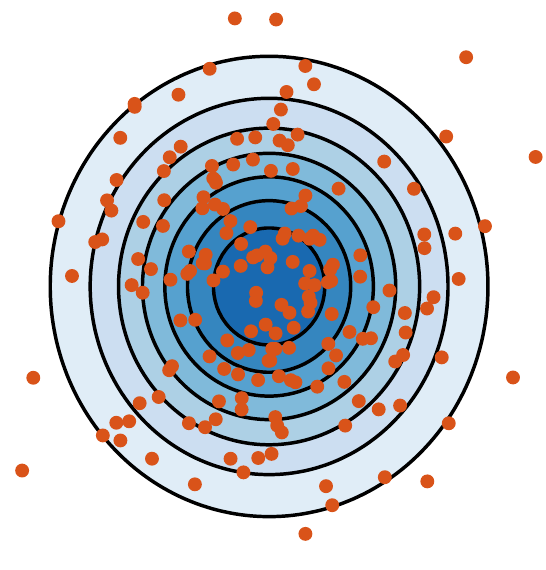}};
	\node[below=1mm of M1] (M2) {\includegraphics[width=7cm]{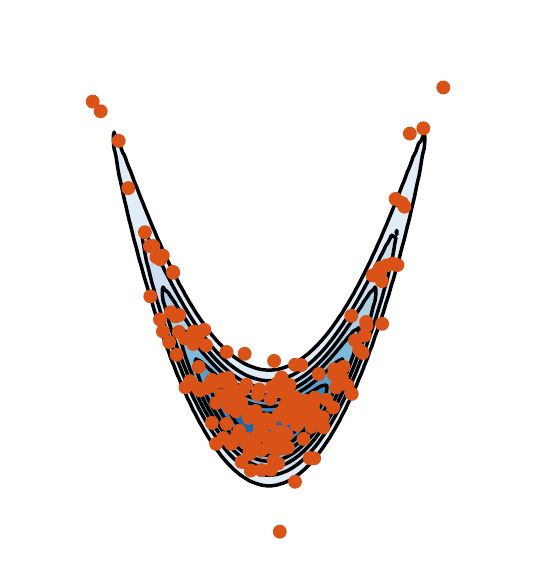}};

	\node[right=-1.3cm of D1.south, yshift=1cm] (D1left) {};
	\node[right=1.3cm of D1.south, yshift=1cm] (D1right) {};
	\node[right=-1.35cm of D2.north, yshift=-2cm] (D2left) {};
	\node[right=1.35cm of D2.north, yshift=-2cm] (D2right) {};
    
	\draw[->,ultra thick] (D1left) edge[bend right] node [left] {\Large $S^{\sharp}\pi_{\mathbf{R}}$} (D2left);
	\draw[<-,ultra thick] (D1right) edge[bend left] node [right] {\Large $S_{\sharp}\pi_{\mathbf{W}}$} (D2right);

	\node[right=-1.3cm of M1.south, yshift=1cm] (M1left) {};
	\node[right=1.3cm of M1.south, yshift=1cm] (M1right) {};
	\node[right=-1.35cm of M2.north, yshift=-2cm] (M2left) {};
	\node[right=1.35cm of M2.north, yshift=-2cm] (M2right) {};

	\draw[->,ultra thick] (M1left) edge[bend right] node [left] {\Large $S^{-1}(\boldsymbol{r}^i)$} (M2left);
	\draw[<-,ultra thick] (M1right) edge[bend left] node [right] {\Large $S(\boldsymbol{w}^i)$} (M2right);

\end{tikzpicture}}
\caption{The target density $\pi_{\bw}$ is represented as the pullback of a standard Gaussian reference density $\pi_{\br}$ through the transport map $S$ (\emph{left}). Samples from $\nu_{\br}$ are generated by mapping reference samples through the inverse of $S$ (\emph{right}). Similarly we can represent the reference density as the pushforward of the target density through $S$ and sample from $\nu_{\bw}$ by mapping target samples through $S$. \label{fig:transport_map}}
\end{figure}
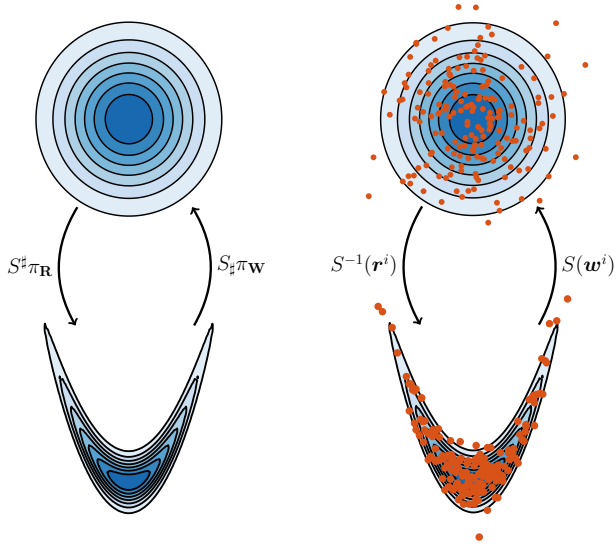

Let $\bw \in \mathbb{R}^{\n}$ be a \emph{target} random variable with measure $\nu_{\bw} \in \mathcal{P}(\mathbb{R}^{\n})$ we would like to characterize, and let $\br \in \mathbb{R}^{\n}$ be a \emph{reference} random variable with measure $\nu_{\br} \in \mathcal{P}(\mathbb{R}^{\n})$ that we can easily sample from, for example, a standard Gaussian. Our goal is to find a measurable function $S\colon \mathbb{R}^{\n} \rightarrow \mathbb{R}^{\n}$ that defines a deterministic coupling between these two variables such that the law of $S(\bw)$ is equal to the law of $\br$. For such a map $S$, we say that $S$ \emph{pushes forward} $\nu_{\bw}$ to $\nu_{\br}$ and we denote this by $S_{\sharp}\nu_{\bw} = \nu_{\br}$. Equivalently, we say that $S$ \emph{pulls back} $\nu_{\br}$ to $\nu_{\bw}$ and we denote this by $S^{\sharp}\nu_{\br} = \nu_{\bw}$. For measures $\nu_{\bw}$ and $\nu_{\br}$ with densities $\pi_{\bw}$ and $\pi_{\br}$, respectively, similar relationships apply, $S_{\sharp}\pi_{\bw} = \pi_{\br}$ and $S^{\sharp}\pi_{\br} = \pi_{\bw}$. If $S$ is a diffeomorphism (i.e., a differentiable bijection with a differentiable inverse), we can evaluate the pullback density $S^{\sharp}\nu_{\br}$ using the change-of-variables formula as 
\begin{equation}
    S^{\sharp}\pi_{\br}(\w) \coloneqq \pi_{\br} \circ S(\w) \vert \det \nabla S(\w) \vert.
\end{equation}
Therefore, having the transport map $S$ that couples $\bw$ and $\br$ and the reference density $\pi_{\br}$ enables us to evaluate the target density $\pi_{\bw}$. Furthermore, we can use this map to generate i.i.d.\thinspace samples from the target distribution by sampling from the reference measure $\rsample^i \sim \nu_{\br}$ and applying the inverse map to generate samples $S^{-1}(\rsample^i) \sim \nu_{\bw}$. In Fig.~\ref{fig:transport_map}, we provide an example of using $S$ to characterize the pullback density and generate samples from a 2-dimensional ``banana'' target distribution. 

The existence of a map $S$ that pushes forward $\pi_{\bw}$ to $\pi_{\br}$ is guaranteed under very mild conditions such as $\nu_{\bw}$ and $\nu_{\br}$ being absolutely continuous with respect to the Lebesgue measure, i.e., the distributions do not contain atoms and their densities exist. We refer the reader to~\cite{santambrogio2015optimal} for approaches that relax these conditions.
\subsection{The Knothe--Rosenblatt rearrangement} \label{subsec:triangular_maps}

In general, many maps $S$ can satisfy the measure transport conditions above~\cite{villani2008optimal}. For instance, the optimal transport literature looks for maps $S$ that couple two distributions and also minimize an integrated transportation distance, such as the Wasserstein metric~\cite{peyre2019computational}. In this work, we consider the Knothe\textendash Rosenblatt (KR) rearrangement for $S$, i.e., the unique monotone and lower triangular map that pushes forward $\pi_{\bw}$ to $\pi_{\br}$~\cite{Rosenblatt1952, knothe1957contributions, bogachev2005triangular}. A monotone and lower triangular map is a multivariate function of the form
\begin{equation}\label{eq:KR_map_vector}
    S(\w) = \begin{bmatrix*}[l] S^1(w_1) \\ S^2(w_1,w_2) \\ \vdots \\ S^{\n}(w_1,\dots,w_{\n}) \end{bmatrix*},
\end{equation}
where $\xi \rightarrow S^{k}(w_1,\dots,w_{k-1},\xi)\colon \mathbb{R} \to \mathbb{R}$, i.e., the restriction of the map component $S^k$ to its last variable $w_k$, by fixing its first $k-1$ variables, is monotone increasing for all $(w_1,\dots,w_{k-1}) \in \mathbb{R}^{k-1}$. We denote the space of triangular and monotone maps on $\mathbb{R}^{\n}$ by $\mathcal{S}_{\triangle}(\mathbb{R}^{\n})$. 

Working with the KR rearrangement, rather than other (non-triangular)  transport maps, has several computational benefits~\cite{Marzouk2017}. First, monotone triangular maps $S \in \mathcal{S}_{\triangle}(\mathbb{R}^{\n})$ can be easily inverted to sample from $\nu_{\bw}$ by solving the nonlinear system $S(\w^i) = \rsample^i,\; \rsample^i\sim \nu_{\br}$ via a sequence of 1-dimensional nonlinear solves. Furthermore, each nonlinear solve has a unique solution due to the injectivity of the map. Second, the Jacobian determinant of a triangular map $\det \nabla S(\w)$ is tractable to compute as the product of the partial derivatives of each component with respect to the last variables, i.e., $\det \nabla S(\w) = \prod_{k = 1}^{\n} \partial_k S^k(\w_{1:k})$. Third, these maps are easier 
to parameterize than maps derived via optimal transport and can be computed quickly using gradient-based optimization techniques~\cite{parno2018transport, Baptista2020}. 

Most importantly in our context, these maps provide direct access to the conditional densities of the target distribution. For a reference distribution with independent components, such as the standard Gaussian, the $k$ map component $S^k$ pushes forward the $k$th marginal conditional density $\pi_{\bw_k|\bw_1,\dots,\bw_{k-1}}$ in the factorization of the target density $\pi_{\bw} = \pi_{\bw_1}\pi_{\bw_2|\bw_1}\dots\pi_{\bw_\n|\bw_1,\dots,\bw_{\n-1}}$, to the $k$th marginal of the reference density $\pi_{\br}$ \cite[App. A]{spantini2018inference}. In Section~\ref{subsec:cde_maps}, we use this property to extract the posterior density, i.e., the conditional density of the parameters given the data, from their joint density.

\subsection{Estimating the Knothe--Rosenblatt rearrangement} \label{subsec:estimating_maps}

The \emph{exact} KR rearrangement $S_{\textrm{KR}}$ that couples two densities is, in general, analytically unavailable. However, we consider approximations of the KR map in this work. We follow the variational approach in ~\cite{Marzouk2017,spantini2018inference,Baptista2020} of minimizing the KL divergence,\footnote{While we choose the KL divergence, one can also consider other divergences or metrics to measure the discrepancy between densities.} from the pullback density $S^{\sharp}\pi_{\br}$ to the target density $\pi_{\bw}$, which has the form
\begin{equation}
    \mathcal{D}_{\textrm{KL}} \left(\pi_{\bw} || S^{\sharp}\pi_{\br} \right)
    = \mathbb{E}_{\nu_{\bw}} \left[\log \pi_{\bw} - \log \pi_{\br} \circ S(\bw) - \log \det \nabla S(\bw) \right] \nonumber.
\end{equation}
The KR rearrangement is given by the global minimizer of the KL divergence over the space of lower triangular and monotone maps on $\mathbb{R}^{\dimw}$. That is, 
\begin{align} \label{eq:min_KLdivergence}
    S_{\textrm{KR}} &= \underset{S \in \mathcal{S}_{\triangle}(\mathbb{R}^{\dimw})}{\argmin} \; \mathcal{D}_{\textrm{KL}}(\pi_{\bw}||S^{\sharp}\pi_{\br}) \nonumber \\
    &= \underset{S \in \mathcal{S}_{\triangle}(\mathbb{R}^{\dimw})}{\argmin} \; \mathbb{E}_{\nu_{\bw}} \left[-\log \pi_{\br} \circ S(\bw) - \log \det \nabla S(\bw) \right].
\end{align}

In all of our numerical experiments, we let the reference distribution be standard Gaussian. Substituting the form of the reference density $\pi_{\br} = \mathcal{N}(\mathbf{0},\identity_{\dimw})$ in~\eqref{eq:min_KLdivergence}, the optimization problem can be written as
\begin{equation} \label{eq:min_KL_gaussianref}
    S_{\textrm{KR}} = \underset{S \in \mathcal{S}_{\triangle}(\mathbb{R}^{\dimw})}{\argmin} \; \sum_{k = 1}^{\dimw} \mathbb{E}_{\nu_{\bw}}\left[\frac{1}{2} S^k(\bw_{1:k})^2 -  \log \partial_k S^k(\bw_{1:k}) \right],
\end{equation}
where each term in the sum only depends on one map component. Thus, optimizing for $S$ can be decoupled into $\dimw$ optimization problems for each map component\footnote{In fact, this holds for any reference density with independent components.}, which can each be solved independently and in parallel to determine each $S^{k}$. More information on the parametrization and 
computation of monotone and triangular maps is provided in~\ref{app:map_details}. 
\subsection{Conditional density estimation via triangular transport maps} \label{subsec:cde_maps}

To estimate the conditional densities $\pi_{\bx|\bq}$ and any posterior $\pi_{\bx|\q^*}$ in particular,  we begin by considering the KR rearrangement $S_{\textrm{KR}}$ that pushes forward the target joint density $\pi_{\bx,\bq}$ to a reference density with independent components
$\pi_{\br_1,\br_2} = \pi_{\br_1}\pi_{\br_2}$, that are both defined on $\mathbb{R}^{\dimx+q}$. Given that $S_{\textrm{KR}}$ is lower triangular, we can partition the map into two blocks as 
\begin{equation} \label{eq:block_KRmap}
    S_{\textrm{KR}}(\q,\x) = \begin{bmatrix*}[l] S^{\mathcal{Q}}(\q) \\ S^{\mathcal{X}}(\q,\x) \end{bmatrix*},
\end{equation}
where $S^{\mathcal{Q}}\colon \mathbb{R}^{q} \rightarrow \mathbb{R}^{q}$ and $S^{\mathcal{X}}(\q,\cdot)\colon \mathbb{R}^{\dimx} \rightarrow \mathbb{R}^{\dimx}$ are monotone functions for all $\q \in \mathbb{R}^{q}$.From the measure transport condition, then $S^{\mathcal{Q}}$ pulls back the marginal $\pi_{\br_1}$ to the marginal for the summary statistics $\pi_{\bq}$, and $S^{\mathcal{X}}(\q,\cdot)$ pulls back the marginal $\pi_{\br_2}$ to the conditional $\pi_{\bx|\bq}(\cdot|\q)$ for each $\q \in \mathbb{R}^{q}$.

From the decoupling of the optimization problem over the map components, we only need to solve $\dimx$ independent optimization problems for the components in $S^{\mathcal{X}}$ to approximate the conditional density $\pi_{\bx|\bq}$. 
Moreover, the conditional density can be estimated given $N$ i.i.d.\thinspace samples from the joint distribution, $\{(\x^i,\q^i)\}_{i=1}^N \sim \nu_{\bx,\bq}$. We use these samples to define a Monte Carlo approximation of the expectation in~\eqref{eq:min_KL_gaussianref} and to find the $k$th component of $S^{\mathcal{X}}$ by solving 
\begin{equation}\label{eq:monte_carlo_map_estimate}
   (\widehat{S}^{\mathcal{X}})^{q + k} = \underset{S^{q + k}}{\argmin} \; \frac{1}{N}\sum_{i=1}^N \left[ \frac{1}{2} S^{q + k}(\q^i,\x^i_{1:k})^2 -  \log \partial_k S^{q + k}(\q^i,\x^i_{1:k}) \right],
\end{equation}
under the monotonicity constraint $\partial_{k}(S^{\mathcal{X}})^k(\q,\x_{1:k}) > 0$ for all $(\q,\x_{1:k}) \in \mathbb{R}^{q+k}$. Then, collecting the components for $k=1,\dots,\dimx$ in $\widehat{S}^{\mathcal{X}}$ produces an estimator of the conditional density $\pi_{\bx|\bq}(\cdot|\q)$ for any $\q \in \mathbb{R}^{q}$ as 
\begin{align}
\x \mapsto \widehat{\pi}_{\bx|\bq}(\x|\q) &\coloneqq  \widehat{S}^{\mathcal{X}}(\q,\x)^{\sharp}\pi_{\br_2}(\x) \nonumber \\
&= \pi_{\br_2}\left(\widehat{S}^{\mathcal{X}}(\q,\x)\right) \det \nabla_{\x} \widehat{S}^{\mathcal{X}}(\q,\x) \label{eq:conditional_density_estimate}.
\end{align}

\begin{figure}[!ht]
\scalebox{0.55}{\begin{tikzpicture}
	\node (J) {\includegraphics[width=12cm]{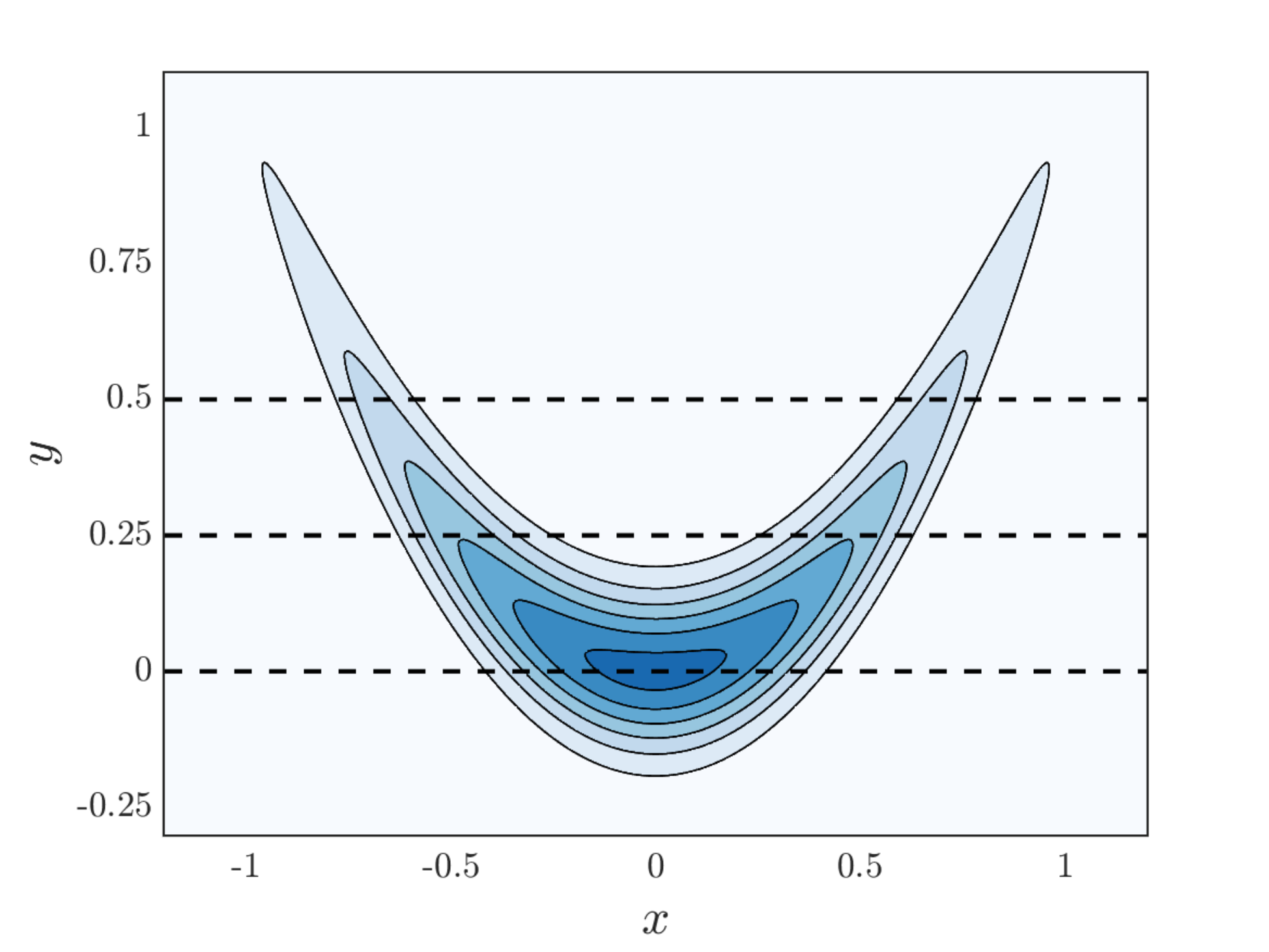}};

	\node[right=1cm of J]  (P2) {\includegraphics[width=7cm]{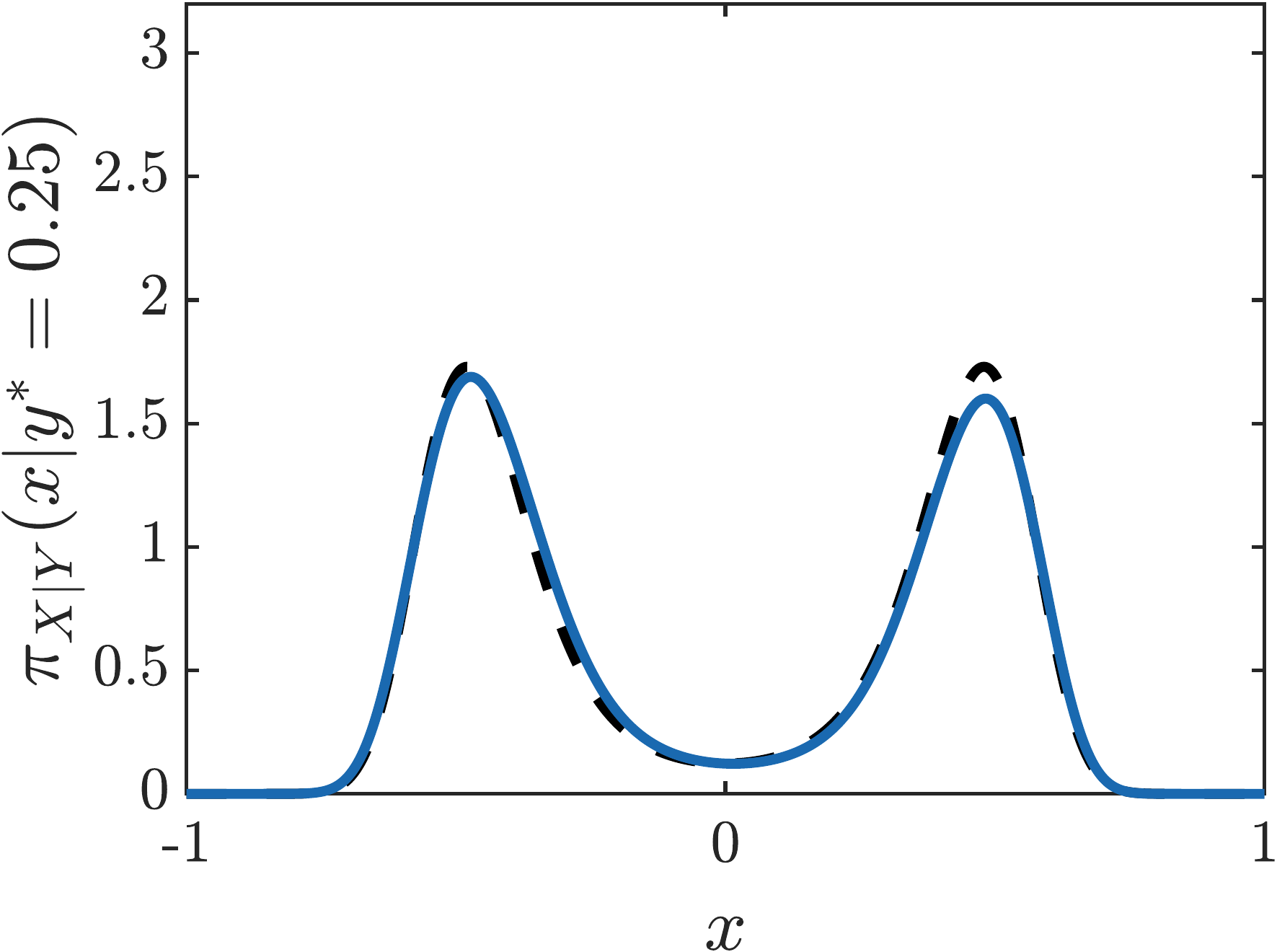}};
	\node[above=1mm of P2] (P1) {\includegraphics[width=7cm]{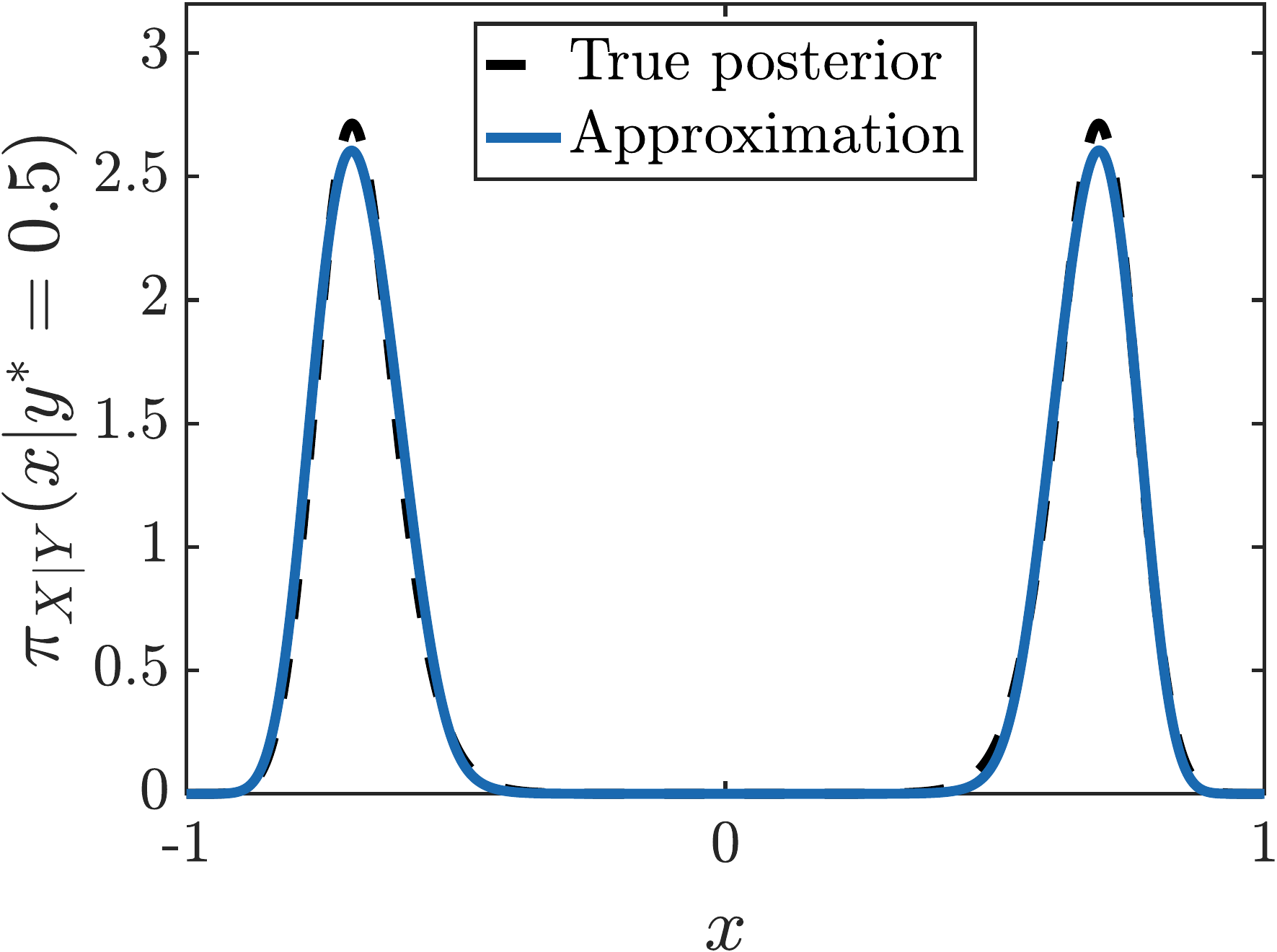}};
	\node[below=1mm of P2] (P3) {\includegraphics[width=7cm]{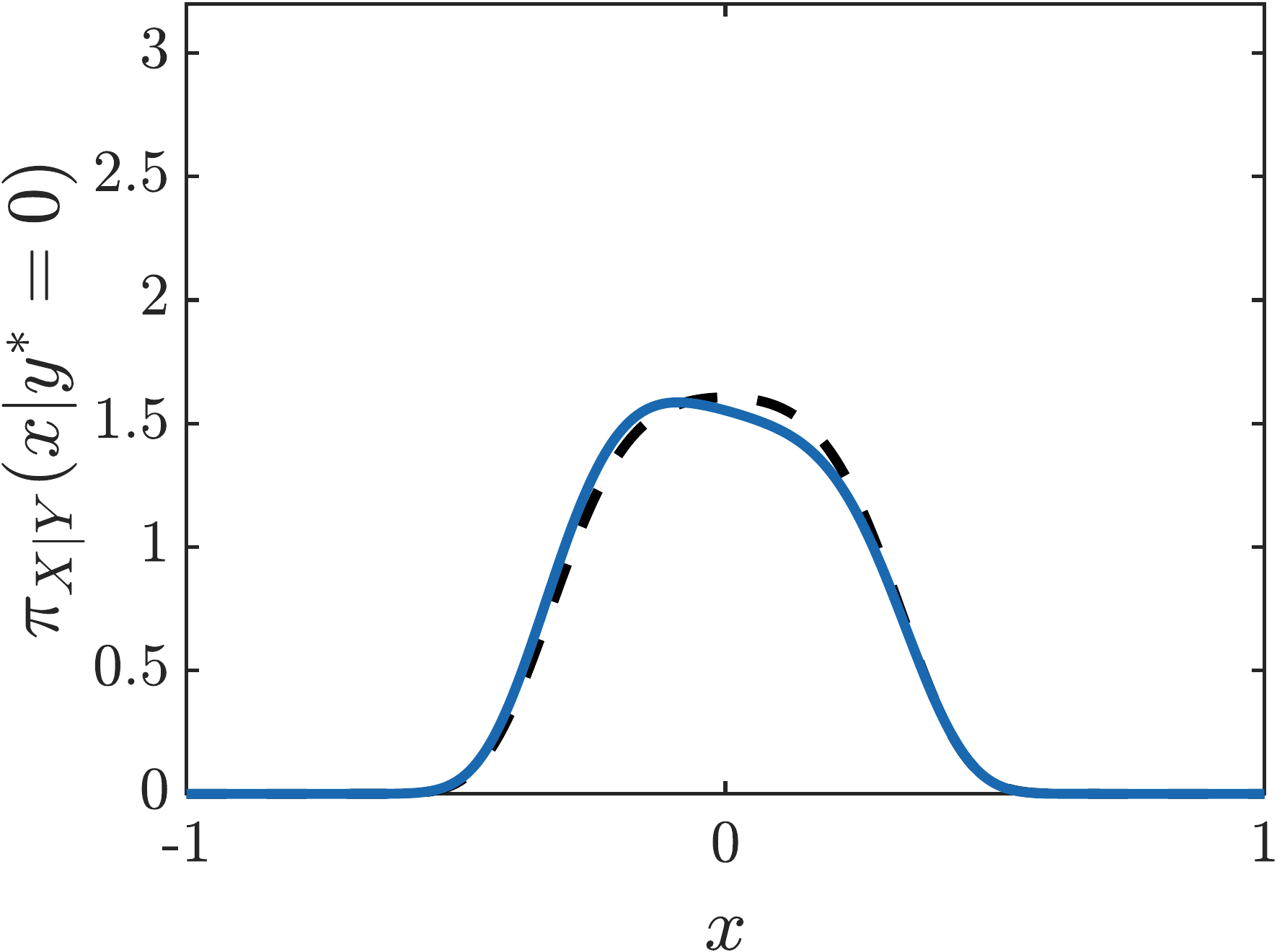}};

	\node[right=5mm of P1] (M1) {\includegraphics[width=7cm]{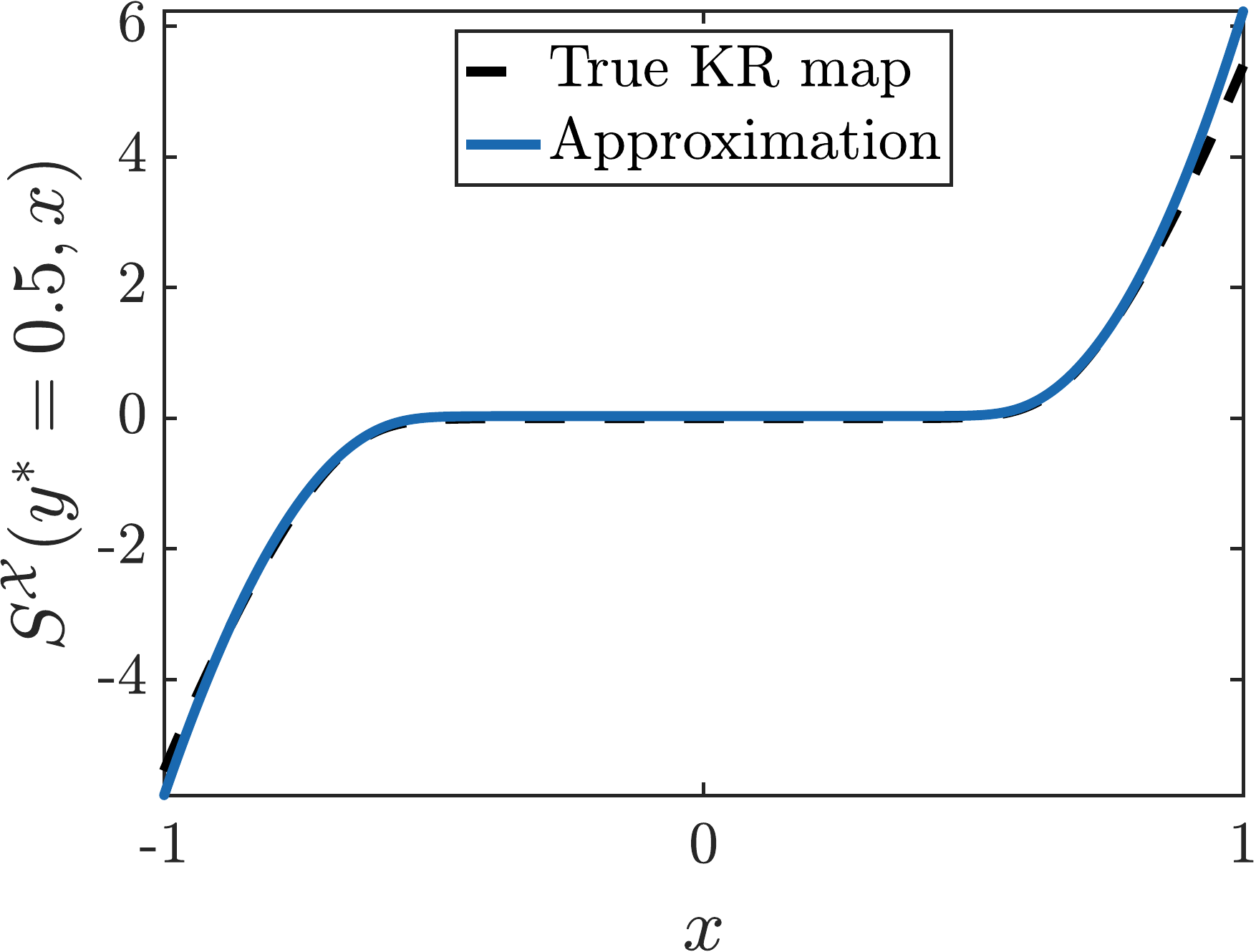}};
	\node[right=5mm of P2] (M2) {\includegraphics[width=7cm]{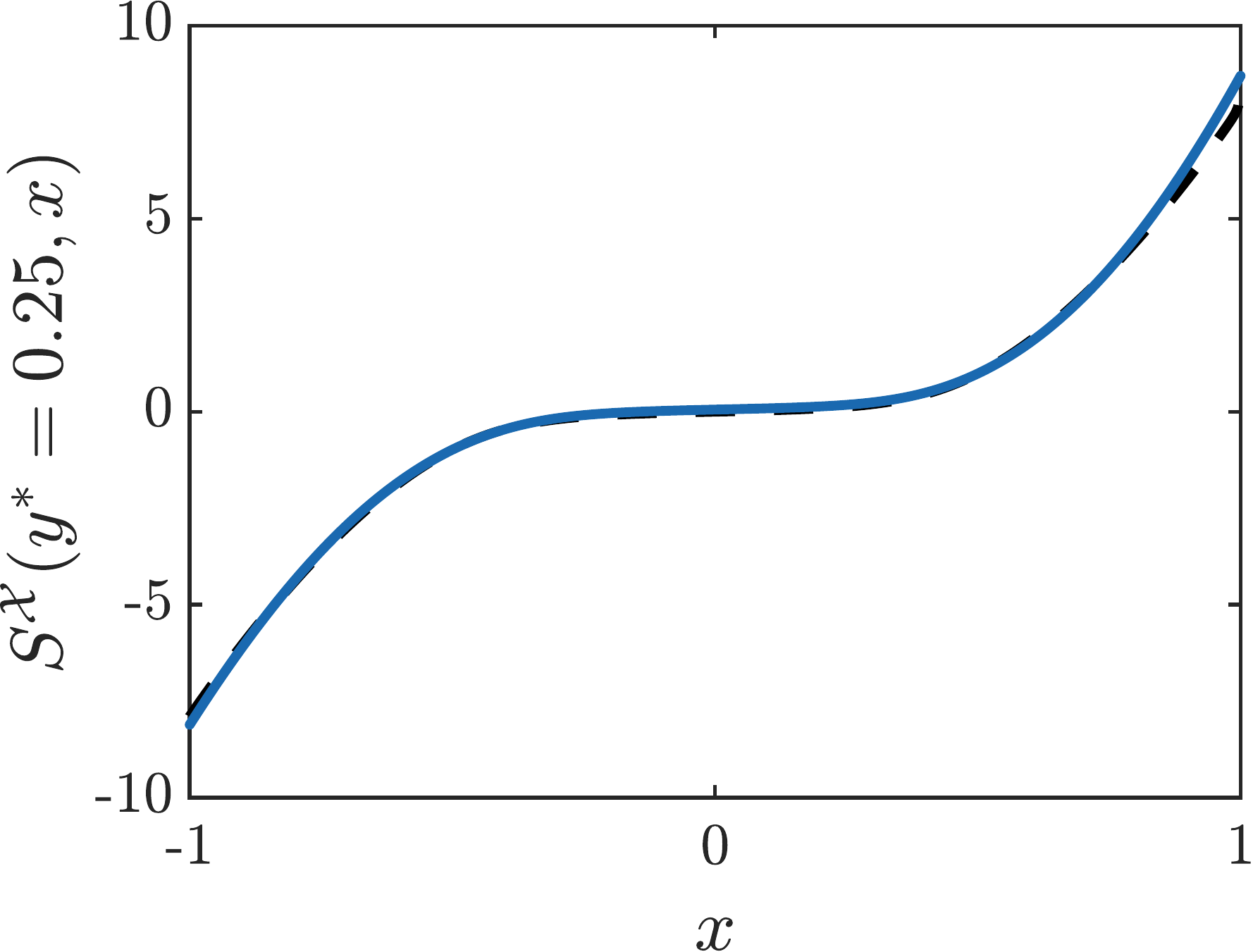}};
	\node[right=5mm of P3] (M3) {\includegraphics[width=7cm]{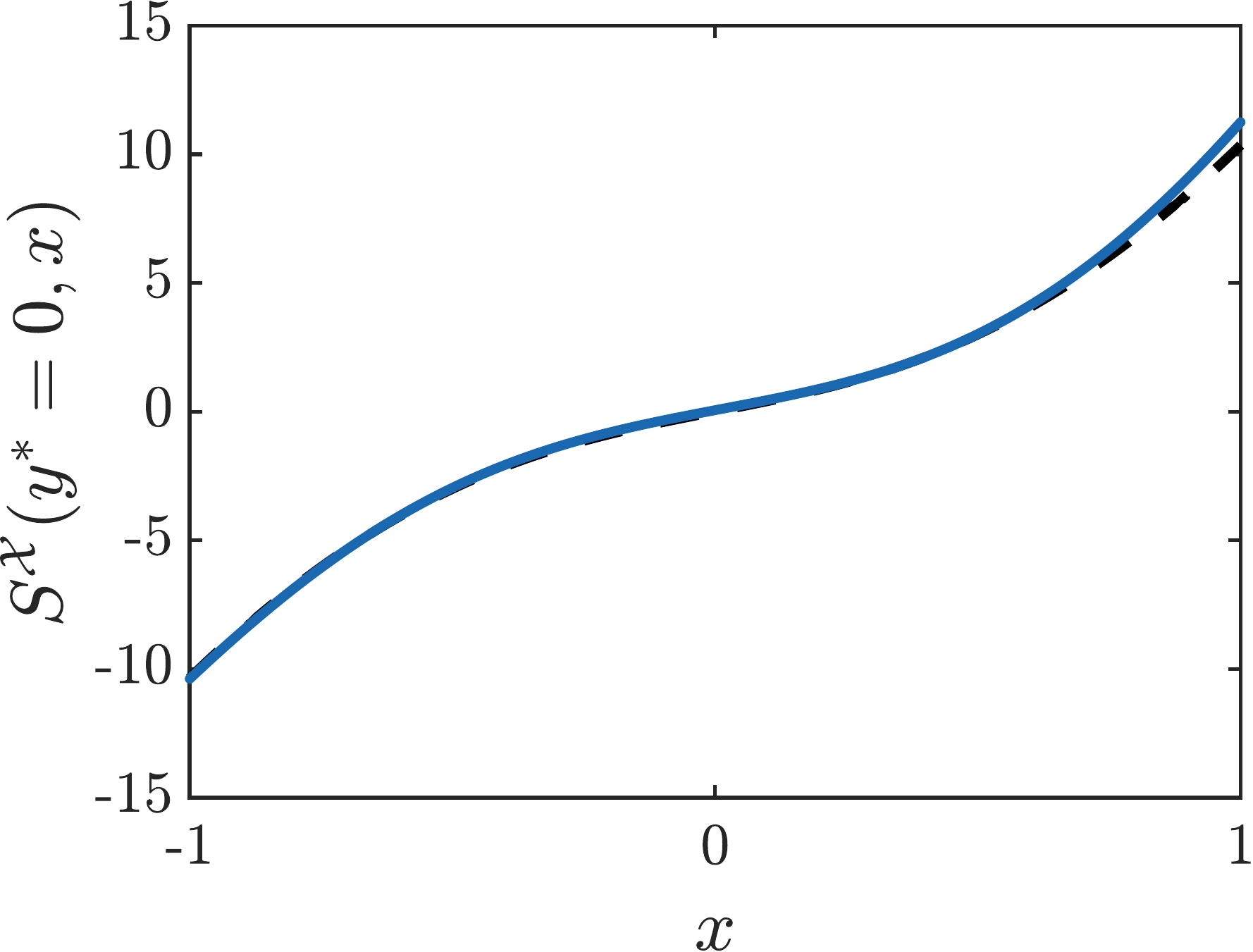}};

	\node[right=-1.49cm of J.east, yshift=-1.79cm] (C3) {};
	\node[right=-1.5cm of J.east, yshift=-0.65cm] (C2) {};
	\node[right=-1.49cm of J.east, yshift=0.575cm] (C1) {};

	\draw[->,ultra thick] (C3) -- (P3.west) {};
	\draw[->,ultra thick] (C2) -- (P2.west) {};
	\draw[->,ultra thick] (C1) -- (P1.west) {};
\end{tikzpicture}}
\caption{Conditional density estimation for the posterior densities with the quadratic likelihood model: $Y^2 = X^2 + N$ for $X \sim \mathcal{N}(0,0.25)$ and $N \sim \mathcal{N}(0,0.01)$. Contours of the joint density of the parameter $x$ and the observable $y$ are on the left, and the posterior densities for three values of the observable $y^* \in \{0,0.25,0.5\}$ and the monotone maps that pull back a standard normal density to each posterior are plotted on the right. \label{fig:example_cde}}
\end{figure}
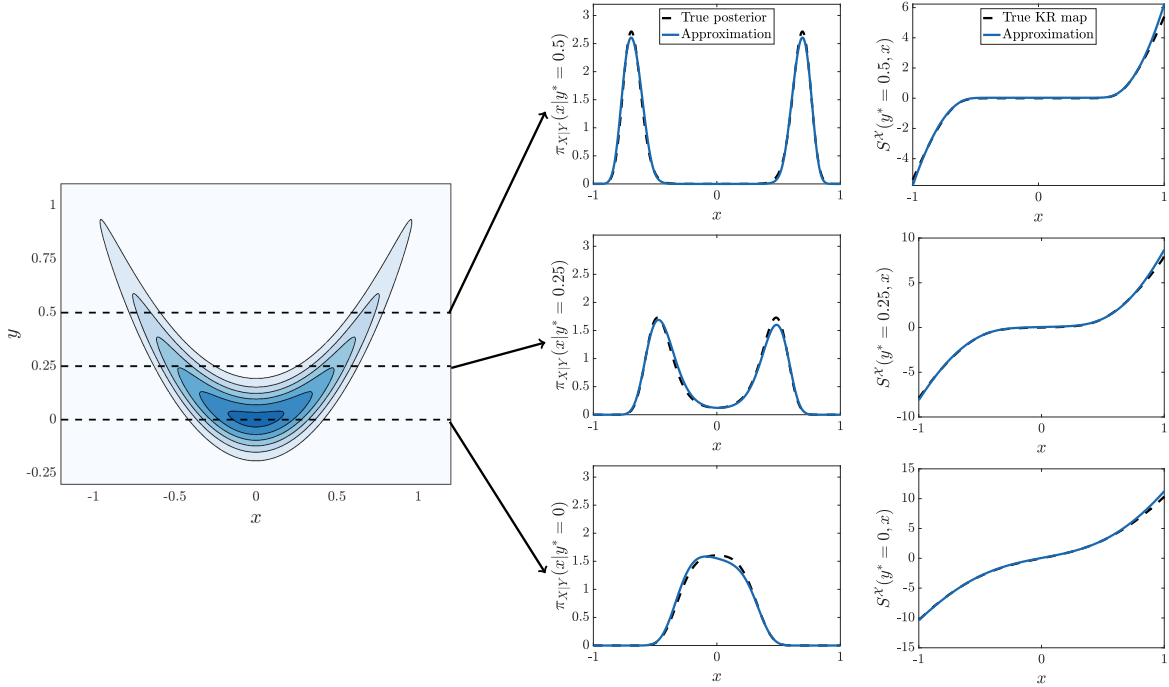

An example of the 
estimated conditional densities $\widehat{\pi}_{\bx|\bq}(\cdot|\q)$ extracted from the transport map $\widehat{S}^{\mathcal{X}}(\q,\cdot)$ is shown in Fig.~\ref{fig:example_cde}. We note that a single map $S^{\mathcal{X}}$ that is a function of the summary statistics $\q$ can characterize conditional densities with different properties across the data space, such as unimodal and bimodal behavior.

In addition to having a functional form for the conditional density $\widehat{\pi}_{\bx|\bq}(\cdot|\q)$, we can also use the estimated map to generate i.i.d.\thinspace samples $\x^i$ from the conditional distribution $\widehat{\nu}_{\bx|\bq}(\cdot|\q)$ by sampling $\rsample^i$ from the reference measure and solving for the parameter $\x^i$ that is mapped to $\rsample^i$ by the function $\widehat{S}^{\mathcal{X}}(\q,\cdot)$, i.e.,
\begin{equation}\label{eq:posterior_sampling}
\begin{rcases}
    \rsample^i\sim\nu_{\br_2}\\
    \widehat{S}^{\mathcal{X}}(\q,\x^i) = \rsample^i
    \end{rcases}
    \implies \x^i \sim \widehat{\nu}_{\bx|\bq}(\cdot|\q).
\end{equation}

To summarize the approach outlined in this section, we use a sample representation $(\x^i,\q^i)$ of the joint distribution $\nu_{\bx,\bq}$ to compute an approximate map $\widehat{S}^{\mathcal{X}}$ that couples a reference density to the conditional densities $\pi_{\bx| \bq}$; this map enables both evaluation of approximate posterior densities $\widehat{\pi}_{\bx|\q^*}$ for any and all summary statistics $\q^*$, via \eqref{eq:conditional_density_estimate}, and simulation of i.i.d.\thinspace samples from their respective distributions $\widehat{\nu}_{\bx|\bq}(\cdot|\q^*)$, via \eqref{eq:posterior_sampling}.
\section{Estimating expected information gain via triangular transport maps} \label{sec:eig} 

\begin{figure}[!ht]
    \centering
    \begin{subfigure}{0.49\textwidth}
        \includegraphics[width=\textwidth]{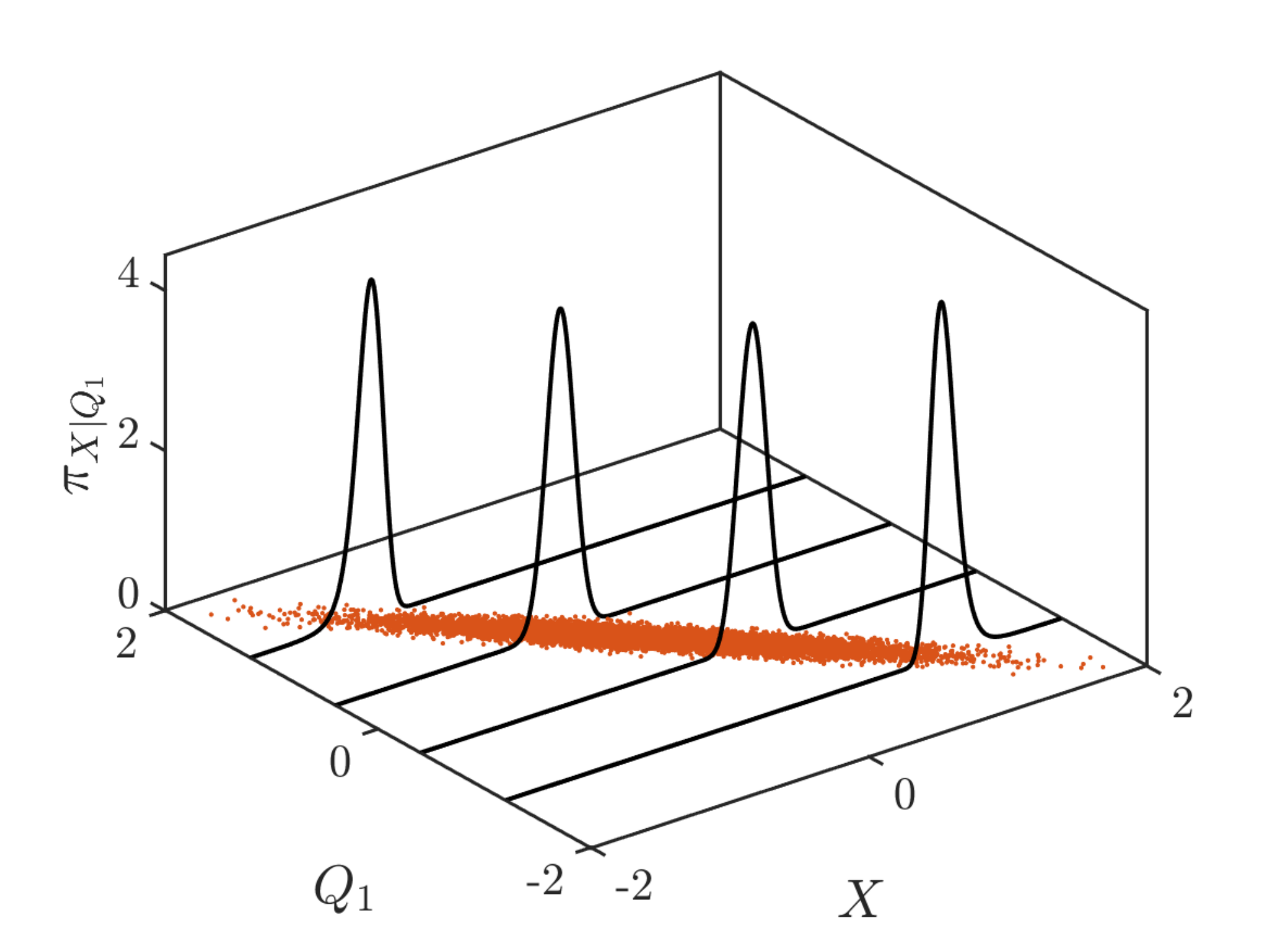}
    \end{subfigure}
    \begin{subfigure}{0.49\textwidth}
        \includegraphics[width=\textwidth]{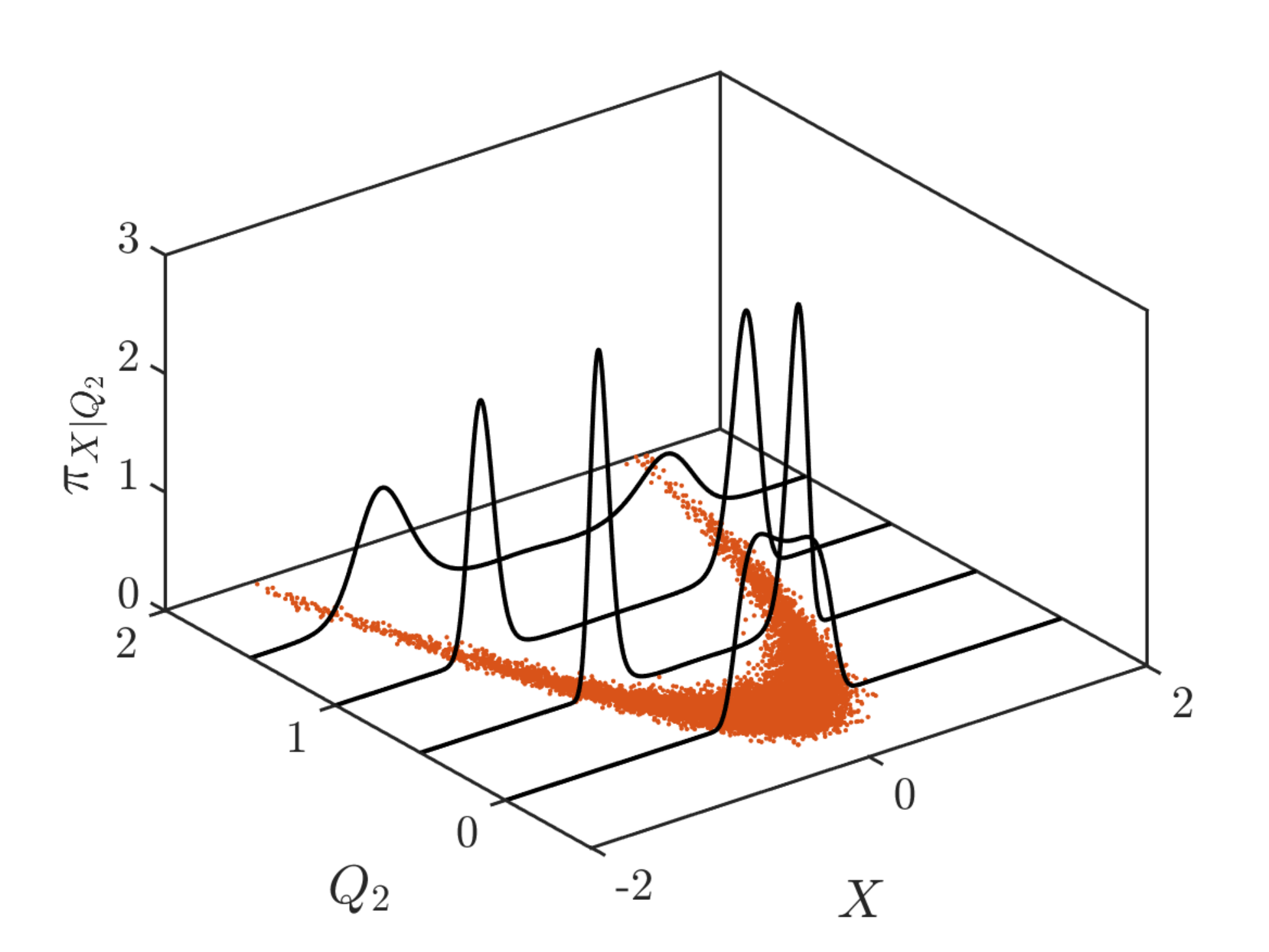}
    \end{subfigure}
    \caption{Posterior densities for the Gaussian $X$ parameter in Fig.~\ref{fig:example_cde} given two summary statistics $Q_1 = -X + N$ (\emph{left}) and $Q_2 = X^2 + N$ (\emph{right}) for $N \sim \mathcal{N}(0,0.01)$. The posteriors for $Q_1$ are concentrated as compared to the bi-modal posteriors for $Q_2$, indicating that $Q_1$ is more informative for predicting $X$. Integrating the KL divergence between the prior for $X$ and the posteriors over the different values of $Q_i$ results in EIGs of $\widehat{I}(X;Q_1) = 1.64 \pm 0.01$ and $\widehat{I}(X;Q_2) = 0.86 \pm 0.01$. This provides a quantitative approach to measure and rank the informativeness of different summary statistics.}
    \label{fig:example_EIG}
\end{figure}

A core advantage of estimating the conditional densities for \emph{all values} of $\q$ is that it enables the computation of expectations over the data random variables. One of these quantities is the EIG defined in Section~\ref{subsec:eig_intro}. An example of using EIG to compare the informativeness of two summary statistics for parameter inference is shown in Fig.~\ref{fig:example_EIG}. In Section~\ref{subsec:related_work_eig} we describe existing estimators for EIG and in Section~\ref{subsec:lfi_EIG} we present a lower bound for the EIG based on the transport maps computed in Section~\ref{sec:transportLFI}.

\subsection{Related work}\label{subsec:related_work_eig} 

Given conditional densities $\pi_{\bq|\bx}$ that are tractable to evaluate, several nested Monte Carlo (NMC) estimators have been proposed to estimate the EIG using samples from the joint distribution of parameters and observables~\cite{rainforth2018nesting, ryan2003estimating, Huan2013}. The EIG in~\eqref{eq:EIG_posterior} and \eqref{eq:EIG_likelihood} contains an expectation over $\nu_{\bx,\bq}$ and an unknown marginal density for the summary statistics that is typically unavailable in closed form. To estimate the EIG, NMC uses $N_1$ i.i.d\thinspace samples $\{(\x^i,\q^i)\}_{i=1}^{N_1} \sim \nu_{\bx,\bq}$ to estimate the outer expectation as 
$I(\bx;\bq) \approx \frac{1}{N_1} \sum_{i=1}^{N_1} \log \pi_{\bq|\bx}(\q^i|\x^i) - \log \pi_{\bq}(\q^i)$ and 
an independent set of $N_2$ i.i.d\thinspace samples $\{\x^j\}_{j=1}^{N_2} \sim \nu_{\bx}$ to approximate the integral for the marginal density of the summary statistics at $\q^i$ as $\pi_{\bq}(\q^i) \approx \frac{1}{N_2} \sum_{j=1}^{N_2} \pi_{\bq|\bx}(\q^i| \x^{j})$. Let us note that estimates for the marginal density are often numerically zero from arithmetic underflow when evaluating the likelihood function in high dimensions. 

To address this, Beck et.\thinspace al~\cite{Beck2018} proposed a Laplace approximation\textendash based importance sampling scheme to estimate the marginal likelihood. Although NMC is asymptotically unbiased for estimating EIG as $(N_1,N_2)\rightarrow \infty$, NMC is often considered too computationally costly. Even with an optimal allocation of samples to the inner and outer estimators, the mean squared error of a NMC estimator for EIG converges at a rate of $\mathcal O(N^{-2/3})$, where $N$ is the total number of samples. However, in order to achieve this rate, one needs to have $N_2 = c\sqrt{N_1}$, and finding what $c$ is involves some tuning~\cite{feng2019}.

To alleviate this cost, Wu et.\thinspace al~\cite{wu2020fast} considered EIG estimation based on Laplace approximations 
that exploit likelihood-informed subspaces and data dimension reduction. The authors numerically demonstrate that various biased estimators are often sufficient to perform BOED with unimodal posteriors that arise from affine or mildly nonlinear forward operators, unimodal priors, and additive Gaussian observation errors with small variance. The downside, however, is that accurate approximations or bounds on EIG are no longer guaranteed, and (gradient) information of the forward operator is required.

In a likelihood-free setting, estimating EIG requires us to characterize two densities, the prior and posterior, or the likelihood and marginal density of the data, which can be computationally expensive. One of the most well-known methods proposed in~\cite{Kraskov2003} is to estimate mutual information (MI) using k-nearest neighbor entropy estimators. Another method is to estimate MI using kernel density estimators (KDE)~\cite{Moon1995}. While the former requires the number of samples to scale exponentially with the MI, the latter KDE often performs poorly in high dimensions. To circumvent these challenges, Suzuki et.\thinspace al~\cite{Suzuki2008} proposed to approximate EIG by directly estimating a ratios of densities. Thomas et.\thinspace al~\cite{Thomas2020} proposed an approach using logistic regression to classify prior and posterior samples in order to estimate the ratio of the prior and posterior densities. Although this method allows us to estimate the density ratio directly, it requires solving an optimization problem for each observation. Alternatively, several approaches have been proposed in the machine learning literature for constructing variational bounds for MI as proposed in~\cite{Foster2019,poole2019}. In~\cite{Foster2019}, the approximate posterior is obtained by optimizing over a parameterized family of distributions, e.g., the mean-field approximation. The choice can introduce bias in the EIG estimation. 
\subsection{Likelihood-free estimators for expectation information gain}\label{subsec:lfi_EIG}

We now describe an approach to estimate the EIG that uses the variational approximation $\widehat{\pi}_{\bx|\bq}$ in~\eqref{eq:conditional_density_estimate} for the posterior density $\pi_{\bx|\bq}$. A simple calculation \cite[Appx. A]{Foster2019} shows that a lower bound for the EIG in~\eqref{eq:EIG_posterior} is:
\begin{align} \label{eq:EIG_lowerbound}
    I(\bx;\bq) \geq \widehat{I}(\bx;\bq) \coloneqq \mathbb{E}_{\nu_{\bx,\bq}}\left[\log \frac{\widehat{\pi}_{\bx|\bq}}{\pi_{\bx}}\right].
\end{align}
Moreover, the gap between the EIG and the lower bound is equal to the expectation of the KL divergence from the variational approximation to the true posterior, i.e., $\mathbb{E}_{\nu_{\bq}}[\mathcal{D}_{\textrm{KL}}(\pi_{\bx|\bq}(\cdot|\q)||\widehat{\pi}_{\bx|\bq}(\cdot|\q))]$. Thus, the lower bound for the EIG provided by the variational approximation approaches the true EIG as $\widehat{\pi}_{\bx|\bq}(\cdot | \q)$ better approximates the true posterior density.

Given $N$ i.i.d.\thinspace samples from the joint density of parameters and summary statistics, an unbiased estimate for the lower bound in~\eqref{eq:EIG_lowerbound} is 
\begin{equation} \label{eq:EIG_lower_bound_estimate}
    \mathbb{E}_{\nu_{\bx,\bq}}\left[\log \frac{\widehat{\pi}_{\bx|\bq}}{\pi_{\bx}}\right] \approx
    \frac{1}{N} \sum_{i=1}^{N} \log \widehat{\pi}_{\bx|\bq}(\x^i|\q^i) - \mathbb{E}_{\nu_{\bx}} [\log \pi_{\bx}],
\end{equation}
where $\{(\x^i,\q^i)\}_{i=1}^N \sim \nu_{\bx,\bq}$. The second term in~\eqref{eq:EIG_lower_bound_estimate} is the entropy of the prior density for the parameters. When the prior density is specified in a parametric family, this quantity is often available in closed form. As an example, for a standard Gaussian prior density $\pi_{\bx} = \mathcal{N}(0,\identity_{\dimx})$ where $\bx \in \mathbb{R}^{\dimx}$, the entropy 
is given by $\mathbb{E}_{\nu_{\bx}} [-\log \pi_{\bx}] = \frac{1}{2}\log(2\pi e\dimx)$.

We note that in the general likelihood\textendash free setting when the prior density $\pi_{\bx}$ is also analytically unavailable, we can also find a variational approximation $\widehat{\pi}_{\bx}$ for $\pi_{\bx}$ and define an estimator for the EIG as
\begin{equation} \label{eq:general_EIGestimator}
   I(\bx;\bq) \approx 
   \mathbb{E}_{\nu_{\bx,\bq}} \left[\log\frac{\widehat{\pi}_{\bx|\bq}}{\widehat{\pi}_{\bx}} \right] \approx
   \frac{1}{N} \sum_{i=1}^N \left[\log \widehat{\pi}_{\bx|\bq}(\x^i|\q^i) - \log \widehat{\pi}_{\bx}(\x^i) \right],
\end{equation}
where $\{(\x^i,\q^i)\}_{i=1}^N \sim \nu_{\bx,\bq}$. However, as shown in \cite{Foster2019}, the estimator in~\eqref{eq:general_EIGestimator} does not
yield a bound on EIG. This is problematic if the goal is to maximize the EIG with respect to different design variables in the context of optimal experimental design.  

An important property of MI, and hence EIG, is its invariance under marginal transformations of the random variables~\cite{cover1999elements}. If we have two bijective functions $f$ and $g$, then $I(\bx;\bq) = I(f(\bx);g(\bq))$. The property allows us to compute the EIG by finding variational approximations on transformed random variables so that the resulting EIG estimator has better statistical properties. In our numerical experiments, we transform the model parameters $\bx$ defined on a bounded subset of $\mathbb{R}^3_+$ to the unbounded domain $\mathbb{R}^3$ using the map described in Section~\ref{subsec:prior_transform}. The transformed random variables have a standard Gaussian prior distribution allowing us to use the EIG estimator in~\eqref{eq:EIG_lower_bound_estimate}. Moreover, the distributions for the transformed parameters satisfy the absolute continuity assumptions required by our approach to compute the transport maps in Section~\ref{sec:transportLFI}. A similar transformation can be defined for the summary statistics. For instance, the function $g(\bq) = \log(\bq)$ transforms variables fully supported on the positive numbers $\mathbb{R}_{+}$ to be fully supported on $\mathbb{R}$. 
\section{Case study: Model definitions} 
\label{sec:model_definitions}
In the next two sections, we perform Bayesian calibration of the OK model for the unguided self-assembly of Di-BCP thin films with top-down microscopy image data. In this section, we focus on defining the data model and the prior distribution for the model parameters used by the LFI and EIG computation methodologies described above. In the next section, we present numerical results of model calibration and EIG computation.

We first introduce the OK model for unguided Di-BCP thin film self-assembly and the model parameters in Section \ref{subsec:ok_model}. The associated forward operator and auxiliary variable are defined in Section~\ref{subsec:ok_forward_operator}. We then define a model for top-down microscopy characterization that contains restriction operators that mimic microscope magnification, a blurring convolution, and additive sensor noise in Section~\ref{subsec:microscopy_model}. Several low-dimensional summary statistics that quantitatively describe relevant features in the image data are proposed in Section~\ref{subsec:summary_statistics}, which are categorized into energy-based statistics and Fourier-based statistics. Lastly, we construct a model-specific prior for the parameters based on assumptions of experimental conditions in Section~\ref{subsec:prior_transform}.

\subsection{The Ohta--Kawasaki model}\label{subsec:ok_model}

Consider a Di-BCP melt, composed of two types of monomers that are labeled A and B respectively, in a physical domain $\mathcal{D}\subset\R^3$. Let $u_A$ and $u_B\colon\mathcal{D}\to[0,1]$, denote the normalized segment densities for the two monomers. These segment densities satisfy the incompressibility constraint $u_A+u_B = 1$. Then, our state variable $u\coloneqq u_A-u_B\in[-1,1]$, known as the order parameter, represents the difference between the normalized local fractions of the two phases. Under the thin film approximation, we may only consider the variation of the order parameter parallel to the substrate; thus $\mathcal{D}\subset \mathbb{R}^2$ is assumed in our case study\footnotemark.
\footnotetext{If one is interested in model calibration with respect to the effects of film thickness and/or polymer-substrate interaction, then the combination of X-ray scattering, which has the ability to probe the internal structure of BCP melts \cite{Sunday2013, Sunday2014}, and the OK model in the full 3D domain or $x_1$-$x_3$ cross-section domain should be considered. We refer to works discussed in Section \ref{subsec:related_work} for setups of such a calibration task.}

The free energy of a Di-BCP melt, $\mathcal{E}^{\textrm{OK}}$, given by the OK model \cite{Ohta1986, Uneyama2005} is assumed to be the sum of three contributing energies: the \emph{double well energy} $\mathcal{D}^{\textrm{OK}}$, the \emph{interfacial energy} $\mathcal{I}^{\textrm{OK}}$, and the \emph{non-local energy} $\mathcal{N}^{\textrm{OK}}$, scaled by the model parameters $(\kappa,\epsilon,\sigma)\in\mathbb{R}_{+}^3$,
respectively. That is,
\begin{align}\label{eq:ok_energy_strong}
\mathcal{E}^{\textrm{OK}}(u; \kappa, m, \epsilon, \sigma) \coloneqq \kappa \mathcal{D}^{\textrm{OK}}(u) + \epsilon^2\mathcal{I}^{\textrm{OK}}(u) + \sigma \mathcal{N}^{\textrm{OK}}(u; m).
\end{align}
The double well energy $\mathcal{D}^{\textrm{OK}}$ quantifies the free energy associated with the mixing of monomers and favors pure monomer phases represented by values $\pm 1$. The interfacial energy $\mathcal{I}^{\textrm{OK}}$ is a $H^1(\mathcal{D})$ seminorm that characterizes the short-range interaction between monomers and controls the interfacial length between different monomer phases. Lastly, the nonlocal energy $\mathcal{N}^{\textrm{OK}}$ characterizes the long-range interaction between monomers and controls the expansion of phase boundaries. The competition between the three energies during free-energy minimization forms a length scale selection mechanism that leads to various structures of equilibrium Di-BCP melts depicted by the experimental phase diagram in Fig.~\ref{fig:experimental_phase_diagram} (\emph{left}). It has been shown numerically that most periodic structures observed in experiments are minimizers of the OK free energy ~\cite{Choksi2009, Choksi2011}. 

\begin{figure}[H]
    \centering
    \begin{minipage}{.5\textwidth}
    \caption{The quartic double well potential, which has $-1$ and $1$ as its minima. The dashed lines delineate physical boundaries. Since $-1$ and $1$ refer to normalized pure densities of $u_{A}$ and $u_{B}$, respectively, when the order parameter is outside of $(-1, 1)$, the model does not represent a physically meaningful quantity. In other words, this double well potential $\mathcal{W}$ is \emph{non-degenerate}. A \emph{degenerate}, or \emph{obstacle}, double well potential, such as described in~\cite{Burkovska2021}, produces an infinite wall at these boundaries. \label{fig:double_well}}
  \end{minipage} \quad
  \begin{minipage}{.45\textwidth}
    \includegraphics[width=\textwidth]{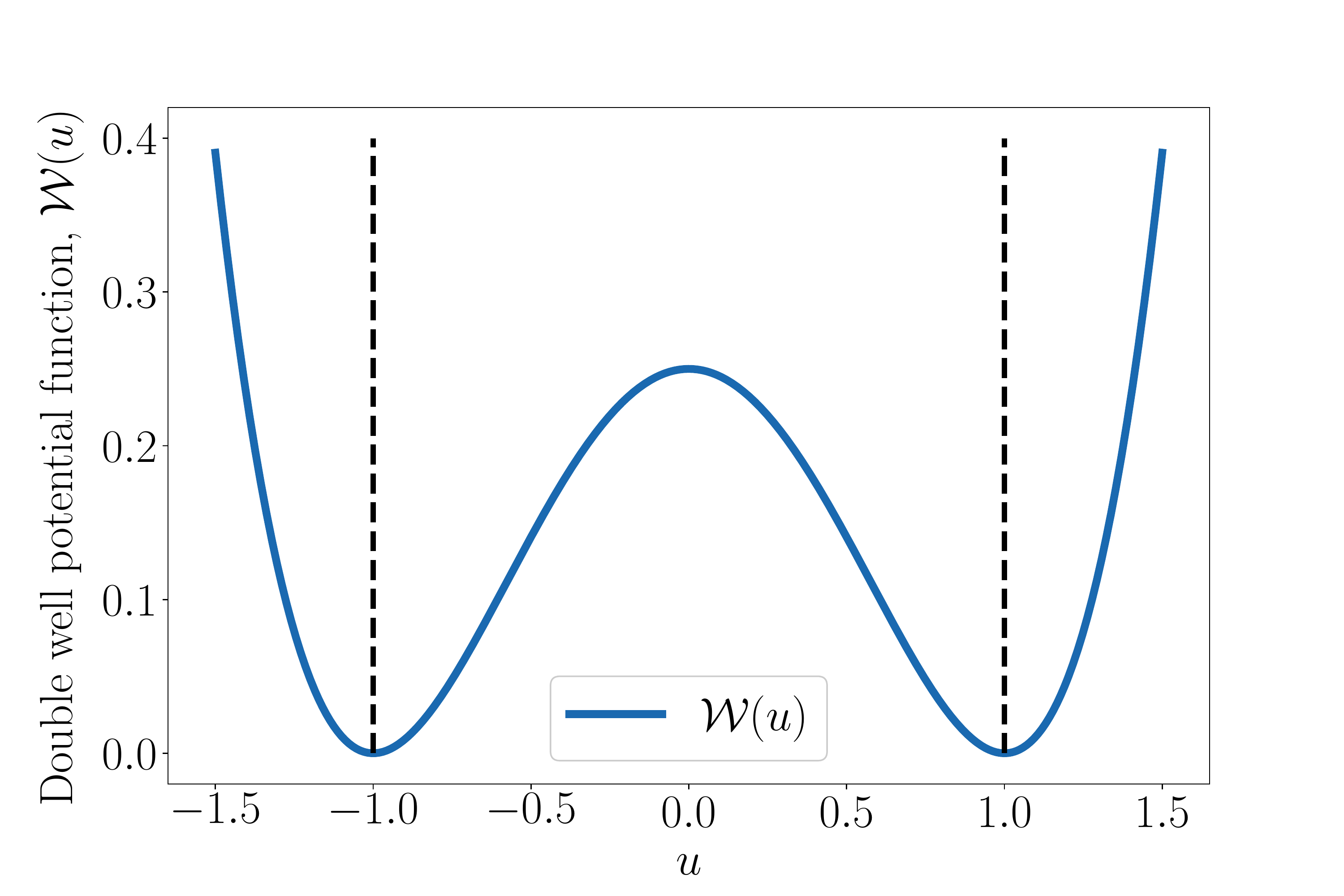}
  \end{minipage}
 \end{figure}

The form of these contributing energies is given by:
\begin{equation}
\begin{gathered}
    \mathcal{D}^{\textrm{OK}}(u)= \int_\mathcal{D} \mathcal{W}(u(\s))\textrm{d}\boldsymbol{s},\quad \mathcal{I}^{\textrm{OK}}(u)=\frac{1}{2}\int_\mathcal{D} |\nabla u(\s)|^2\textrm{d}\boldsymbol{s},\\
    \mathcal{N}^{\textrm{OK}}(u; m)= \frac{1}{2}\int_\mathcal{D}(u(\s)-m)(-\triangle_N)^{-1}(u(\s)-m)\textrm{d}\boldsymbol{s}\;,
\end{gathered}
\end{equation}
where $m\in[-1,1]$ is the spatial mean $\mathcal{M}$ of the order parameter, i.e., \begin{equation}\label{eq:spatial_average_constraint}
m = \mathcal{M}(u) = |\mathcal{D}|^{-1}\int_\mathcal{D} u(\s)\textrm{d}\boldsymbol{s},
\end{equation}
and $\mathcal{W}(u) =(1-u^2)^2/4$ is the double well potential that is plotted in Fig.~\ref{fig:double_well}.
The action of the inverse Laplacian is defined by the solution of a Poisson equation with homogeneous Neumann boundary conditions\footnote{The periodic boundary condition is often used for square or cubic domains, mainly for computational convenience and the minimization of edge effects~\cite[p. 96]{Fredrickson2006}. The Neumann boundary additionally models the effect of a bounded (confined) domain by assuming that polymer chains reflect at the boundary~\cite{Hsu1985}.}:
\begin{equation*}\label{eq:lap_inv}
    w = (-\triangle_N)^{-1}(u-m)\iff 
    \begin{cases}
        -\triangle w(\s) = u(\s)-m &\text{in }\mathcal{D}\;;\\
        \nabla w(\s) \cdot \boldsymbol{n} = 0 &\text{on }\partial\mathcal{D}\;,
    \end{cases}
\end{equation*}
where $\boldsymbol{n}$ is the unit vector normal to $\partial\mathcal{D}$.\\

The parameters in the OK model are connected to the commonly used material parameters of diblock copolymers. In particular, an appeal to the higher-fidelity SCFT model in \cite{Choksi2003} reveals the relationships:
\begin{equation} \label{eq:ok_parameter_maps}
    \kappa = 1\;,\quad \epsilon^2 = \frac{l^2}{3f(1-f)\chi}\;,\quad\sigma = \frac{36}{f^2(1-f)^2l^2\chi N^2}\;, \quad m = 2f-1,
\end{equation}
where $N$ is the degree of polymerization of the polymer, $\chi$ is the Flory-Huggins parameter describing the strength of repulsion between the two monomers, $l$ is the statistical segment length of the polymer\footnotemark, and $f\in(0,1)$ represents the volume fraction of monomer $A$. In this study, we take the common assumption of $\kappa = 1$ and calibrate the collection of model parameters $\x = (m,\epsilon,\sigma)$. We refer to the three parameters as the \emph{spatial average parameter}, the \emph{interfacial parameter}, and the \emph{nonlocal parameter}, respectively.
\footnotetext{The statistical segment length is assumed to be the same for the two types of monomers in a Di-BCP, thus we refer to the OK model as a model of conformationally symmetric Di-BCP melts.}

\subsection{The forward operator of the Ohta--Kawasaki model}\label{subsec:ok_forward_operator}
The forward operator $\mathcal{F}$ is defined as the map from parameters $\x$ and a random initial state $z$ to a local minimizer of the OK energy, $\mathcal{E}^{\textrm{OK}}_{\x}(u) \coloneqq \mathcal{E}^{\textrm{OK}}(u; 1, m, \epsilon,\sigma)$, for the given parameters. That is,
\begin{equation}\label{eq:ok_energy_solutions}
u = \mathcal{F}(\x,z) \in \argmin_{u\in\statespace} \mathcal{E}^{\textrm{OK}}_{\x}(u)
\end{equation}
where the state space is defined as
\begin{equation}\label{eq:state_space}
    V^u\coloneqq\left\{v\in H^1(\mathcal{D}):\int_\mathcal{D} \left(v(\boldsymbol{s})-m\right)\textrm{d}\boldsymbol{s} = 0\right\}.
\end{equation}
As a result of the non-convexity of the double well potential $\mathcal{W}$, the OK energy is non-convex and does not have a unique minimizer. Thus, the local minimizers in~\eqref{eq:ok_energy_solutions} consist of an ensemble of numerous achievable equilibrium patterns for Di-BCP thin film self-assembly. We note that the states  satisfying the first order condition for the minimization problem in~\eqref{eq:ok_energy_solutions} can be found by solving the following mixed problem \cite{Cao2022}.
\begin{align}\label{eq:mixed_first_oder_condition}
    \text{Find } (u,\mu) \in \left( H^1(\mathcal{D})\right)^2 \text{ such that: } \begin{cases}
    (\nabla \mu, \nabla\tilde{u})_{L^2} + \left(\sigma (u-m),\tilde{u}\right)_{L^2} = 0, &\forall \tilde{u}\in H^1(\mathcal{D})\;;\\
    (\mu, \tilde{\mu})_{L^2}-\left(\mathcal{W}'(u), \tilde{\mu}\right)_{L^2} - (\epsilon^2\nabla u, \nabla\tilde{\mu})_{L^2} = 0, &\forall\tilde{\mu}\in H^1(\mathcal{D})\;.
    \end{cases}
\end{align}
where $(\cdot,\cdot)_{L^2}$ denotes $L^2(\mathcal{D})$ or $(L^2(\mathcal{D}))^2$ inner product.
    
The evaluation of the forward operator $\mathcal{F}(\x,\z)$ is accomplished by minimizing the energy $\mathcal{E}^{\textrm{OK}}_{\x}$ via a variational procedure, starting from an initial state $\z$. Conventionally, an $H^{-1}$ gradient flow approach is used, which involves mimicking a time-dependent process by solving the nonlocal Cahn--Hilliard equation \cite{Choksi2009,Choksi2011,Qin2013}, a fourth-order time-evolving partial differential equation (PDE), whose steady-state solutions satisfy \eqref{eq:mixed_first_oder_condition}. According to the dynamical DFT \cite{TeVrugt2020}, the gradient flow approach may faithfully represent the kinetic pathway of the self-assembly if an appropriate mobility operator \cite{Schmid2020} and/or a stochastic noise that satisfies the fluctuation\textendash dissipation theorem \cite{Fraaije1993} are/is incorporated. Instead, the OK model is more commonly treated as an equilibrium model, i.e., the thermal fluctuations and dynamics of Di-BCPs are not modeled.

Several authors have considered a diverse set of strategies to evaluate the forward operator via more efficient optimization schemes\footnotemark. Although there are few studies on the direct optimization of the OK energy, quasi-Newton methods, such as Anderson acceleration~\cite{Thompson2004} and semi-implicit Seidel schemes~\cite{Ceniceros2004}, are widely used in context of the SCFT model, which is also formulated as a free energy minimization problem. Similarly, a modified Newton method for the direct minimization of the OK energy was recently proposed \cite{Cao2022} and shown to be typically three orders of magnitude faster than the gradient flow approach. The latter Newton method is adopted as the optimization procedure in our numerical case study.
\footnotetext{We note that, the forward operator defined by these optimization schemes are often \emph{non-continuous}\textemdash the resulting solutions are not guaranteed to vary smoothly with changes in the initial state. This makes the forward operator strictly \textit{generative}.}

Given the presence of many local minimizers for OK energy, an approach is needed to model the random initial state. A commonly adopted initial state is the realization of a uniform white noise random field~\cite{Qin2013,Choksi2009,Choksi2011}. A more rigorous approach that uses transformed Gaussian random fields is proposed in \cite{Cao2022}. These random fields play the role of the~\emph{auxiliary variable} $\bz$ in the forward operator. When coupled with the optimization procedure, the random fields produce a distribution of equilibrium patterns. As described in Section~\ref{subsec:statistical_models}, a physically consistent model should satisfy the equality in distribution in~\eqref{eq:state_variable}, meaning that the probability associated with each model solution matches the probability of realizing the corresponding equilibrium pattern in an experiment. The validation of these model assumptions underlying the forward operator is an important topic that we leave for future work.

\begin{figure}
    \centering
    \includegraphics[height=0.18\textwidth]{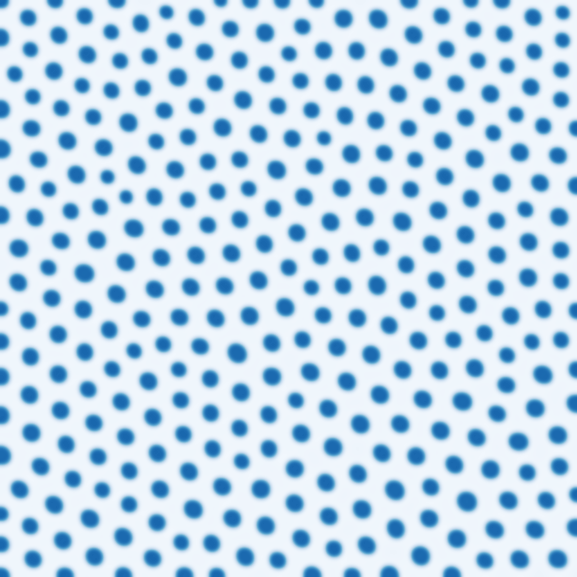}    \includegraphics[height=0.18\textwidth]{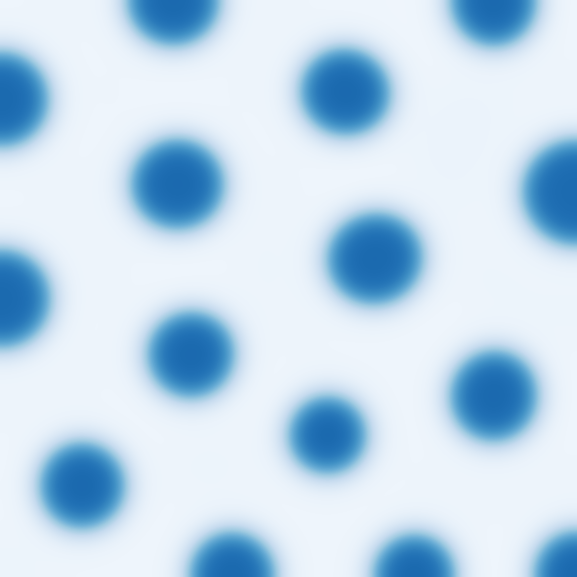}
\includegraphics[height=0.18\textwidth]{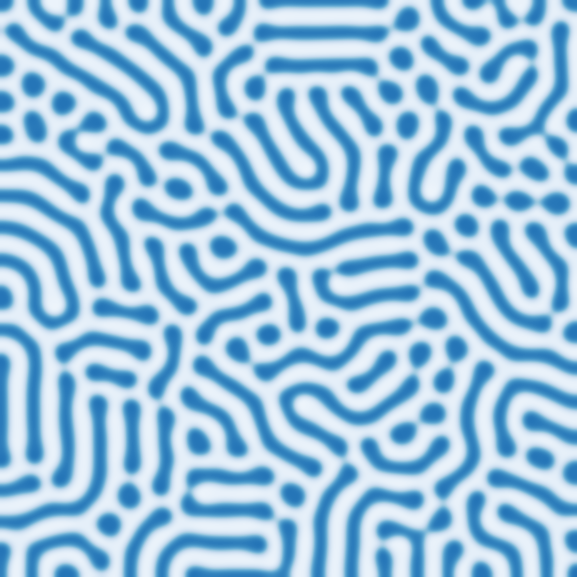}
    \includegraphics[height=0.18\textwidth]{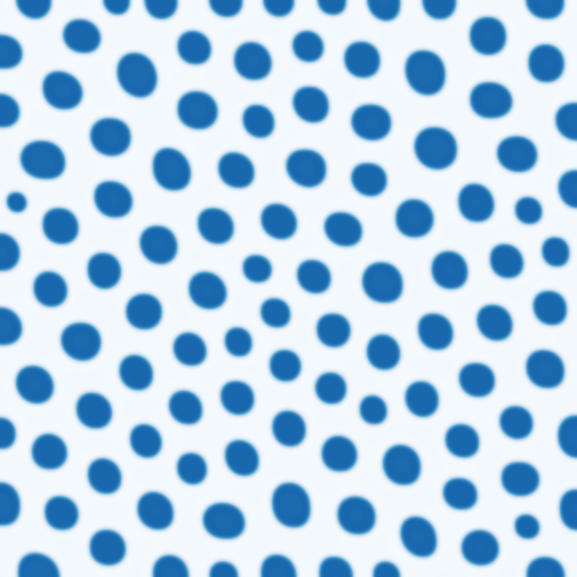}    
    \includegraphics[height=0.18\textwidth]{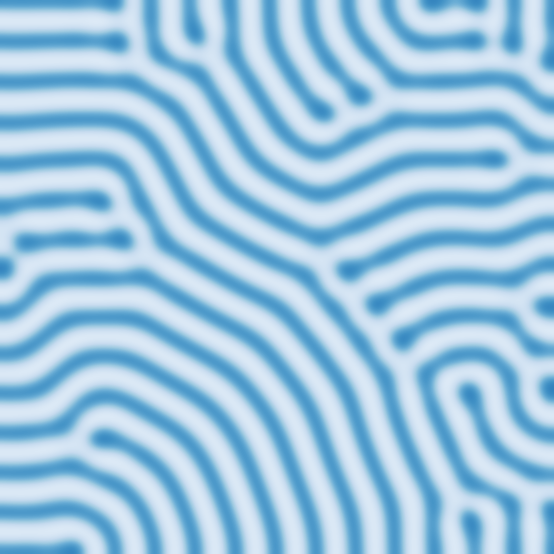}   
    \includegraphics[height=0.18\textwidth]{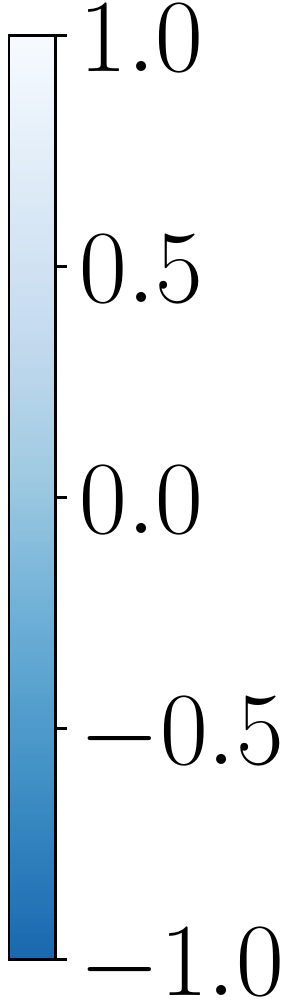}  
    \caption{Examples of Di-BCP pattern simulated via various choices of model parameters ($\m, \epsilon, \sigma)$ of the OK model. The model can characterize a range of morphological phases, interface thicknesses, and strengths of phase separation. \label{fig:Di-BCP melts}}
\end{figure}

We provide a description of our model for the auxiliary variable and finite element discretization of the state space in \ref{app:our_z_model} and \ref{app:function_space}, respectively. In conjunction with the variational minimization procedure for the OK model and its numerical implementation detailed in~\cite{Cao2022}, this completes the definition of the forward operator in our case study. See Fig.~\ref{fig:Di-BCP melts} for examples of Di-BCP patterns generated by the forward operator. 

\subsection{A model of top-down microscopy characterization }\label{subsec:microscopy_model}

In this subsection, we define the state-to-observable map $\mathcal{J}$ for top-down microscopy characterization of Di-BCP thin films. We present the restriction operator $\mathcal{R}$, the imaging operator $\mathcal{I}$, and the statistical sensor operator $\mathcal{S}$ used in our case study. Recall that $\mathcal{J} = \mathcal{I} \circ \mathcal{R}$, as defined in Section~\ref{subsec:characterization_model}.

\begin{figure}[h]
    \centering
    \begin{tabular}{|c|c|c|c|}\hline
     \textbf{reference field of view}  & $70\%$ & $40\%$ & $10\%$\\\hline\hline
     \includegraphics[width = 0.205\linewidth]{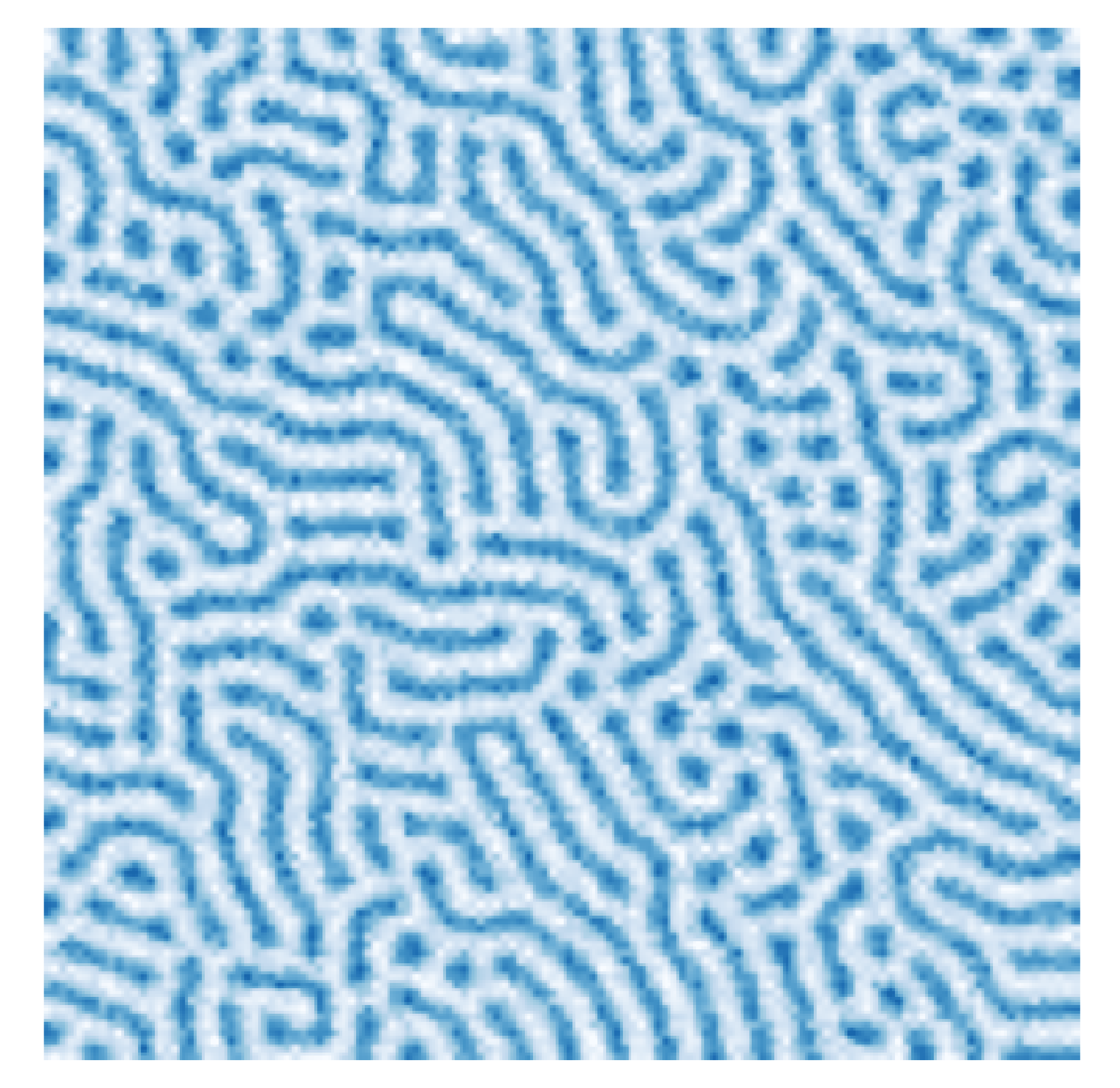} & \includegraphics[width = 0.2\linewidth]{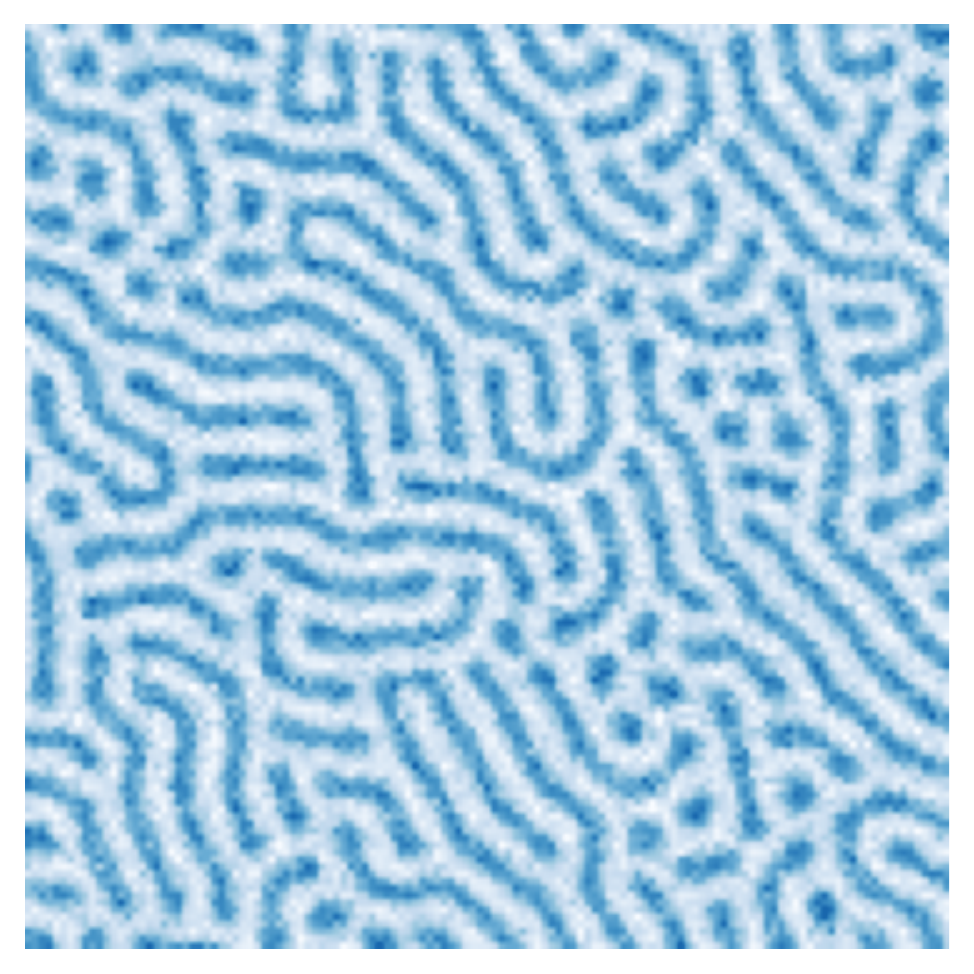} &
     \includegraphics[width = 0.2\linewidth]{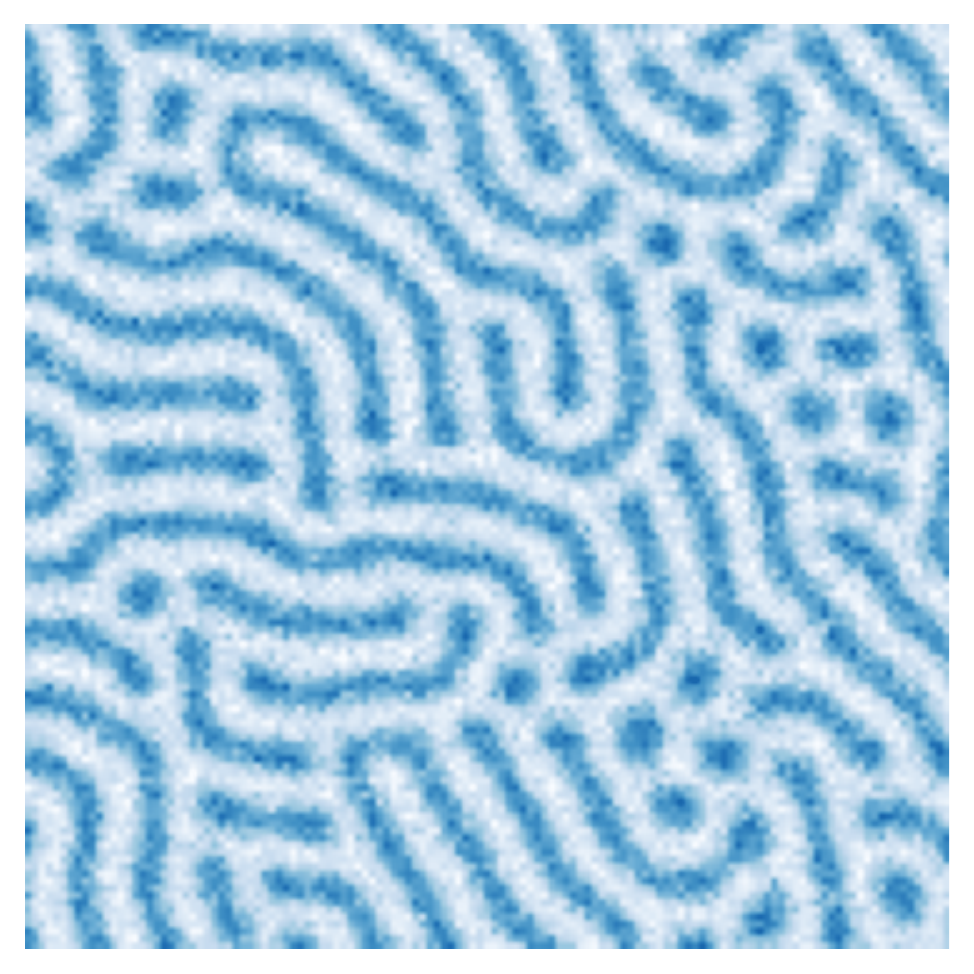} &
     \includegraphics[width = 0.2\linewidth]{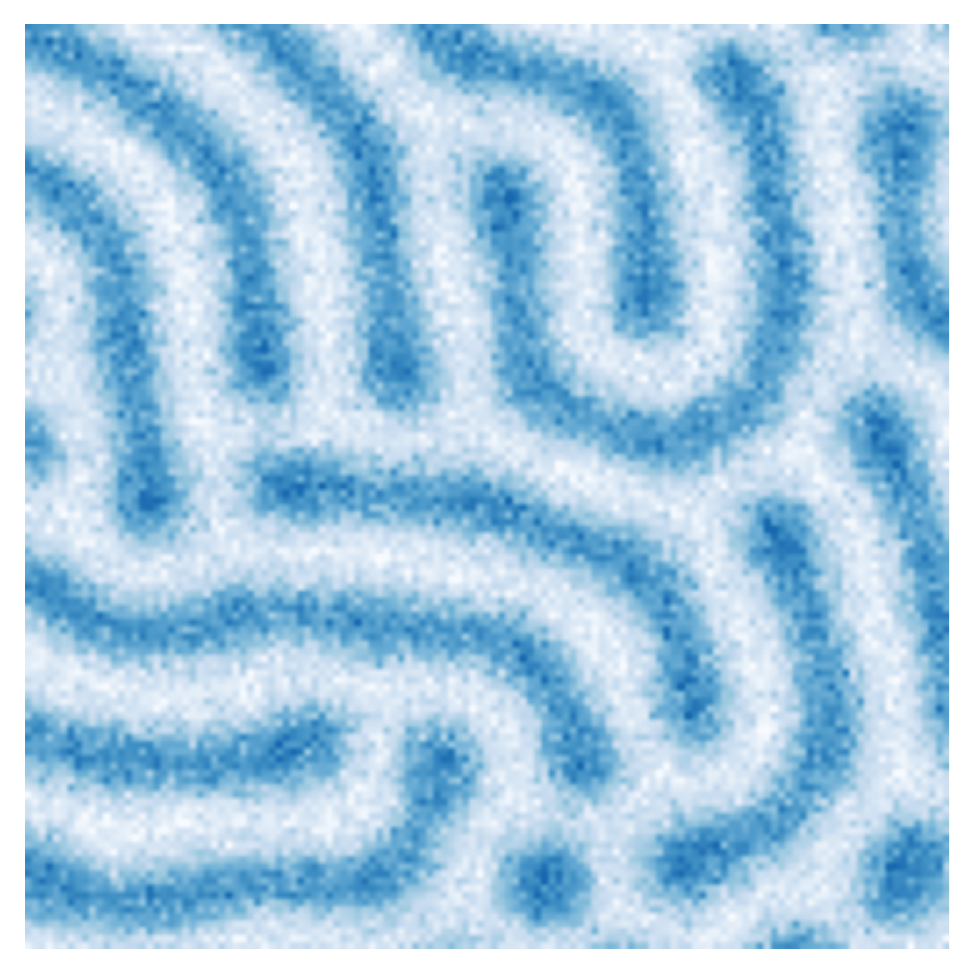}\\\hline
    \end{tabular}
    \caption{A set of synthetic top-down microscope images of a Di-BCP thin film at different levels of magnification, specified by the restriction operator $\mathcal{R}_j$ to the centered subdomain $R_j(\mathcal{D})$ with $j\%$, $j = 100, 70, 40, 10$ (from \emph{left} to \emph{right}), of the area of the reference field of view $\mathcal{D}$.}
    \label{fig:restriction_operator}
\end{figure}

We assume a set of experimental scenarios, in which microscope images are taken in fields of view of various sizes, mimicking the effect of varying levels of magnification. In terms of the characterization model, we first define a reference field of view $\mathcal{D} = [0,L]^2$, where $L$ denotes its side length, corresponds to a relatively large region of the specimen top surface. Then, we define the restrictions to its centered subdomains $R_j(\mathcal{D})$, where  $j = 2, 4,\dots,10, 20,\dots 90$ denotes the percentage of the area of $\mathcal{D}$ retained by the restriction. The restriction operator is thus defined by $\mathcal{R}_j:H^1(\mathcal{D})\to H^1(R_j(\mathcal{D}))$. The effect of these restriction operators, i.e., varying levels of magnification, on model calibration will be quantitatively studied in the next section. The numerical implementation of the restriction operators is discussed \ref{app:function_space} and a numerical example of the synthetic data generated by these restriction operators is provided in Fig.~\ref{fig:restriction_operator}.

For the microscopy imaging operator $\mathcal{I}$, we only consider the blurring effect, which is often modeled as a \textit{point-spread function} (PSF) in the context of electron or optical microscopy. The PSF is a linear integral operator which degrades the fine-scale features in the observed specimen to resemble lens-aberrations. Numerous PSFs for microscopy imaging have been considered in the literature~\cite{Zhang2007, Roels2016, Zotta2018}, such as Airy disks, Debye diffraction integral, and Gaussian convolution integral. For our numerical case study, we employ the convolution with a Gaussian kernel\footnotemark as a representative prototype:
\footnotetext{We apply zero-padding to extend $u$ beyond the boundary of the reference field of view $\partial\mathcal{D}$.}
\begin{equation} \mathcal{I}(\state) = \int_{\mathcal{R}(\mathcal{D})} \frac{1}{\sqrt{2}\sigma_{\textrm{blur}}}\exp{\bigg(-\frac{1}{2\sigma_{\textrm{blur}}^2}\|\s_1 -  \s_2\|^2\bigg)} \state(\s_1)\textrm{d}\s_1,
\end{equation}
where $\mathcal{I}\colon H^1(\mathcal{R}(\mathcal{D}))\to H^1(\mathcal{R}(\mathcal{D}))$ and $\sigma_{\textrm{blur}}$ is a blurring strength parameter.

Since the sensor noise in microscope images are generally `fine-grained', we assume there is i.i.d.\ additive noise for each pixel. Other authors have considered Poisson, Gaussian or mixed Poisson--Gaussian noise random variable in a combination of additive and multiplicative form~\cite{Meiniel2018, Roels2014, Roels2016}; we employ additive Gaussian noise as a prototype for simplicity, although more complex and accurate models could have been employed within the introduced inference framework without additional difficulty. 

In our study, we assume a $256 \times 256$ grid array of sensors. The sensor signal is mathematically defined as local averages in grid cells,  denoted as $\mathcal{D}_{jk}$, centered at the sensor positions:
\begin{equation}\label{eq:sensor_discretization}
    \left(\boldsymbol{\mathcal{D}}(o)\right)_{jk} = \left|\mathcal{D}_{jk}\right|
    ^{-1}\int_{\mathcal{D}_{jk}} o(\boldsymbol{s})\:\textrm{d}\boldsymbol{s},\quad j,k \in (1,2,\ldots,256).
\end{equation}
We thus arrive at the following characterization model:
\begin{equation}
\mathcal{S}(\mathcal{J}(\state),\noise) = \boldsymbol{\mathcal{D}}(\mathcal{J}(\state)) + \noise, \quad \noise \overset{\textrm{i.i.d.}}{\sim} \mathcal{N}_{jk}(0, \sigma^2_{\noise}),
\end{equation}
where $\mathcal{J} = \mathcal{I}\circ \mathcal{R}$ as defined above.

In the next section, we examine, jointly and separately, the effect of various levels of blurring, $\sigma_{\text{blur}}= (\text{None}, 5L\times 10^{-3}, L\times 10^{-2})$, and/or the effect of various sensor noise variances $\sigma^2_{\noise}$, based on a signal-to-noise ($\textrm{SNR}$) ratio of $(\infty, 5.0, 3.0, 2.5)$, on the informativeness of the data. $\textrm{SNR}$ is related to the sensor noise variance $\sigma^2_{\noise}$ by, naively, $\sigma_{\noise}^{-1} = \textrm{SNR}$.
\subsection{Summary statistics for top-down microscope images of diblock copolymer thin films}\label{subsec:features}

In this section, several summary statistics for top-down microscopy characterization images of Di-BCP thin films, i.e. $\bq = \mathcal{Q}(\by)\in\R^q$, are proposed. First, in Section~\ref{subsubsec:energy_features} we introduce summary statistics defined by the composition of energy functionals and an $H^1$-interpolation operator. Then, in Section~\ref{subsubsec:fourier_features}, we present summary statistics based on the magnitude spectrum of these images obtained by discrete Fourier transform. For both cases, we discuss connections between these summary statistics and features in the image data, and provide reasoning for why we expect these summary statistics to be informative of the model parameters. 

\subsubsection{Energy-based summary statistics} \label{subsubsec:energy_features}
We introduce five summary statistics, $Q_i = \mathcal{Q}_i(\by)$, $i = 0, 1,\dots, 4$, where $\mathcal{Q}_i$ is defined by the composition of an energy functional and an $H^1$ interpolation operator. The $H^1$ interpolation operator, denoted as $\Pi$, interpolate the image $\y\in\R^y$ into the function space $H^1(\mathcal{D}_u)$ using a linear finite element basis, where $\mathcal{D}_u$ is a unit square domain\footnotemark. The details of the interpolation operator for our numerical case study are given in \ref{app:function_space}. The energy functionals act on the $H^1$-interpolated image data. Their names, notations, and mathematical forms are listed in Tab.~\ref{tab:energy-based_statistics}. 

\footnotetext{We note that scalings of the summary statistics do not change the EIG between $\bx$ and $Q_i$, since EIG is invariant under invertible transformations of $\by$. The scalings can be helpful to adjust the magnitude of summary statistics to the same order.}

\begin{table}
    \centering
    \begin{tabular}{|l|l|l|}\hline
    \textbf{Notation}&\textbf{Name} & \textbf{Functional} \\\hline\hline     \textbf{$Q_0, \widehat{M}$}& Empirical spatial average & $ \displaystyle \mathcal{M}(\widehat{u}) = \int_{\mathcal{D}_u}\widehat{u}\;\textrm{d}\boldsymbol{s}$ \\\hline
    \textbf{$Q_1$}& Double well & $\displaystyle\mathcal{D}^{\textrm{OK}}(\widehat{u}) = \int_{\mathcal{D}_u} (1-\widehat{u}^2)^2/4\;\textrm{d}\boldsymbol{s}$ \\\hline
    \textbf{$Q_2$}& Interfacial & $\displaystyle\mathcal{I}^{\textrm{OK}}(\widehat{u}) = \int_{\mathcal{D}_u} |\nabla \widehat{u}|^2/2\;\textrm{d}\boldsymbol{s}$ \\\hline
    \textbf{$Q_3$}& Nonlocal & $\displaystyle\mathcal{N}^{\textrm{OK}}(\widehat{u}) = \int_{\mathcal{D}_u} (\widehat{u}-\mathcal{M}(\widehat{u}))(-\triangle_N)^{-1}(\widehat{u}-\mathcal{M}(\widehat{u}))/2\;\textrm{d}\boldsymbol{s}$ \\\hline
    \textbf{$Q_4$}&Total variation & $\displaystyle\mathcal{TV}(\widehat{u}) = 2(1-|\mathcal{M}(\widehat{u})|)^{-1}\int_{\mathcal{D}_u} (\nabla \widehat{u}\cdot \nabla \widehat{u})^{1/2}\;\textrm{d}\boldsymbol{s}$ \\\hline
    \end{tabular}
    \caption{The notations, names, and corresponding energy functionals of the proposed energy-based summary statistics.}
    \label{tab:energy-based_statistics}
\end{table}

Three summary statistics, $Q_1,Q_2,Q_3$, are inspired by the OK free energy. They are defined using the functionals in the OK energy, i.e.,  $\mathcal{D}^{\textrm{OK}}$, $\mathcal{I}^{\textrm{OK}}$, and $\mathcal{N}^{\textrm{OK}}$, evaluated at the interpolated image data random variable $\Pi(\by)$. The motivation behind choosing these summary statistics are our perceptions that (i) the magnitude of the three energies are sensitive to the model parameters and relatively insensitive to the auxiliary variable, and (ii) they characterize informative geometric features of Di-BCP patterns, such the total area of phase interface, degree of phase separation, etc. Since the microscopy characterization visually preserves the geometry of phase interfaces, we suspect that the negative impact of sensor blurring and noise on the informativeness of these summary statistics for inferring the model parameters is relatively minor. Therefore, we hypothesize that the parameters are well-informed by these energy-based summary statistics.

We also define two additional energy-based summary statistics, given by the empirical spatial average, $Q_0 = \mathcal{Q}_0(\by) \coloneqq (\mathcal{M}\circ\Pi)(\by)$, and a total-variation-like functional, $Q_4=\mathcal{Q}_4(\by) \coloneqq(\mathcal{TV}\circ\Pi)(\by)$. We expect the empirical spatial average to be informative of the material parameter $f$ that represents the segment number ratio of monomer A. For the OK model, $f$ is directly related to the model parameter $m$ through \eqref{eq:ok_parameter_maps}. The total-variation functional $\mathcal{TV}$ formally captures the average ratio between the interfacial length and the inclusion domain size, such as the radius of spots for spot patterns, of Di-BCP patterns. It is also relatively insensitive to the sensor noise with edge preserving property, which makes it popular in image denoising problems~\cite{Rudin1992}. 

\begin{figure}[h]
    \centering
    \footnotesize
    \begin{tabular}{|c|c|c|c|c|}\hline
        \textbf{Synthetic image $\y^*$} & $\mathcal{Q}_1(\y^*)$ & $\mathcal{Q}_2(\y^*)$ & $\mathcal{Q}_3(\y^*)$ \textbf{(log scale)} & $\mathcal{Q}_4(\y^*)$\\\hline\hline
        \includegraphics[width = 0.13\linewidth]{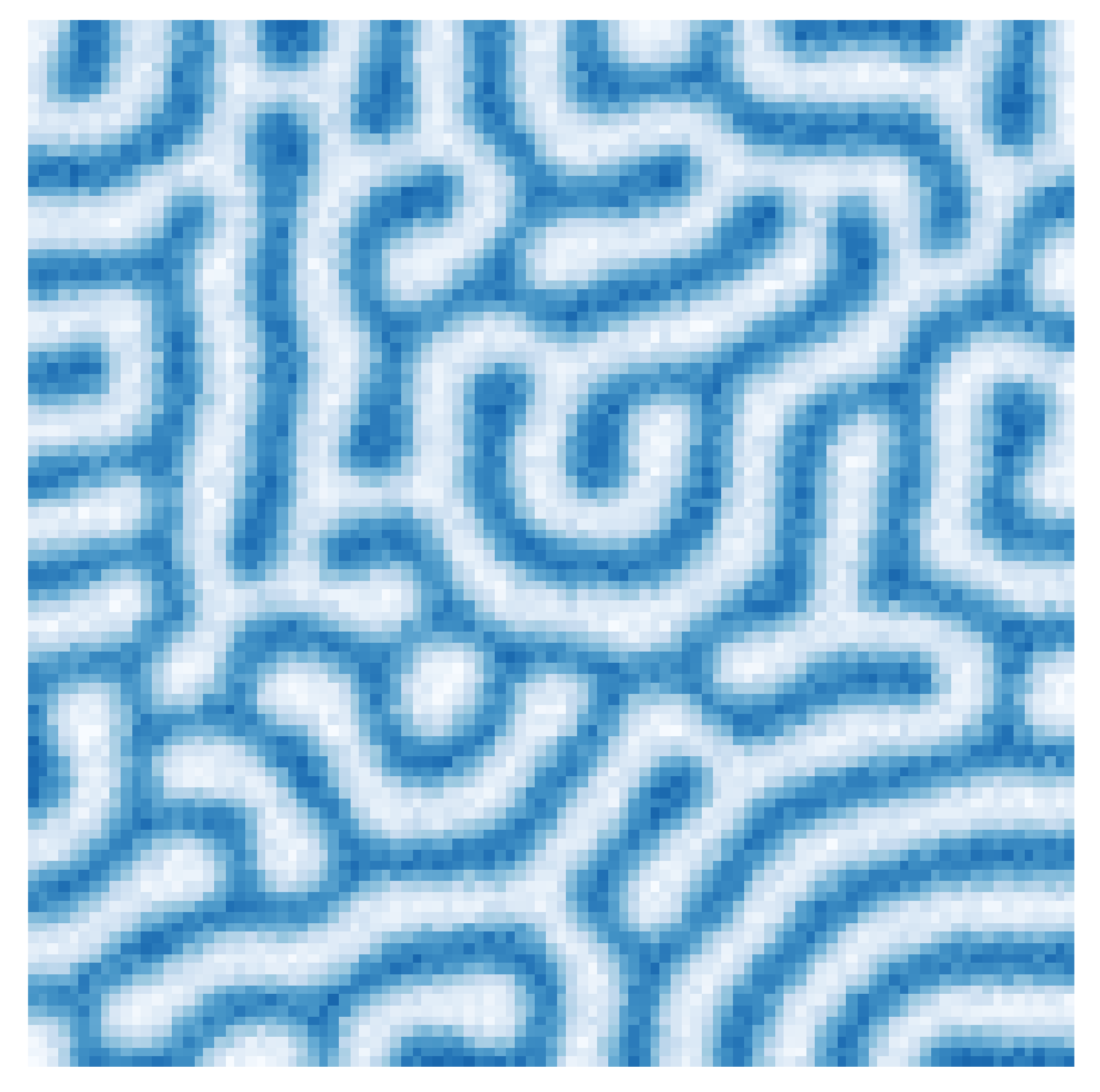}& \includegraphics[width = 0.17\linewidth]{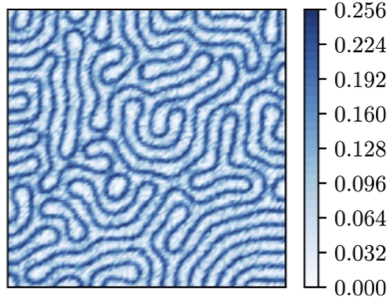} & \includegraphics[width = 0.18\linewidth]{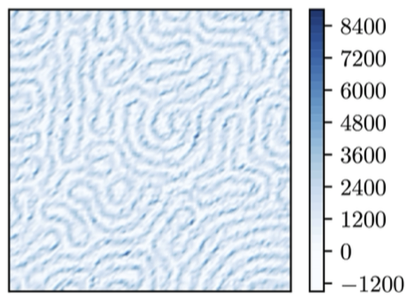} & \includegraphics[width = 0.175\linewidth]{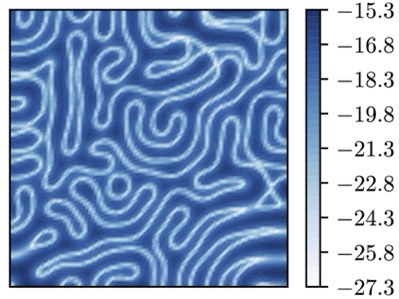}  & \includegraphics[width = 0.17\linewidth]{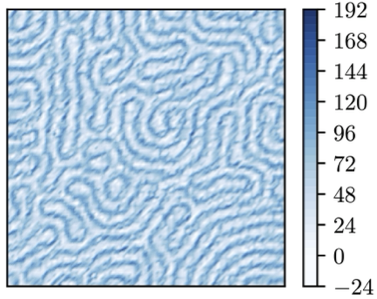}
        \\\hline
        \includegraphics[width = 0.13\linewidth]{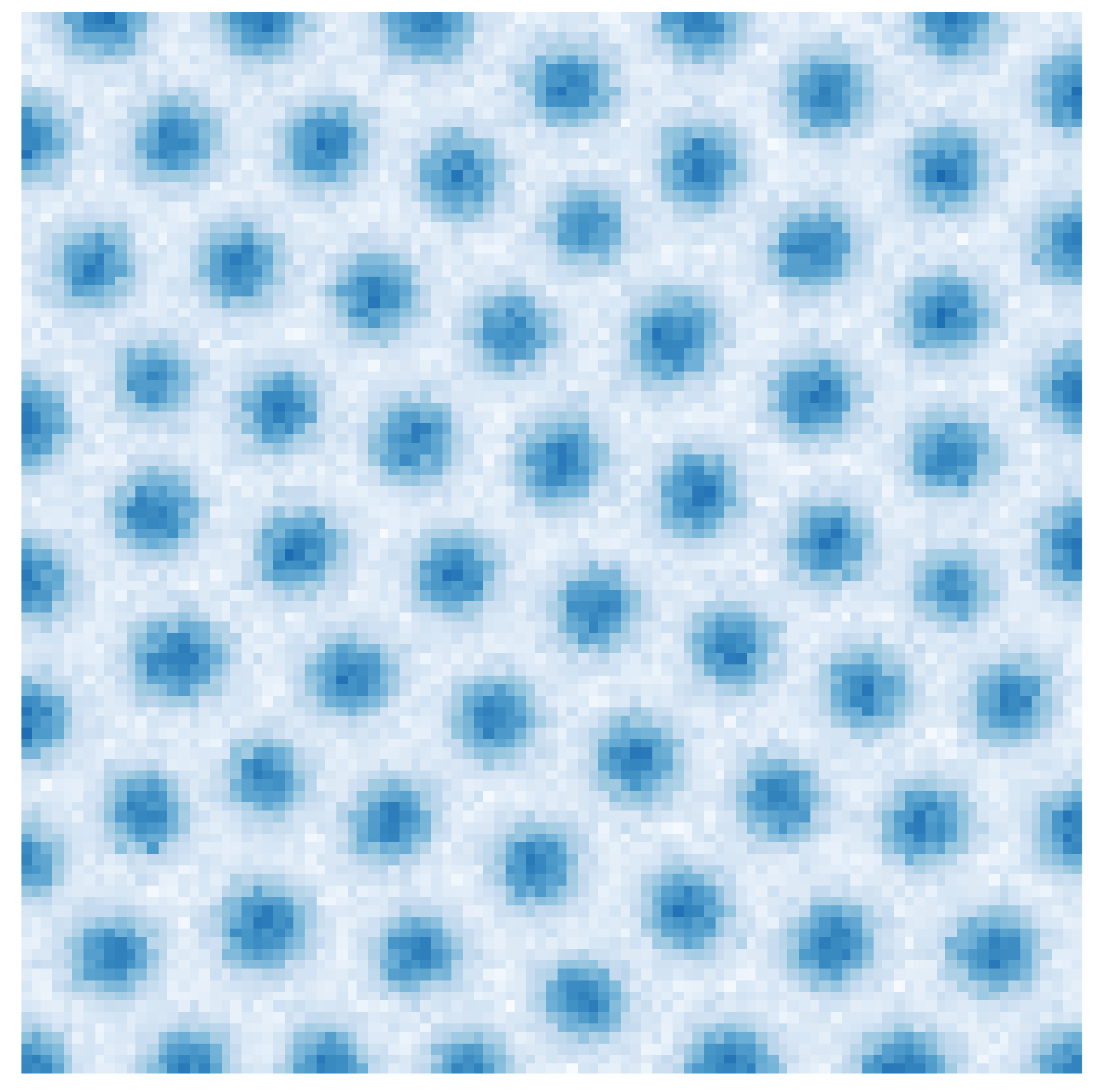}& \includegraphics[width = 0.17\linewidth]{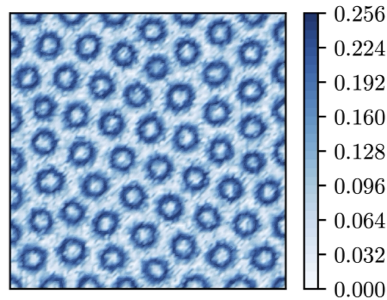} & \includegraphics[width = 0.18\linewidth]{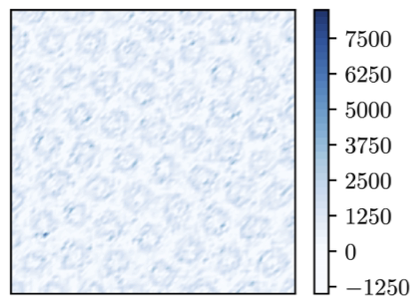} & \includegraphics[width = 0.175\linewidth]{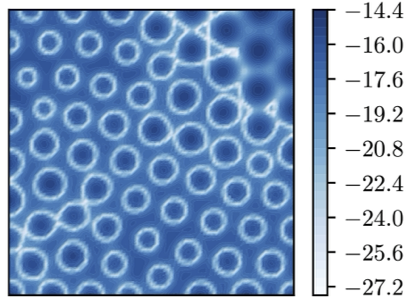}  & \includegraphics[width = 0.17\linewidth]{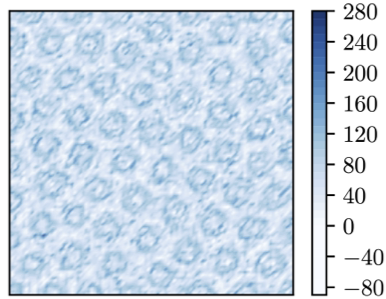}\\\hline
    \end{tabular}
    \caption{Two synthetic top-down microscopy characterization images, $\y^*$, of Di-BCP thin films and the integrands for functionals, $\mathcal{Q}_i$, $i = 1,\dots,4$, evaluated at the images. These functionals are the energy-based summary statistics shown in Tab.~\ref{tab:energy-based_statistics}.}
    \label{fig:energy-based_statistics}
\end{figure}

We visualize integrands of the functionals defining the energy-based summary statistics for synthetic data with lamella and spot latent patterns in Fig.~\ref{fig:energy-based_statistics}.

\subsubsection{Fourier-based summary statistics} \label{subsubsec:fourier_features}
Di-BCP thin film equilibrium patterns are well characterized by their intrinsic length scales. For example, the \emph{interfacial thickness} and the \emph{periodicity length} of Di-BCP patterns are often extracted from microscopy images for quantitative analysis \cite{Burger2020,Murphy2015}. In particular, the \textit{azimuthally-averaged magnitude spectrum} is widely used to process and analyze the microscopy characterization images of BCP equilibrium patterns \cite{Harrison2000, Edwards2005}. In the absence of external guidance or boundary geometry effects, there is no orientational preference in the latent patterns, which suggests that the orientational averaging of the magnitude spectrum preserves most of the information for inferring the model parameters. We thus hypothesize that the azimuthally-averaged magnitude spectrum of microscopy image data and its summary statistics may be effective for model parameter inference. 

We note that the azimuthally-averaged magnitude spectrum is similar to \emph{scattering intensity profiles} in small angle X-ray scattering. The scattering intensity profiles are modeled as noisy Fourier domain representations of the latent states~\cite{Harkless1990, Lee2005}. In this sense, the use of azimuthally-averaged magnitude spectrum and its related summary statistics for parameter inference and EIG computation can be seen as a rough prototype for the same task performed with small angle X-ray scattering data\footnotemark. 
\footnotetext{However, we acknowledge several important differences. First, the X-ray scattering imaging operator involves a Fourier transform and the sensor noise is applied on top of the observables, whereas the magnitude spectrum of microscope images already includes the sensor noise and the blurring effect. Secondly, the imaging operator for the X-ray scattering experiment depends highly on the experimental set up. Changes in the experimental design variables, such as beam stop positions and X-ray incident angles, can lead to dramatic changes in the observables.}

We now define the procedure for obtaining the azimuthally-averaged magnitude spectrum from an image $\y^* \in \mathbb{R}^{N\times M}$. The discrete Fourier transform of the image outputs a complex Fourier coefficient matrix $\widehat{\y^*}\in\mathbb{C}^{N\times M}$. We first translate the entries to center the zero frequency and examine the magnitude spectrum:
\begin{equation}
    \mathcal{P}(\y^*) =  |\widehat{\y^*}| \in \R^{N\times M}.
\end{equation}
For azimuthal averaging of the magnitude spectrum, a change-of-variables to radial coordinates centered at the zero frequency is performed. In the radial coordinate, the zero-centered magnitude spectrum has $N_r$ unique radial frequencies, $(r_1,\dots, r_{N_r})\in{(0, r_{\textrm{max}}]}^{N_r}$. We only study the magnitude spectrum for frequencies no larger than some right-closed intervals, as higher radial frequencies are not well-resolved and typically uninformative. Finally, we obtain the azimuthally-averaged magnitude spectrum, $\mathcal{P}_a\colon\R^{N\times M}\to \R^{N_r}$, by averaging over orientations $\theta$. Specifically, the average magnitude $\mathcal{P}_a(\y^*)(r_k)$ is computed for $N_\theta(r_k)$ number of entries in $\mathcal{P}(\y^*)$ with the identical radial frequency $r_k$:
\begin{equation}\label{eq:aaps}
    \mathcal{P}_a(\y^*)(r_k) = \frac{1}{N_{\theta}(r_k)} \sum_{j = 1}^{N_{\theta}(r_k)} \mathcal{P}(\y^*)(r_k,\theta_j),\quad k = 1, 2, \dots, N_r.
\end{equation}
We hypothesize that not much, if any, information that is relevant toward inferring the length scale dependent parameters is lost after the azimuthal averaging. 

The number of unique radial frequencies $N_r$ is typically large. To further reduce the dimension of the azimuthally-averaged magnitude spectrum data, we also apply averaging over each unit interval of the radial frequency to obtain $\floor{r_{\text{max}}}$-dimensional averaged spectrum $\tilde{\mathcal{P}}_a(\y^*)$:
\begin{equation}\label{eq:aaps_interval}
    \left(\tilde{\mathcal{P}}_a(\y^*)\right)_j = \frac{1}{\sum_{j-1< r_k\leq j}N_{\theta}(r_k)} \sum_{j-1< r_k\leq j}\sum_{i=1}^{N_{\theta}(r_k)} \mathcal{P}(\y^*)(r_k,\theta_i)\;,\quad j = 1, 2, \dots, \floor{r_{\text{max}}}.
\end{equation}

\begin{figure}
\centering
\begin{tabular}{|c|c|c|c|}\hline
\small
  \textbf{Synthetic image} $\y^*$ & $\ln (\mathcal{P}(\y^*))$ & $\mathcal{P}_a(\y^*)$ & $\tilde{\mathcal{P}}_a(\y^*)$ \\\hline\hline
  \includegraphics[width=0.16\linewidth]{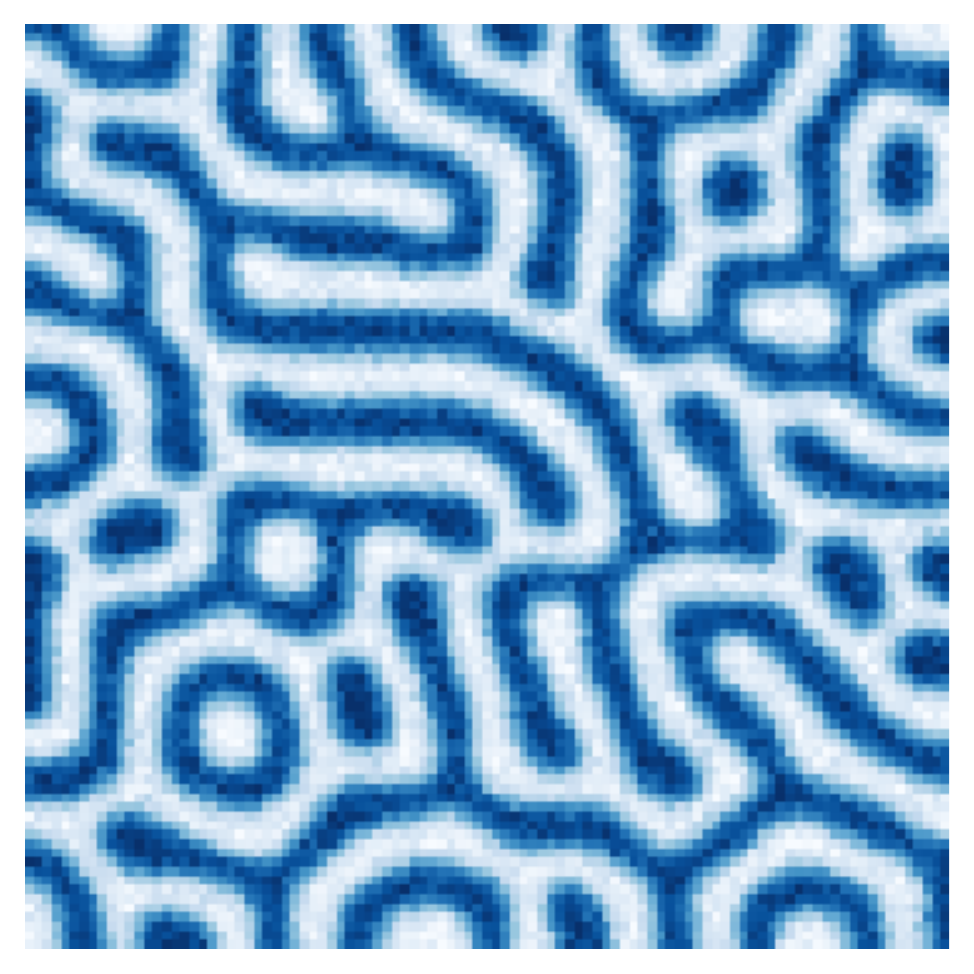}  & \includegraphics[width=0.2\linewidth]{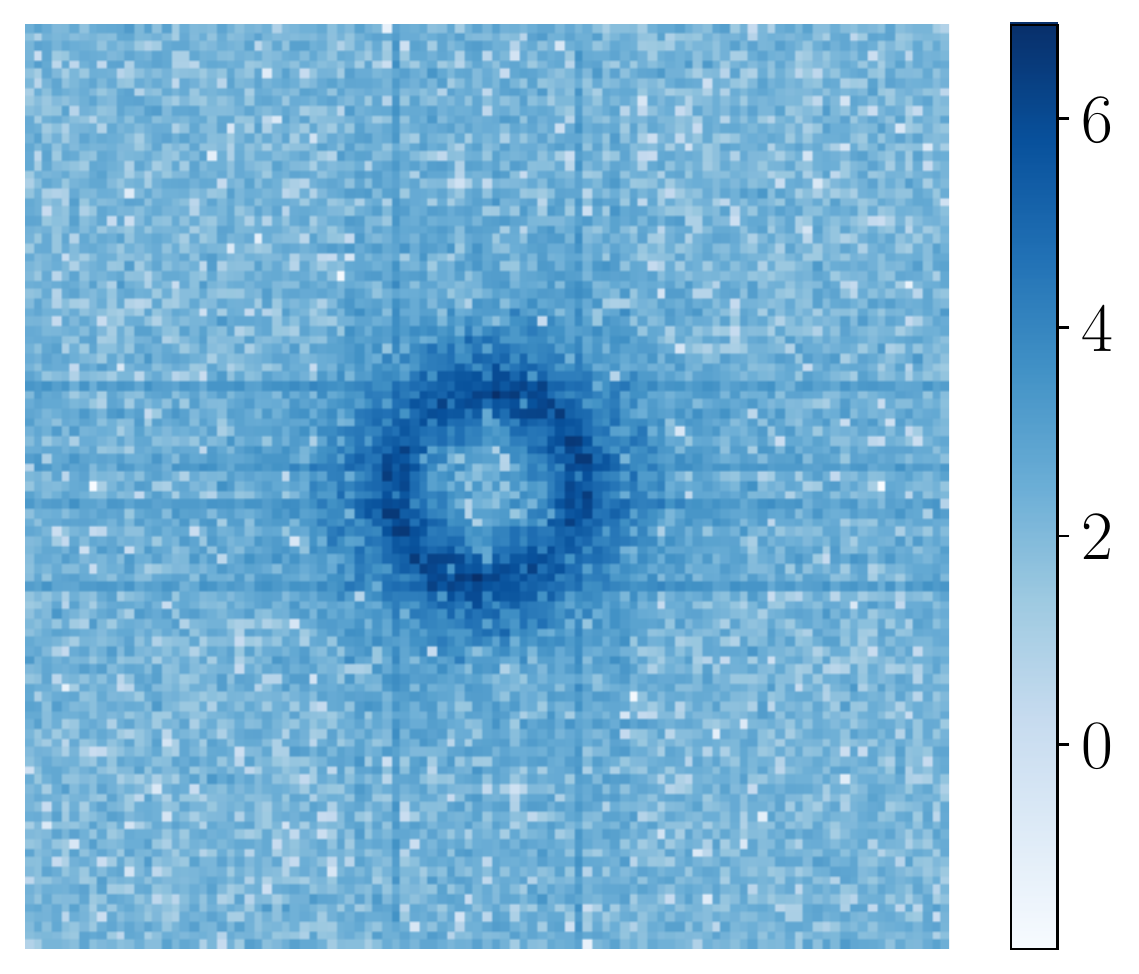} & \includegraphics[width=0.21\linewidth]{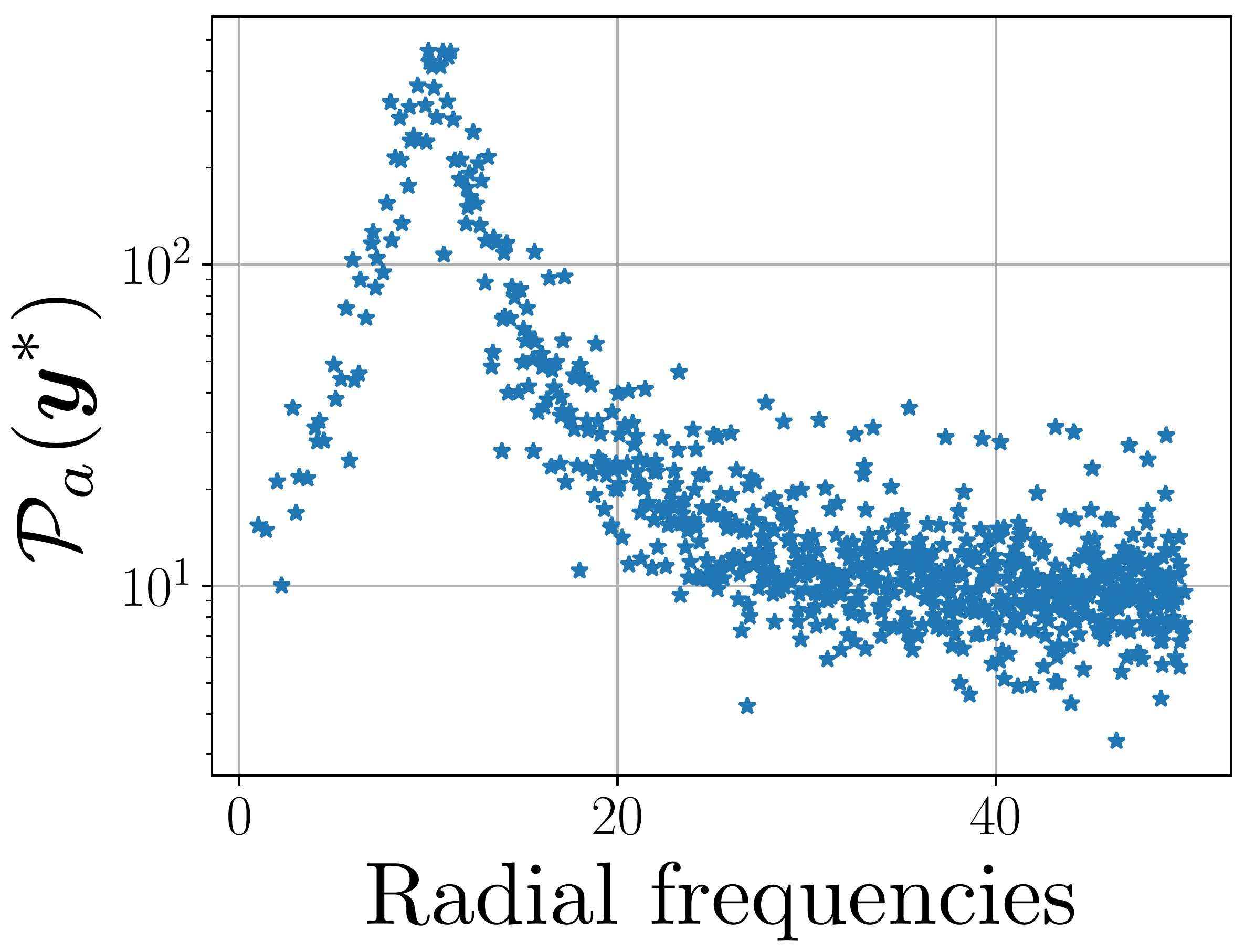} &         \includegraphics[width=0.20\linewidth]{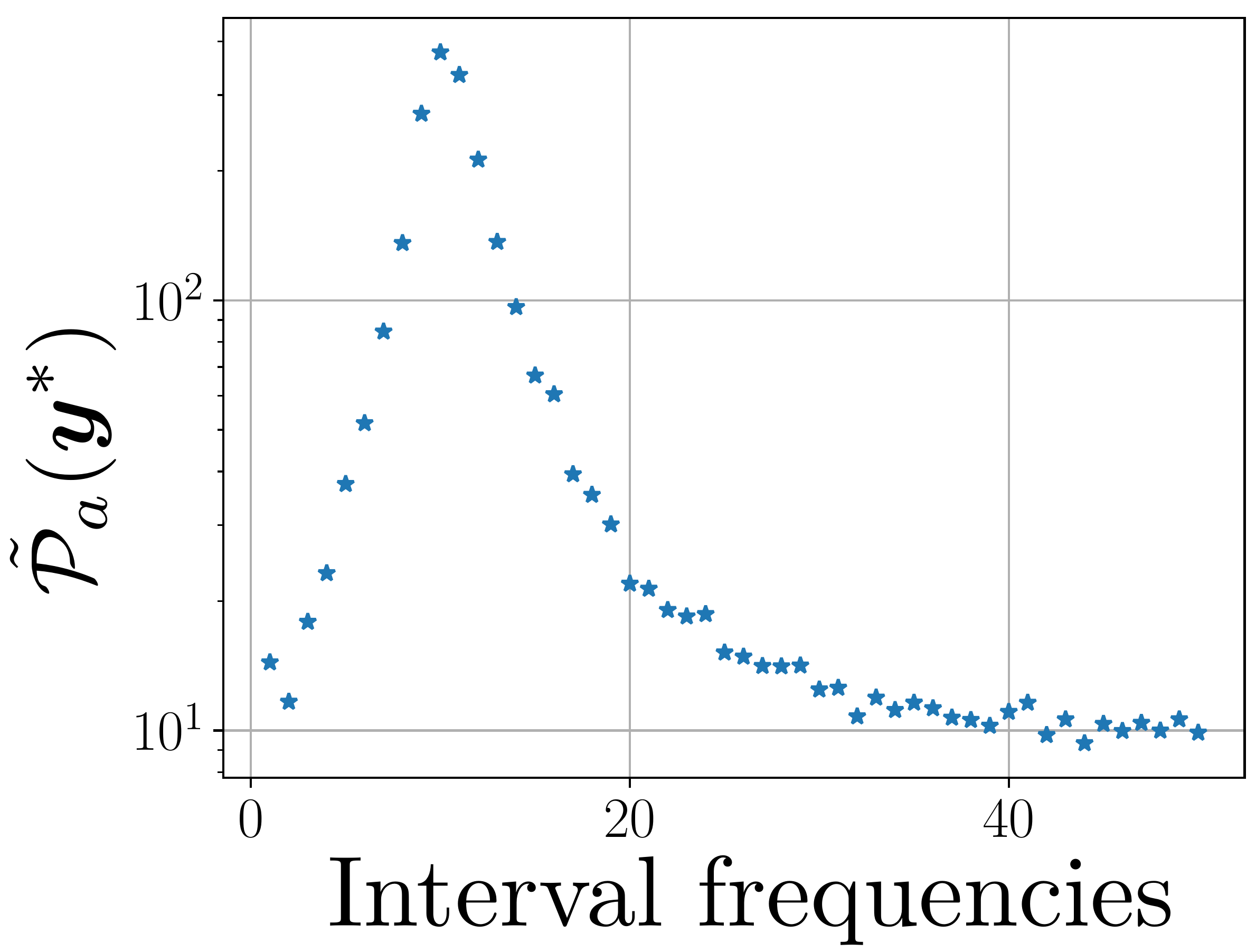}\\\hline
    \includegraphics[width=0.16\linewidth]{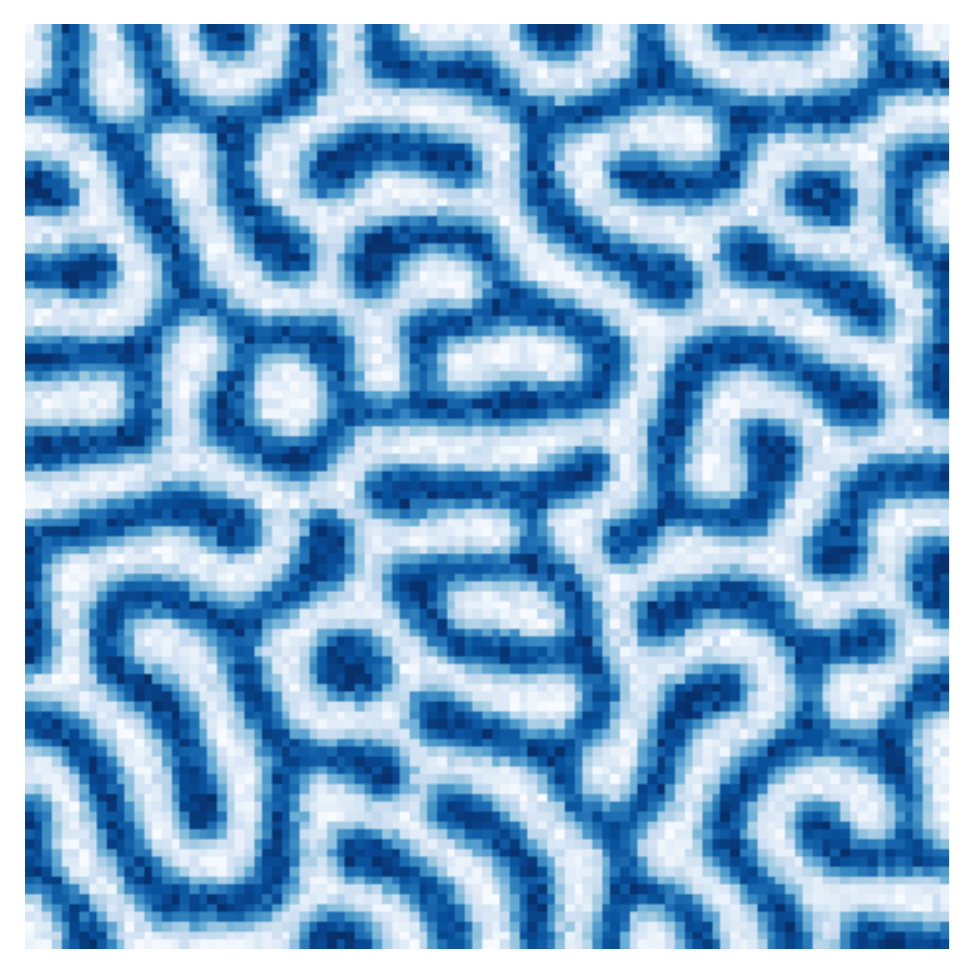}  & \includegraphics[width=0.19\linewidth]{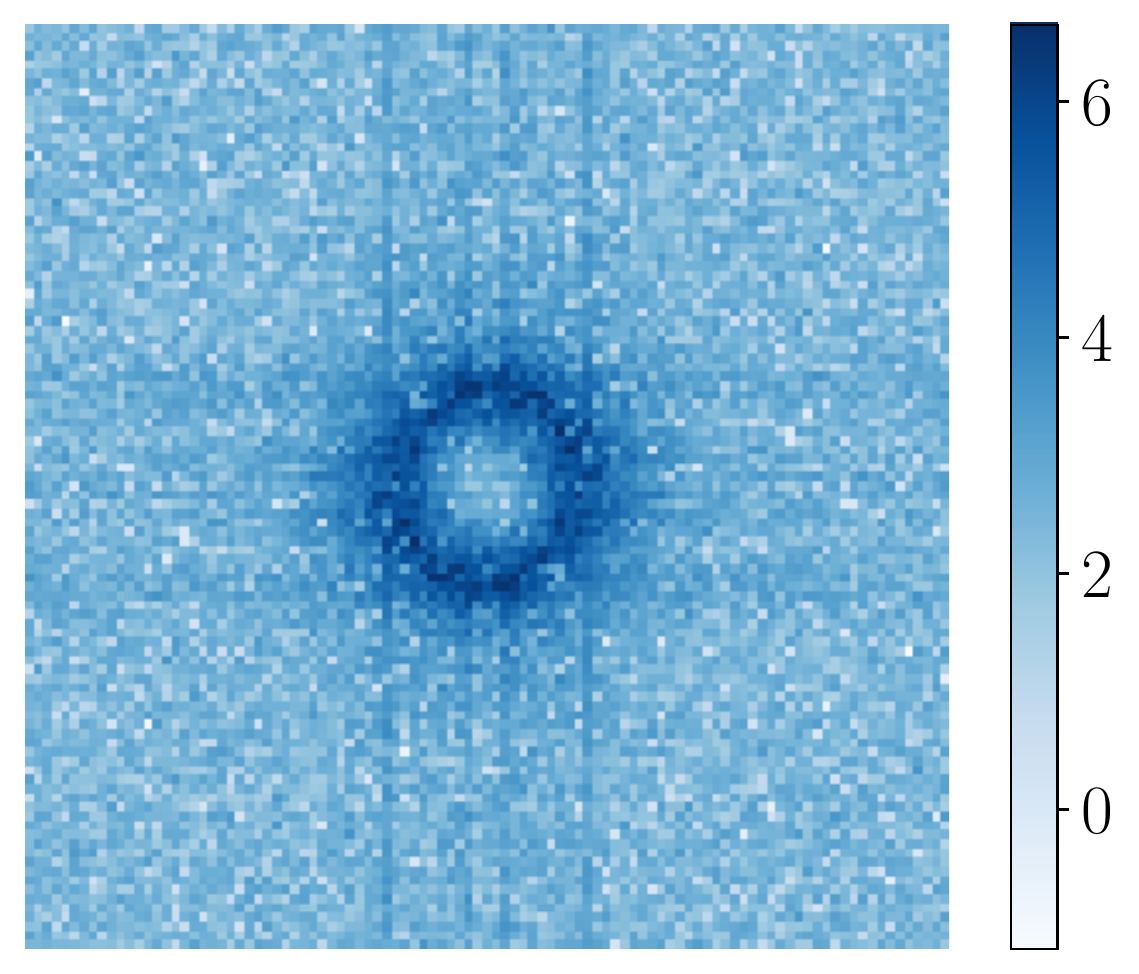} & \includegraphics[width=0.21\linewidth]{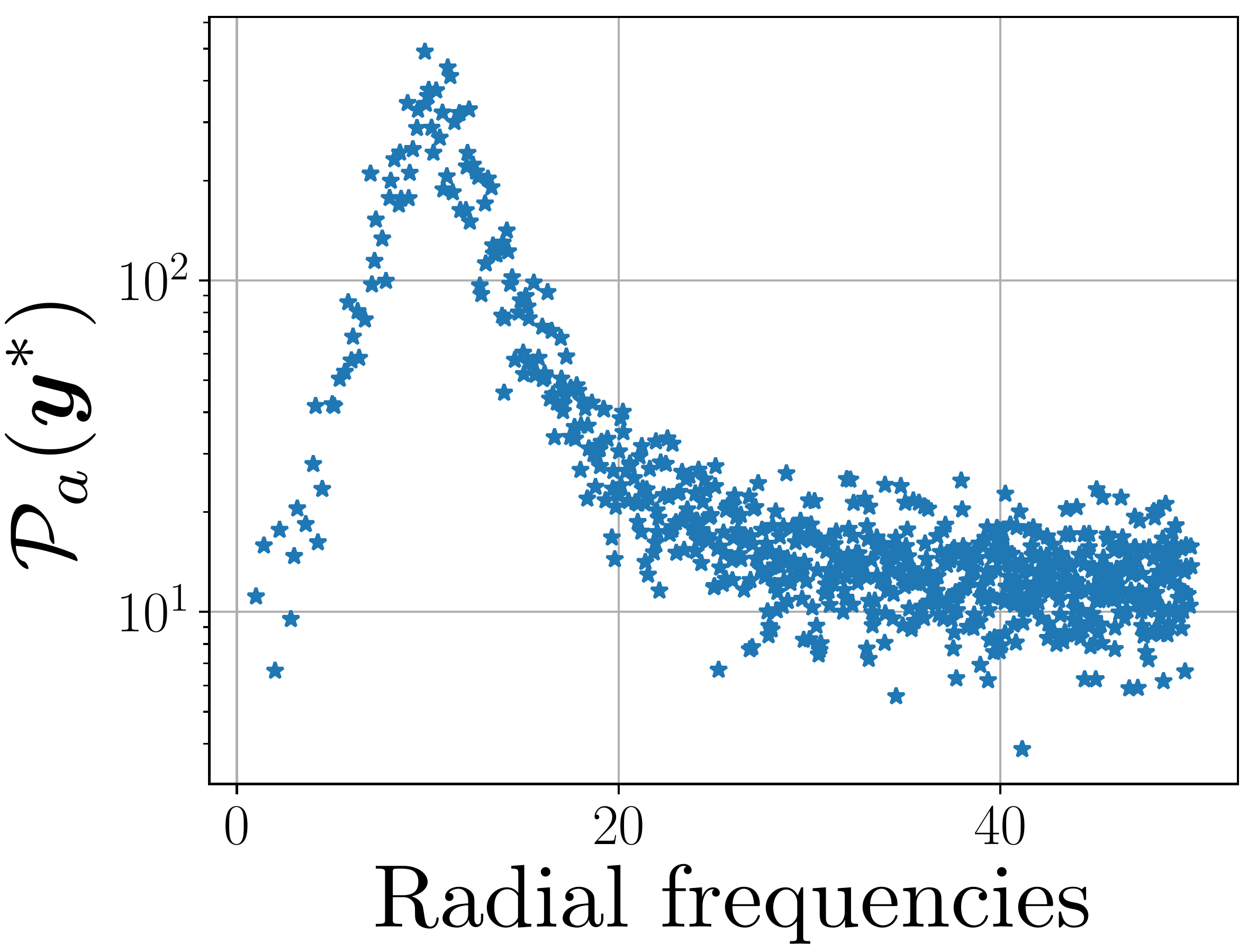} &         \includegraphics[width=0.2\linewidth]{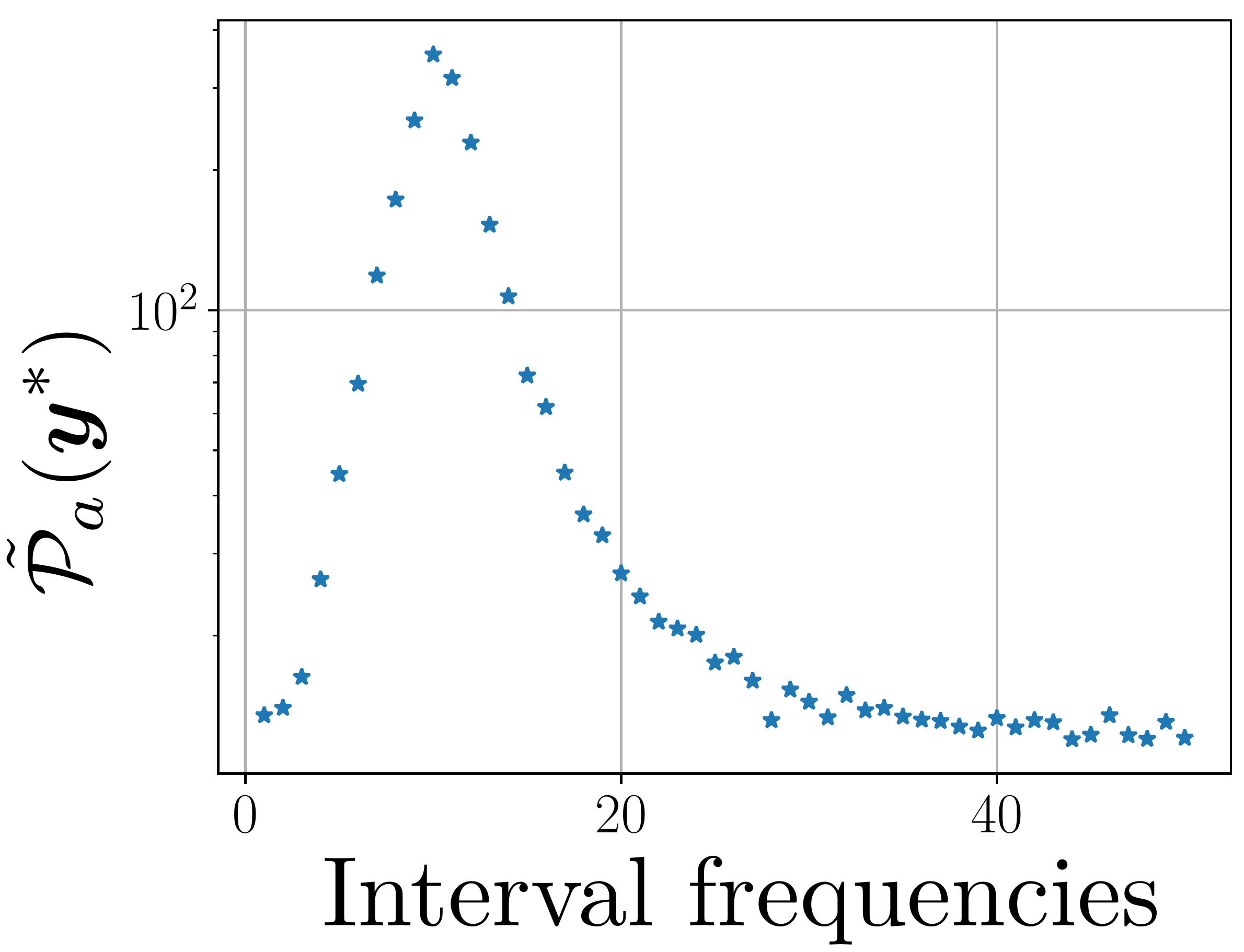}\\\hline
  
\end{tabular}
    \caption{Two synthetic top-down microscopy characterization images, $\y^*$, of Di-BCP thin films, their log-scale magnitude spectrum, $\ln(\mathcal{P}(\y^*))$, their azimuthally-averaged magnitude spectrum, $\mathcal{P}_a(\y^*)$, with $r_{\text{max}} = 50$, and their interval-averaged magnitude spectrum, $\tilde{\mathcal{P}}_a(\y^*)$. The synthetic images are generated by the OK model at the same model parameter.}
    \label{fig:magnitude_spectrum_examples}
\end{figure}

As shown in Fig.~\ref{fig:magnitude_spectrum_examples}, the zero-centered magnitude spectrum of synthetic microscope images have `rings' of frequencies with dominant magnitude values. They represent the periodicity length of the latent Di-BCP pattern. Furthermore, the overall shape of the magnitude spectrum distribution is sensitive to the change in model parameters and relatively insensitive to the auxiliary variable. Therefore, we further extract summary statistics such as \textit{dominant radial frequency}, or some quantitative descriptions of the shape of the azimuthally-averaged magnitude spectrum, such as \textit{moments of various orders}. Moreover, using $\boldsymbol{Q}_a = \tilde{\mathcal{P}}_a(\by)$ as summary statistics of microscopy data allows us to further investigate and compare the informativeness of the averaged magnitude spectrum in different ranges of the radial frequency for learning the model parameters.

\subsection{Prior distribution} \label{subsec:prior_transform}

We now define the prior distribution $\nu_{\bx}$, with density $\pi_{\bx}$, for the parameters used in our case study. The parameters of interest, again, are $\x = (m, \epsilon, \sigma)$, where $m \in (-1,1)$, $\epsilon \in \mathbb{R}_+$, and $\sigma \in \mathbb{R}_{+}$. We label the individual parameter random variables as $\bx = (M, E, \Sigma) \sim \nu_{\bx}$. 

$\nu_{\bx}$ is constructed based on hypotheses on the experimental setting related to the parameters of the OK model. These assumptions put restrictions on the feasible or admissible portions in parameter space. We denote this admissible set of parameters by $\mathcal{A}\subsetneq \mathbb{R}_{+}^3$. Then, under the principal of maximum entropy~\cite{caticha2004maximum}, we consider a uniform prior distribution\footnote{We note that if one's experimental setup provides additional information about the model parameters, the prior distribution may be much more concentrated in regions of parameter space.} over this admissible set, whose density is given by: 
\begin{equation}\label{eq:admissible_set_uniform_prior}\pi_{\bx}(\x) = \begin{cases} 1/|\mathcal{A}| \quad &\x \in  \mathcal{A} \\ 
0 \quad &\x 
\not\in  \mathcal{A}.\end{cases}\end{equation}

Our uniform prior is defined \emph{hierarchically}, that is, the parameters are defined conditionally on the others given the variable ordering $(M,
E, \Sigma)$. This hierarchical distribution implies a factorization of the prior density:
\begin{align}\label{eq:factorized_prior}
    \pi_{\bx}(m,\epsilon,\sigma) &= \pi_{M}(m)\pi_{E|M}(\epsilon | m)\pi_{\Sigma|M,E}(\sigma | m, \epsilon).
\end{align}
The marginal density $\pi_{M}$, for the spatial average parameter $M$, known as the \emph{hyper-parameter prior} or \emph{hyper-prior}, is uniform and is independent of the other two parameter random variables $(E, \Sigma)$; the conditional densities $\pi_{E|M}(\cdot | m)$ and $\pi_{\Sigma|M,E}(\cdot | m, \epsilon)$ are also uniform.
We implicitly define the factors in \eqref{eq:factorized_prior} by describing the constraints that form the boundary $\partial\mathcal{A}$ of the admissible set $\mathcal{A}$.

First, we consider constraints on the spatial average parameter $M$. In our experimental setting, we assume that the latent material states of observed images exhibit phase separation with a high probability. Note that the homogeneously mixed state are trivial critical points of the OK energy. In this case, the length scale related parameters $E$ and $\Sigma$ do not play a role in the forward operator, and thus not identifiable from data. To define constraint boundaries for the hyper-prior $\pi_{M}$ based on this assumption, we consider a result from the analysis of the OK energy functional: 

\begin{observation}[Constant state: Global minimum or metastable]\label{obs:disordered_state}
The constant state $u = m$ with no phase seperation is the global minimum of $\mathcal{E}_{\x}^{\textrm{OK}}$~\eqref{eq:ok_energy_strong} whenever $|m| > 1/\sqrt{3}$. Otherwise, it is a local minimum, i.e. a metasable state.~\cite{Choksi2009, Choksi2011, Cao2022}.
\end{observation} 

It is often assumed that global minimum states of $\mathcal{E}_{\x}^{\textrm{OK}}$ are relatively more stable configurations, i.e., more probable. Therefore, we exclude portion of parameter space in which the constant states are global minima, and define the admissible set for the hyper-prior for $M$ as $(0, 1/\sqrt{3})$, and the marginal prior density for $M$ as
\begin{equation}\label{eq:m_prior}
    \pi_{M} = \mathcal{U}(0, 1/\sqrt{3}),
\end{equation}
where we consider only positive numbers in our study, without loss of generality, due to symmetry of the forward operator about $m= 0 $.

To define admissibility constraints for the conditional density of $E$ and $\Sigma$ given the parameter $M$, we employ two assumptions on the equilibrium patterns encountered in our experimental scenario. These assumptions put restrictions on the size of the field of view of the microscope relative to different length scales of latent Di-BCP patterns. We note that these assumptions are based on a rescaled computational domain $\mathcal{D}_c = [0, L]^2$, $L = 40$ , with the material and model parameters rescaled according to the relations in \eqref{eq:ok_parameter_maps}. See a discussion on the computational domain in \ref{app:function_space}.

\begin{assumption}[Resolving phase interface]\label{ass:interfacial_energy}
We assume the phase interface of the latent Di-BCP patterns can be sufficiently resolved if given little or no image corruptions, which correspond to $\epsilon > \epsilon_{\textrm{min}} = 0.01$.
\end{assumption}

\begin{assumption}[Observing relatively large surfaces]\label{ass:domain_size}
We assume the periodicity of Di-BCP patterns can be sufficiently captured in the image data~\cite{Choksi2009, Choksi2011}, if given little or no image corruptions:
\begin{equation}
    L \gg O\left(\left(\epsilon/\sigma\right)^{1/3}\right) \implies \sigma > \epsilon C^3/L^3,\quad C \gg 1,
\end{equation}
and we thus assume $\sigma > \sigma_{\textrm{min}} = 0.01$.
\end{assumption} 

To complete the admissibility constraints, we emulate the work of~\cite{Choksi2009,Choksi2011}, in which the authors characterize a \emph{phase diagram} for the OK model that describes the morphology of equilibrium states for given model parameters $(\epsilon, \sigma)$. By analyzing the (linear) stability of global minimizers of the OK energy under periodic boundary conditions through Fourier expansions, they derive ``order/disorder transition (ODT) curves'' that partition the parameter space of $(\epsilon, \sigma)$ into: (i) a region in which the OK energy functional possess the constant states as the global minimizers, (ii) a region in which the constant states are \emph{not} the global minimizers, and  (iii) a transition region where the analysis is not applicable. It is hypothesized in~\cite{Choksi2011}, based on numerical evidence, that the \emph{actual} region in parameter space for which the constant states are \emph{not} the global minimizers is actually larger, therefore expanding the ODT curve further into the transition region.

\begin{figure}[!ht]
    \centering
    \begin{minipage}{.6\textwidth}
  \caption{(\emph{left}) The admissible set $\mathcal{A}\subset \mathbb{R}^3$. (\emph{Right}) The boundary $\partial \mathcal{A}$ of the admissible set for $(\epsilon, \sigma)$ for three choices of the order parameter spatial average $m = (0, 0.3, 1/\sqrt{3})$. The prior distribution for the parameters $(E, \Sigma)$ conditioned on the hyper-parameter $M$, $\nu_{E,\Sigma | M}(\cdot,\cdot | m)$, is uniformly distributed over the admissible set. We also plot 10,000 independently drawn samples from each conditional distribution.}
  \label{fig:prior_slice}
   \includegraphics[width=\textwidth]{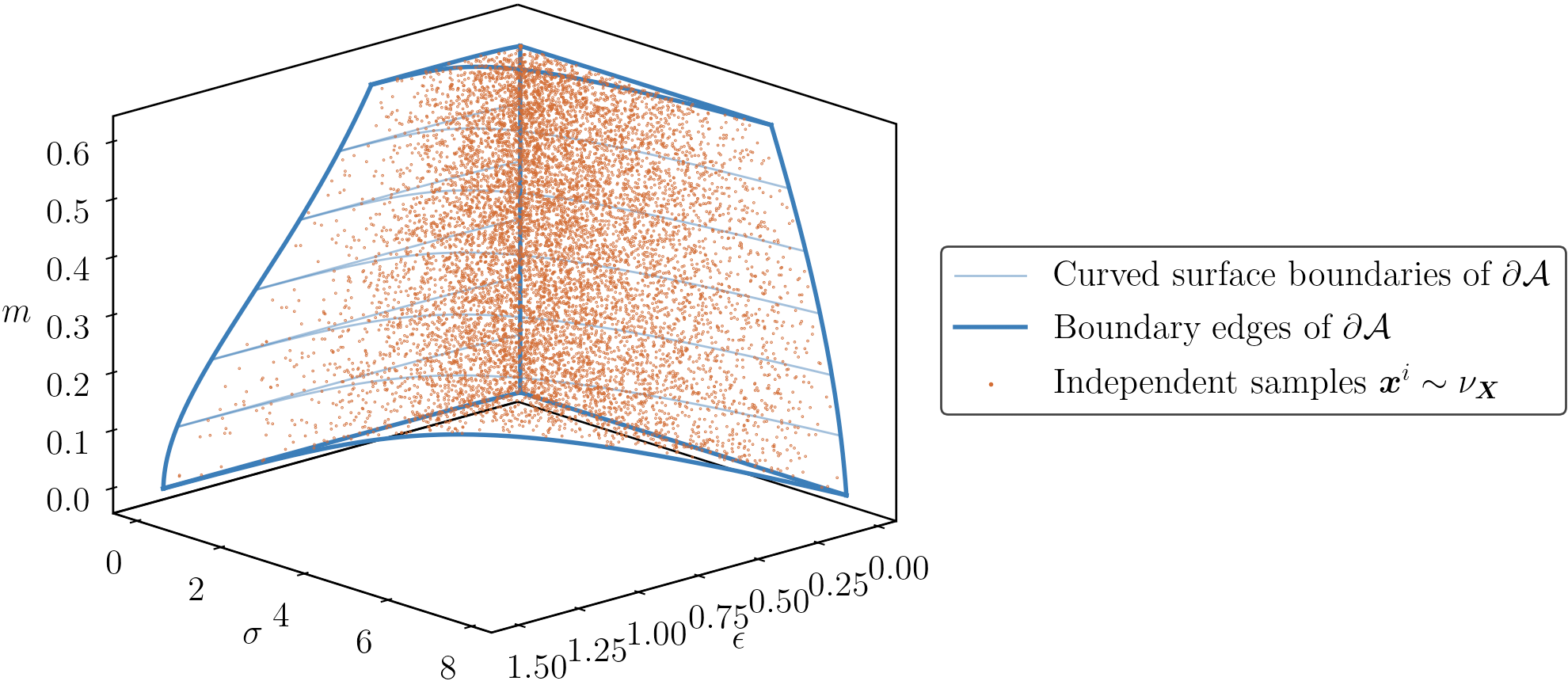}
  \end{minipage}
  \begin{minipage}{0.39\textwidth}
    \includegraphics[width=\textwidth]{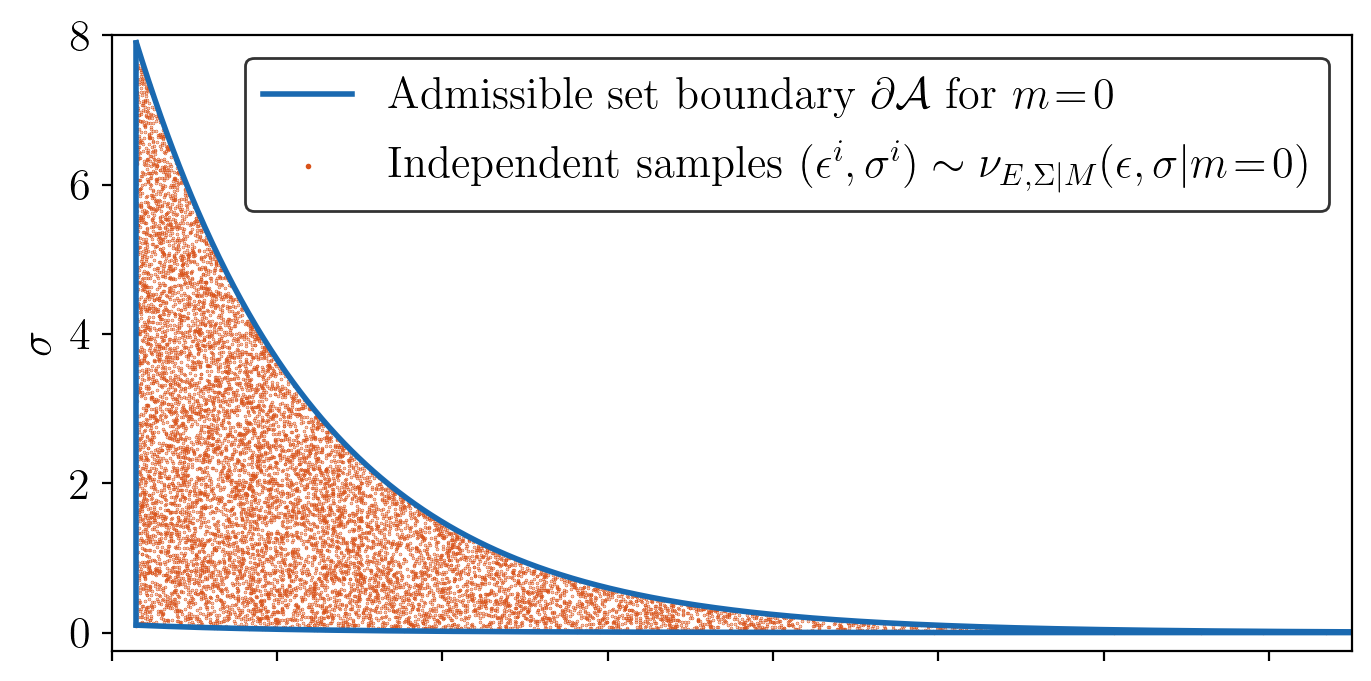}\\
    \includegraphics[width=\textwidth]{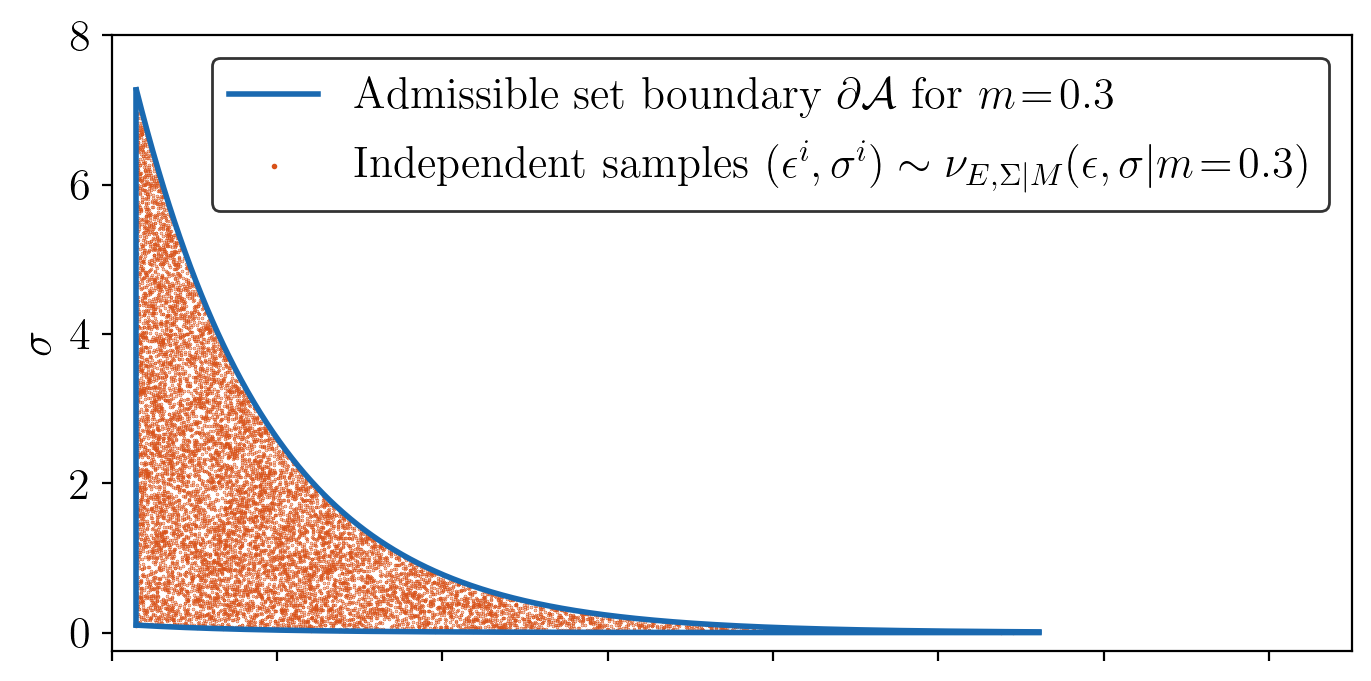}\\
    \includegraphics[width=\textwidth]{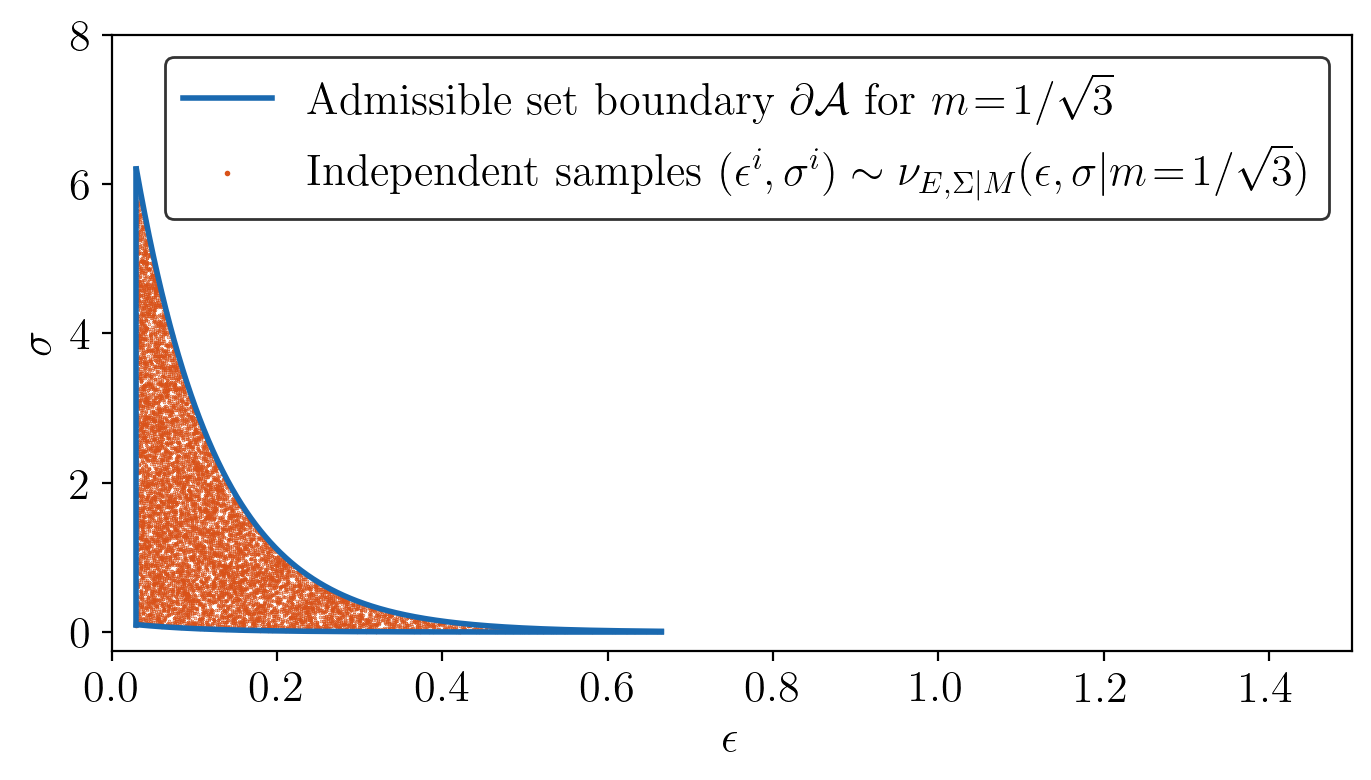}
  \end{minipage}
\end{figure}

We follow these authors in producing a \emph{numerically discovered empirical boundary} for the ODT curve based on the OK model.\footnote{While we consider the homogeneous Neumann boundary conditions \eqref{eq:lap_inv}, for which the analysis in~\cite{Choksi2011} does not directly apply\textemdash since the Fourier expansion arguments only apply to the case of periodic boundary conditions\textemdash our empirical ODT curves have similar qualitative behavior as in the authors' work.} 
The result of the numerical procedure is two parameterized curved boundary surfaces $\sigma_{\textrm{max}}(\epsilon; m)$ and $\sigma_{\textrm{min}}(\epsilon; m)$ which define the upper and lower boundary for the parameter $\sigma$ conditioned on the values for m and $\epsilon$. A visualization of the curved line ``slices" of the surfaces, for three values of $m$ are provided in Fig.~\ref{fig:prior_slice}. Intersecting these surfaces with the fixed bounds $\epsilon_{\textrm{min}} = 0.01$, $\sigma_{\textrm{min}} = 0.01$ from Assumptions \ref{ass:interfacial_energy}, \ref{ass:domain_size}, and $M \in (0, 1/\sqrt{3})$ from Observation \ref{obs:disordered_state}, produces the boundary $\partial \mathcal{A}$.

\section{Case study: Numerical results} \label{sec:numerical_results}
In this section, we present the numerical results for our case study. These results demonstrate the insights one can extract from performing calibration tasks via the measure transport approach. In Section~\ref{subsec:posterior_predictive}, we provide representative posterior samples to visually demonstrate the reduction of the model parameter uncertainty after performing the model calibration with the energy-based summary statistics. Samples of the observables, i.e., the latent Di-BCP patterns, at these parameter values are also generated and plotted to qualitatively validate the inference results. In Section~\ref{subsec:EIG_energy}, we present results on estimating EIG of model parameters from various combinations of the energy-based summary statistics with varying levels of noise and blurring corruptions. The similar EIG estimation results from the Fourier-based summary statistics are provided in Section~\ref{subsec:EIG_fourier}. In Section~\ref{subsec:EIG_specimen_size}, we study the impact of various levels of magnification for microscopy imaging on the calibration results. Finally, in Section~\ref{subsec:EIG_spatial_average_fixed}, we study the EIG estimation for the interfacial and nonlocal parameters $(E,\Sigma)$ when the spatial average parameter $M = m$ is known.

To perform our numerical experiments, we first collect $n = 10^5$ samples from the joint distribution of parameters and the model-generated data $\{(\x^i,\y^i)\}_{i=1}^{n}$, as described in Section~\ref{subsec:computation_wo_likelihoods}. Then, the samples of model-generated data are used to compute samples of the proposed summary statistics $\{(\x^i,\q^i = \mathcal{Q}(\y^i))\}_{i=1}^{n}$. We use $60\%$ of the samples as a training set for estimating the posterior probability densities using the triangular transport map approach introduced in Section~\ref{sec:transportLFI} and use the remaining samples to estimate the EIG using the likelihood-free estimator introduced in Section~\ref{sec:eig}.  

A very small subset ($\sim\thinspace 200$) of the samples are numerically problematic due to (i) failure to resolve phase interface geometries in the observable and (ii) failure to evaluate the forward operator within a reasonable residual tolerance. In both cases, the numerical anomalies primarily occur near the boundary of the admissible set $\partial \mathcal{A}$ and are caused by relatively coarse spatial discretization of the state space. In our numerical case study, we choose to discard these samples. To detect these numerically problematic samples, an empirical approach is adopted based on the value of the KL divergence between $\tilde{\mathcal{P}}_a(\y^i)$ and a reference set of values. 
We observe that the numerically problematic samples invariably possess a large value of such a KL-divergence. This ansatz is used an automated criterion to discard a majority of these numerically problematic samples. Additionally, to deal with outlier samples, we compute the $0.05$ and $0.995$ quantiles of the log posterior evaluations $\widehat{\pi}_{\bx|\bq}(\x^i|\q^i)$ and discard samples that are outside of these quantiles for the EIG estimation.

\subsection{Posterior and posterior predictives} \label{subsec:posterior_predictive}

To visualize posterior predictive samples, we first generate representative posterior quadrature points. In particular, we take Gauss--Hermite quadrature points $\rsample_{\textrm{GH}}$ which are used for accurately integrating functions with respect to the standard normal distribution, i.e, the reference distribution $\nu_{\br_2}$, and transform them through the inverse of the posterior transport map, i.e. $\widehat{S}^{\mathcal{X}}(\q^*, \cdot)^{-1}(\rsample_{\textrm{GH}})$, as described in~\eqref{eq:conditional_density_estimate}. These points are depicted within the admissible set for the prior in Tab.~\ref{table:posteriors}. While these points, paired with appropriate weights, can be used for both visualizing and accurately integrating functions with respect to the posterior distribution, we only present numerical results on the visualization of the model predictions at posterior samples in this study. We note that the points include the inverse transformation of the center quadrature point, $\x_{\boldsymbol{0}} = \widehat{S}^{\mathcal{X}}(\q^*, \cdot)^{-1}(\boldsymbol{0})$, as a representative sample that is typically in the region with highest posterior probability\footnotemark. In addition, we include the transformation of quadrature points in the tails of each reference marginal $\mathcal{N}(0,1)$, in order to study the `tails', or regions of low probability, of the posterior. These points are labeled as $\x_{\pm\delta m} \coloneqq \widehat{S}^{\mathcal{X}}(\q^*, \cdot)^{-1}(\boldsymbol{0}\pm \delta m)$, $\x_{\pm\delta \epsilon}:=\widehat{S}^{\mathcal{X}}(\q^*, \cdot)^{-1}(\boldsymbol{0}\pm \delta \epsilon)$, and $\x_{\pm\delta \sigma}:=\widehat{S}^{\mathcal{X}}(\q^*, \cdot)^{-1}(\boldsymbol{0}\pm \delta \sigma)$ where $\delta m = \Delta \mathbf{e}_1$, $\delta \epsilon = \Delta \mathbf{e}_2$ and $\delta \sigma = \Delta \mathbf{e}_3$ for some scalar $\Delta > 0$. In our studies we set $\Delta = 2$, which corresponds to the $0.977$ quantile of a univariate standard normal random variable.

In Tab.~\ref{table:posteriors}, we present the posterior quadrature points from the parameter posterior distribution that conditioned on the energy-based summary statistics of three different synthetic top-down microscopy characterization images. They are generated at three different sets of model parameters, with their latent Di-BCP patterns shown in the left column. For this example, we consider the characterization model with the reference field of view ($\mathcal{J} = \identity_{H^1(\mathcal{D})}$), a low level of the additive noise corruption ($\text{SNR} = 5$), and an insignificant level of the blurring corruption ($\mathcal{I} = \identity_{H^1(\mathcal{D})}$). In the top row, the case of symmetric Di-BCPs ($m = 0$) is considered, which exhibit lamella patterns. The other two are chosen to have patterns consist of primarily spots (\emph{middle row}) and a mixture of lamellae and spots (\emph{bottom row}). For each of the synthetic image data $\y^*$, we consider two sets of posteriors that are conditioned on $\q^* = \boldsymbol{\mathcal{Q}}(\y^*)$ with (i) $\boldsymbol{\mathcal{Q}} = (\mathcal{Q}_2, \mathcal{Q}_4)$  only (\emph{middle column}), and (ii) $\boldsymbol{\mathcal{Q}}= \boldsymbol{\mathcal{Q}}_{1:4}\coloneqq(\mathcal{Q}_1, \mathcal{Q}_2,\mathcal{Q}_3, \mathcal{Q}_4)$ (\emph{right column}). When comparing the two sets of posterior samples, we observe a strong concentration of the posterior achieved by conditioning on the set of four energy-based summary statistics---all seven posterior quadrature points are essentially indistinguishable in the plots. This observation is quantified via EIG calculations that are presented in the following subsections.

\begin{table}[!ht]
\centering
\centerline{\begin{tabular}{|@{}c@{}c@{}|} \hline
\textbf{Synthetic latent pattern} $\state^*$  & $\pi_{\bx | \boldsymbol{q}^*}$\\ \hline
& \\[-14 pt] 
\begin{tabular}{c}\Centerstack{\includegraphics[width=0.25\textwidth]{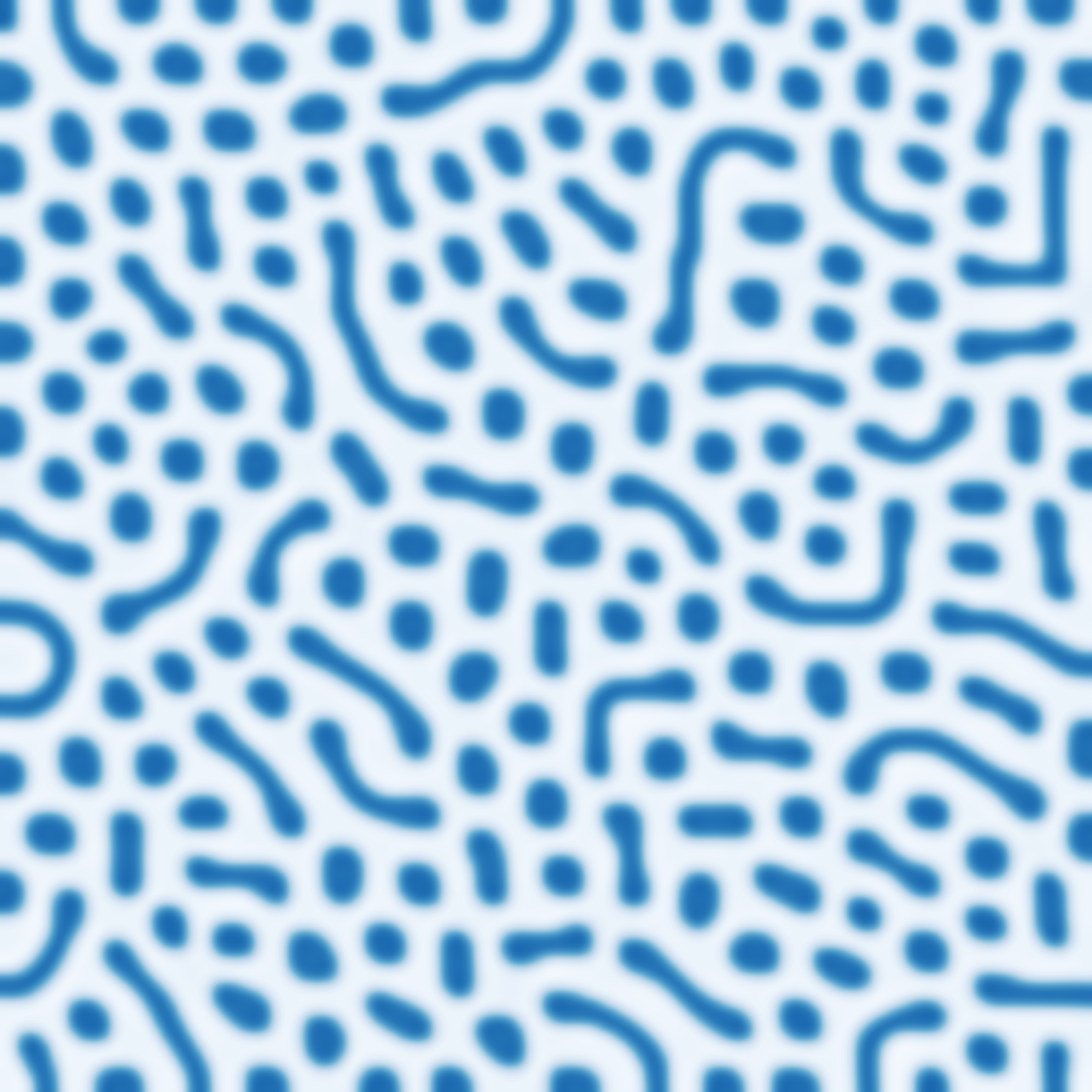}}
\end{tabular} &
\begin{tabular}{cc} &\\
             $\boldsymbol{\mathcal{Q}}=(\mathcal{Q}_2,\mathcal{Q}_4)$ & $\boldsymbol{\mathcal{Q}}=\boldsymbol{\mathcal{Q}}_{1:4}$ \\
 \Centerstack{\includegraphics[width=0.335\linewidth]{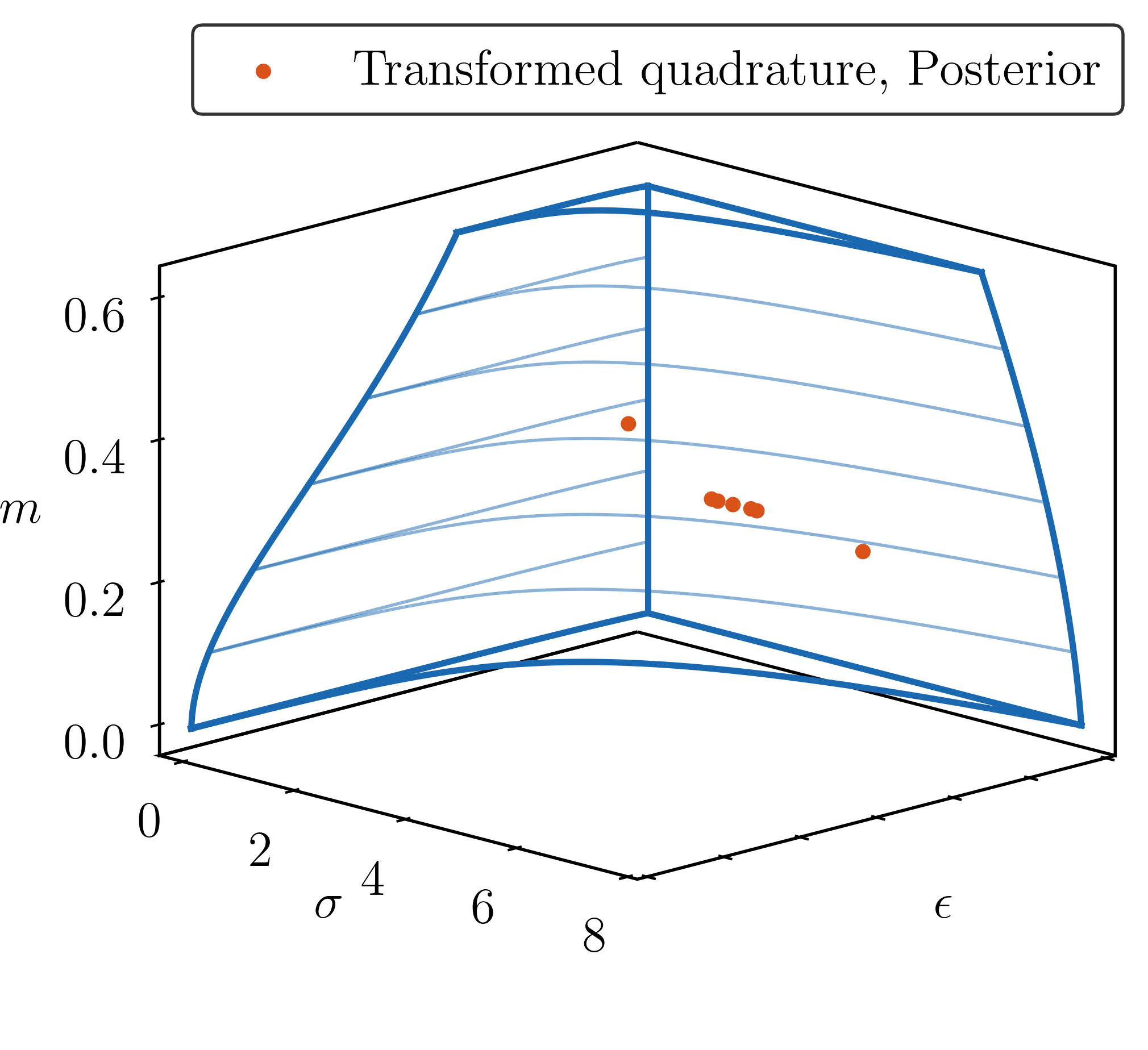}} &
  \Centerstack{\includegraphics[width=0.335\linewidth]{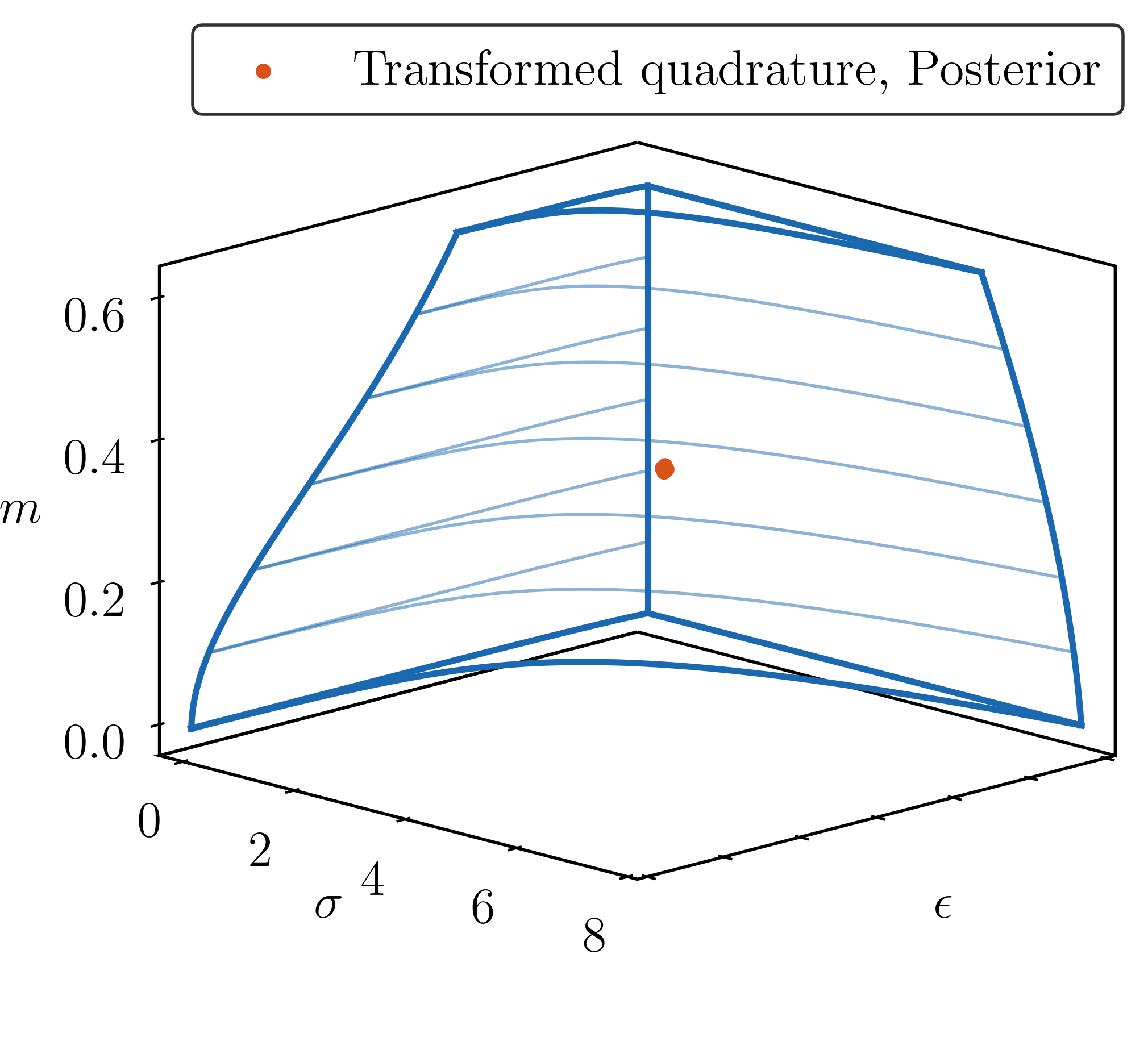}} 
  \end{tabular}
\\[-18pt] & \\\hline
& \\[-14 pt] 
    \begin{tabular}{c}\Centerstack{\includegraphics[width=0.25\textwidth]{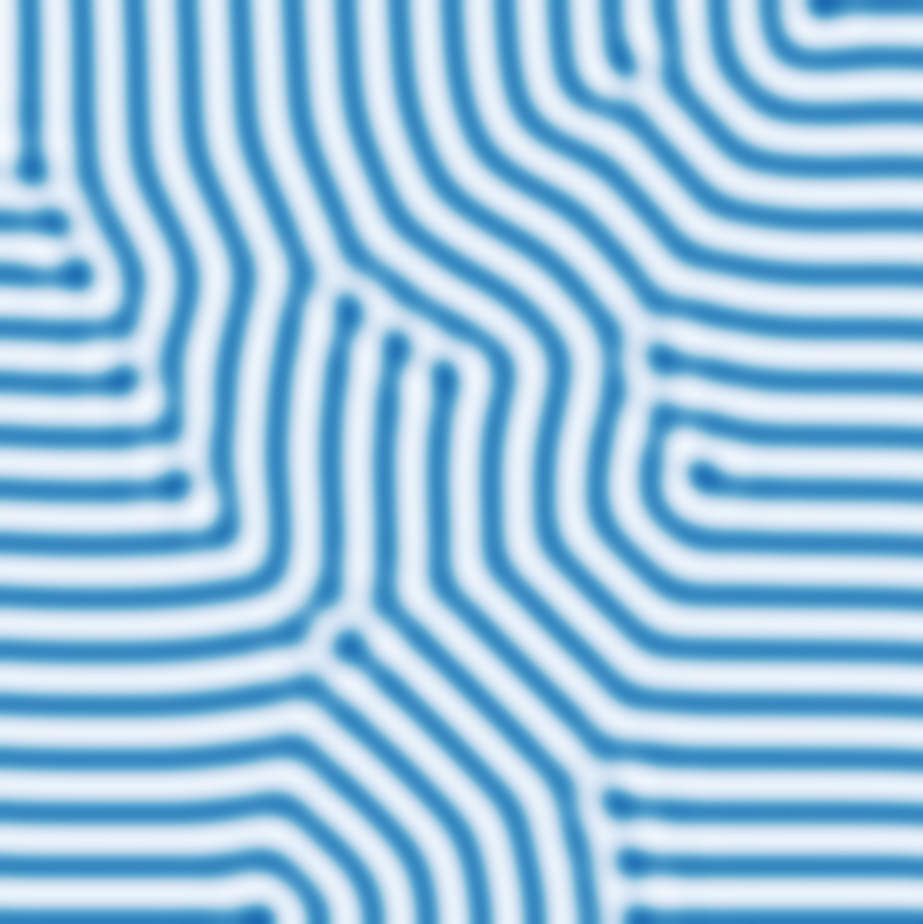}}\end{tabular} &
             \begin{tabular}{cc}&\\
         $\boldsymbol{\mathcal{Q}}=(\mathcal{Q}_2,\mathcal{Q}_4)$ & $\boldsymbol{\mathcal{Q}}=\boldsymbol{\mathcal{Q}}_{1:4}$ \\
 \Centerstack{\includegraphics[width=0.335\linewidth]{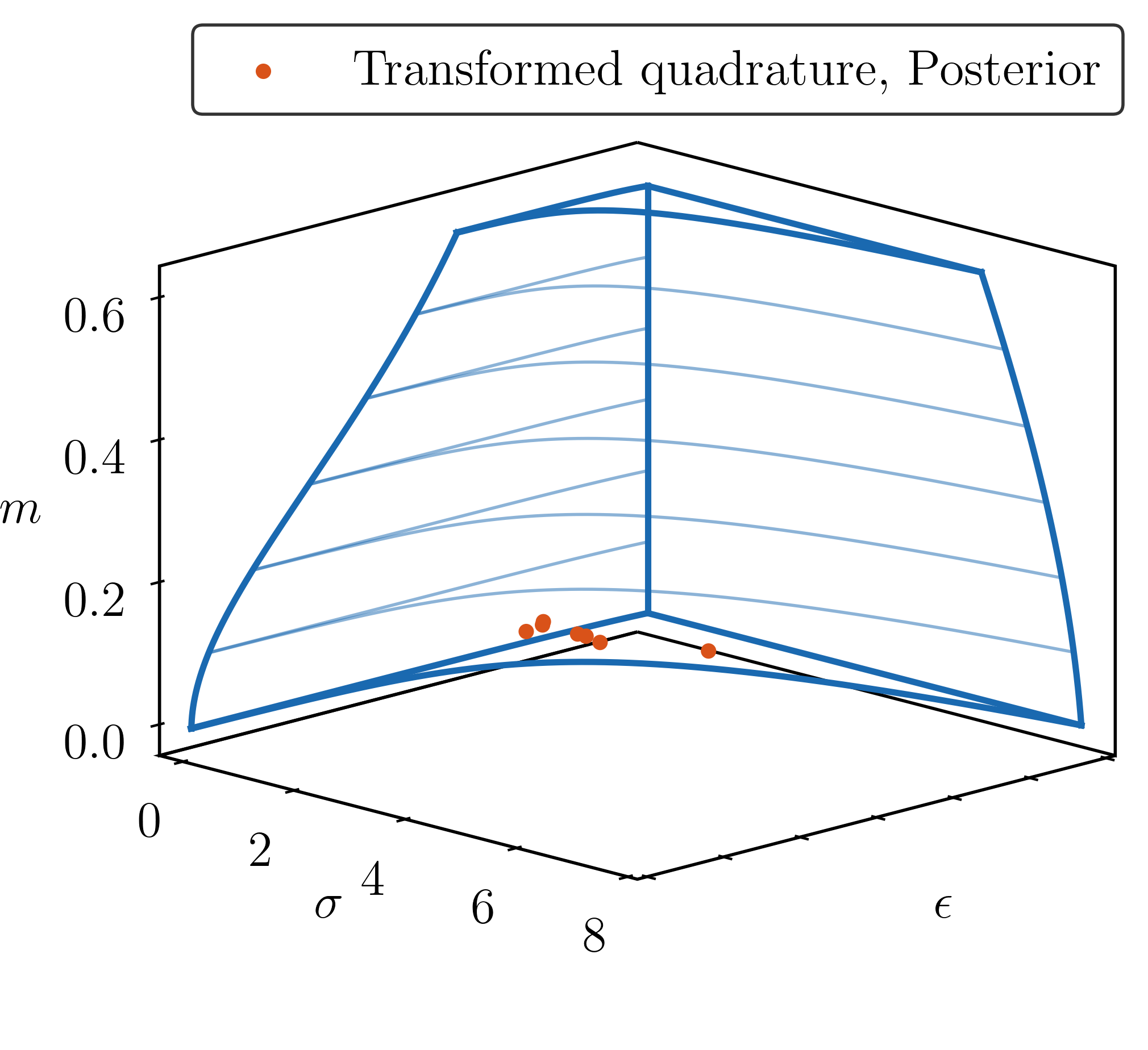}} &
  \Centerstack{\includegraphics[width=0.335\linewidth]{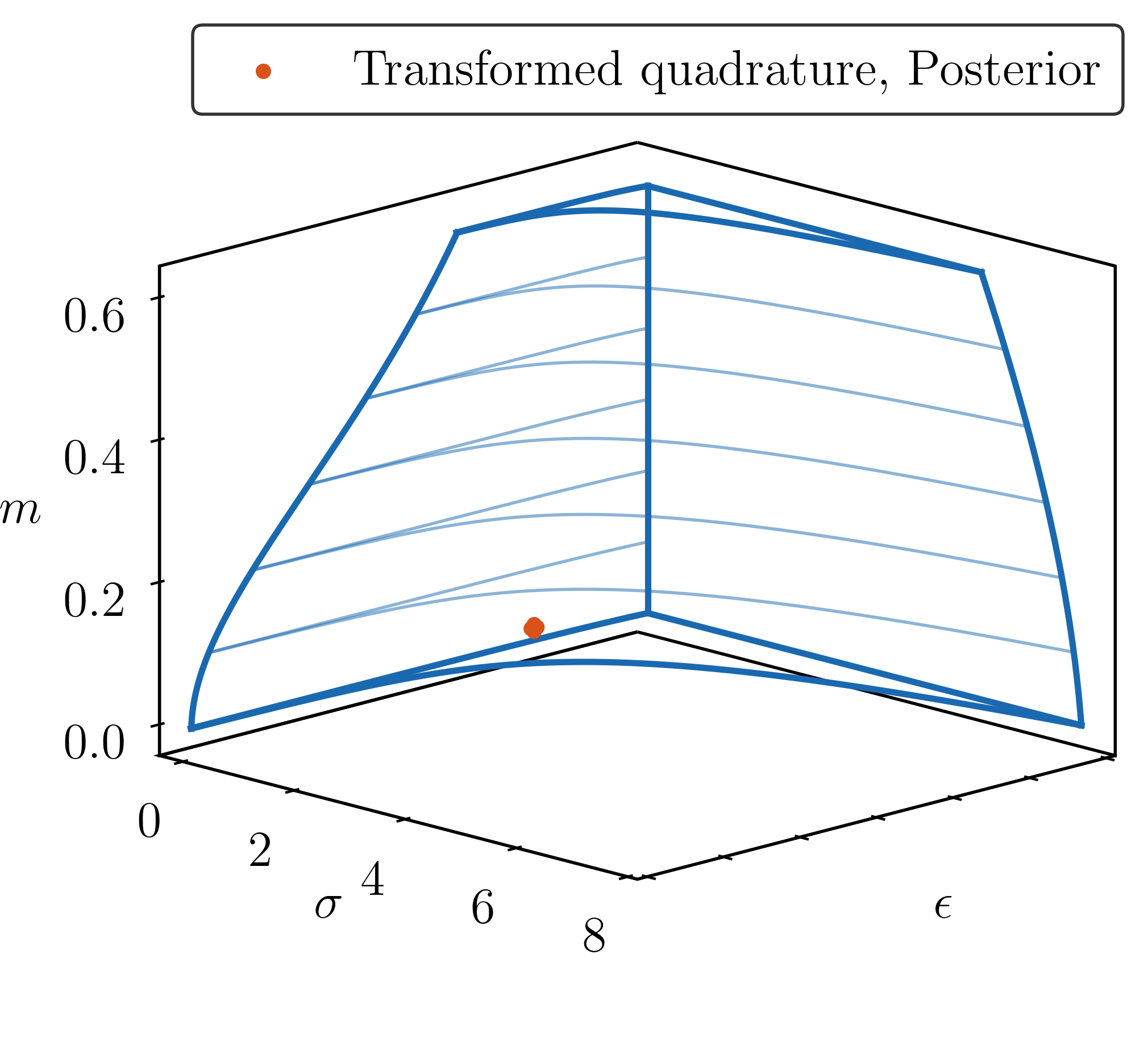}}
  \end{tabular}
            \\[-18pt] & \\\hline
& \\[-6 pt] 
            \begin{tabular}{c}\Centerstack{\includegraphics[width=0.25\textwidth]{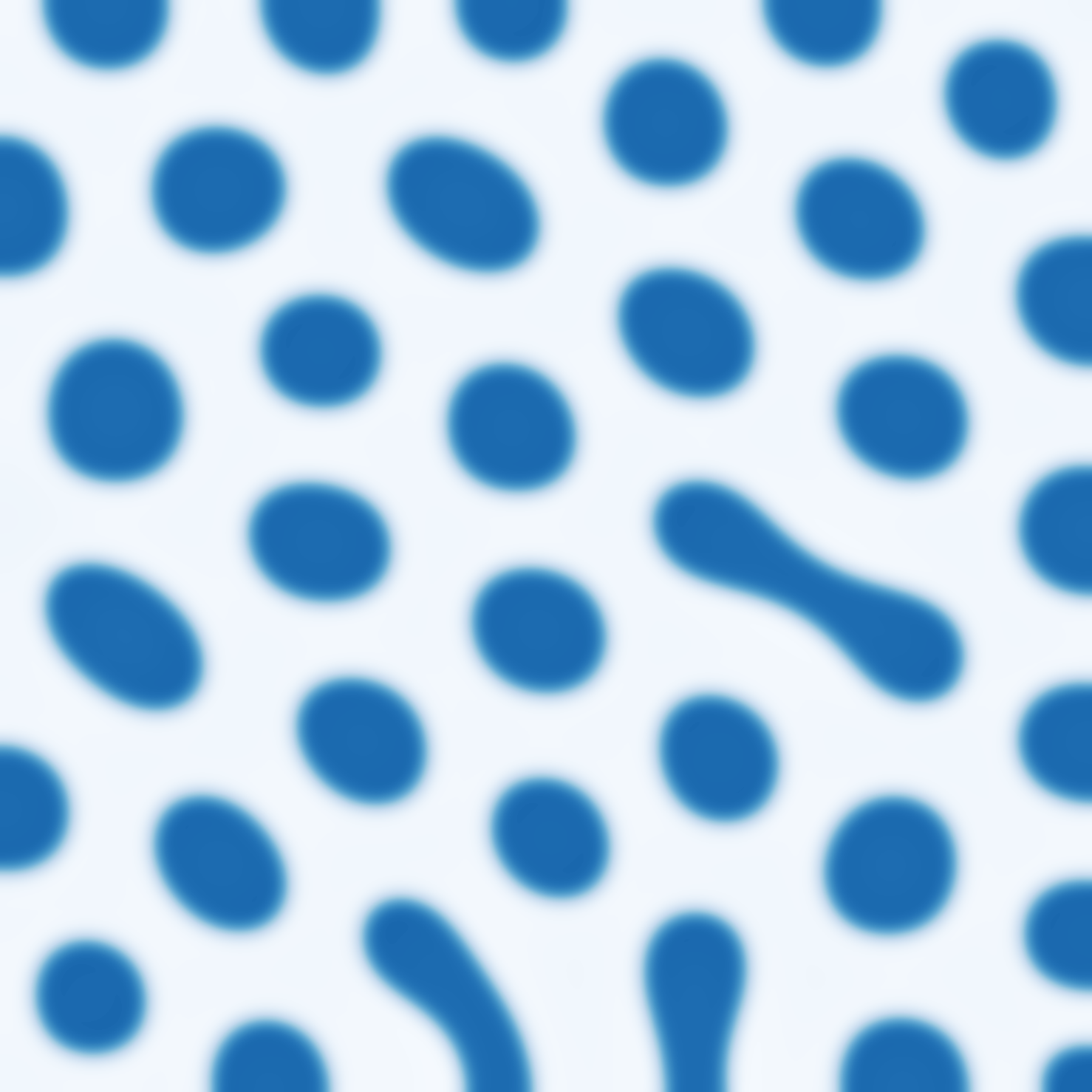}}
            \end{tabular} &
               \begin{tabular}{cc}
         $\boldsymbol{\mathcal{Q}}=(\mathcal{Q}_2,\mathcal{Q}_4)$ & $\boldsymbol{\mathcal{Q}}=\boldsymbol{\mathcal{Q}}_{1:4}$ \\
 \Centerstack{ \adjustbox{trim={0.0\width} {0.09\height} {0.0\width} {0.0\height},clip}%
  {\includegraphics[width=0.335\linewidth]{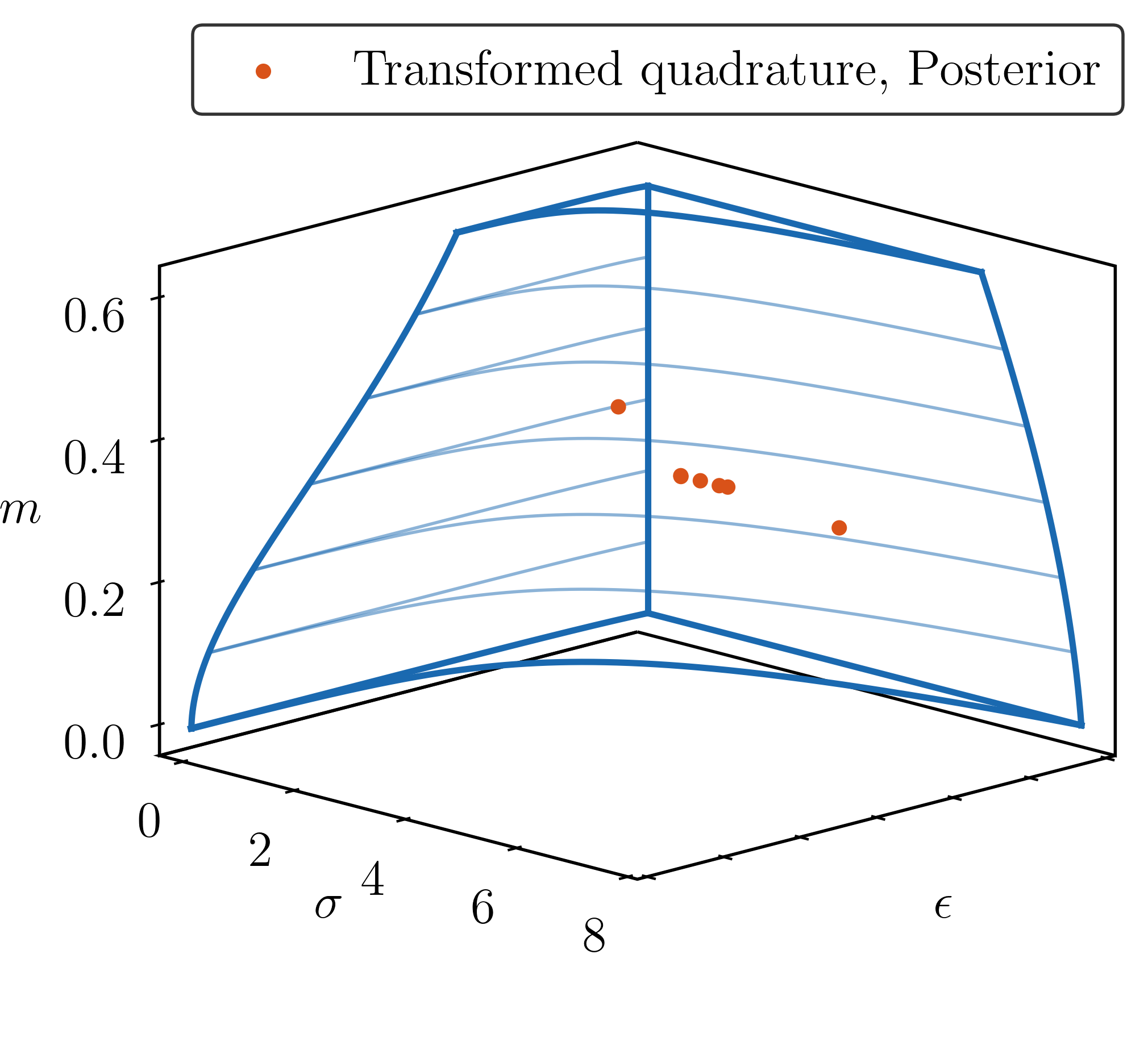}}} &
  \Centerstack{\adjustbox{trim={0.0\width} {0.09\height} {0.0\width} {0.0\height},clip}%
  {\includegraphics[width=0.335\linewidth]{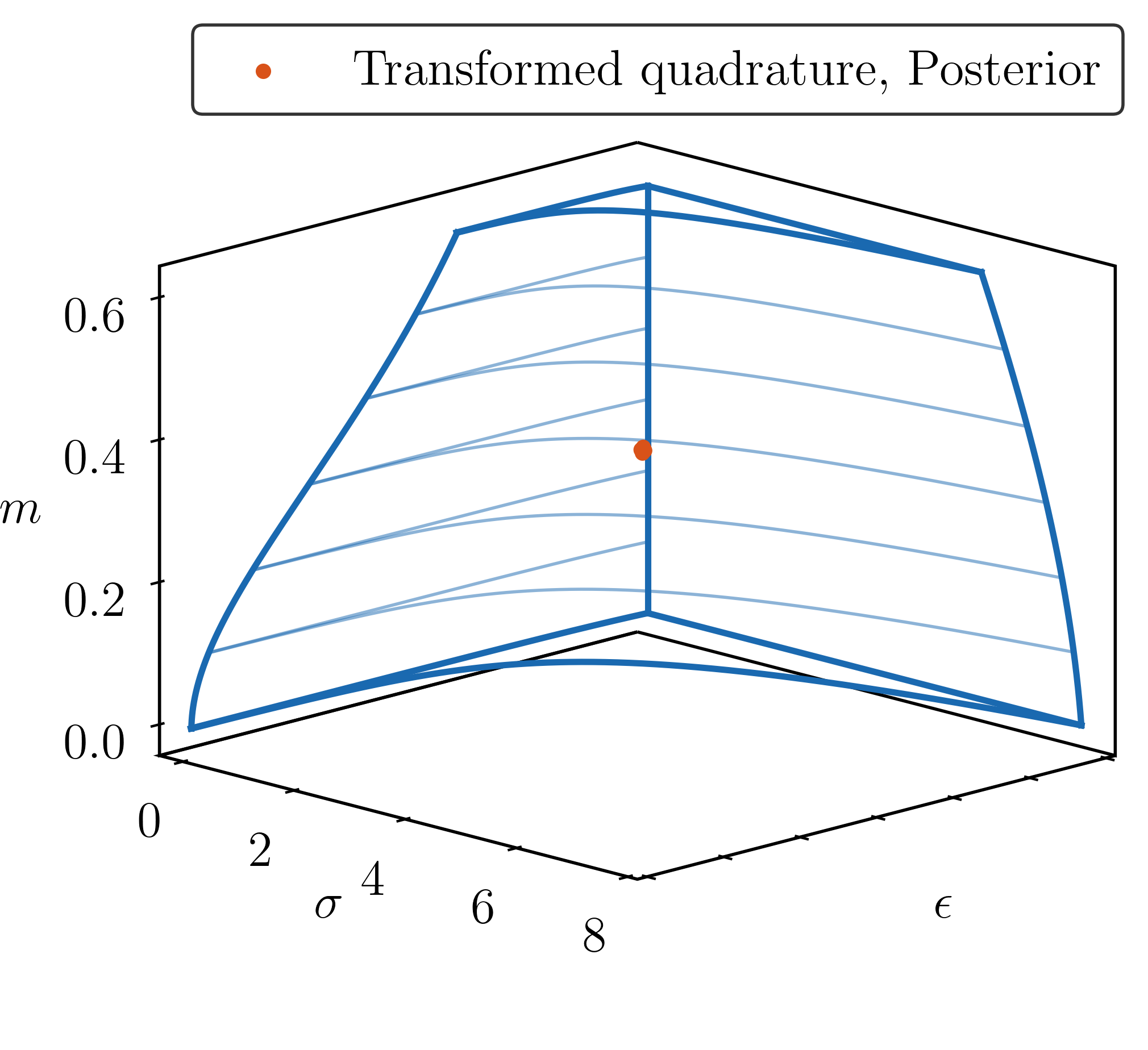}}}\end{tabular}
  \\[-10pt] & \\\hline 
  \end{tabular}}
    \caption{(\emph{left column}) The latent material states $u^*$, described by order parameters, of three synthetic microscopy image $\y^*$ data of Di-BCP thin films. (\emph{middle column}) The Gauss--Hermite quadrature points of the reference density pulled back to the model parameter posterior, illustrating posterior concentration conditioned on $\q^* = (\mathcal{Q}_2(\y^*), \mathcal{Q}_4(\y^*))$. (\emph{right column}) The same as the middle column, but conditioned on $\q^*=(\mathcal{Q}_1(\y^*),\dots, \mathcal{Q}_4(\y^*))$.}
    \label{table:posteriors}
\end{table}

\begin{table}[!hp]
\centering
\centerline{\begin{tabular}{|@{}c@{}c@{}c@{}c@{}c@{}c@{}|} \hline
\textbf{Synthetic latent pattern} $\state^*$  & $\boldsymbol{\mathcal{Q}}$ & $\mathcal{F}(\x_{\boldsymbol{0}},\z^i)$ & $\mathcal{F}(\x_{\pm \delta m},\z^i)$ & $\mathcal{F}(\x_{\pm \delta \epsilon},\z^i)$  & $\mathcal{F}(\x_{\pm \delta \sigma},\z^i)$ \\ \hline
&&&&& \\[-6 pt]
\begin{tabular}{c}\Centerstack{\includegraphics[width=0.375\textwidth]{figure/ricardo/posterior_predictives/data/state_3.png}}\end{tabular} &
            
             \begin{tabular}{c}$(\mathcal{Q}_2,\mathcal{Q}_4)$ \\ \\\\\\\\\\\\ $\boldsymbol{\mathcal{Q}}_{1:4}$\end{tabular} &
              \begin{tabular}{c}\Centerstack{\includegraphics[width=0.1835\linewidth]{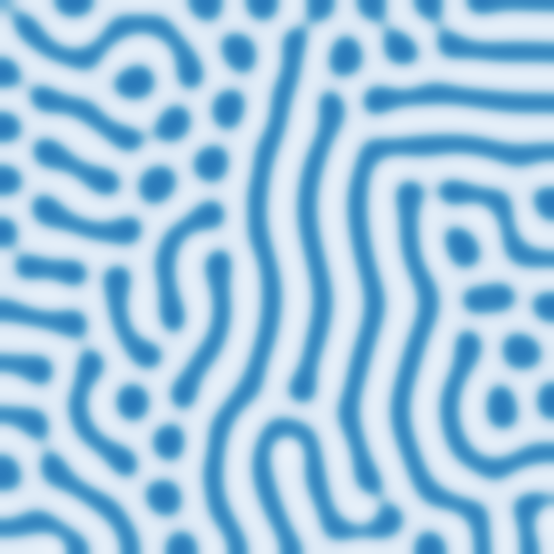}} \\
             \Centerstack{\includegraphics[width=0.1835\linewidth]{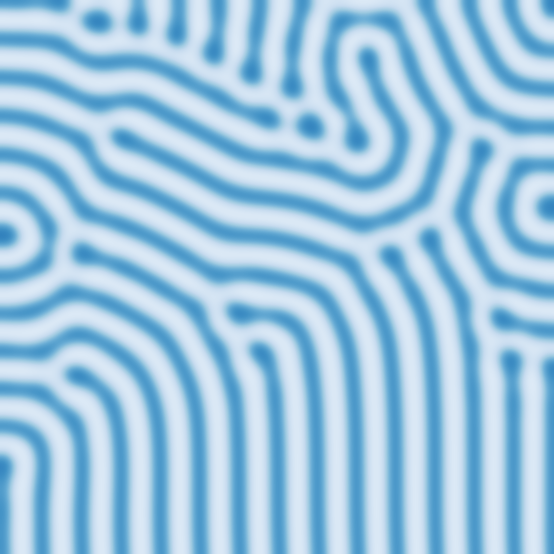}}
            \end{tabular}&
             \begin{tabular}{c} \Centerstack{\includegraphics[ width=0.088\linewidth ]{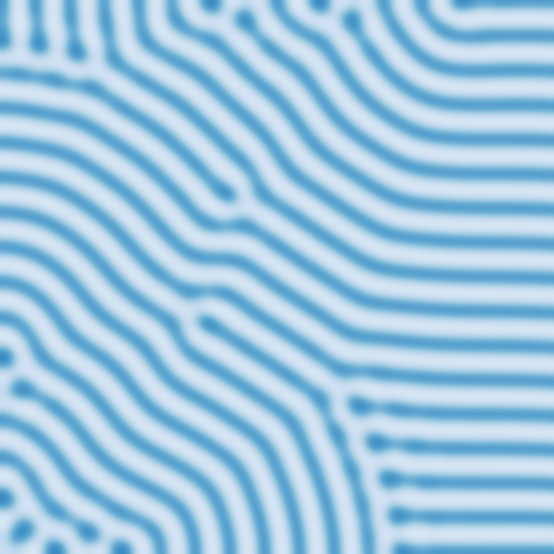}} \\
             \Centerstack{\includegraphics[ width=0.088\linewidth ]{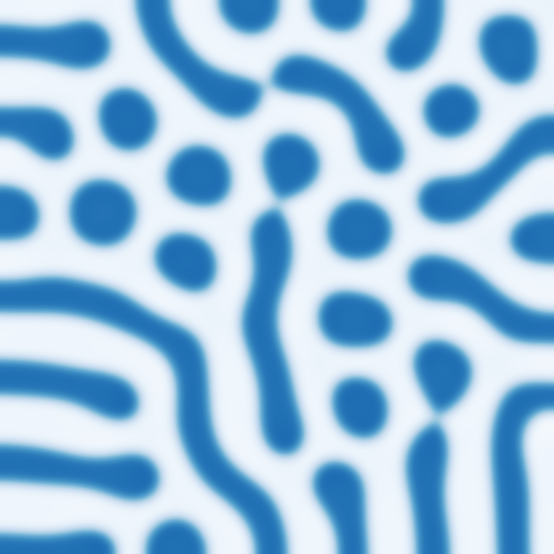}}
             \\
               \Centerstack{\includegraphics[ width=0.088\linewidth ]{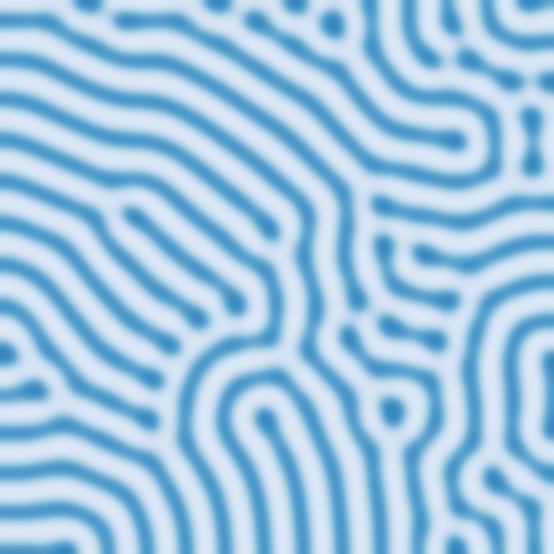}} \\
                 \Centerstack{\includegraphics[ width=0.088\linewidth ]{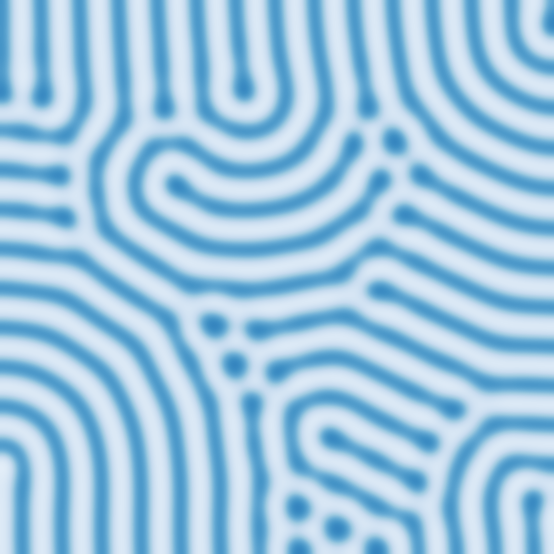}}

            \end{tabular} &
            \begin{tabular}{c} \Centerstack{\includegraphics[ width=0.088\linewidth ]{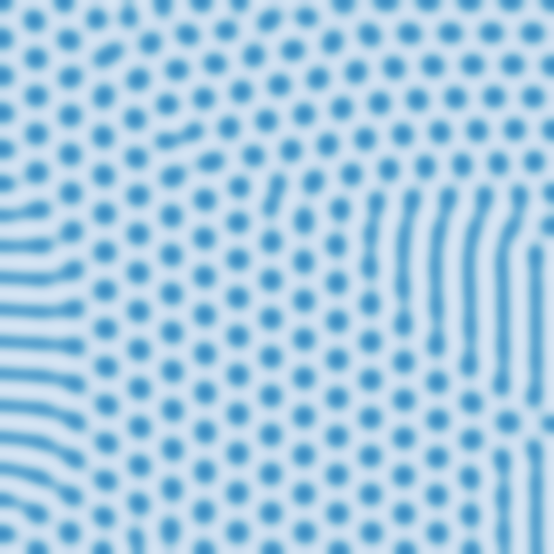}} \\
             \Centerstack{\includegraphics[ width=0.088\linewidth ]{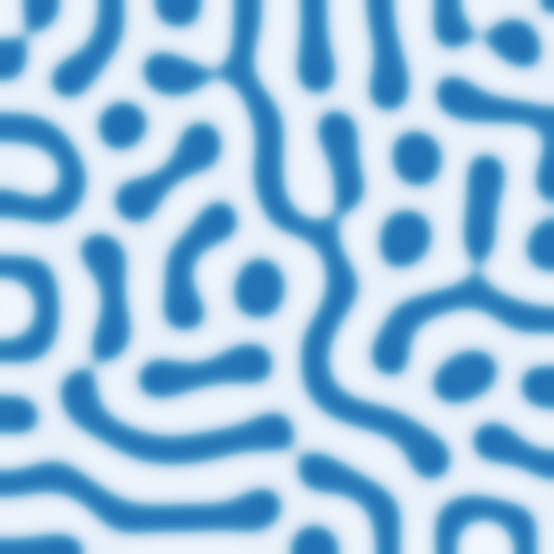}} \\
                 \Centerstack{\includegraphics[ width=0.088\linewidth ]{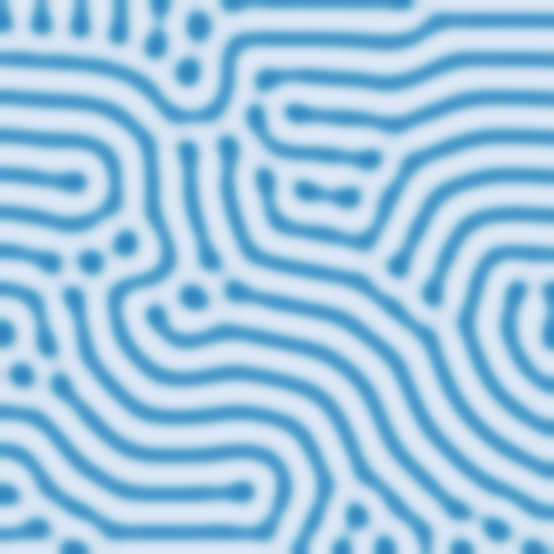}} \\
                 \Centerstack{\includegraphics[ width=0.088\linewidth ]{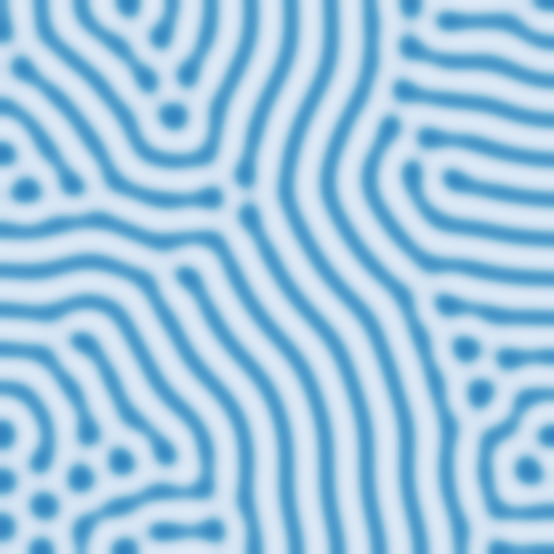}}
            \end{tabular} &
            \begin{tabular}{r} \Centerstack{\includegraphics[ width=0.088\linewidth ]{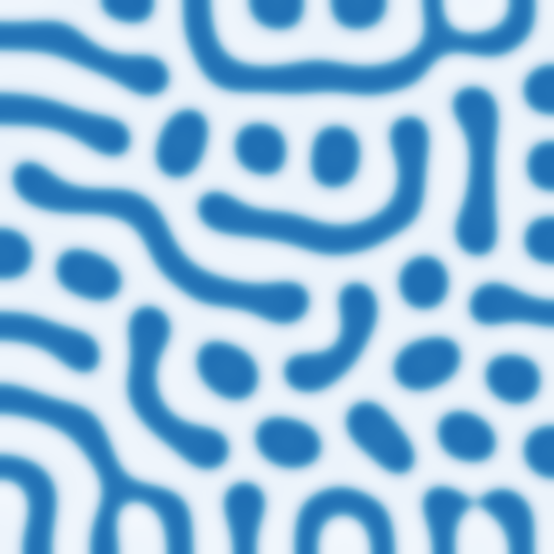}} \\
             \Centerstack{\includegraphics[ width=0.088\linewidth ]{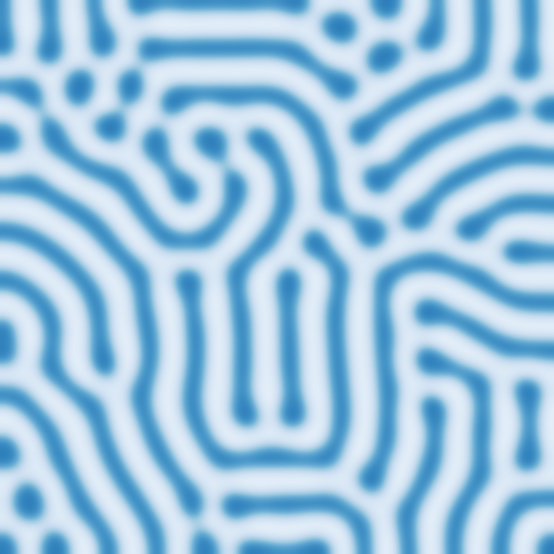}} \\
             \Centerstack{\includegraphics[ width=0.088\linewidth ]{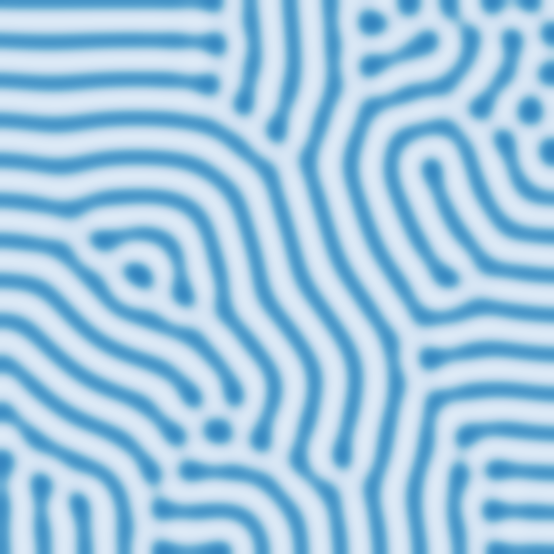}} \\
                 \Centerstack{\includegraphics[ width=0.088\linewidth ]{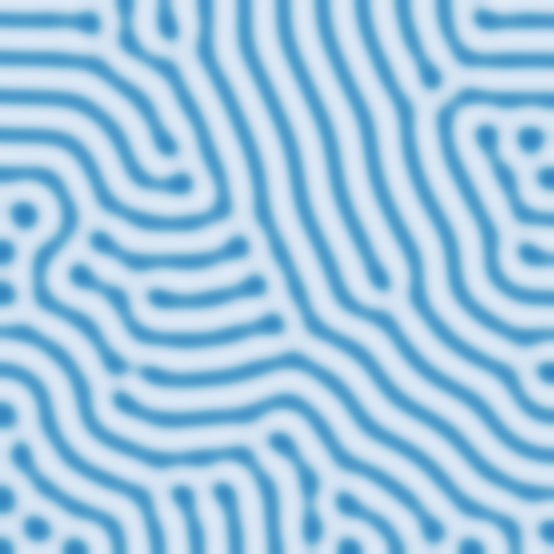}}
            \end{tabular} \\[-10pt] &&&&& \\\hline
&&&&& \\[-6 pt]
        \begin{tabular}{c}\Centerstack{\includegraphics[width=0.375\textwidth]{figure/josh/true_state_150_8.png}}\end{tabular} &
            
             \begin{tabular}{c}$(\mathcal{Q}_2,\mathcal{Q}_4)$ \\ \\\\\\\\\\\\ $\boldsymbol{\mathcal{Q}}_{1:4}$\end{tabular} &
              \begin{tabular}{c}\Centerstack{\includegraphics[width=0.1835\linewidth]{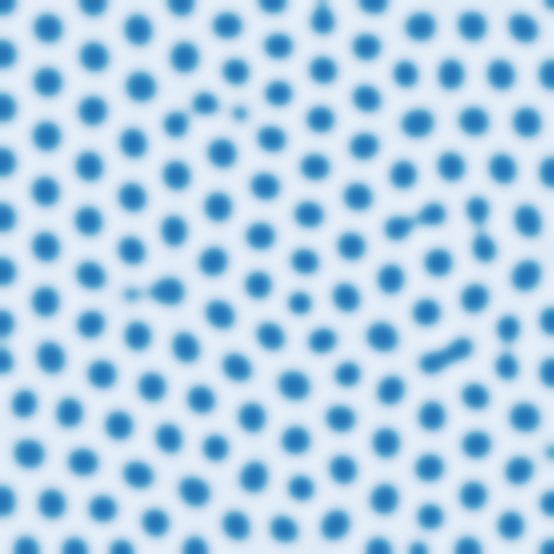}} \\
             \Centerstack{\includegraphics[width=0.1835\linewidth]{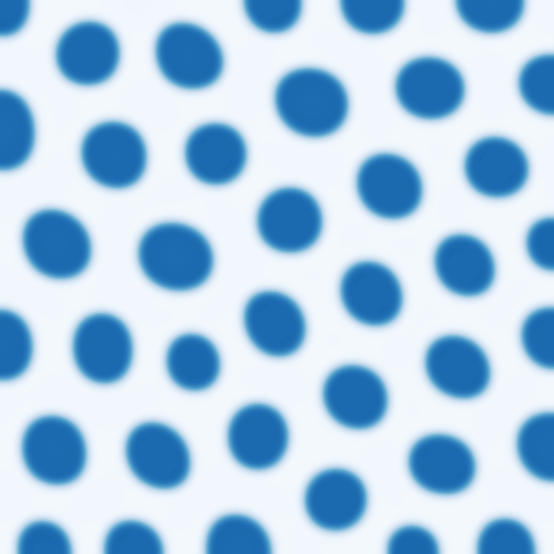}}
            \end{tabular}&
             \begin{tabular}{c} \Centerstack{\includegraphics[ width=0.088\linewidth ]{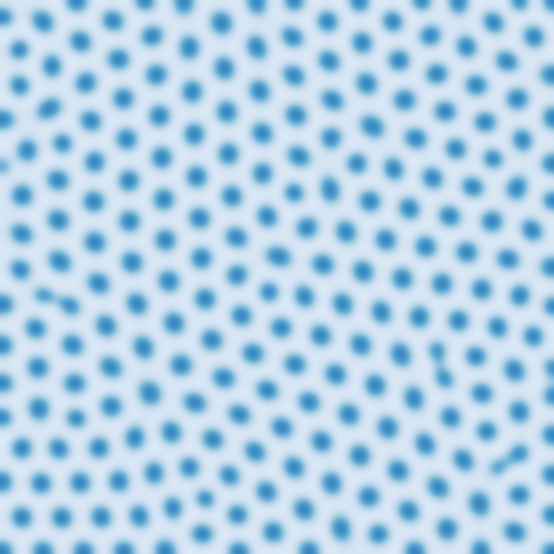}} \\
             \Centerstack{\includegraphics[ width=0.088\linewidth ]{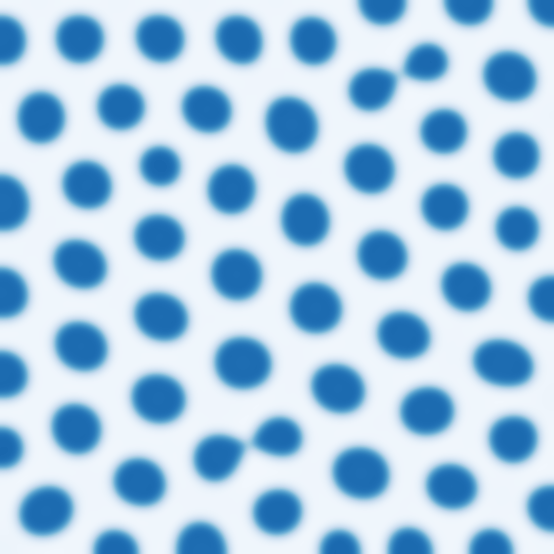}}
             \\
               \Centerstack{\includegraphics[ width=0.088\linewidth ]{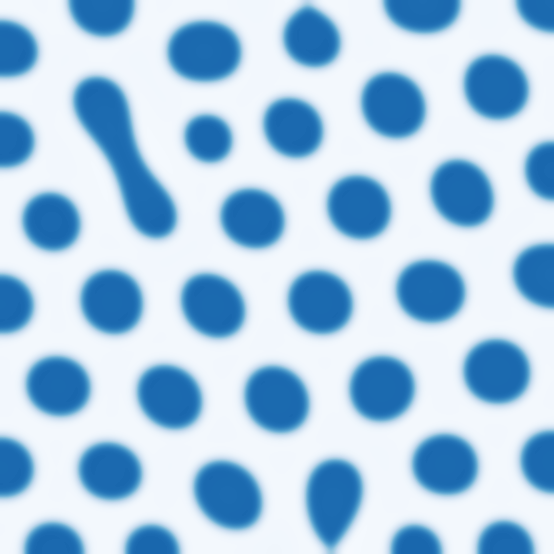}} \\
                 \Centerstack{\includegraphics[ width=0.088\linewidth ]{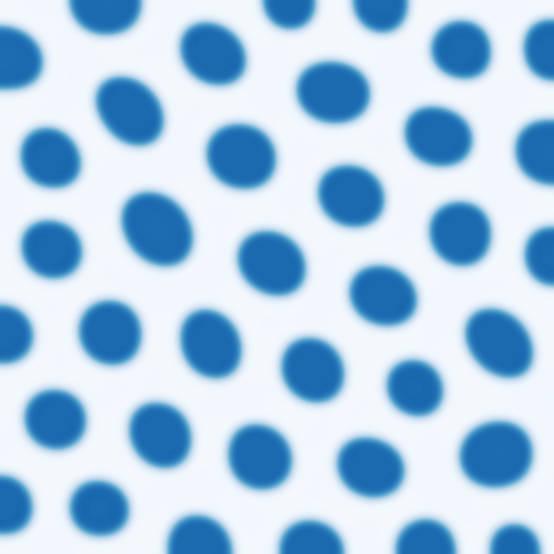}}

            \end{tabular} &
            \begin{tabular}{c} \Centerstack{\includegraphics[width=0.088\linewidth]{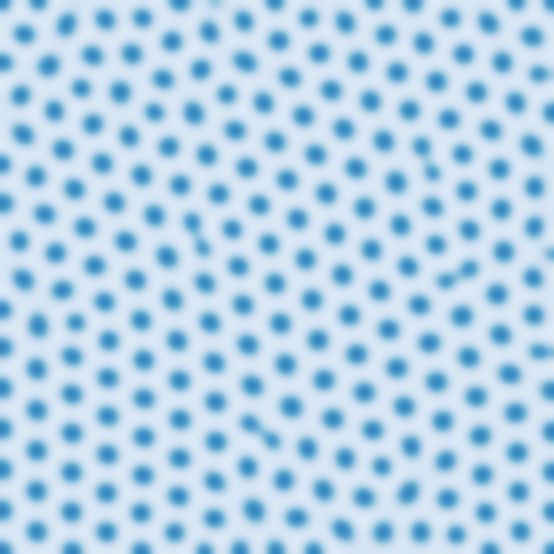}} \\
             \Centerstack{\includegraphics[width=0.088\linewidth]{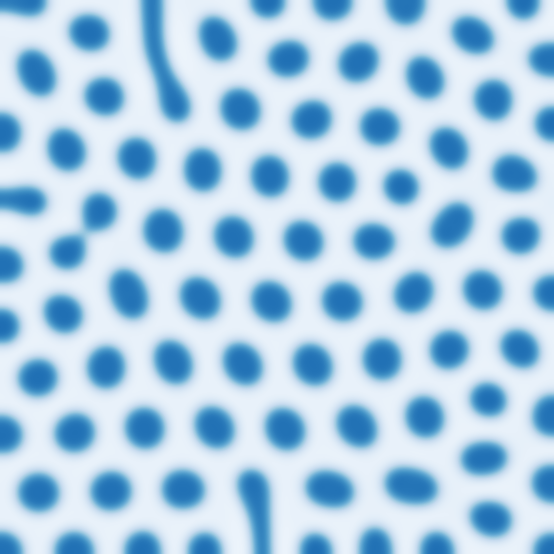}} \\
                 \Centerstack{\includegraphics[width=0.088\linewidth]{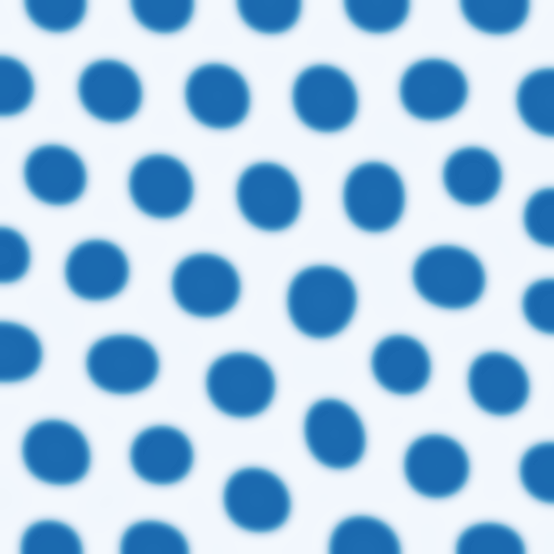}} \\
                 \Centerstack{\includegraphics[width=0.088\linewidth]{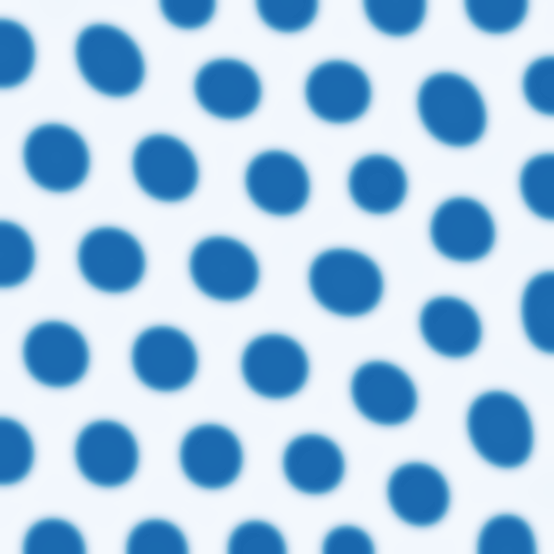}}
            \end{tabular} &
            \begin{tabular}{r} \Centerstack{\includegraphics[width=0.088\linewidth]{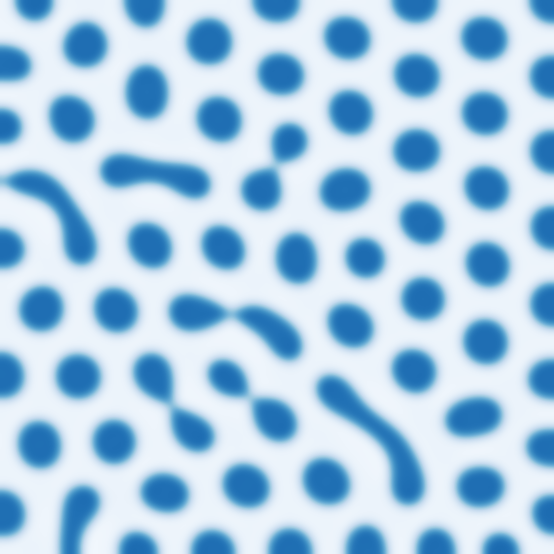}} \\
             \Centerstack{\includegraphics[width=0.088\linewidth]{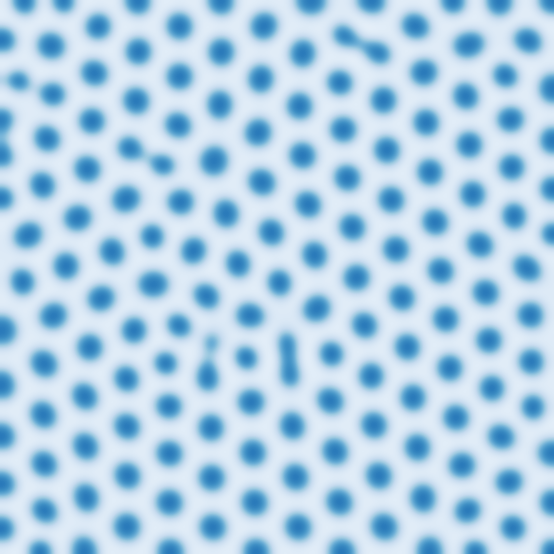}} \\
             \Centerstack{\includegraphics[width=0.088\linewidth]{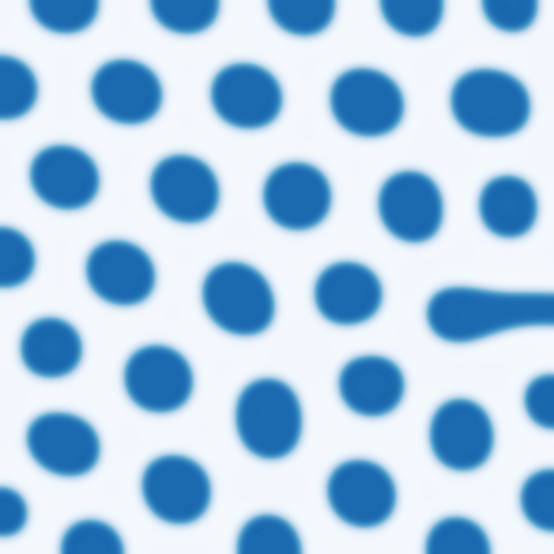}} \\
                 \Centerstack{\includegraphics[width=0.088\linewidth]{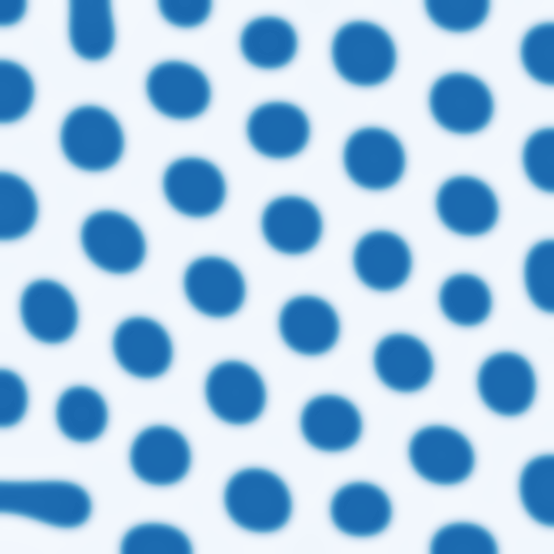}}
            \end{tabular} 
            \\[-10pt] &&&&& \\\hline
&&&&& \\[-6 pt]
        \begin{tabular}{c}\Centerstack{\includegraphics[width=0.375\textwidth]{figure/josh/true_state_352_27.png}}\end{tabular} &
            
             \begin{tabular}{c}$(\mathcal{Q}_2,\mathcal{Q}_4)$ \\ \\\\\\\\\\\\ $\boldsymbol{\mathcal{Q}}_{1:4}$\end{tabular} &
              \begin{tabular}{c}\Centerstack{\includegraphics[width=0.1835\linewidth]{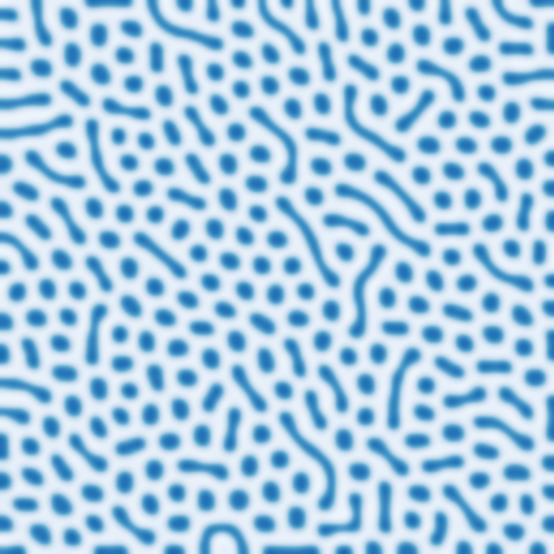}} \\
             \Centerstack{\includegraphics[width=0.1835\linewidth]{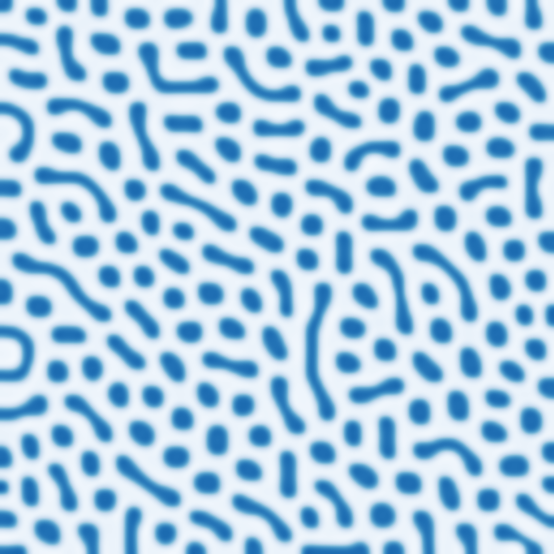}}
            \end{tabular}&
             \begin{tabular}{c} \Centerstack{\includegraphics[ width=0.088\linewidth ]{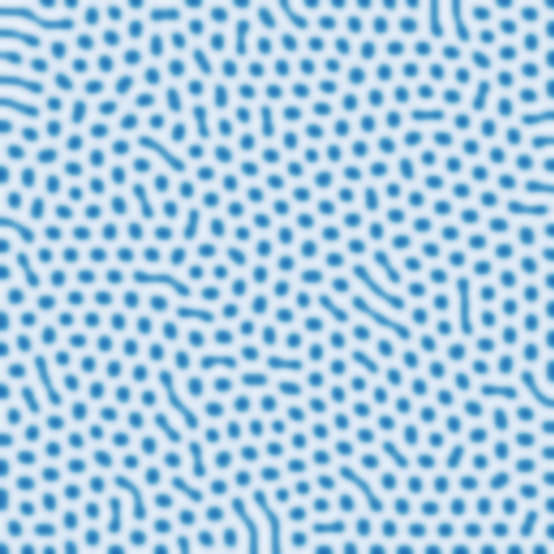}} \\
             \Centerstack{\includegraphics[ width=0.088\linewidth ]{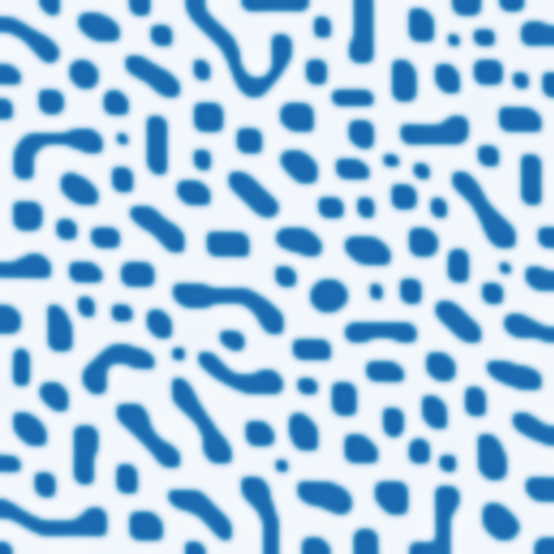}}
             \\
               \Centerstack{\includegraphics[ width=0.088\linewidth ]{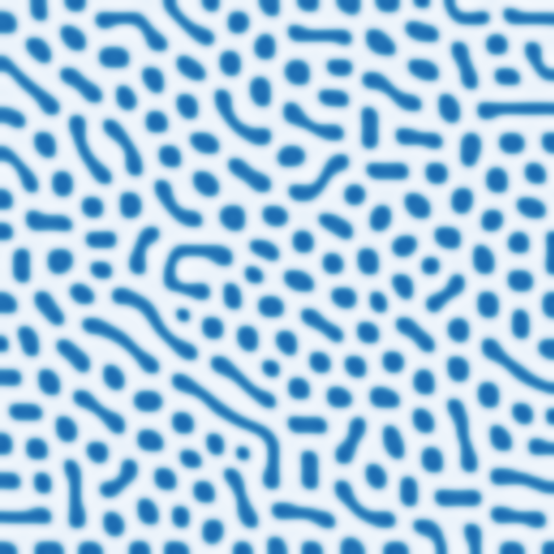}} \\
                 \Centerstack{\includegraphics[ width=0.088\linewidth ]{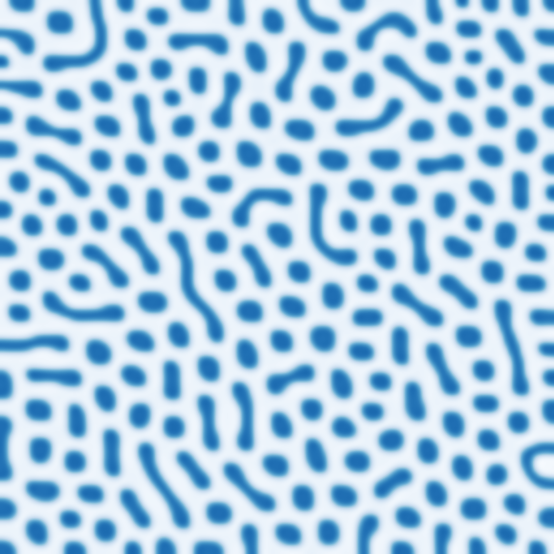}}

            \end{tabular} &
            \begin{tabular}{c} \Centerstack{\includegraphics[ width=0.088\linewidth ]{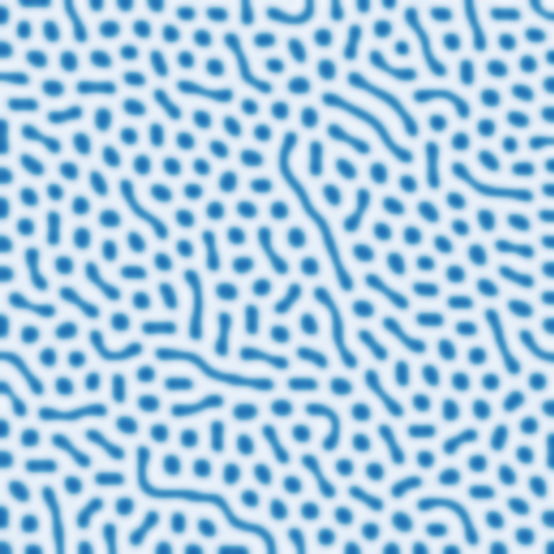}} \\
             \Centerstack{\includegraphics[ width=0.088\linewidth ]{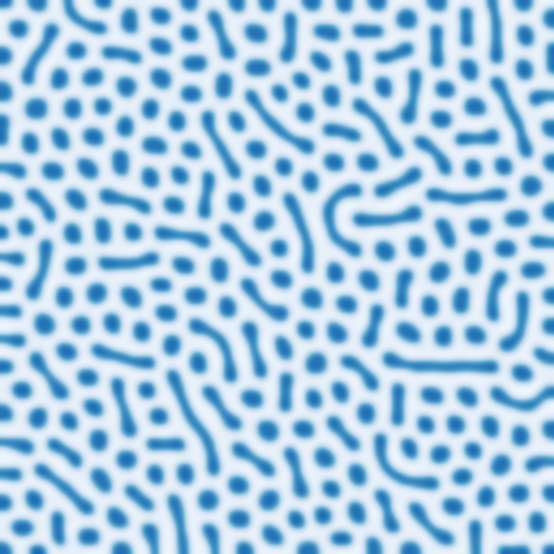}} \\
                 \Centerstack{\includegraphics[ width=0.088\linewidth ]{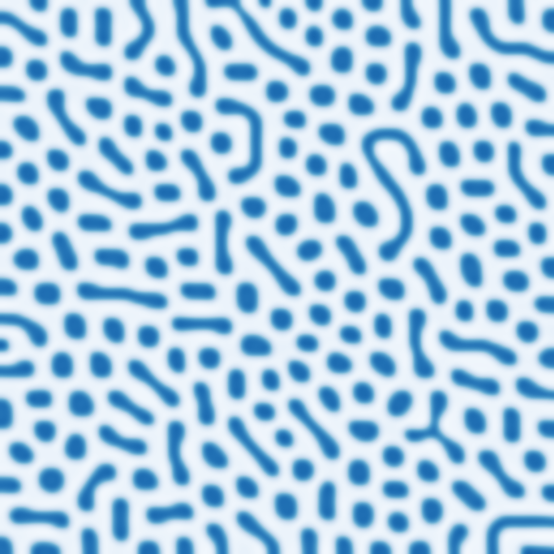}} \\
                 \Centerstack{\includegraphics[ width=0.088\linewidth ]{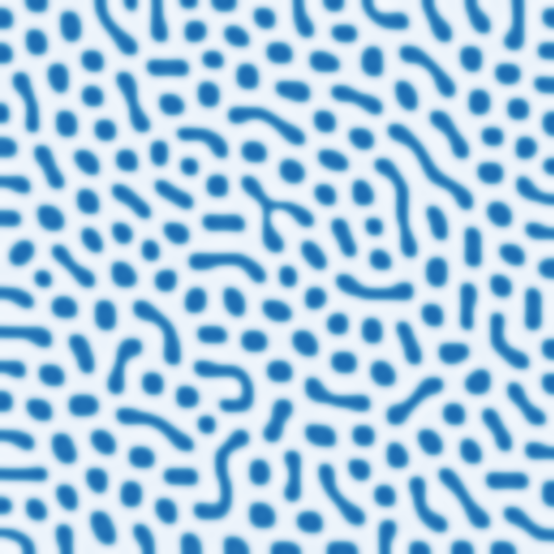}}
            \end{tabular} &
            \begin{tabular}{r} \Centerstack{\includegraphics[ width=0.088\linewidth ]{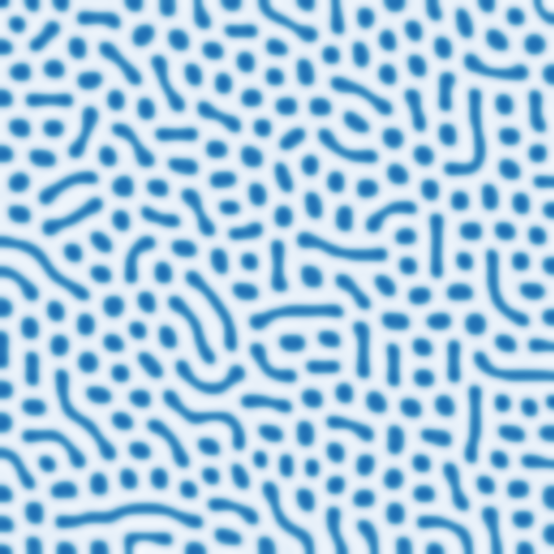}} \\
             \Centerstack{\includegraphics[ width=0.088\linewidth ]{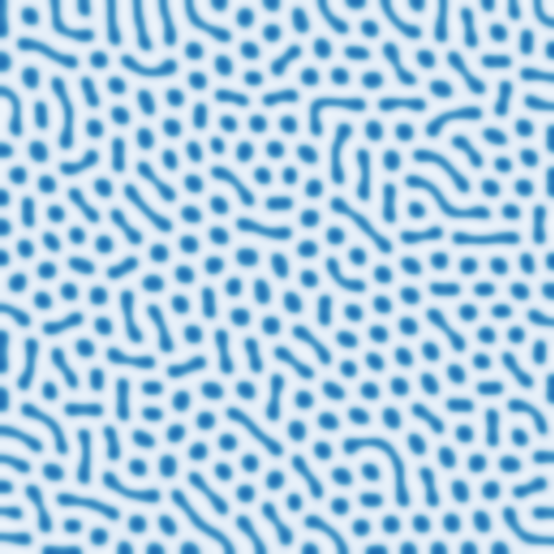}} \\
             \Centerstack{\includegraphics[ width=0.088\linewidth ]{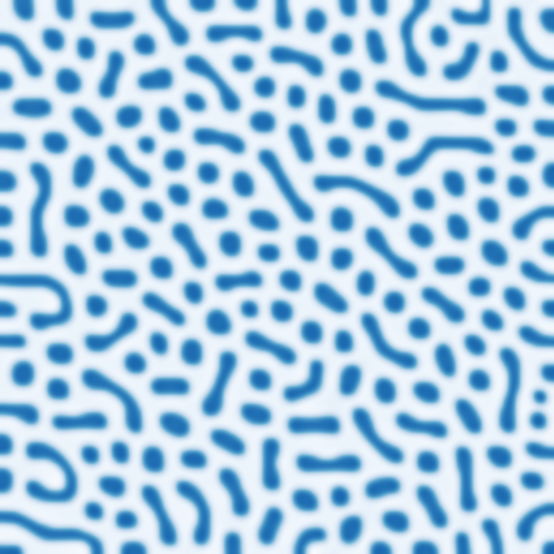}} \\
                 \Centerstack{\includegraphics[ width=0.088\linewidth ]{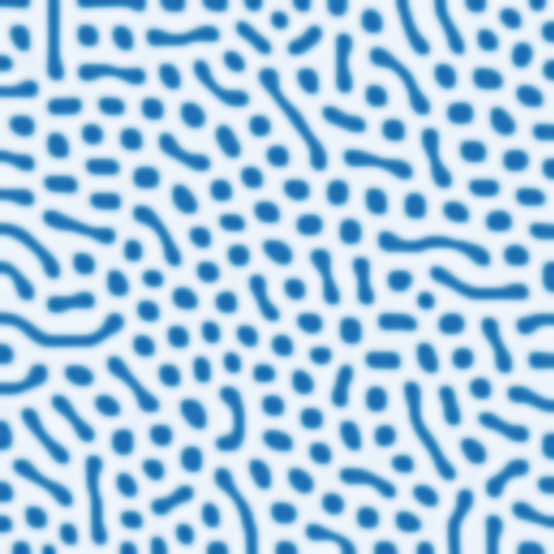}}
    \end{tabular} \\[-10pt] &&&&& \\ \hline\end{tabular}}
    \caption{Posterior predictive samples, $\mathcal{F}(\x, z^i)$, where each image on the right column is generated by independent draws from $\z^i \sim\nu_{Z|M}(\cdot |m)$, The model parameter samples $\x_0$, $\x_{\pm \delta m}$, $\x_{\pm \delta \epsilon}$, and $\x_{\pm\delta\sigma}$ are transformed quadrature points shown in Table~\ref{table:posteriors}.}
    \label{table:posterior_predictions}
\end{table}

Next, we present posterior predictive samples in Tab.~\ref{table:posterior_predictions} in order to examine the downstream effects of Bayesian calibration on model predictions. That is, we generate samples from the posterior predictive distribution 
\begin{equation}
    \nu_{\bstate^{\textrm{pred}} | \bq}(\cdot | \q^*) = \int \nu_{\bstate^{\textrm{pred}} | \bx}(\cdot|\x) \pi_{\bx|\q^*}(\x) \textrm{d}\x.
\end{equation}
The random variable $\bstate^{\textrm{pred}}= \mathcal{F}(\bx |\q^*, \bz)\sim  \nu_{\bstate^{\textrm{pred}} | \bq}(\cdot | \q^*)$ represents the Di-BCP patterns predicted by the forward operator $\mathcal{F}(\cdot, \bz)$ given samples of the calibrated parameters. These posterior predictive samples help us to visually understand the gain in prediction confidence from model calibration. 
 
Generating these samples is accomplished by simulating from the forward operator $\x \mapsto \mathcal{F}(\x, z^i)$ where $z^i \sim \nu_{Z|M}(\cdot |m)$ and $\x$ are the quadrature points in Tab.~\ref{table:posteriors}. The Di-BCP pattern predictions of models calibrated according to the summary statistics $\boldsymbol{Q} = (Q_2, Q_4)$ (\emph{top row} for each $\state^*$) are much more visually dissimilar than the predictions of models calibrated according to the summary statistics $\boldsymbol{Q} = \boldsymbol{Q}_{1:4}$  (\emph{bottom row} for each $\state^*$), regardless of the phases of latent Di-BCP patterns.

Overall, we see that the energy-based summary statistics $\boldsymbol{Q}_{1:4}$ are highly informative of the OK model parameters, despite having a much lower dimension than the image data. We note that sampling from all three posterior distributions in Tab.~\ref{table:posteriors} and~\ref{table:posterior_predictions} is easily accomplished after finding the map $\widehat{S}^{\mathcal{X}}$. Furthermore, evaluating the posterior densities can also be easily accomplished using the map, as discussed in Section \ref{sec:transportLFI}.
\subsection{Expected information gain: Energy-based summary statistics}\label{subsec:EIG_energy}

In this subsection, we present results on the estimation of EIGs for inferring the model parameters from the energy-based summary statistics introduced in Section~\ref{subsec:summary_statistics}.

\begin{figure}[!h]
    \centering
    \includegraphics[width=0.32\textwidth]{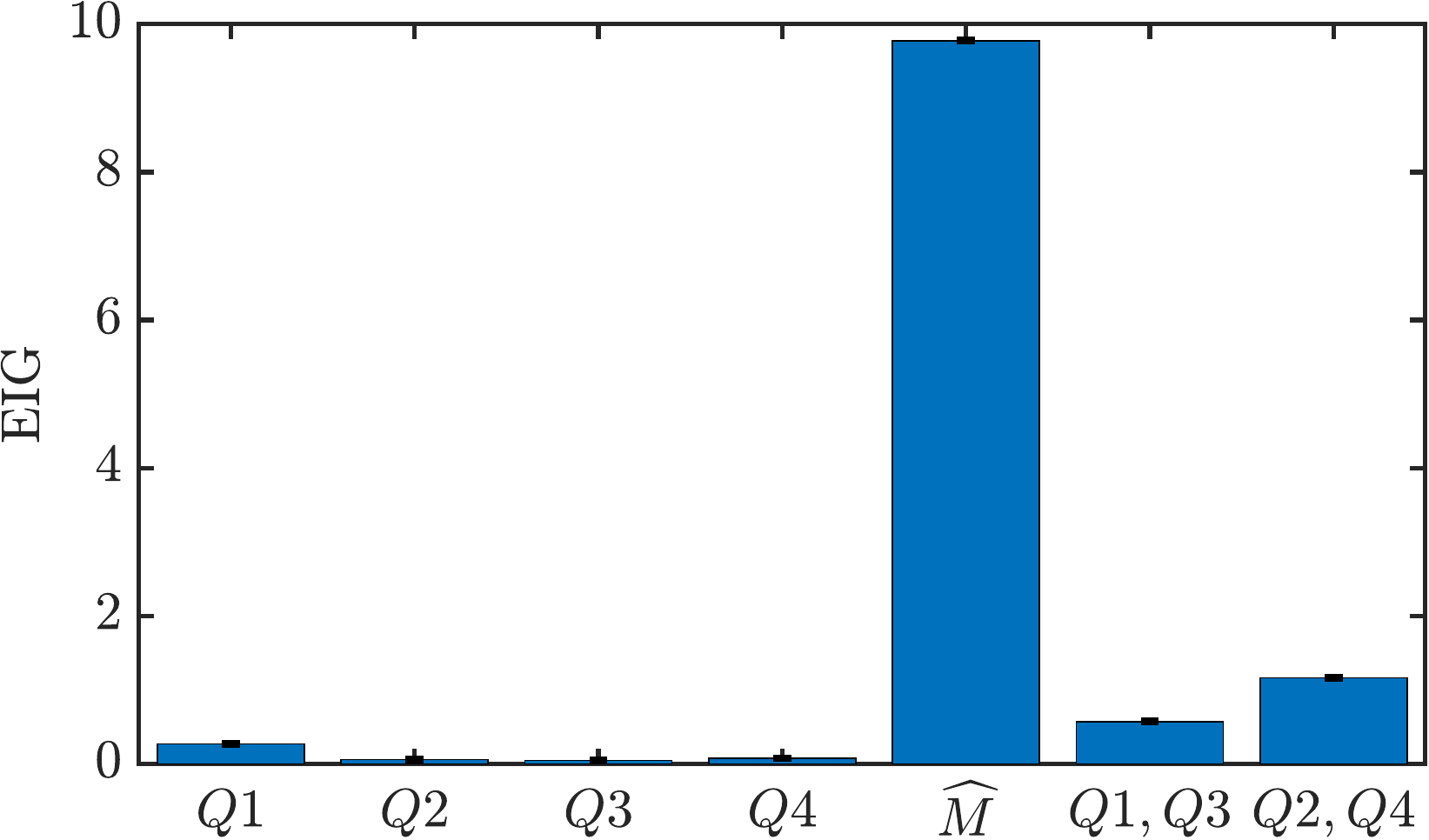}
    \hspace{2pt}
    \includegraphics[width=0.32\textwidth]{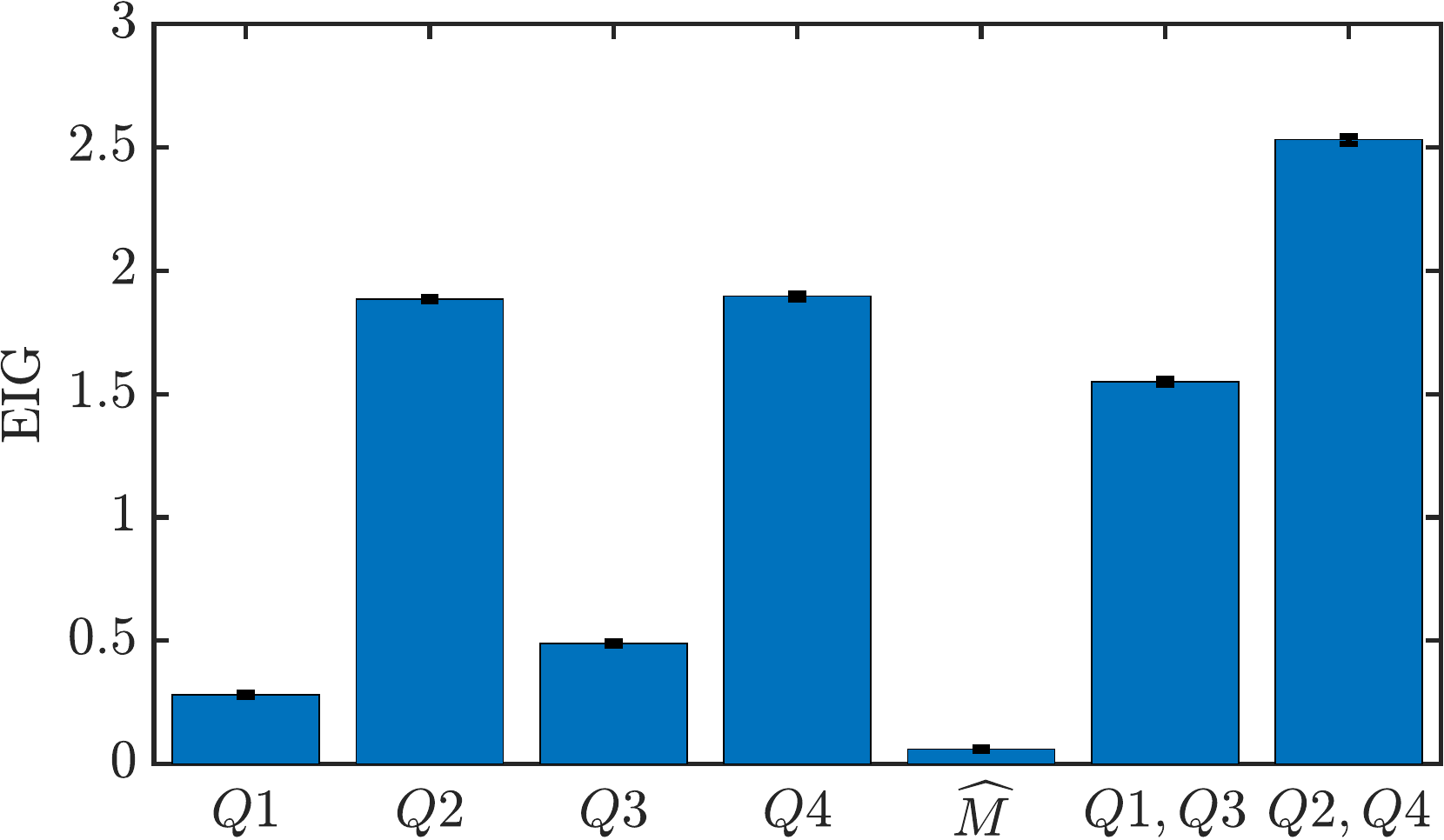}
    \hspace{2pt}
    \includegraphics[width=0.32\textwidth]{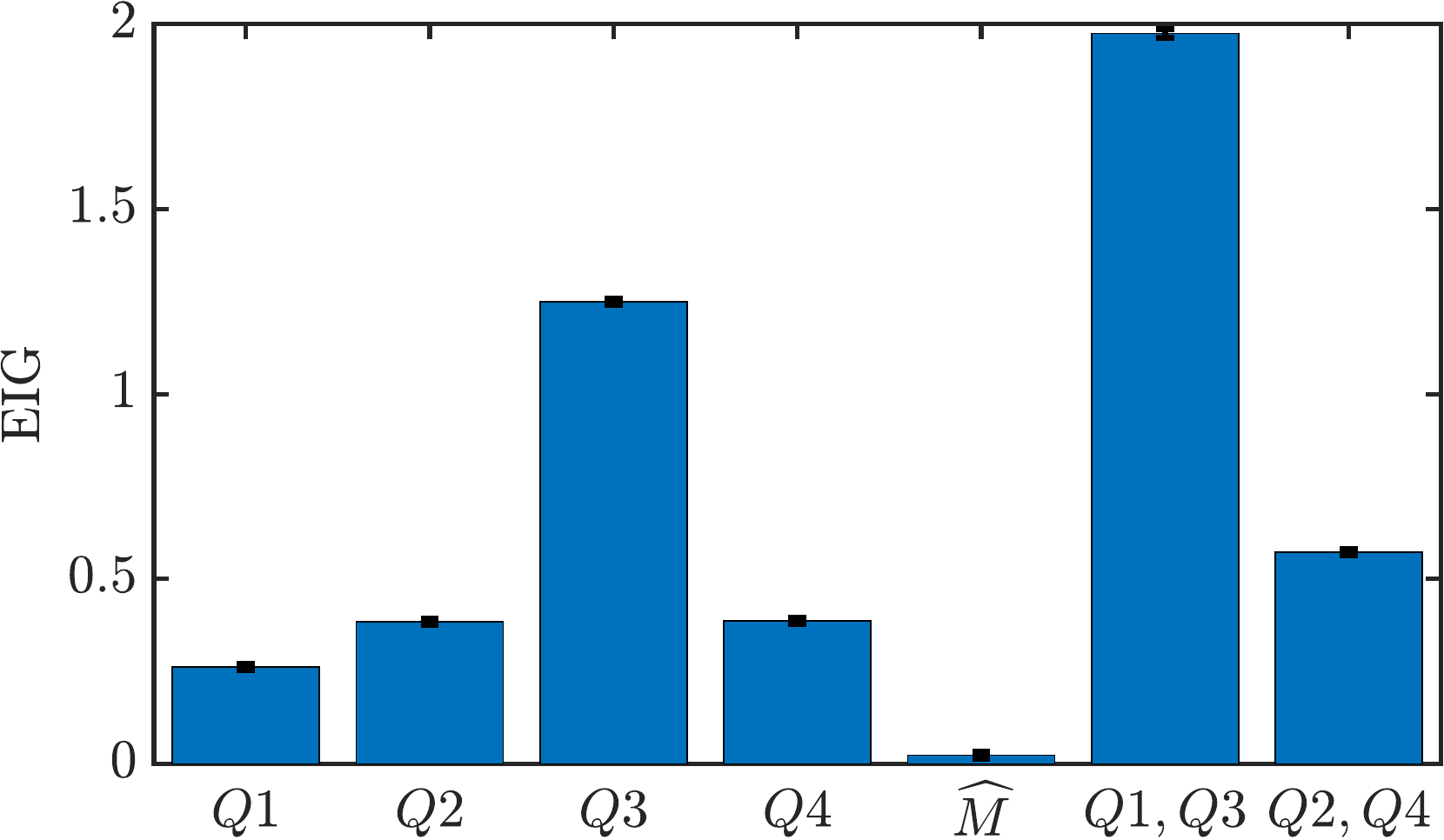}
    \caption{The estimated EIGs for inferring each of the three parameters $\bx = (M,E,\Sigma)$ (\emph{left} to \emph{right}) from (i) each of the energy-based summary statistics $Q_i$, $i = 1,\dots,4$, and $\widehat{M}$, and (ii) pairs of summary statistics, $(Q_1, Q_3)$ and $(Q_2, Q_4)$. We assume no additive noise or blurring corruption in the characterization process. \label{fig:EIG_marginal_parameters_energy_QoI}}
\end{figure}

In Fig.~\ref{fig:EIG_marginal_parameters_energy_QoI},
we plot the estimation of EIGs for inferring each marginal parameter random variable in $\bx =  (M, E, \Sigma)$ from the energy-based summary statistics when observing \emph{raw images} in the reference field of view, i.e., assuming no noise or blurring corruption in the characterization process. The EIG calculations imply that the empirical spatial average summary statistics $\widehat{M}$ strongly informs the spatial average parameter $M$, as expected (\emph{left}). Furthermore, other energy-based summary statistics also jointly inform the spatial average parameter. This corroborates what we learn from the phase diagram in Fig.~\ref{fig:experimental_phase_diagram}: the spatial average parameter, or the volume ratio parameter $f$, can be learned from the morphological phase of the Di-BCP material. The interfacial energy $Q_2$ and/or the total variation energy $Q_4$ strongly inform the interfacial parameter $E$ (\emph{center}), and, finally, the double well energy $Q_1$ and nonlocal energy $Q_3$ jointly strongly inform the nonlocal parameter $\Sigma$ (\emph{right}). The EIG calculations also reveal that even though, marginally, the double well energy $Q_1$ does not strongly inform $\Sigma$, $Q_1$ and $Q_3$ are jointly very informative of $\Sigma$. In fact, the estimated EIG from these two summary statistics, $\widehat{I}(\Sigma; (Q_1,Q_3))$, is larger than the sum of the estimated EIGs from each summary statistics marginally, $\widehat{I}(\Sigma; Q_1) + \widehat{I}(\Sigma; Q_3)$\footnotemark. Similarly, we observe that $Q_1$ and $Q_3$ jointly inform the interfacial parameter $E$ `almost' as much as $Q_2$ or $Q_4$ each informs $E$, even though $Q_1$ and $Q_3$ marginally each provide very little information for learning $E$.
\footnotetext{From information theory, we know that extracting multiple summary statistics jointly is always at least as informative as extracting fewer summary statistics~\cite{cover1999elements}. However, the EIG from multiple summary statistics is not generally the sum of the EIGs from each summary statistics. In fact, the former is sometimes much more that the latter, as is the case here with $Q_1$ and $Q_3$ for informing $\Sigma$.}

\begin{figure}[!ht]
    \centering
    \includegraphics[width=0.32\textwidth]{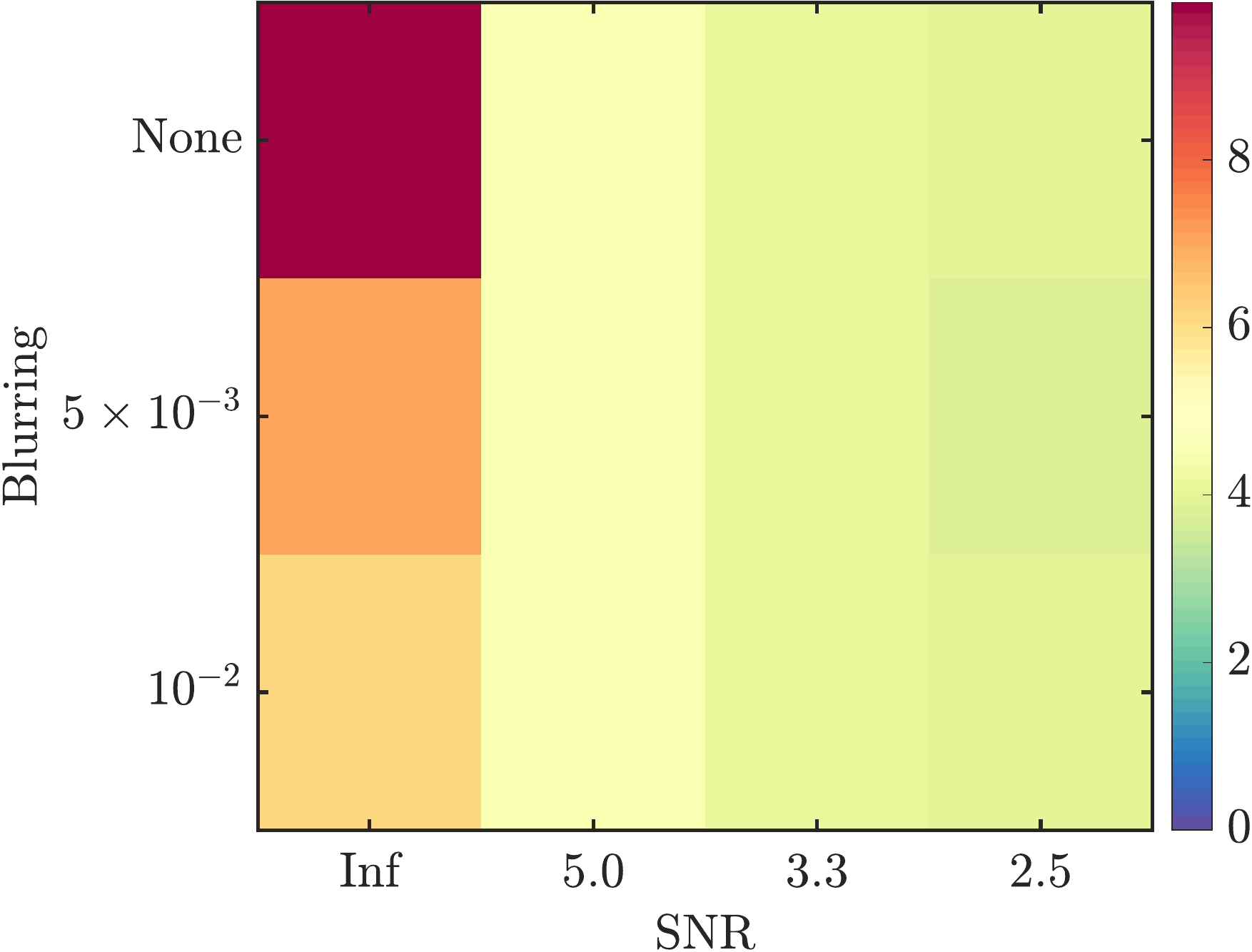}
    \hspace{2pt}
    \includegraphics[width=0.32\textwidth]{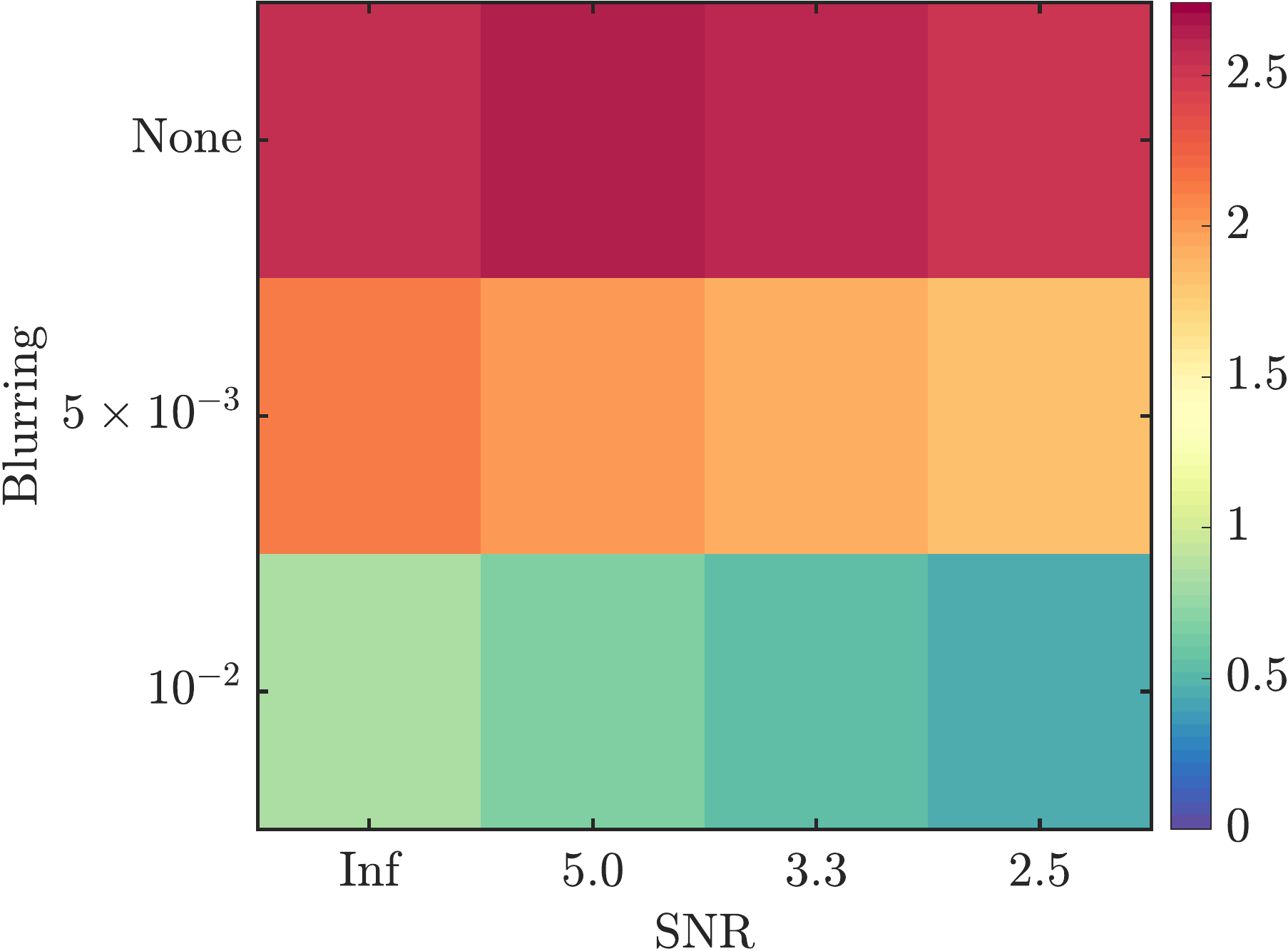}
    \hspace{2pt}
    \includegraphics[width=0.32\textwidth]{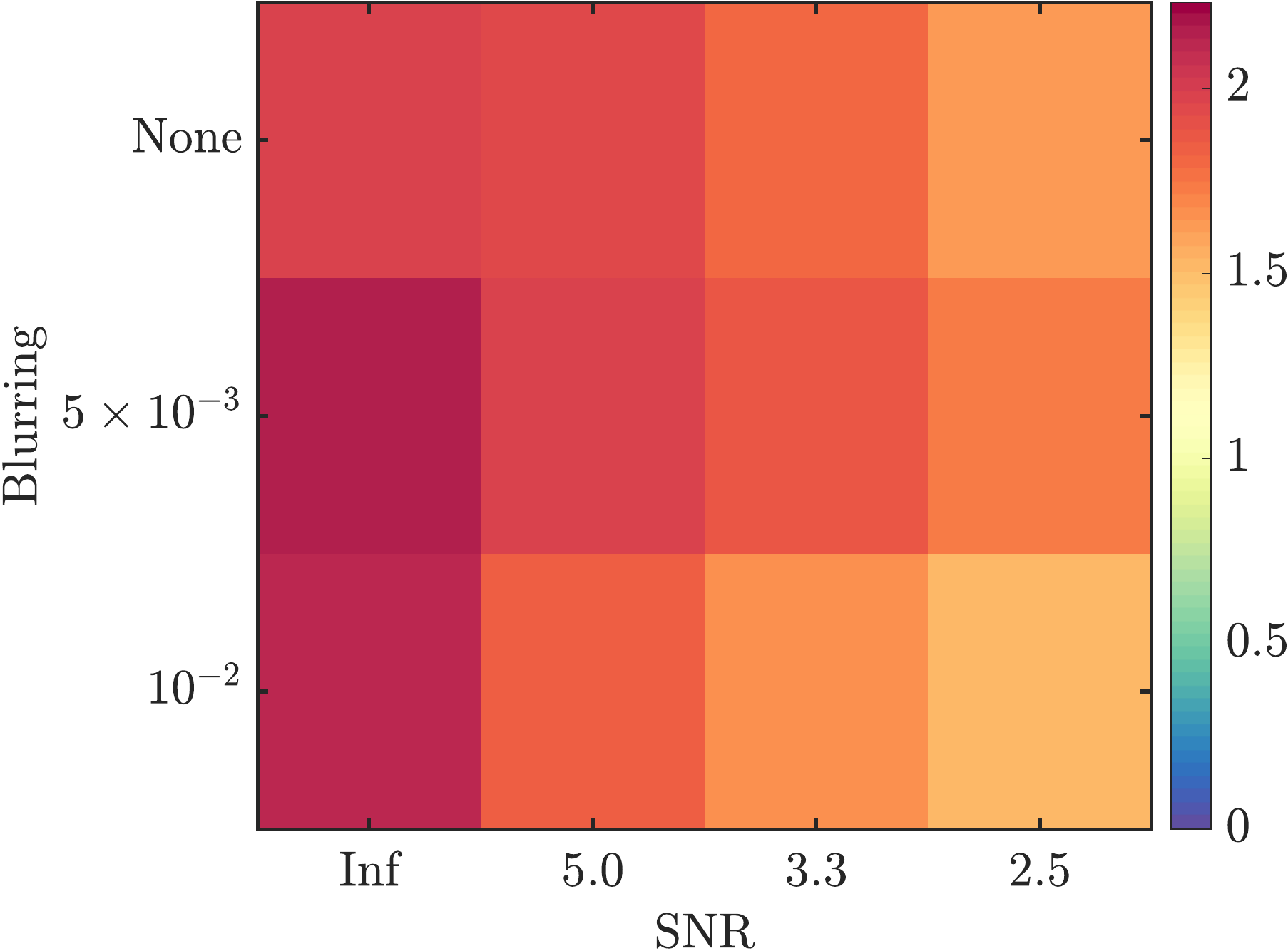}
    \caption{The estimated EIGs for inferring each of the three parameters $\bx = (M,E,\Sigma)$ from its most informative summary statistics according to Fig.~\ref{fig:EIG_marginal_parameters_energy_QoI}, with varying levels of additive noise and blurring corruptions. In particular, we have $\widehat{I}(M;\widehat{M})$ (\emph{left}), $\widehat{I}(E;(Q_2, Q_4))$ (\emph{center}), and $\widehat{I}(\Sigma;(Q_1, Q_3))$ (\emph{right}) . \label{fig:EIG_energyQoI_sensitivity}}
\end{figure}

Next, to investigate the effects of additive noise and blurring corruptions on the informativeness of the energy-based summary statistics, we plot in Fig.~\ref{fig:EIG_energyQoI_sensitivity} the estimated EIGs for inferring each model parameter from its most informative summary statistics according to Fig.~\ref{fig:EIG_marginal_parameters_energy_QoI}, with different levels of additive noise and blurring corruptions specified in Section~\ref{subsec:microscopy_model}. First, we observe that while the empirical spatial average summary statistics $\widehat{M}$ is highly informative of the spatial average parameter in the absence of the noise and blurring corruptions, its informativeness deteriorates by half with the presence of additive noise (\emph{left}), despite the noise being zero mean. Thus, we may have difficulty directly pinpointing the spatial average parameter from the empirical spatial average statistics $\widehat{M}$ of noisy characterization images. On the other hand, the informativeness of $(Q_1,Q_3)$ for inferring $\Sigma$ is not significantly degraded by the various degrees of noise and blurring corruptions (\emph{right}), while the informativeness of $(Q_2, Q_4)$ for inferring $E$ is more significantly degraded by the blurring corruption than by the additive noise corruption (\emph{center}).

The variables of these grids conceptually represent the specifications of the characterization instrument one might be able to choose or control. Ultimately, the EIG computation can help one determine the design of characterization experiments, trading-off the cost of experiments and model calibration accuracy. While some of these quantitative results can sometimes be theoretically deduced without numerical simulations, often these relationships are complex and intractable. Thus it is highly useful to be able to \emph{empirically} extract these relationships with numerical simulations instead. Using the framework for EIG computation introduced in Section~\ref{sec:eig}, each EIG value in the entries of the $3 \times 4$ grids is estimated by a map estimate $\widehat{S}^{\mathcal{X}}$, given the parameter and summary statistics joint samples $(\x^i,\q^i)\sim \nu_{\bx,\bq}$.

Lastly, we note that the EIGs for learning model parameters are examined marginally, although the same computations could have been conducted to estimate the EIG for the three parameters jointly. However, since we were mostly interested in the impact of various summary statistics on learning each individual parameter, we leave EIG studies involving the joint parameter $\bx = (M, E, \Sigma)$ for future study.
\subsection{Expected information gain: Fourier-based summary statistics} \label{subsec:EIG_fourier}

\begin{figure}[!ht]
    \centering
    \includegraphics[width=0.32\textwidth]{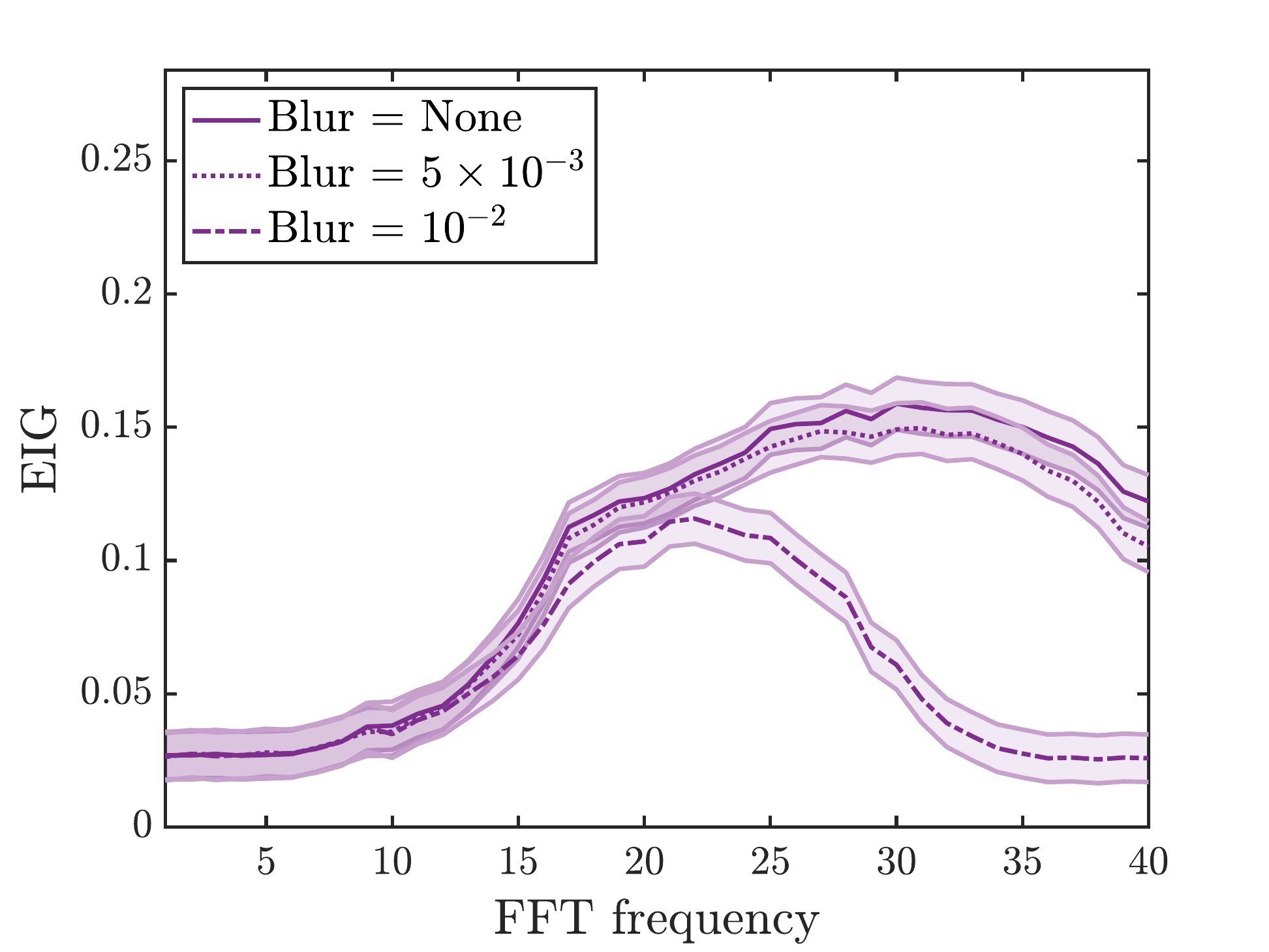}
    \includegraphics[width=0.32\textwidth]{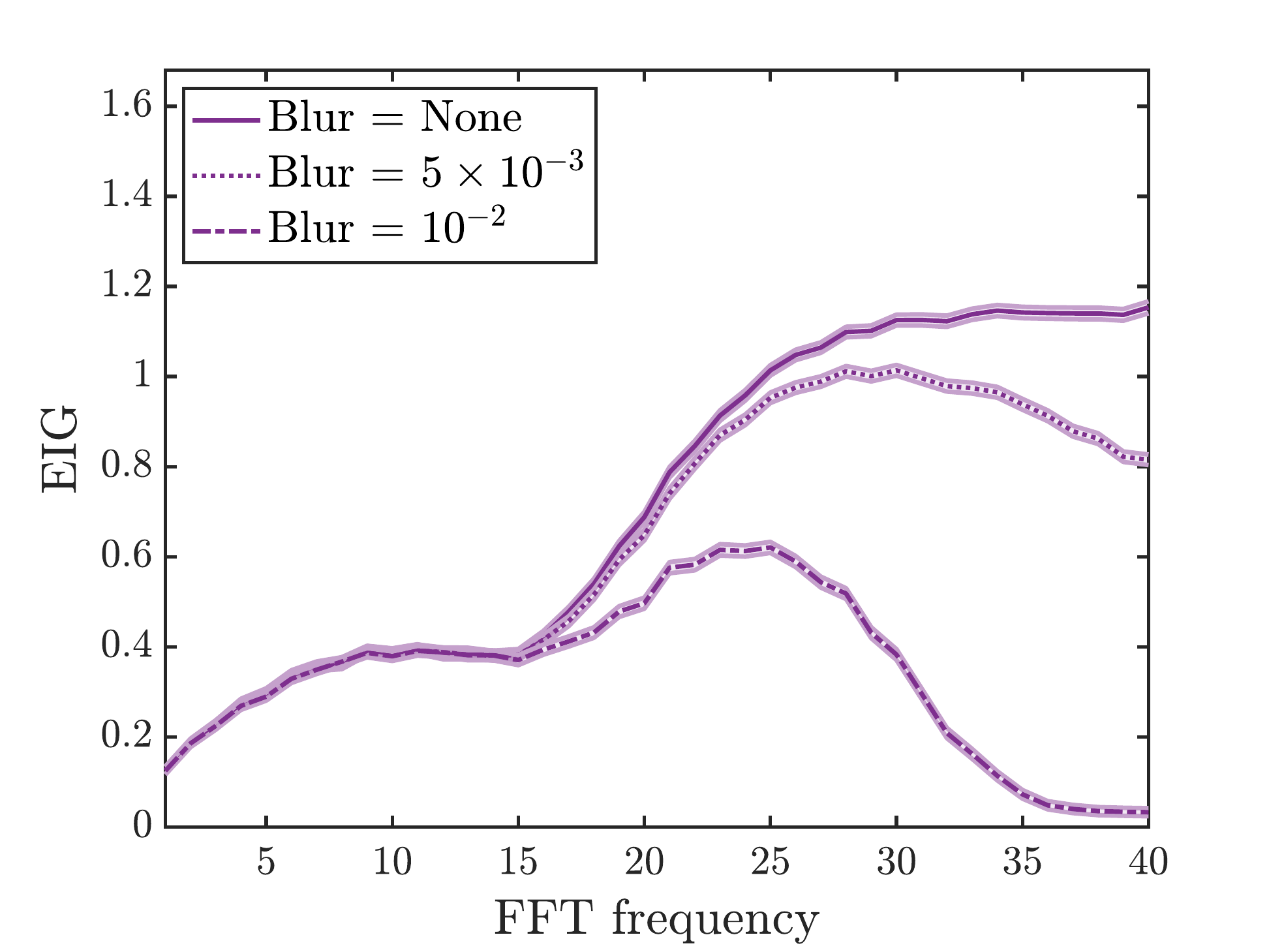}
    \includegraphics[width=0.32\textwidth]{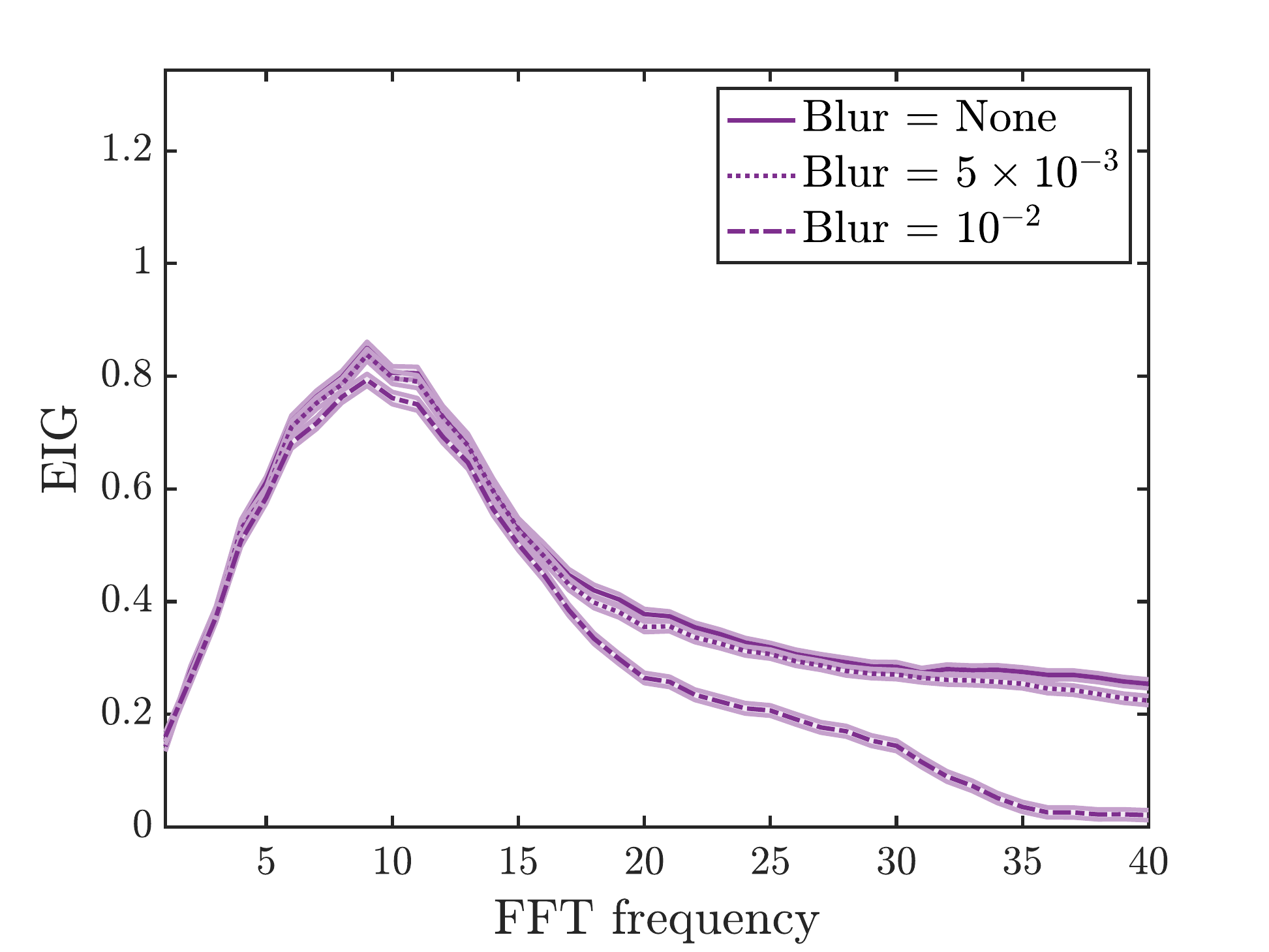}
    \caption{The estimated EIGs for inferring each of the three parameters $\bx = (M,E,\Sigma)$ (from \emph{left} to \emph{right}) from $\mathcal{Q}_{a,i}$, $i = 1,\dots,40$, with increasing levels of the blurring corruption and a fixed level of the additive noise corruption at $\text{SNR} = 5.0$. \label{fig:EIG_maps_FFT}}
\end{figure}

\begin{figure}[!ht]
    \centering
    \includegraphics[width=0.32\textwidth]{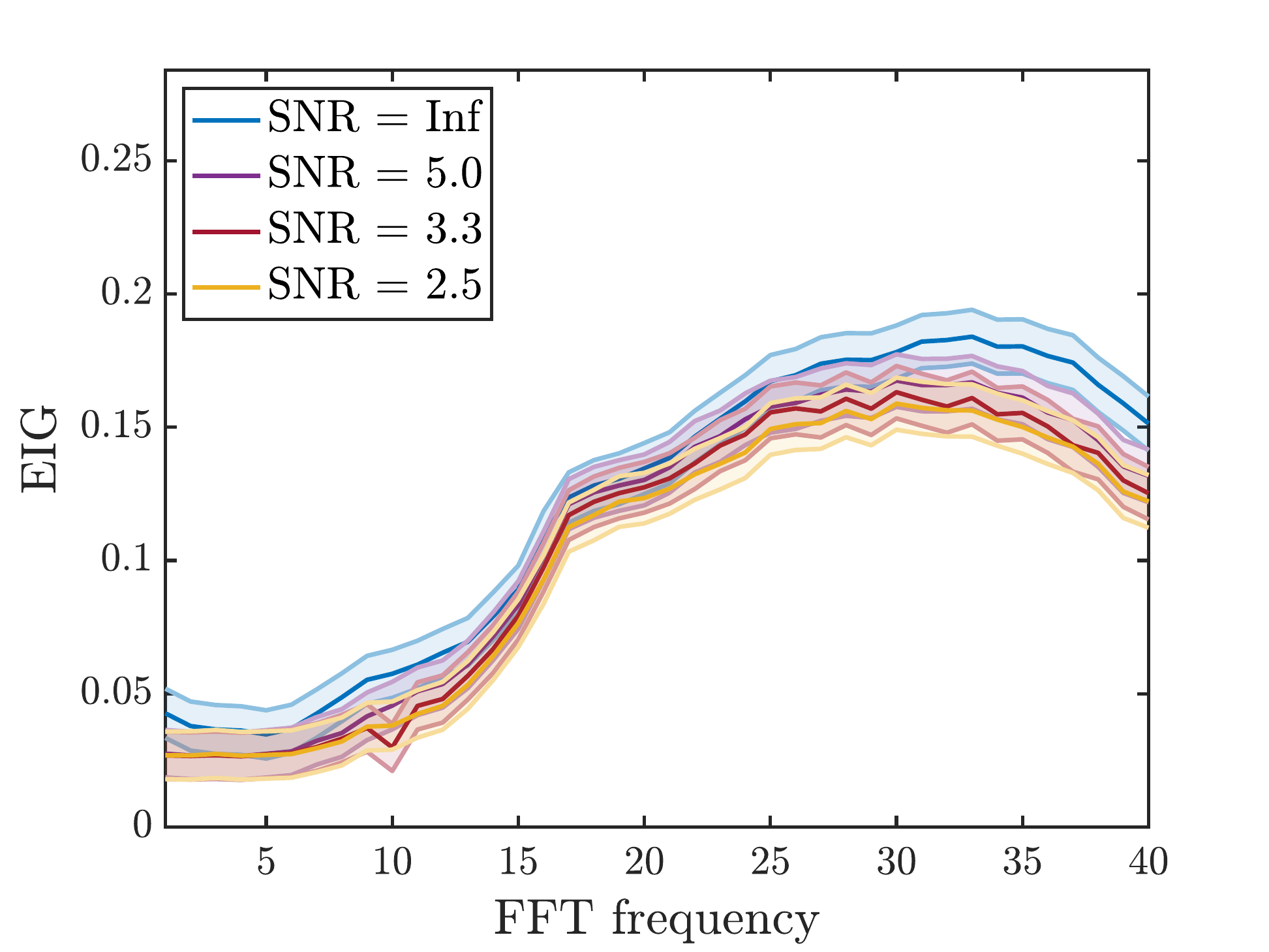}
    \includegraphics[width=0.32\textwidth]{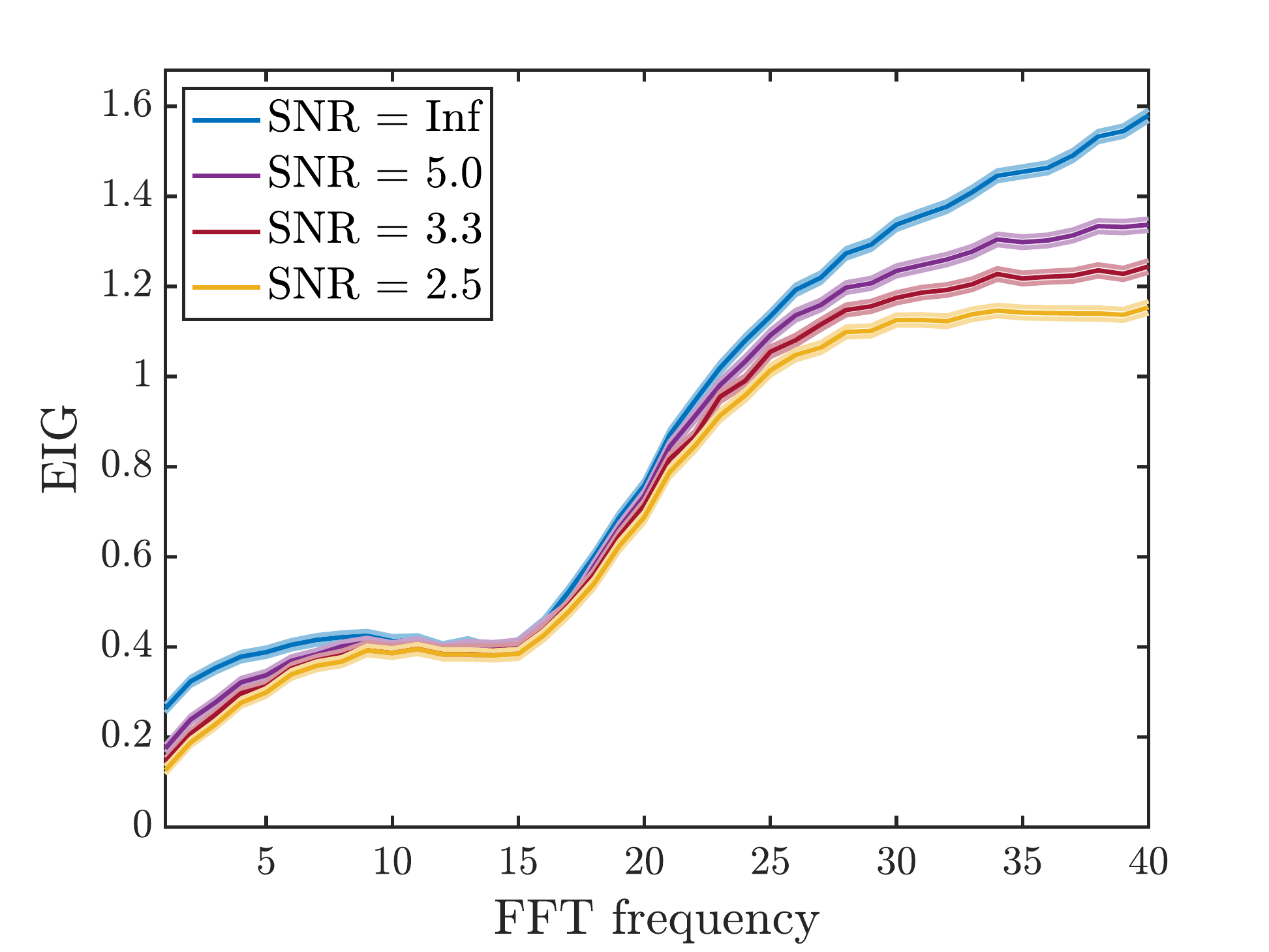}
    \includegraphics[width=0.32\textwidth]{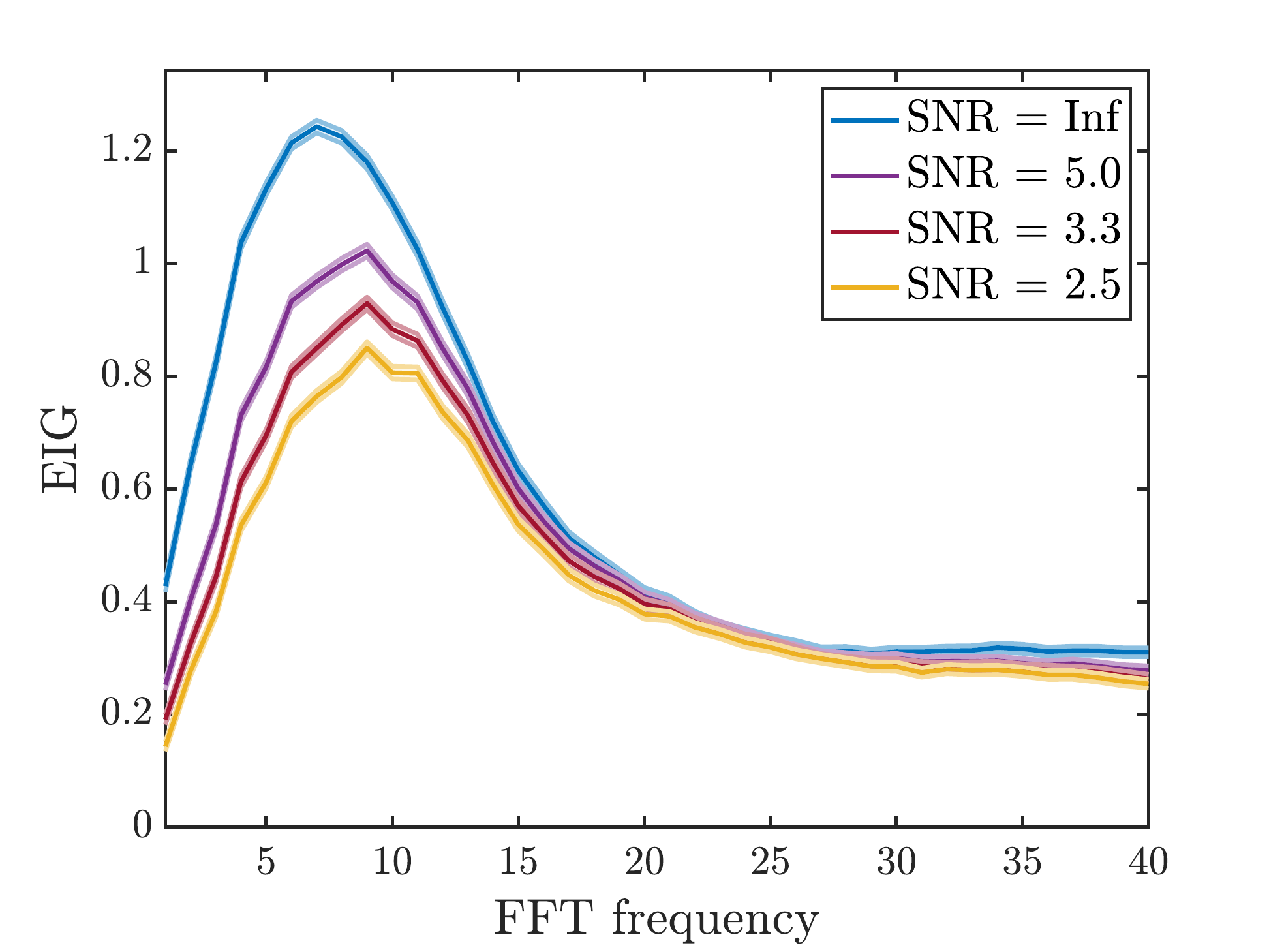}
    \caption{The estimated EIGs for inferring each of the three parameters $\bx = (M,E,\Sigma)$ (from \emph{left} to \emph{right}) from $\mathcal{Q}_{a,i}$, $i = 1,\dots,40$, with decreasing levels of the additive noise corruption and no blurring corruption. \label{fig:EIG_FFT_vs_snr}}
\end{figure}

Next, we computed the EIG from the Fourier-based summary statistics for learning each of the three parameters. In particular, we consider $\bq_a = \tilde{\mathcal{P}}_a(\by)\in \R^{40}$, as defined in \eqref{eq:aaps_interval}, where each entry corresponds to the averaged magnitude spectrum value in a unit interval of the radial frequency. Let $Q_{a,i}$ denote the $i$th component of $\bq_a$. In Fig.~\ref{fig:EIG_maps_FFT} and~\ref{fig:EIG_FFT_vs_snr}, we show the estimated EIGs for inferring marginally each of the three parameters from each component of $\boldsymbol{Q}_a$, with various degrees of additive noise and blurring corruptions.

With the presence of the additive noise corruption, the informativeness of the averaged magnitude spectrum decreases, while the blurring corruption primarily reduces the information contained in the averaged magnitude spectrum at higher radial frequencies. Both the blurring and noise corruptions tend to more strongly impact the EIG of the spectrum at frequencies that are \emph{most informative} for each parameter, with the exception of the EIG for $\Sigma$ subjects to blurring corruptions (Fig.~\ref{fig:EIG_maps_FFT}, \emph{right}). Moreover, the spectrum at higher frequencies generally tend to inform $M$ and $E$ more while the spectrum at lower frequencies ($\sim 10$) inform $\Sigma$ more, so that the spectrum at both higher and lower frequencies play a non-negligible role in informing the model parameters.

\begin{figure}[H]
    \centering
    \includegraphics[width=0.4\textwidth]{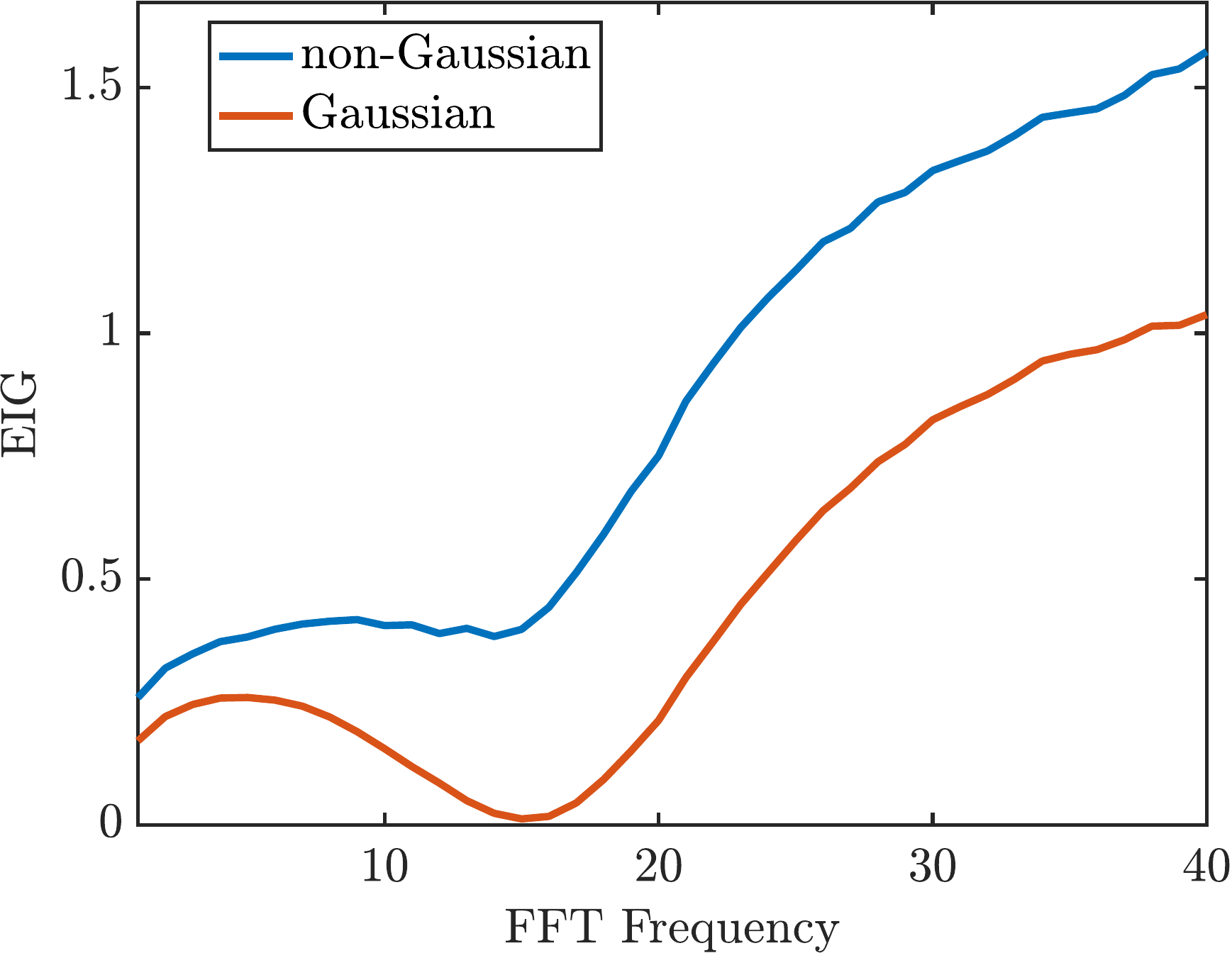}
    \hspace{0.5cm}
    \includegraphics[width=0.4\textwidth]{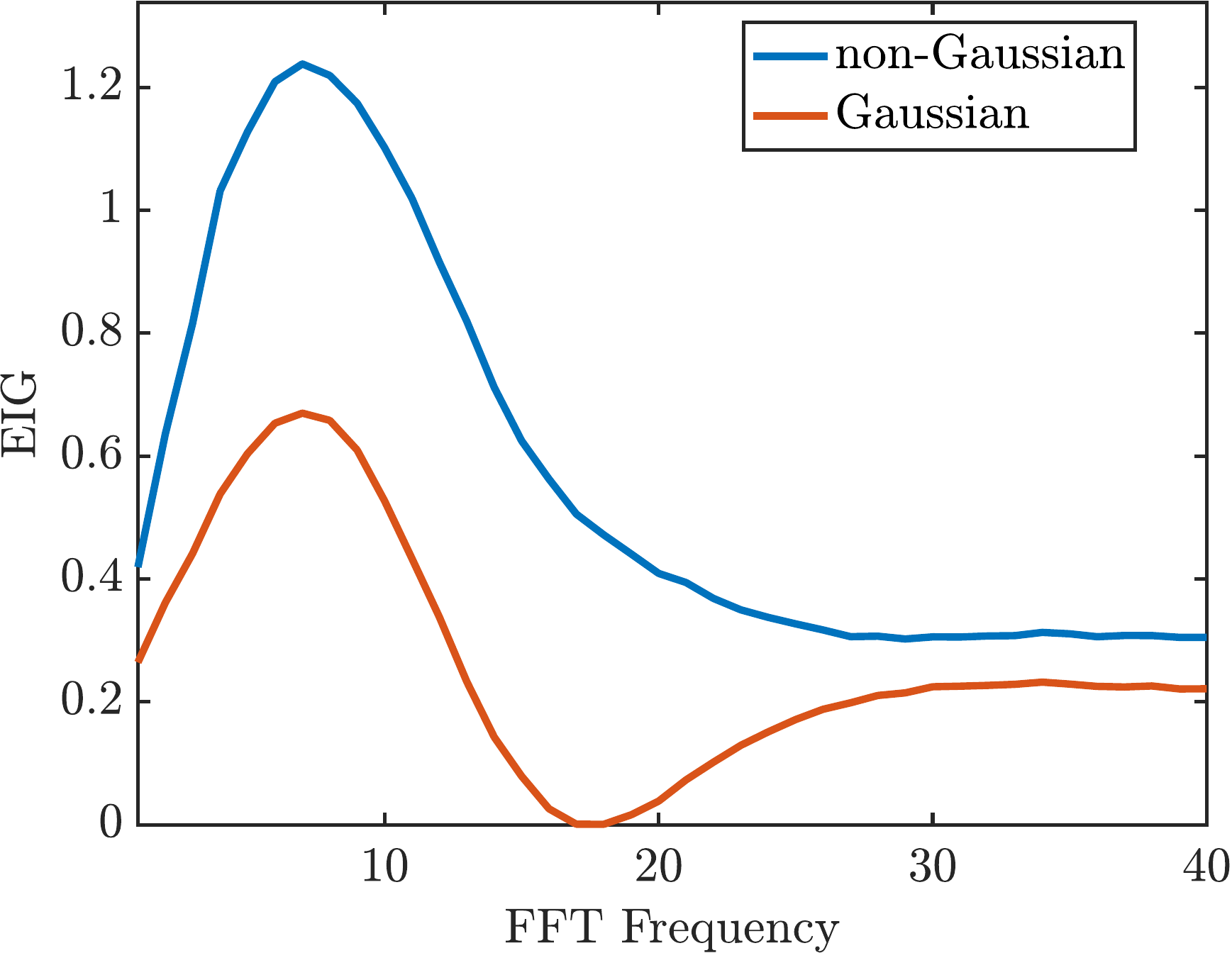}
    \caption{The estimated EIGs for inference of $E$ (\emph{left}) and $\Sigma$ (\emph{right}) from each entry of the averaged magnitude spectrum $Q_{a,i}$, $i=0,\dots, 40$. The Gaussian approximations to the joint density lead to under-predictions of informativeness. \label{fig:Gaussian_vs_nonGaussian_FFT}}
\end{figure}

For the remainder of this subsection, we address the difference in EIG approximation quality when using the parametric transport framework laid out in Section~\ref{sec:eig}, compared to a traditional Gaussian approximation based EIG approximation. In Fig.~\ref{fig:Gaussian_vs_nonGaussian_FFT}, the EIG computed with our non-Gaussian parametric approach is compared to the EIG computed via a Gaussian approximation of the posterior densities. As expected from expression~\eqref{eq:EIG_lowerbound}, the EIG estimator with a variational approximation to the posterior density is negatively biased and this bias increases prediction for an approximate density that has a larger discrepancy with respect to the true posterior (see Section~\ref{subsec:lfi_EIG}). For some components of the averaged magnitude spectrum, this results in an estimated EIG based on the Gaussian approximation (\emph{red}) that is near zero, while the actual information gain in those components (\emph{blue}) is quite larger as seen from the non-Gaussian approximation. This kind of bias can hamper experimental design decision-making that relies on EIG estimations. 

\begin{figure}[!ht]
    \centering
    \includegraphics[width=0.31\textwidth]{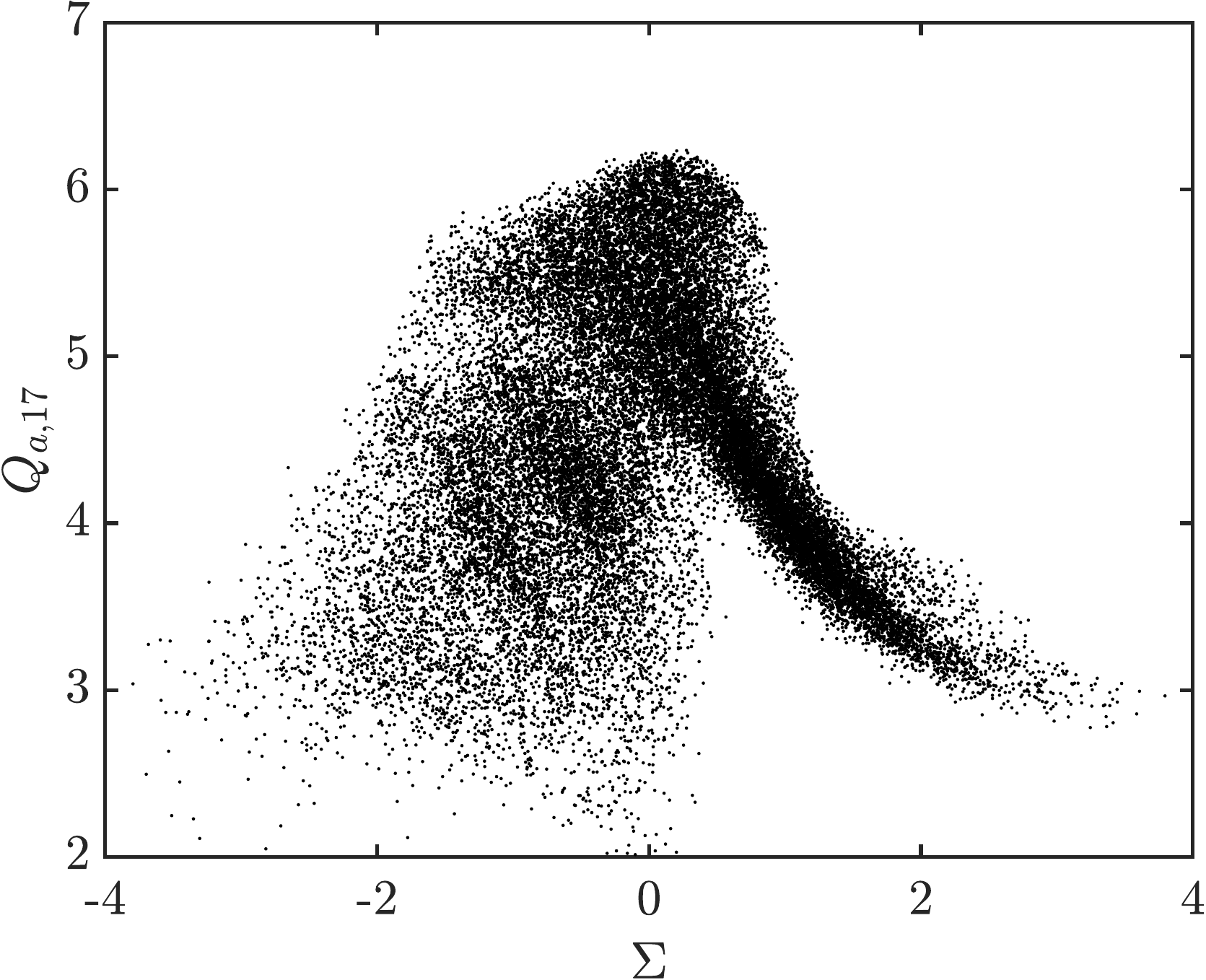}
    \hspace{2pt}
    \includegraphics[width=0.32\textwidth]{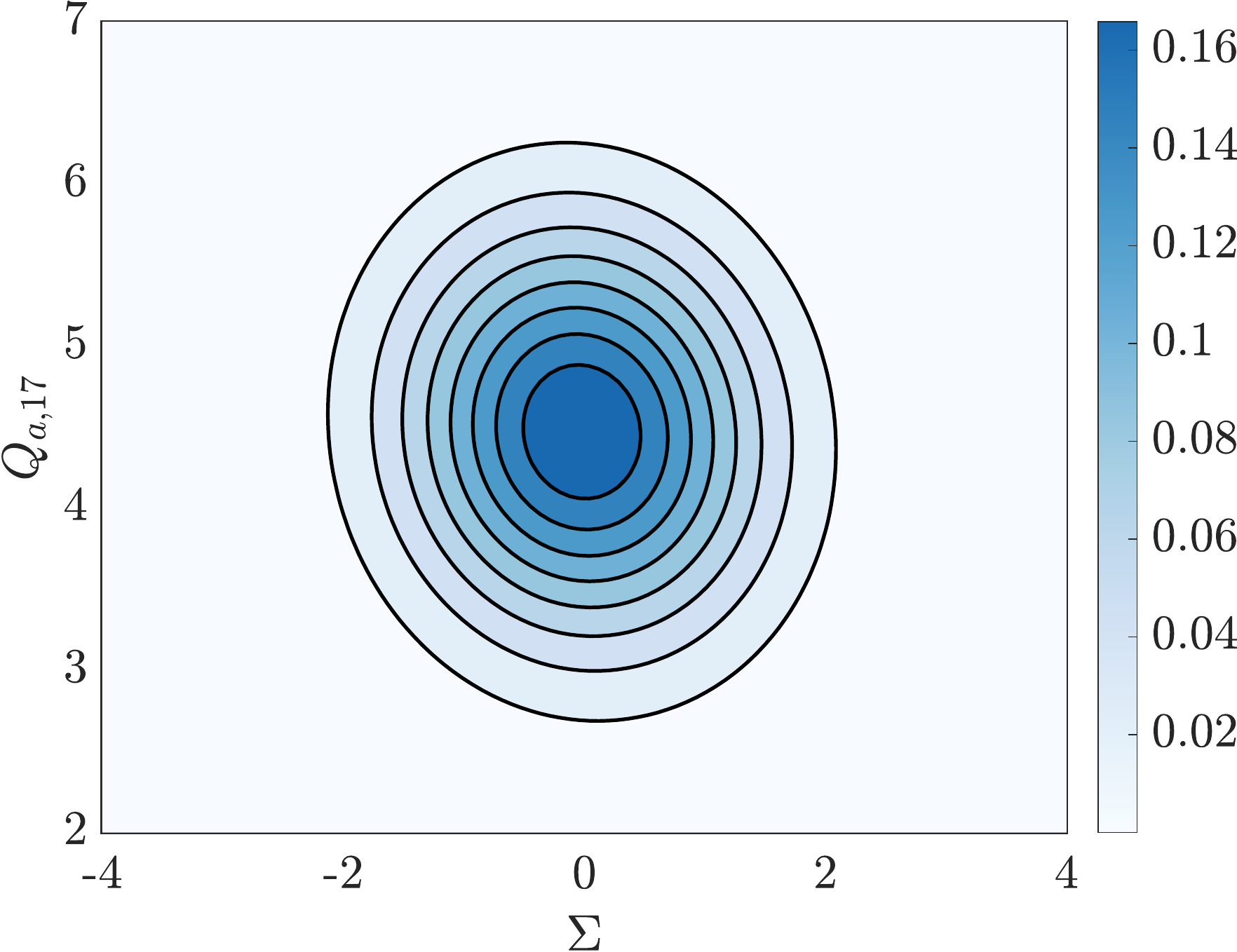}
    \hspace{2pt}
    \includegraphics[width=0.32\textwidth]{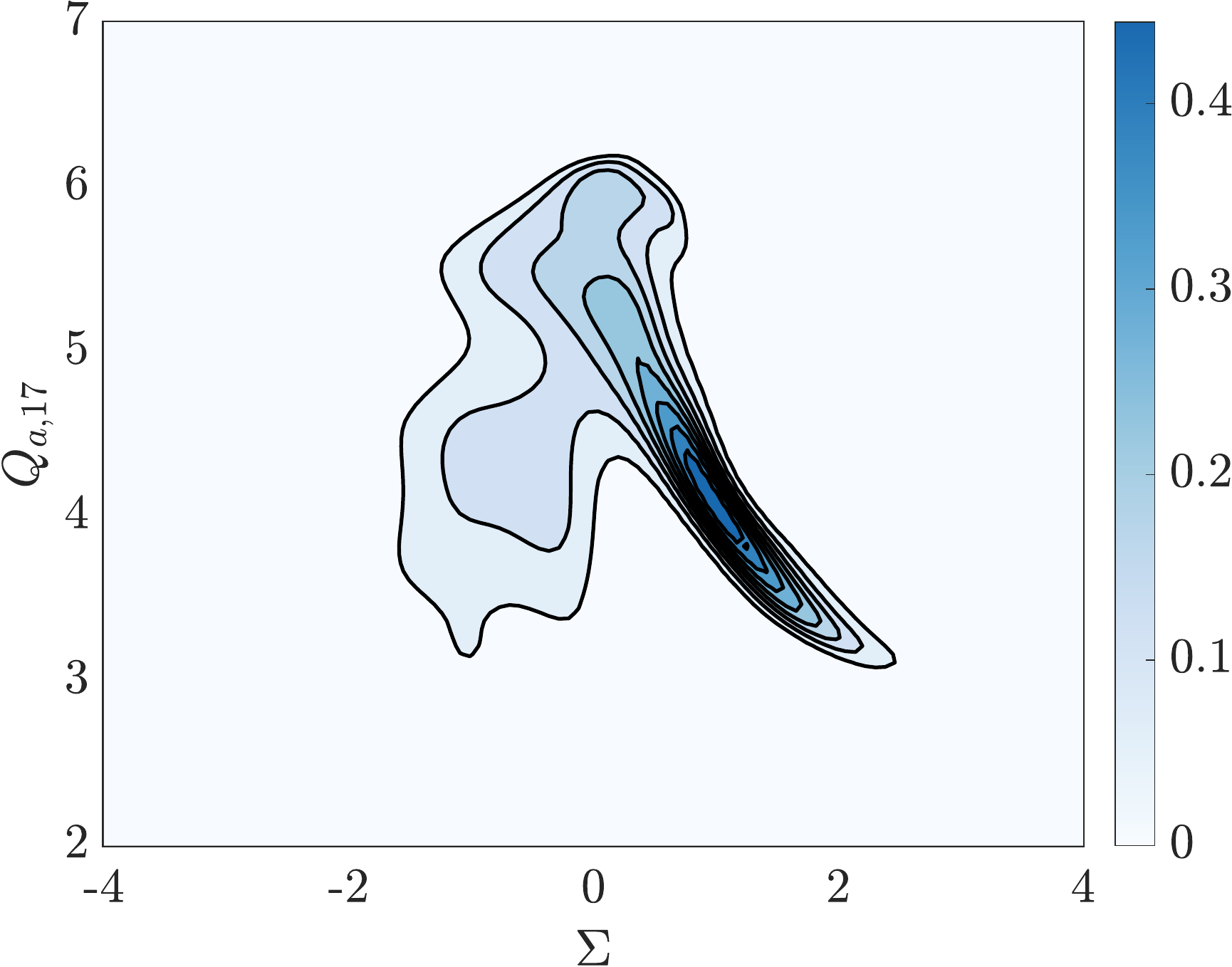}
    \caption{Gaussian vs. non-Gaussian joint densities for $(\Sigma, Q_{a,17})$. The Gaussian approximation incorrectly predicts near independence between the random variables resulting in a poor estimate of $\widehat{I}(\Sigma;Q_{a, 17}) \approx 0$. \label{fig:samples_density_approx_FFT_example}}
\end{figure}

We further demonstrate how poor the approximation can be by visualizing the Gaussian approximation of the joint density that underrepresents the mutual information between two random variables. Specifically, we consider the joint density of $\Sigma$ and $Q_{a,17}$. In Fig.~\ref{fig:samples_density_approx_FFT_example},
we plot the joint samples from this distribution on the left along with a Gaussian and non-Gaussian approximation to the joint density. 

While we considered the EIG from the individual components of the averaged magnitude spectrum for inferring each parameter, interesting future work would be to determine the subset of spectrum data that jointly inform the parameters most. Techniques for doing so are found within the field of active learning or optimal experimental design\textemdash these fields can benefit from the framework laid out in the paper, since they often rely on EIG estimates in order to evaluate optimization objectives. 
\subsection{Expected information gain: Levels of magnification} \label{subsec:EIG_specimen_size}

\begin{figure}[!ht]
    \centering
    \includegraphics[width = 0.32\linewidth]{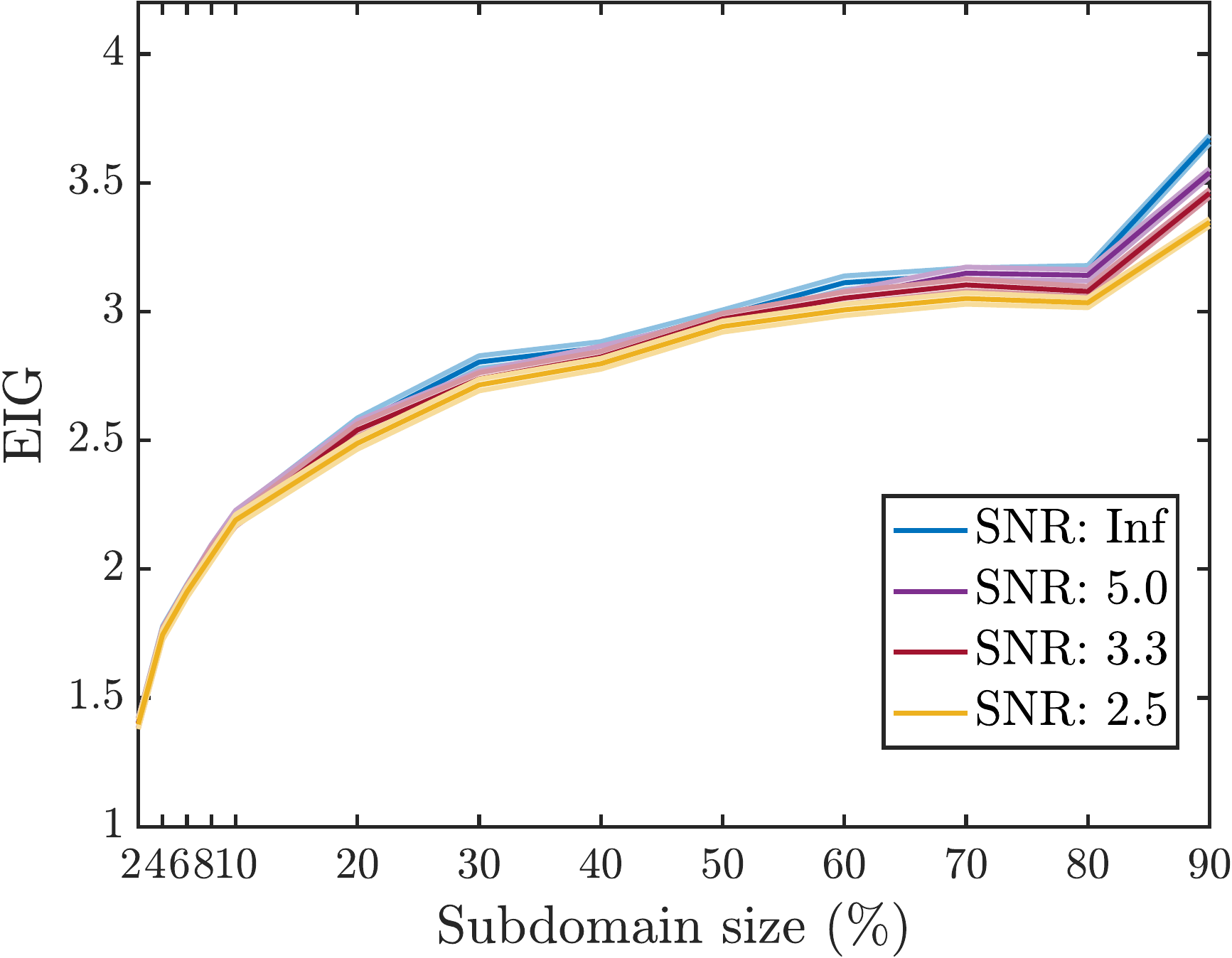}
    \includegraphics[width = 0.32\linewidth]{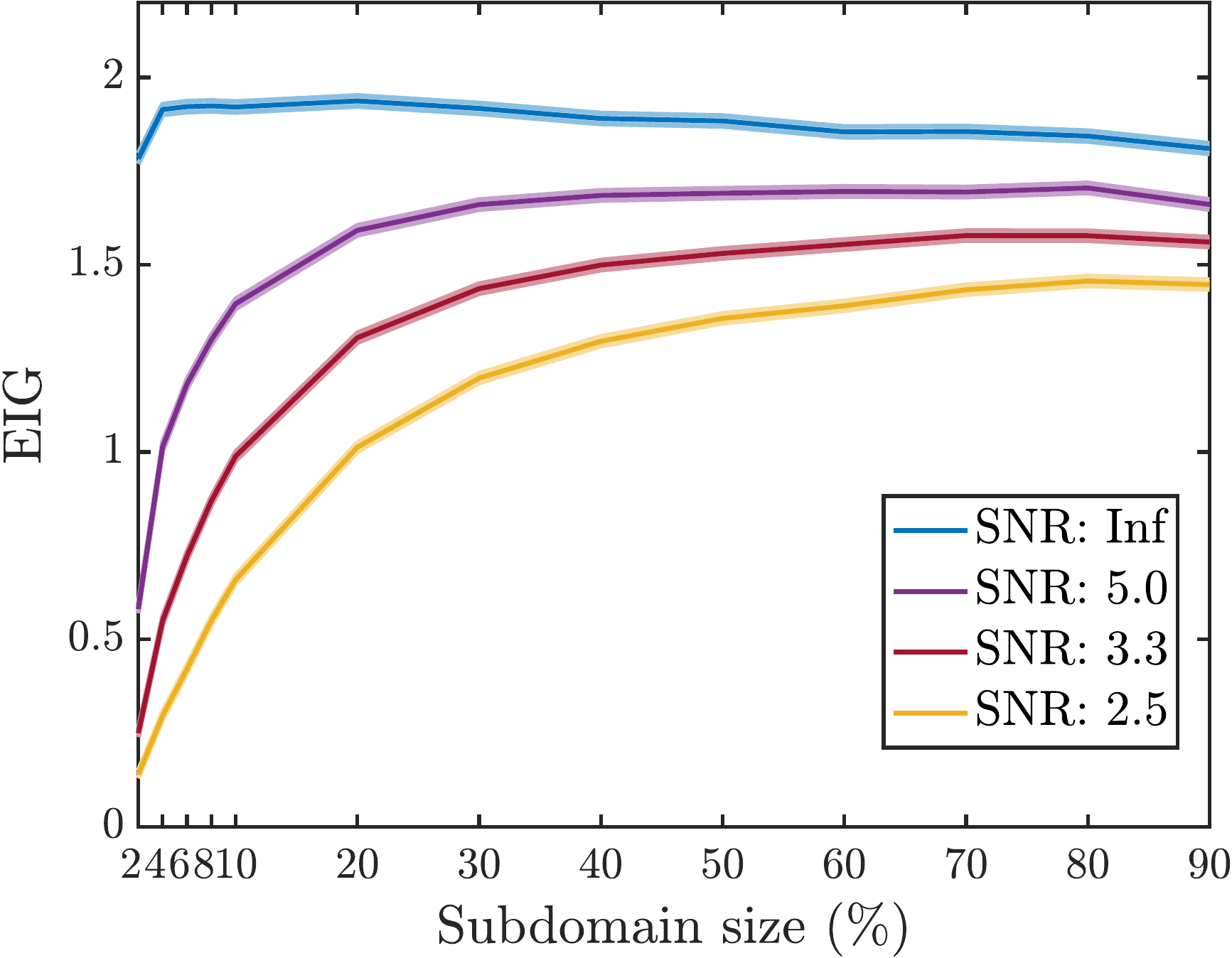}
    \includegraphics[width = 0.32\linewidth]{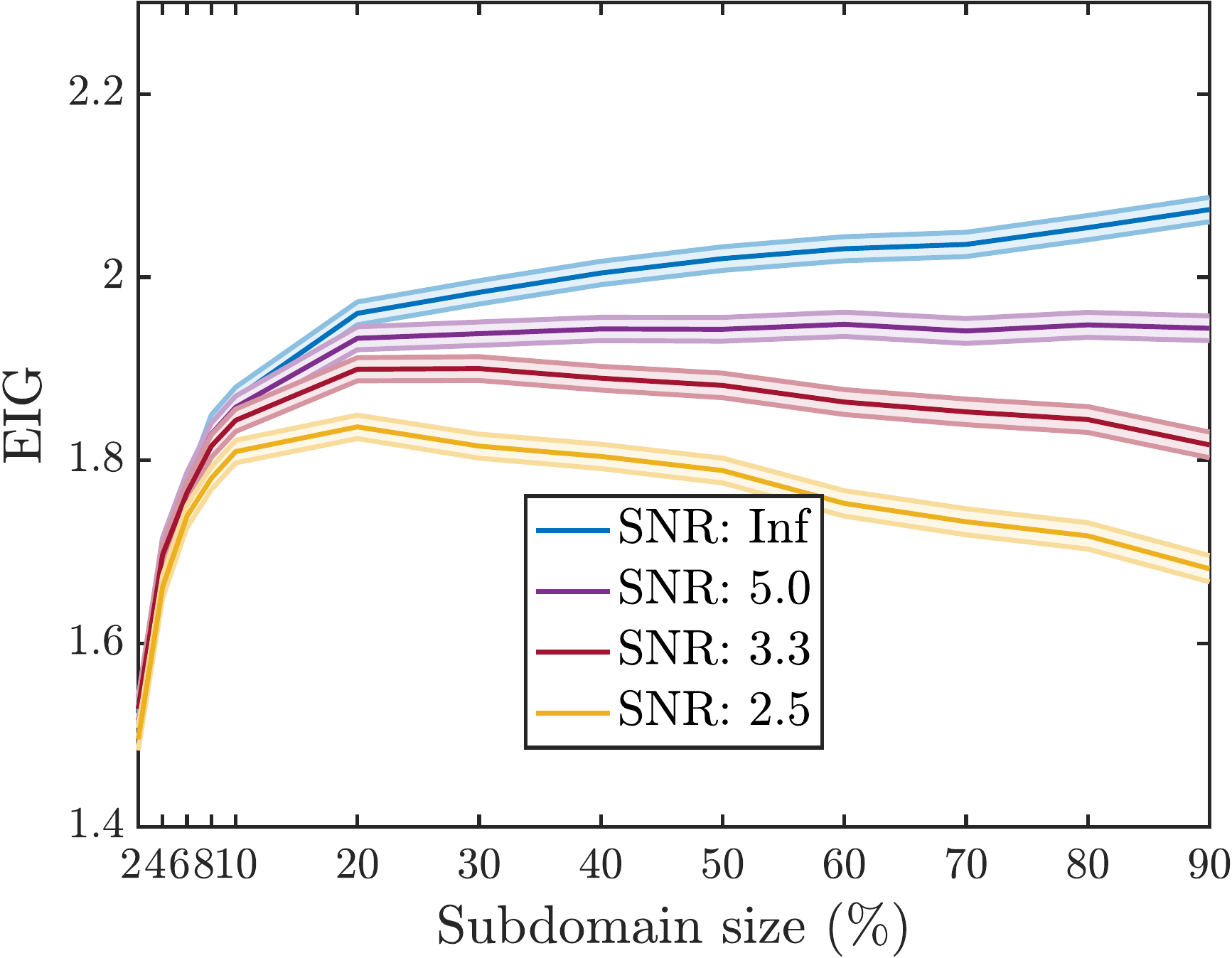}
    \caption{The estimated EIG for inferring each of the three parameters $\bx = (M,E,\Sigma)$ (\emph{left} to \emph{right}) from the observables $(\widehat{M},(Q_1,Q_3),(Q_2,Q_4))$ with different magnification levels, i.e., the specimen subdomain size, varying noise corruption and a fixed blurring corruption of strength $\sigma_{blur} = 5 \times 10^{-3}$.}
    \label{fig:EIG_subdomain_fixedblur}
\end{figure}

\begin{figure}[!ht]
    \centering
    \includegraphics[width = 0.32\linewidth]{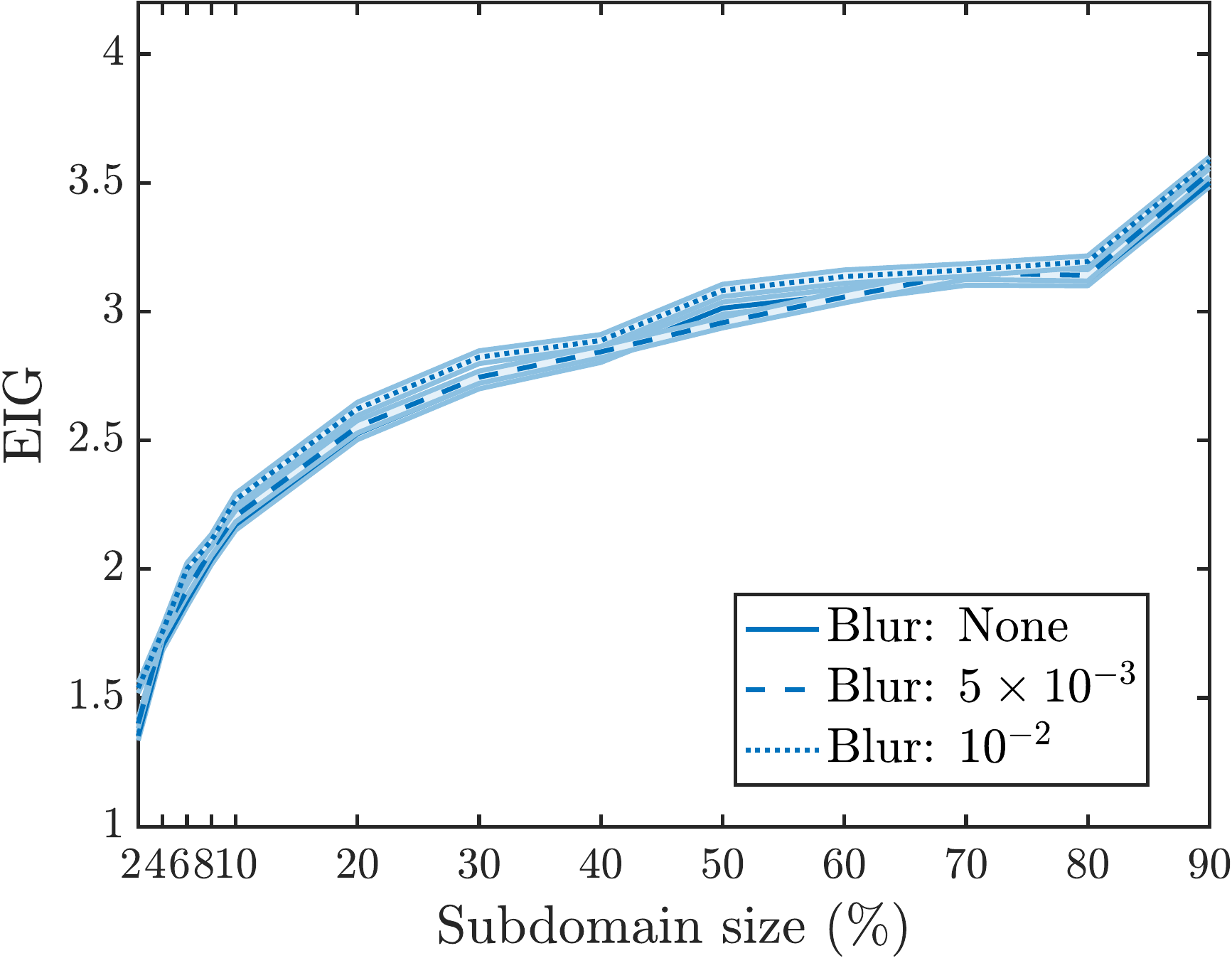}
    \includegraphics[width = 0.32\linewidth]{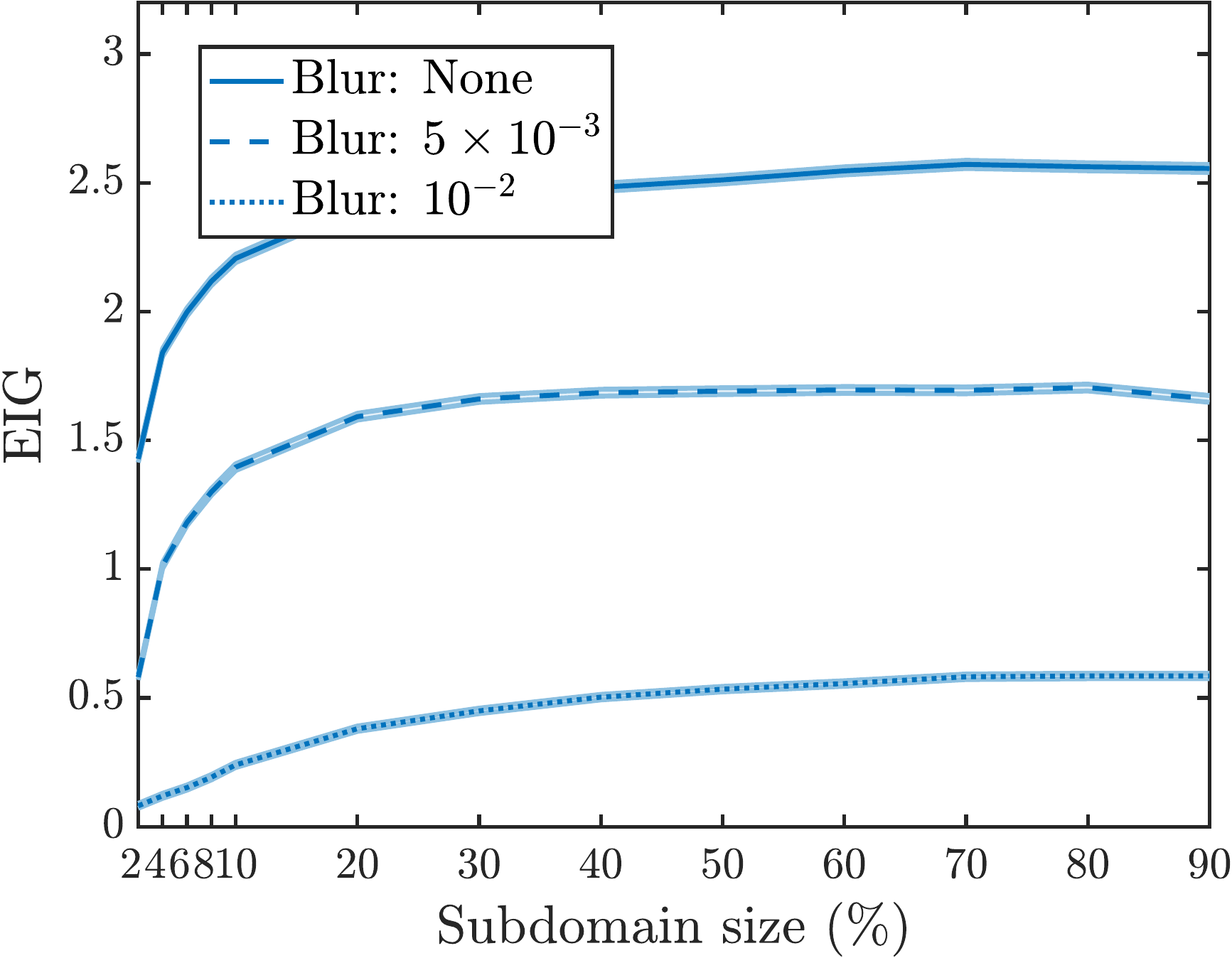}
    \includegraphics[width = 0.32\linewidth]{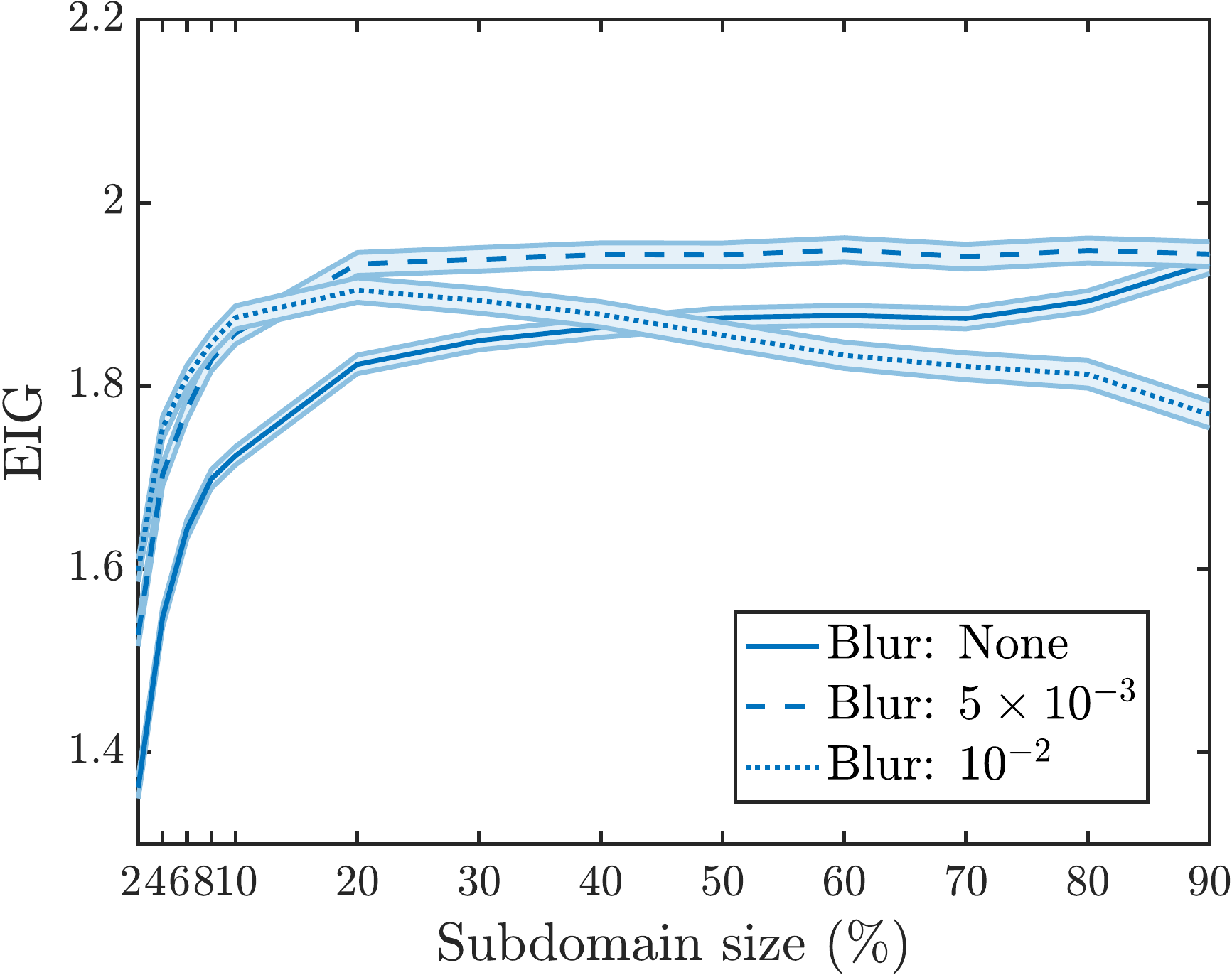}
    \caption{The estimated EIG in Fig.~\ref{fig:EIG_subdomain_fixedblur} with a fixed noise corruption of SNR = 5 and varying blurring corruption.}
    \label{fig:EIG_subdomain_fixedsnr}
\end{figure}

In this subsection, we consider the effects of varying levels of magnification, specified by the restriction operator $\mathcal{R}$ defined in Section~\ref{subsec:microscopy_model}, on the informativeness of energy-based summary statistics, $\widehat{M}$, $(Q_1, Q_3)$ or $(Q_2, Q_4)$, about $\bx = M$, $E$, or $\Sigma$, respectively. In particular, we plot the estimated EIG as a function of magnification levels for a fixed level of blurring corruption and varying levels of noise corruptions in Fig.~\ref{fig:EIG_subdomain_fixedblur}. Similarly, we plot the same quantities in Fig.~\ref{fig:EIG_subdomain_fixedsnr} for a fixed level of noise corruption and varying levels of blurring corruptions.

For learning the spatial average parameter $M$ (\emph{left} in Fig.~\ref{fig:EIG_subdomain_fixedblur} and~\ref{fig:EIG_subdomain_fixedsnr}), the estimated EIG from the empirical mean summary statistics $\widehat{M}$ decreases as the magnification level increases. Moreover, the estimated EIG is insensitive to changes in noise and blurring corruptions, caused by the spatial average--preserving property of the Gaussian blur and the zero-mean property of the noise. We expect this behaviour to change for more complex characterization models that does not satisfy these properties. These numerical results suggest that, under the assumptions of the data model and parameter prior distribution, the action of zooming in will not lead to a decrease in uncertainty in $M$ on average. These result may be interpreted by the loss of phase information related to the material parameter $f$ (see Fig.~\ref{fig:experimental_phase_diagram}) when we only observe a small portion of the top surface of a Di-BCP thin film. For example, one might only observe several spots in a pattern that contains a mixture of spots and strips.

For learning the interfacial parameter $E$ (\emph{middle} in Fig.~\ref{fig:EIG_subdomain_fixedblur} and~\ref{fig:EIG_subdomain_fixedsnr}), the estimated EIG from the energy-based summary statistics $(Q_2, Q_4)$ decreases as the level of magnification increases in the presence of noise and blurring corruptions. This also suggests that the action of zooming in will not lead to a decrease in uncertainty in the $E$ parameter on average. We see that in an idealistic noise-free setting, however, the estimated EIG increases slightly when the level of magnification increases. We hypothesize that the increase in image pixel density via magnification leads to better resolved phase interfaces only in the presence of the blurring corruption, and thus leads to a small increase in the estimated EIG for learning $E$.

For learning the nonlocal parameter $\Sigma$ (\emph{right} in Fig.~\ref{fig:EIG_subdomain_fixedblur} and~\ref{fig:EIG_subdomain_fixedsnr}), we observe an interesting response of the estimated EIG from the energy-based summary statistics $(Q_1, Q_3)$ with respect to changes in the levels of magnification and image corruptions. In the high SNR setting, the estimated EIG decreases as the magnification level increases. On the contrary, we see an increase in the estimated EIG in the low SNR setting. Similarly, in the low blurring setting, the estimated EIGs decreases as the magnification level increases. We also observe an increase in the estimated EIG in the high blurring setting. These results suggest that whether the action of zooming in is beneficial for inferring $\Sigma$ depends on the level of image corruptions.

Overall, while some of the figures behave as one might expect, the unexpected behavior of the EIG for certain summary statistics and model parameters would otherwise be difficult to determine quantitatively without these numerical experiments. These empirical findings could aid and inspire further theoretical investigations about the effects of various image corruptions and experimental designs on model calibration and prediction.
\subsection{Expected information gain: Known volume ratio of monomer blocks} \label{subsec:EIG_spatial_average_fixed}

In many experimental settings, the volume ratio of the two monomer blocks in Di-BCPs can be determined with high certainty. In particular, we have shown in Section \ref{subsec:EIG_energy} that the spatial average summary statistics is exceptional in learning the OK parameter $m$. Therefore, in this subsection, we consider the inference problem for two length scale related parameters in the OK model $(E,\Sigma)$ given image data of Di-BCP patterns with a fixed, or known, spatial average parameter.

\begin{figure}[!ht]
    \centering
    \includegraphics[width=0.3\textwidth]{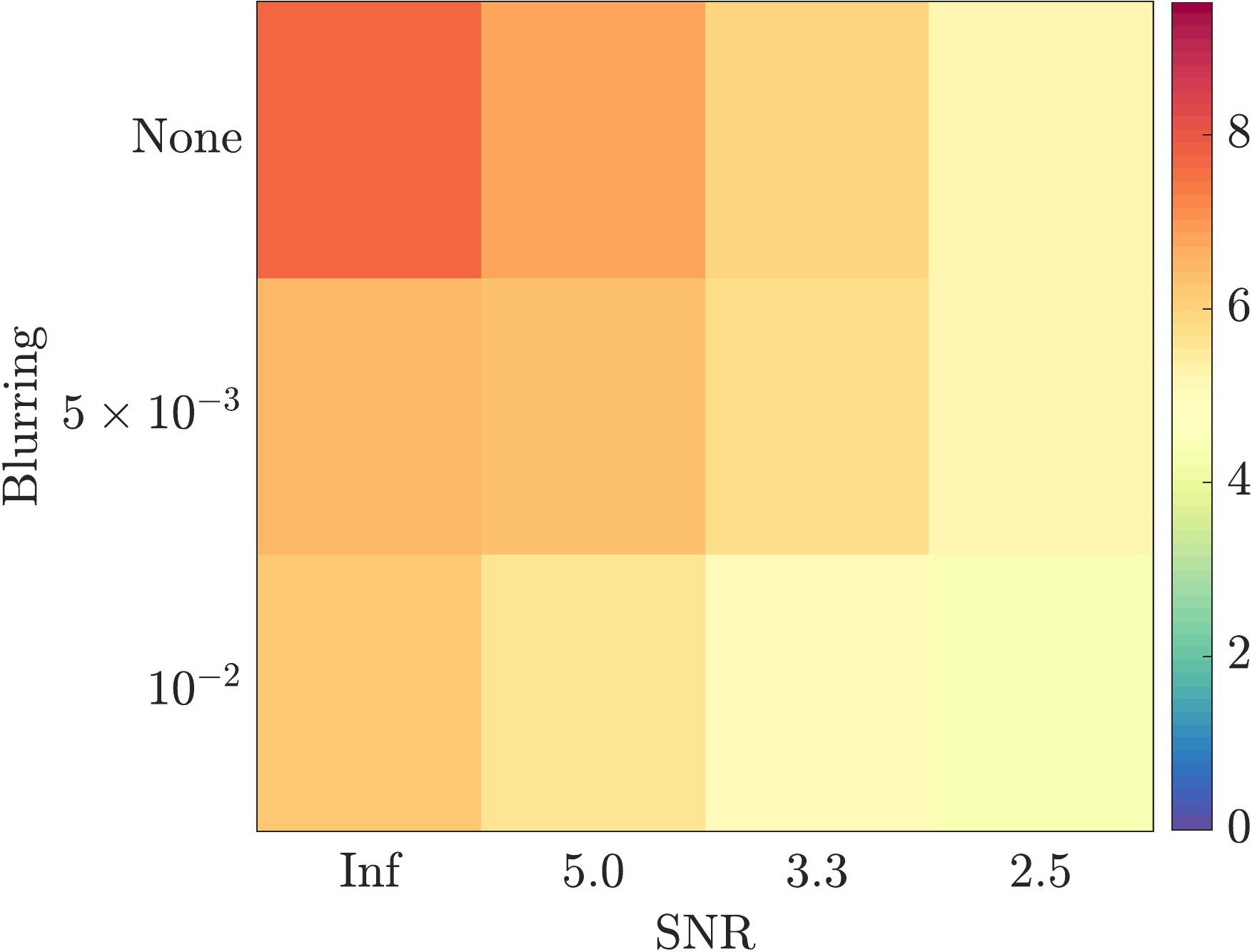}
    \hspace{2pt}
    \includegraphics[width=0.3\textwidth]{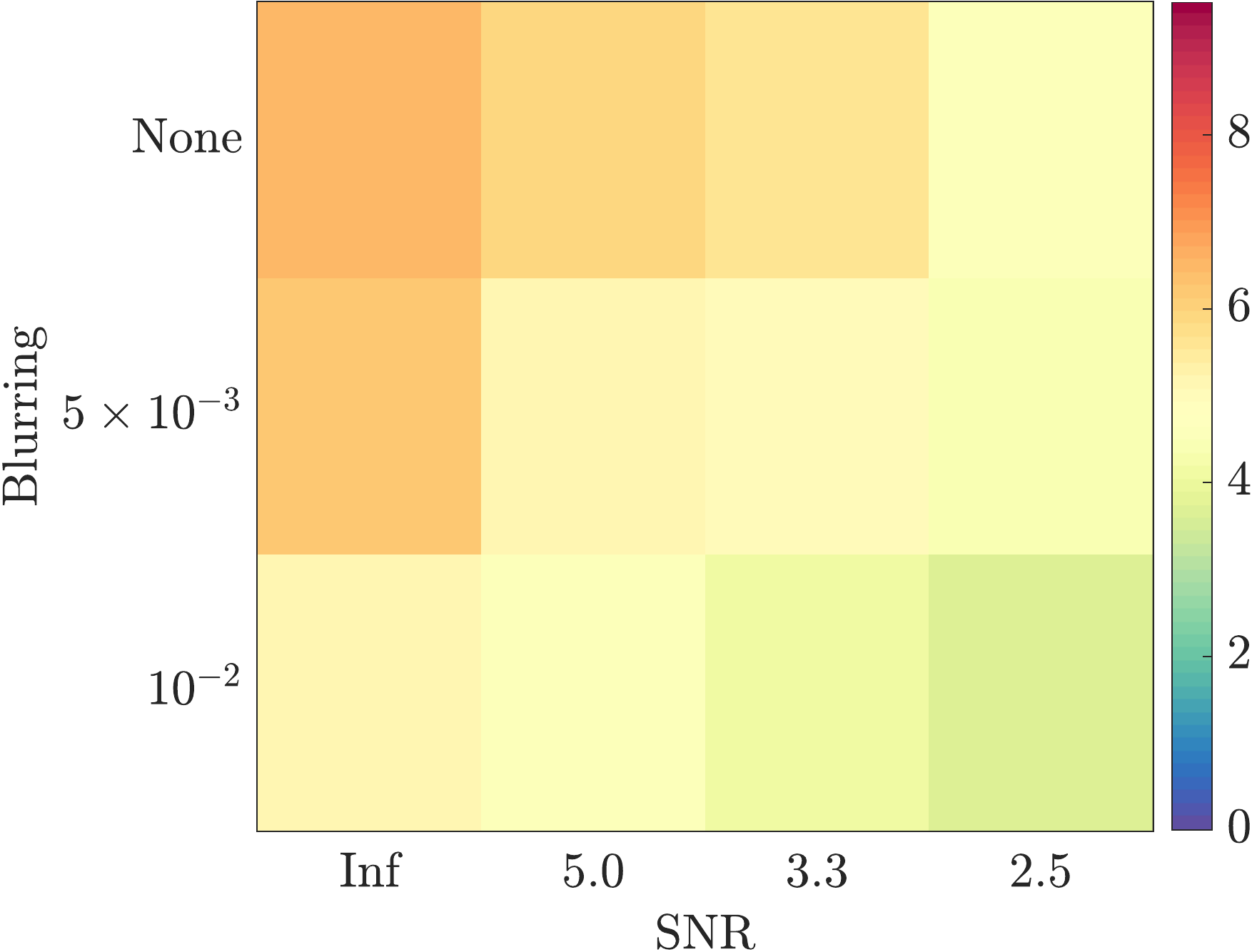}
    \hspace{2pt}
    \includegraphics[width=0.3\textwidth]{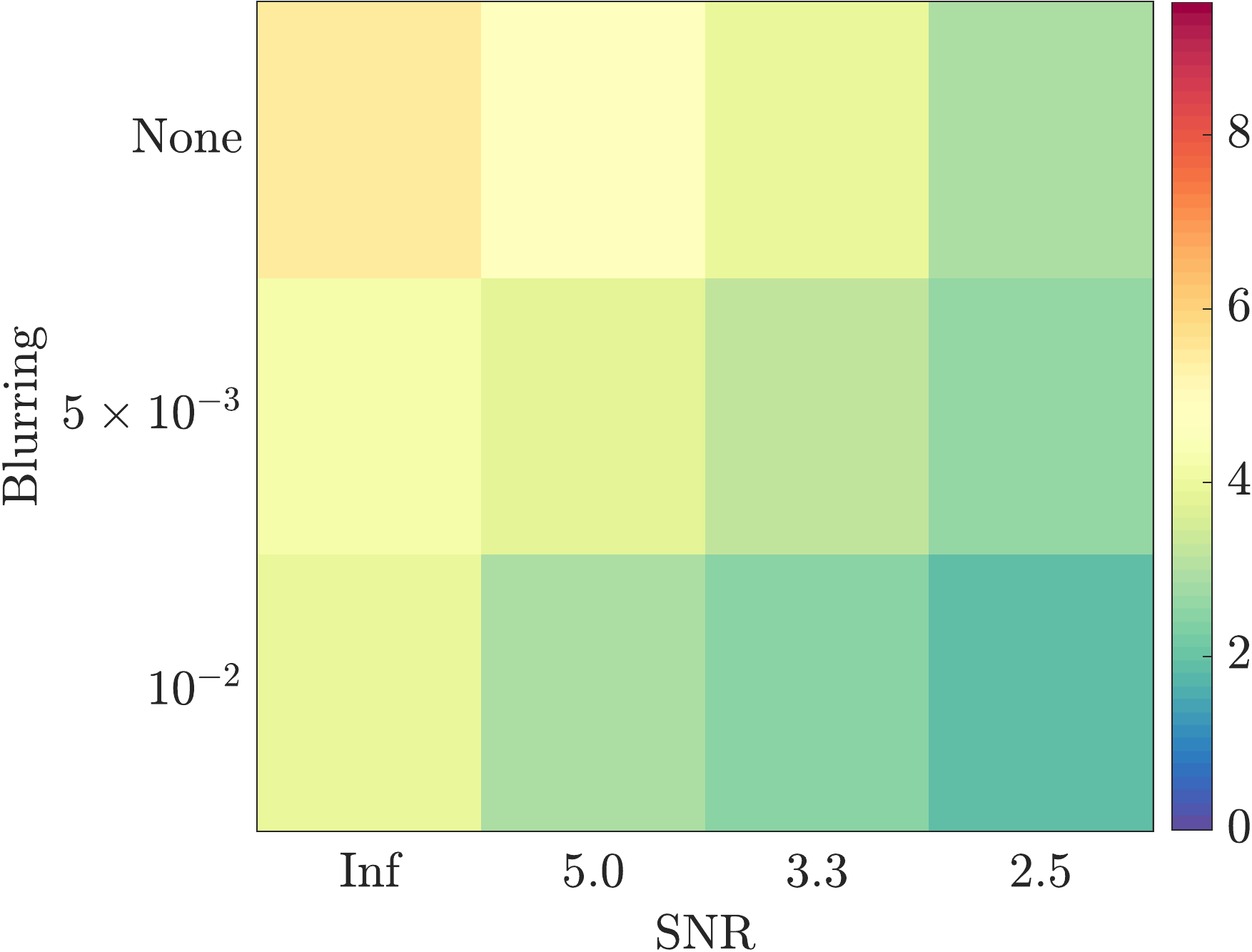}
    \caption{The estimated EIGs for inferring the joint posterior of the two length-scale dependent parameters $(E,\Sigma)$ from energy-based summary statistics $(Q_1,Q_3)$ for $m = 0$ (\it{left}), $m = 0.3$ (\it{center}), and $m = 1/\sqrt{3}$ (\it{right}). \label{fig:EIG_Q13}}
\end{figure}

\begin{figure}[!ht]
    \centering
    \includegraphics[width=0.3\textwidth]{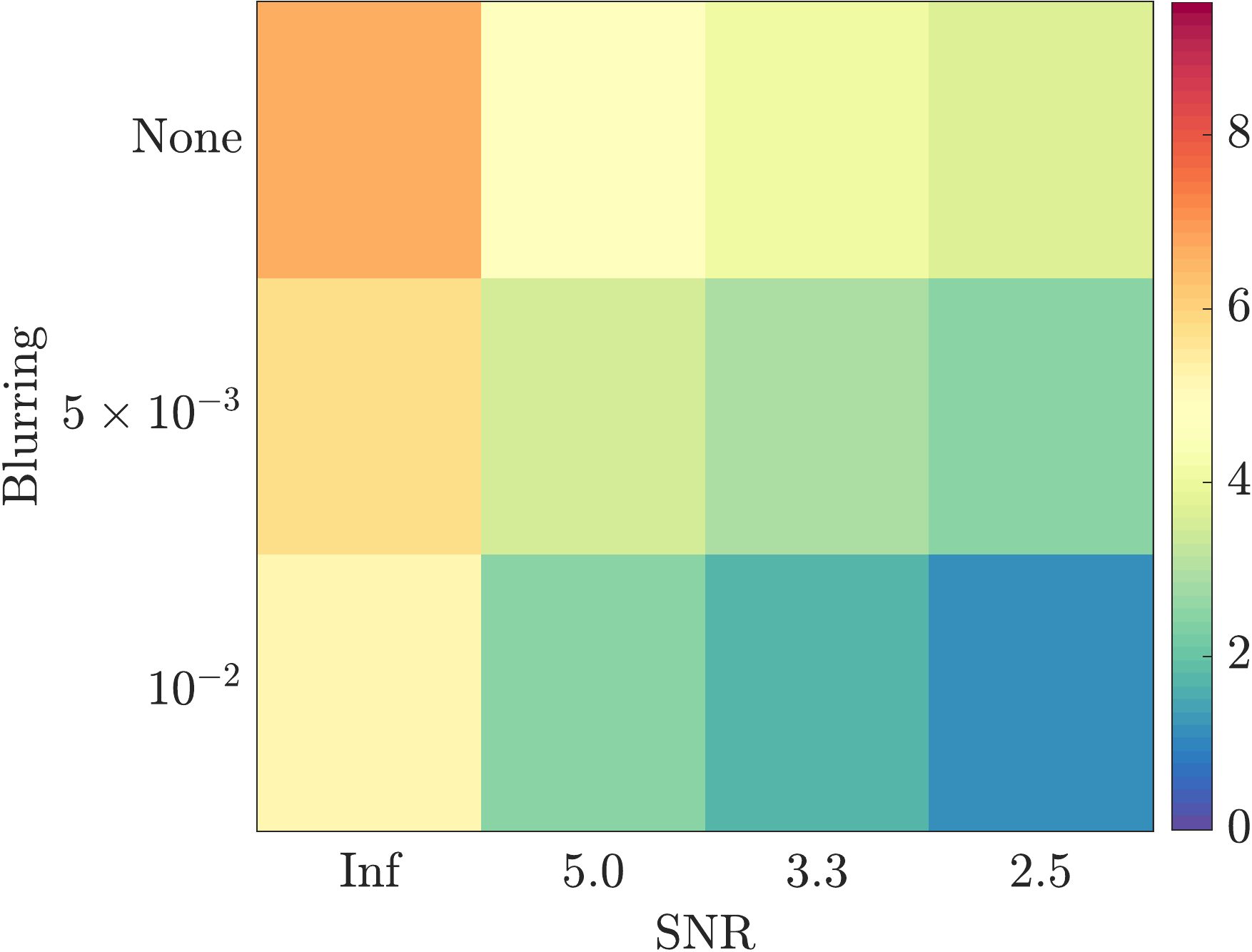}
    \hspace{2pt}
    \includegraphics[width=0.3\textwidth]{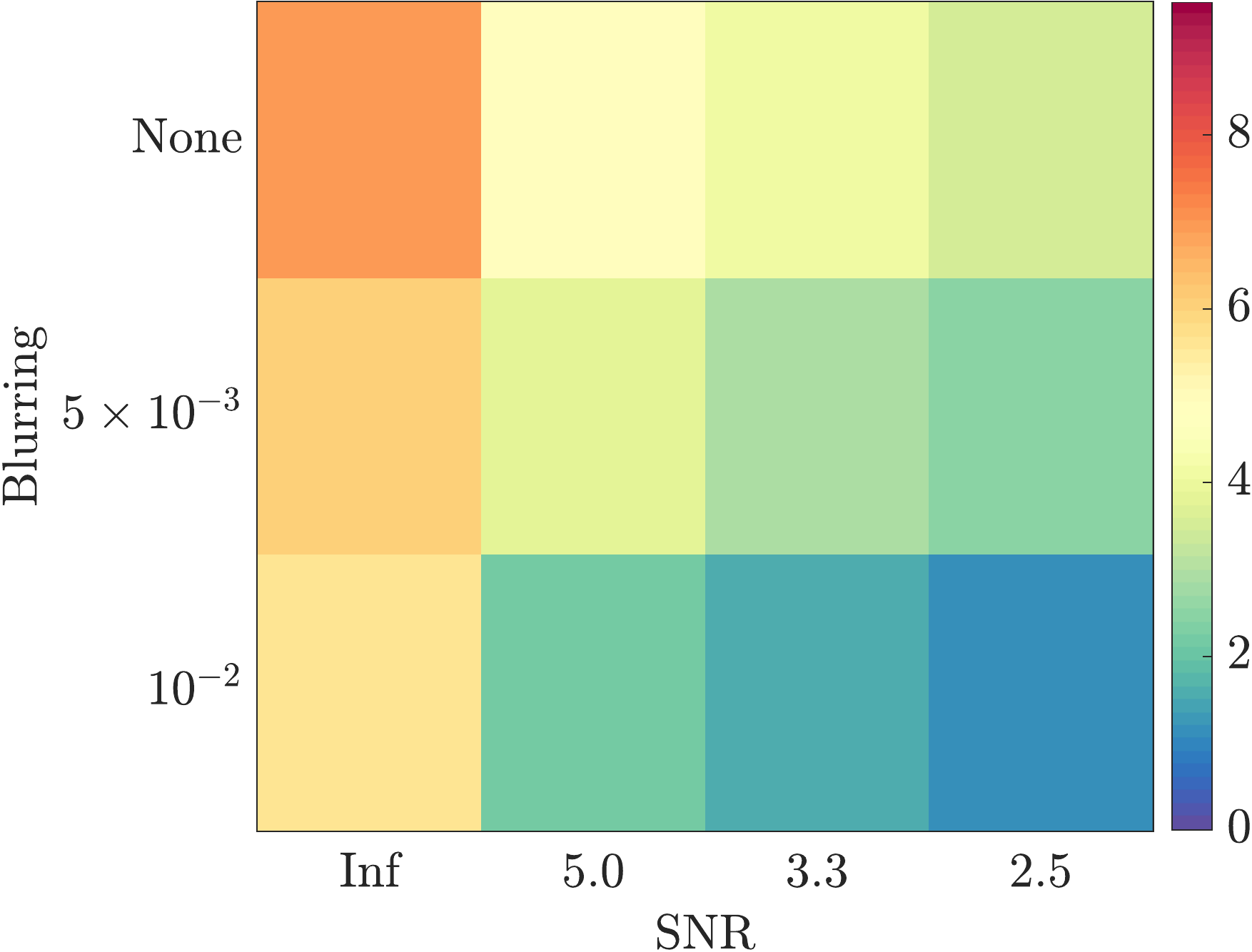}
    \hspace{2pt}
    \includegraphics[width=0.3\textwidth]{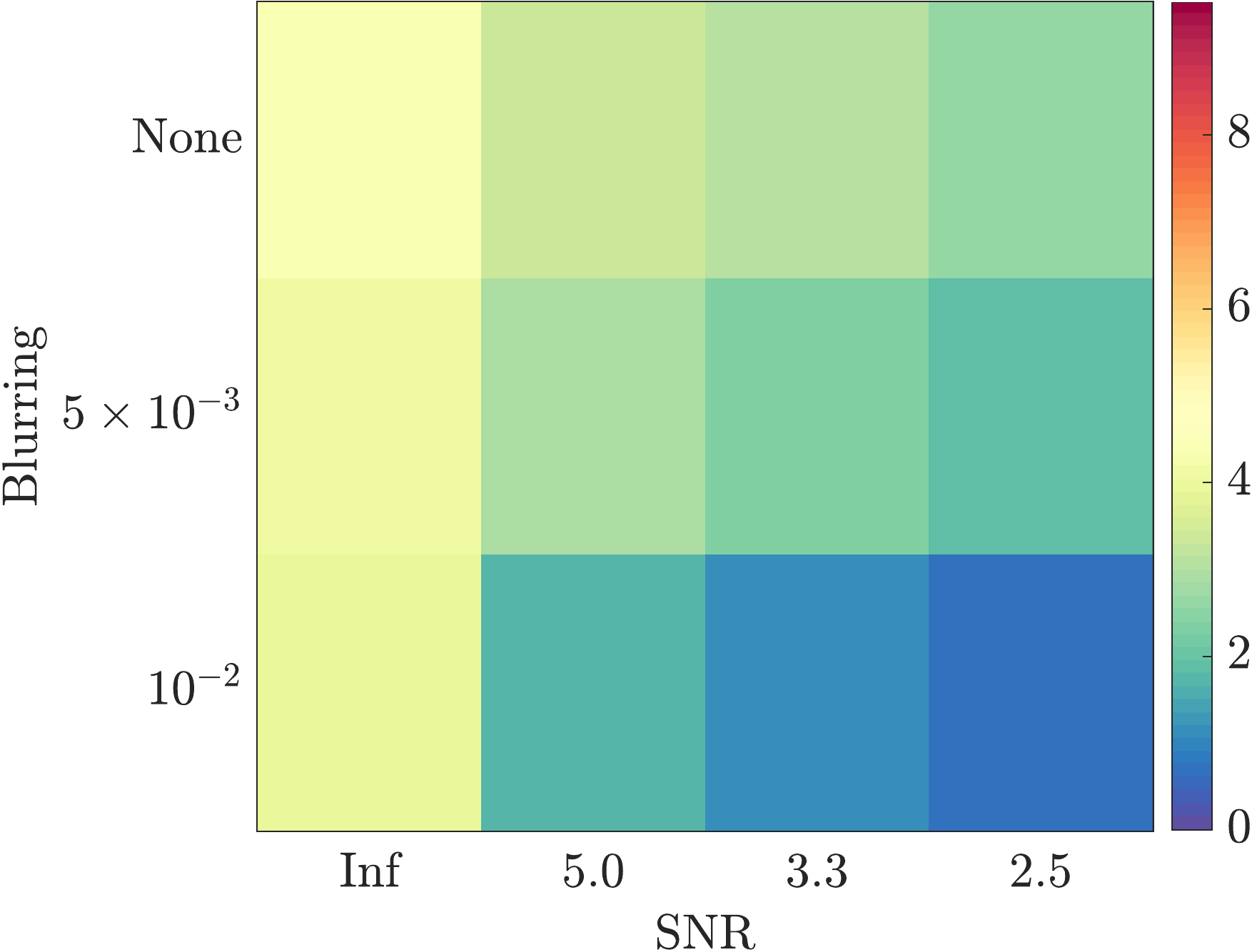}
    \caption{The estimated EIGs for inferring the joint posterior of the two length-scale dependent parameters $(E,\Sigma)$ from energy-based summary statistics $(Q_2,Q_4)$ for $m = 0$ (\it{left}), $m = 0.3$ (\it{center}), and $m = 1/\sqrt{3}$ (\it{right}). \label{fig:EIG_Q24}}
\end{figure}

\begin{figure}[!ht]
    \centering
    \includegraphics[width=0.3\textwidth]{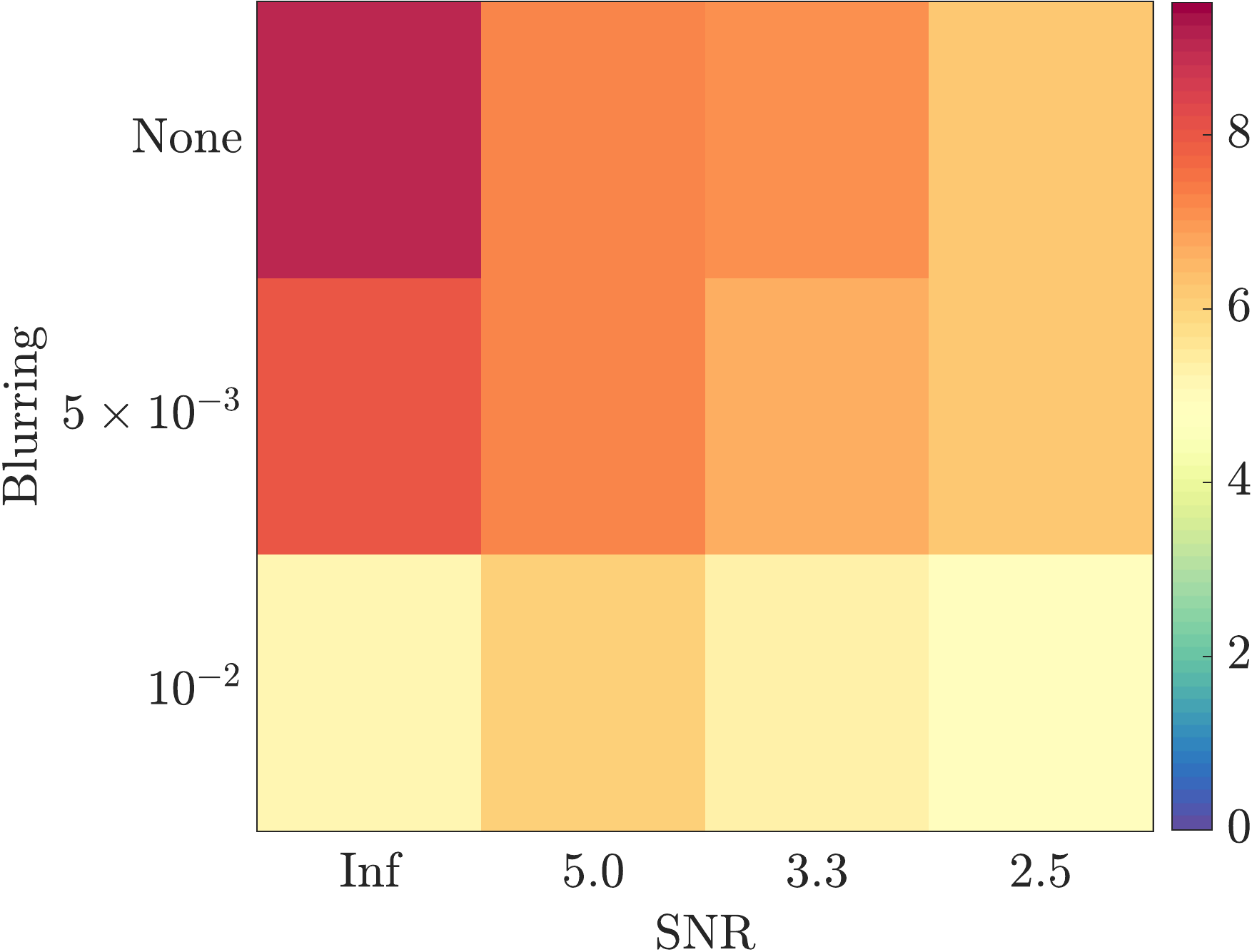}
    \hspace{2pt}
    \includegraphics[width=0.3\textwidth]{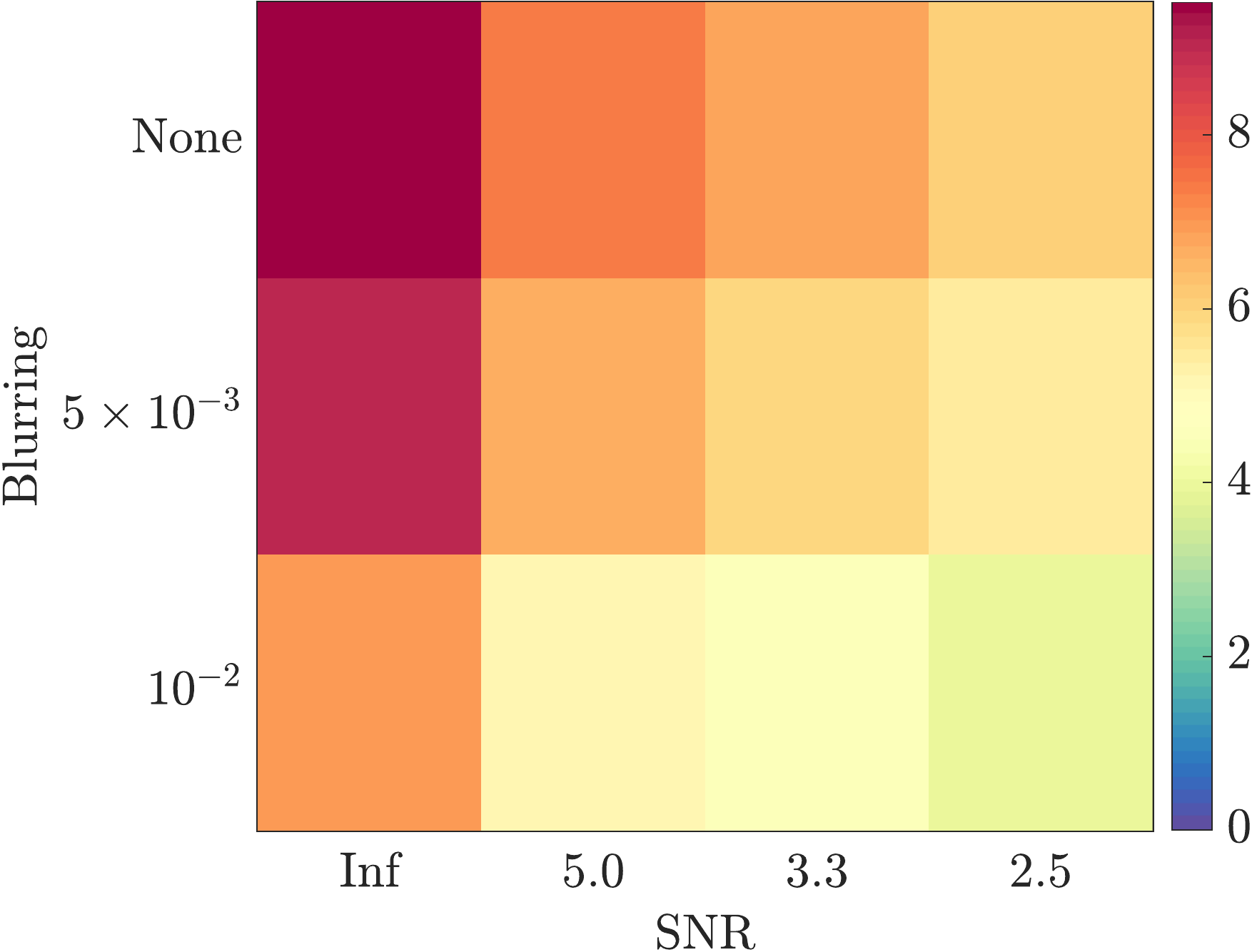}
    \hspace{2pt}
    \includegraphics[width=0.3\textwidth]{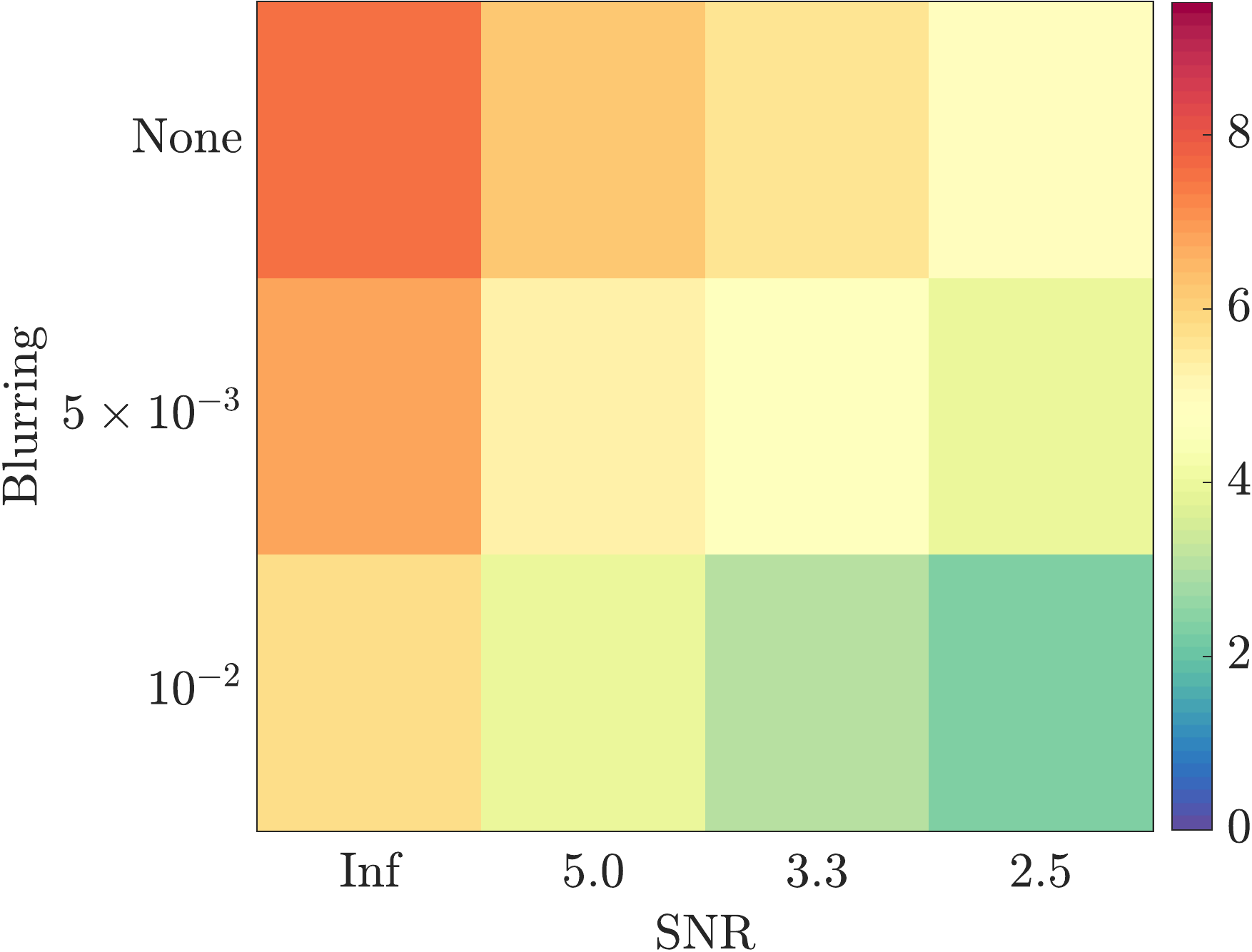}
    \caption{The estimated EIGs for inferring the joint posterior of the two length-scale dependent parameters $(E,\Sigma)$ from energy-based summary statistics $\boldsymbol{Q}_{1:4}$ for $m = 0$ (\it{left}), $m = 0.3$ (\it{center}), and $m = 1/\sqrt{3}$ (\it{right}). \label{fig:EIG_Q1-4}}
\end{figure}

Figures~\ref{fig:EIG_Q13}, \ref{fig:EIG_Q24}, and \ref{fig:EIG_Q1-4} plot the estimated EIGs for learning the parameters $(E,\Sigma)$ jointly from $(Q_1,Q_3)$, $(Q_2,Q_4)$, and $(Q_1, \dots,Q_4)$, respectively, conditioned on three different spatial averages $m = (0, 0.3, 1/\sqrt{3})$. In general, we observe a decrease in the estimated EIGs with with increasing level of noise and blurring corruptions. We also note that the EIGs from $(Q_2,Q_4)$ are more sensitive to noise than those from $(Q_1,Q_3)$ or conditioning on all observables, particularly for the spatial averages $m = 0$ and $m = 0.3$.

\section{Conclusion}
In this work, we explored the Bayesian model calibration problem for block copolymer self-assembly using imaging data. To set up the problem, we introduced a generic probabilistic data model that includes both the measurement uncertainty in the characterization process and the aleatoric uncertainty in the self-assembly procedure. To calibrate the model parameters we characterize their posterior distribution, where the likelihood function accounts for the aleatoric uncertainty by marginalizing over high-dimensional auxiliary variables. 
Moreover, we quantify the expected information gain to measure the informativeness of data for calibrating the parameters.

We argue and demonstrate that, as a result of the generally intractable integrated likelihood evaluations and the high dimensional data, many conventional Bayesian inference methodologies are unsuitable for performing the proposed calibration task. To overcome these challenges, we propose to (i) employ a likelihood-free inference method  that only requires joint samples of data and model parameters, and (ii) construct informative low-dimensional summary statistics for the data. 
In particular, we use an inference methodology based on triangular transport maps, which allows us to (i) consistently characterize the joint distribution of model parameters and summary statistics with limited samples, (ii) perform amortized inference for any realization of the data with a single collection of joint samples, and (iii) consistently compute expected information gain with no significant additional cost, which enables us to study the effects of different summary statistics and corruptions in the data on the calibration results.

Lastly, we demonstrated the power of the proposed inference approach and the computation of expected information gain on 
a numerical case study of calibrating the Ohta--Kawasaki model parameters using top-down microscopy characterization images. We introduced several energy- and Fourier-based summary statistics to reduce the dimension of the image data. A hierarchical uniform prior distribution for the model parameters is constructed by analyzing the forward operator. First, we observe the concentration of the posterior distribution in the parameter space and visualize posterior predictive samples, which show that the proposed energy-based summary statistics are effective for calibrating the parameters with low uncertainty. We then estimate the expected information gains from different summary statistics 
for learning the model parameters, under various experimental scenarios with different levels of (i) noise and blurring corruptions, and (ii) magnifications of the microscope. These numerical results allow us to investigate the effects of experimental settings on the calibration results for this representative model, and enable further optimal experimental design studies of interest.
   

\section*{Acknowledgement}
The work of the authors is supported by \href{https://aeolus.oden.utexas.edu}{the AEOLUS center}, funded by the U.S. Department of Energy, Office of Science, Office of Advanced Scientific Computing Research, Mathematical Multifaceted Integrated Capability Centers (MMICCS), under Award DE‐SC0019303.

\addcontentsline{toc}{section}{Reference}
\bibliographystyle{model1-num-names}
\biboptions{sort,numbers,comma,compress}
\bibliography{main}

\begin{appendices}
\addtocontents{toc}{\protect\setcounter{tocdepth}{0}}
\appendix

\section{The auxiliary variable: a transformed Gaussian random field}\label{app:our_z_model}

In this appendix section, we define the auxiliary random field $\bz$ used in the forward operator $\mathcal{F}(\cdot,\bz)$~\eqref{eq:ok_energy_solutions} to define a random initial state $\z$ in state space when minimizing the OK energy functional. Rather than employing the typically used i.i.d.\thinspace uniform random fields for $\bz$, which are infinitely non-smooth and not amenable to discretization of the PDE equations, we use a transformation $\bz = \mathcal{Z}(\bx, A)$ of a Gaussian random field $A$, that is also known as a \emph{Gaussian related random field}~\cite{hu2000gradual,adler2008some}.

We follow an approach similar to that in in~\cite{Cao2022} by taking $A\sim\nu_{A}$ to be a Gaussian random field with zero-mean
and covariance operator $\mathcal{C}_{A}:L^2(\mathbb{R}^2)\to L^2(\mathbb{R}^2), \mathcal{C}_{A} =(\kappa \mathrm{Id} -\triangle)^{-2}$, which possesses the covariance function
\begin{equation}
    c_{\kappa}(|\s_1-\s_2|) = \frac{\sigma^2}{\Gamma(1)}\kappa|\s_1-\s_2| K_1(\kappa|\s_1-\s_2|),
\end{equation}
where $K_1$ is the order $1$ modified Bessel function of the second kind and $\sigma^2 =\frac{\Gamma(1)}{\Gamma(2)4\pi \kappa}$~\cite{daon2016mitigating}. In this work, $\kappa=200$ is chosen such that the spatial correlation is below the \emph{length scales} of the features of all equilibrium states in our experimental scenario modeled by the OK model \eqref{eq:ok_energy_strong}. $A$ is isotropic and stationary since the function $\boldsymbol{c}$ depends on Euclidean distance $|\s_1-\s_2|$ alone. For this choice of covariance, we have almost surely continuous realizations \cite[Sec. 6.5]{Stuart2010}. Realizations $A^i \sim \nu_{A}$ can be numerically approximated given a sufficiently fine spatial discretization. After discretization, these fields become Gauss--Markov random fields~\cite{Lindgren2011}, which are efficiently implemented using the \texttt{hIPPylib} python package~\cite{Villa2018}, where finite element approximations of these PDEs are solved in the \texttt{FEniCS} library~\cite{AlnaesBlechta2015a, Logg2012}.. A description of finite element discretizations for similar random fields can be found in~\cite{Bui-Thanh2013}.

\begin{figure}[H] \label{fig:realizations_of_random_fields}
    \centering
    \begin{minipage}{.4\textwidth}
  \caption{The Whittle\textendash Mat\'{e}rn covariance function describing the correlation structure of the random fields. For $\kappa = 200$, the correlation decays to near zero for distances $|\s_1-\s_2| > 0.05$. Recall that the computational domain is $\mathcal{D}_c=[0,40]^2$.}
  \end{minipage} \quad
  \begin{minipage}{.5\textwidth}
    \includegraphics[width=\textwidth]{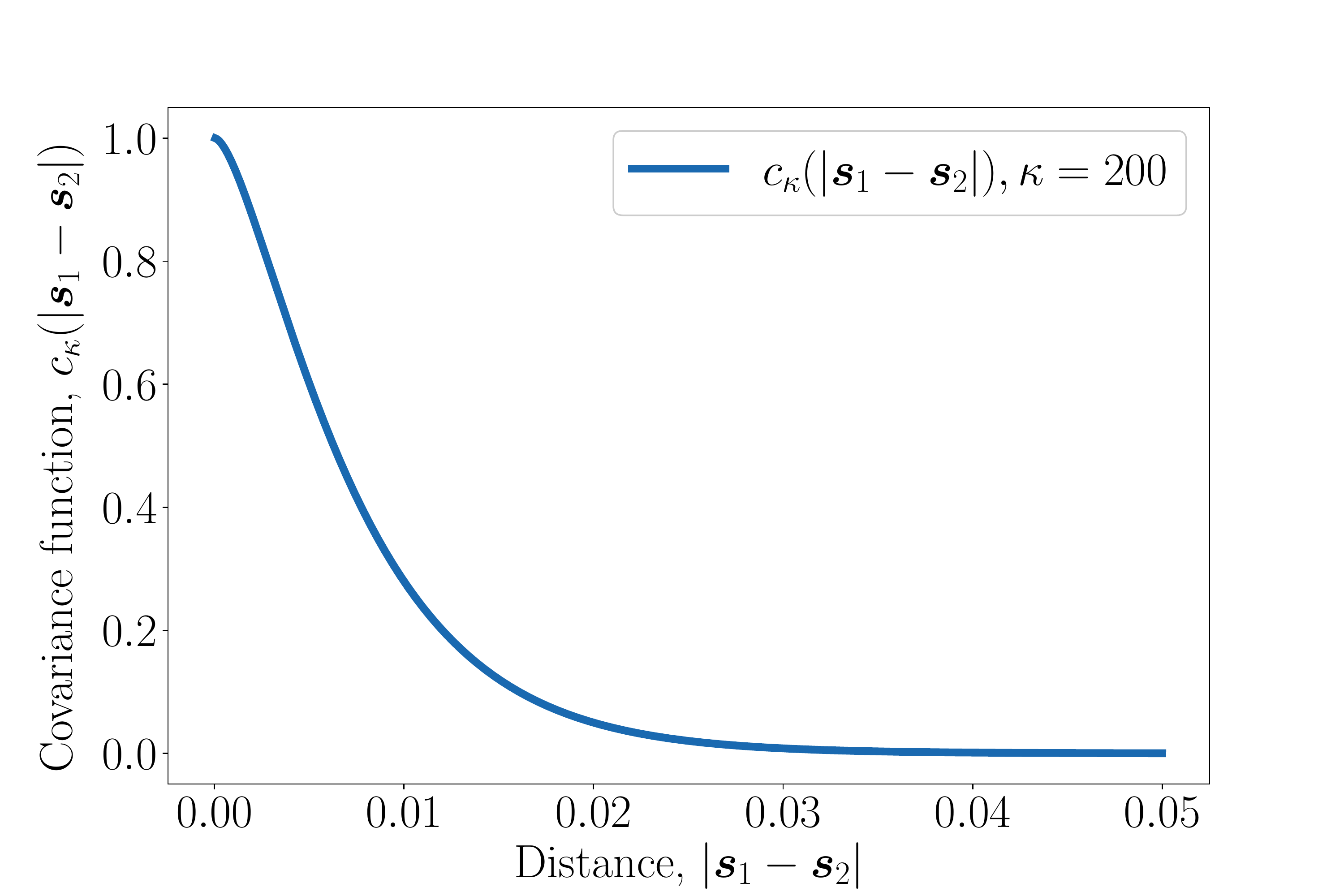}
  \end{minipage}
 \end{figure} 

Finally, we define the transformation $\bz = \mathcal{Z}(\x,A)$ such that the random field $\bz$ is a pointwise uniform a.s.\thinspace continuous field that satisfies certain physical constraints. Specifically, given the point\textendash wise variance $\sigma^2$, the pointwise transformation

\begin{equation}
    \mathcal{Z}(A; \x) = \m + (1-\m)\textrm{erf}\left(\frac{A}{\sigma\sqrt{2}}\right)
\end{equation}
ensures that the pointwise values of the auxiliary random field realizations 
$\z^i = \mathcal{Z}(\bx,A^i)$ for $A^i \sim \nu_{A}$ are within the physical order parameter bounds $(-1,1)$ and have the pointwise mean $m$. Furthermore,
its pointwise marginal densities are $\pi_{\bz(\s) | \boldm}(\cdot | \m) = \mathcal{U}(-1+2\m,1) \;\forall \s\in\mathbb{R}^2.$ Thus, they are physically reasonable random fields to employ as starting points $\bz$ for the variational procedure that defines $\mathcal{F}(\x, \bz)$. 

To exactly match the specified spatial average $m$ for the order parameter~\eqref{eq:spatial_average_constraint}, an additional transformation is employed, in which we choose the closest point satisfying the correct spatial average within the chosen finite element function space. Since this projection does not account for the pointwise constraint $\z^i(\s)\in(-1,1)$, transformed random field samples may actually violate the constraint in regions of the domain\textemdash therefore, our samples only \emph{approximately satisfy} this physical constraint. However, this has a negligible effect on forward solutions, since the double well potential $\mathcal{D}^{\textrm{OK}}$ heavily penalizes deviation from  non-physical regions in state space, so that any variational procedure will correct for this within a few iterations.


\begin{figure}[!htbp]\label{fig:point_wise_variability_of_random_fields}
    \centering
    \includegraphics[height=0.2\linewidth]{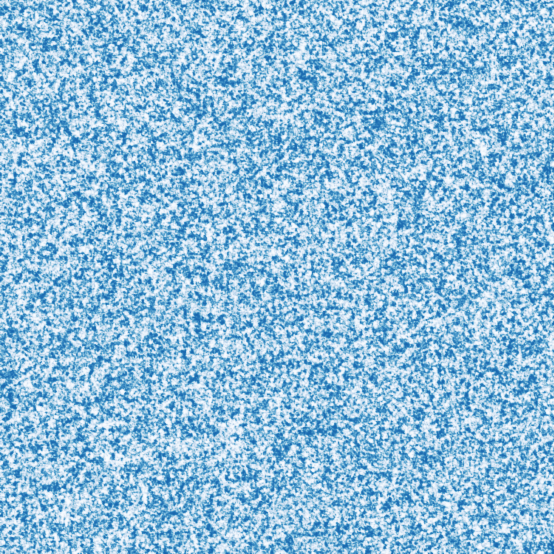} \includegraphics[height=0.2\linewidth]{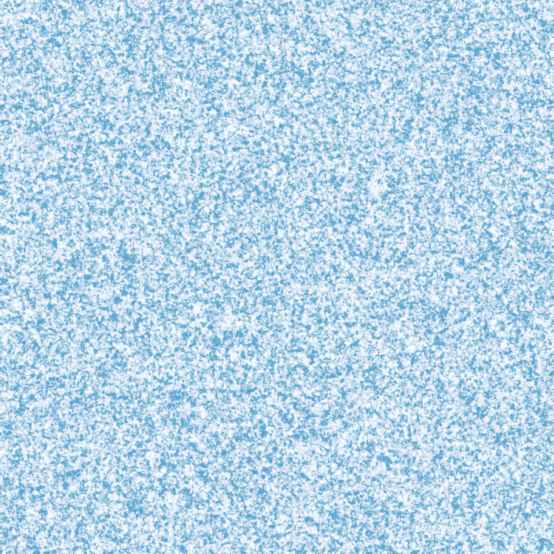}
    \includegraphics[height=0.2\linewidth]{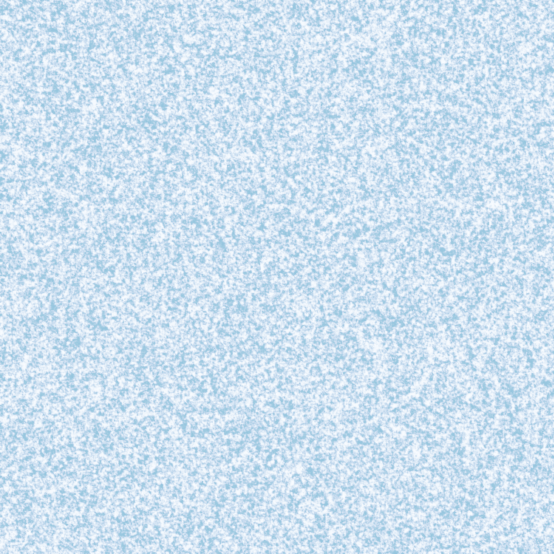}
    \includegraphics[height=0.2\linewidth]{figure/lianghao/colorbar.pdf} 
    \caption{Realizations of $\nu_{\bz | M}(\cdot | m)$ for $\m = 0, 0.3. 0.5$, which have pointwise marginals $\mathcal{U}(-1+2\m,1)$.}
\end{figure}
\section{The discretization of the state space, the restriction operators, and image interpolation}\label{app:function_space}

The state space $V^u$ for th OK model is $\mathring{H}_m^1\big(\mathcal{D}\big)$, as in \eqref{eq:state_space}, with $\mathcal{D}$ corresponds to the reference field of view, as presented in Section~\ref{subsec:microscopy_model}, and $m$ corresponds to the spatial average model parameter. However, when evaluating the forward operator, we solve the equivalent mixed problem as in \eqref{eq:mixed_first_oder_condition}, thus the state space is transformed into the product space $(H^1(\mathcal{D}))^2$. Furthermore, we transform the physical domain $\mathcal{D}$ into a computational domain $\mathcal{D}_c$, with a side length of $40$, thus equivalently rescale the model parameters according to \eqref{eq:ok_parameter_maps}. The purpose of the domain scaling is to improve the numerical conditioning of the forward operator by balancing the relative magnitude of the three terms in the OK free energy \eqref{eq:ok_energy_strong}.

The computational domain $\mathcal{D}_c$ is discretized with uniform triangular mesh, denoted by $\mathcal{P}^h$, such that $\mathcal{D}_c = \bigcup_{K\in\mathcal{P}^h} \bar{K}$. Each element is the image of a scaling map $F_K$ of the reference triangular element $\hat{K}$. We consider a linear finite element basis: let $P^1(\hat{K})$ denote the space of complete linear polynomials, then the finite-dimensional spaces $S^{h,1}\coloneqq S^{h,1}(\mathcal{P}^h)$ by:
\begin{equation}\label{eq:finite_element_basis}
 S^{h,1}\coloneqq \left\{v^h\in H^1(\mathcal{D}_c):v^h|_K=\hat{v}\circ F_K^{-1}, \hat{v}\in P^1(\hat{K})\right\}.
\end{equation}
In our case study, we consider $\mathcal{P}^h$ with same number of cells as the number of sensors, which makes the evaluation of the sensor signal discretization operator $\mathcal{D}$ \eqref{eq:sensor_discretization} relatively straight forward.

The restriction operators $\mathcal{R}_j$, as introduced in Section~\ref{subsec:microscopy_model}, are applied by projecting the space $S^{h, 1}$ to spaces defined with denser uniform triangular meshes $\mathcal{P}^{h_j}$. The denser meshes have the same cells arrangement as $\mathcal{P}^h$ in the subdomain domains $R_j(\mathcal{D}_c)$. We then apply the imaging operator $\mathcal{I}$ and the statistical sensor operator $\boldsymbol{\mathcal{S}}$ to the order parameter restricted to $R_j(\mathcal{D}_c)$.

The interpolation of images into $H^1(\mathcal{D}_u$) is implemented through the similar finite element basis as define in \eqref{eq:finite_element_basis}. The DoFs of the finite element basis match the the dimension of the image and the spatial arrangement of the pixels is preserved.

The finite element discretization, the restriction operators, and image interpolation are implemented through the \texttt{FEniCS} library.
\section{Parameterization and computation of transport maps} \label{app:map_details}

In this section, we discuss our parametrization of triangular and monotone transport maps and an adaptive algorithm for approximating the map given samples from the target density $\pi$. As shown in Section~\ref{subsec:triangular_maps}, the optimization problem for the map $S$ decouples into separate optimization problems for each map component. Thus, this section focuses on the individual optimization problem for each component.

To learn a monotone function for approximating $S^k$, we consider maps that are parameterized to be monotone by construction. Following~\cite{Baptista2020}, we represent each map component $S^k:\mathbb{R}^k \to \mathbb{R}$ as  
\begin{equation} \label{eq:monotone_comp_parameterization}
    S_k(\w_{1:k}) = \mathcal{M}_k(f)(\w_{1:k}) \coloneqq f(\w_{1:k-1},0) + \int_0^{w_k} g(\partial_k f(\w_{1:k-1},t)) \textrm{d}t,
\end{equation}
where $g\colon \mathbb{R} \rightarrow \mathbb{R}_{>0}$ is a positive function so that $\partial_{k} S_k(\w_{1:k}) \ge 0$ for all $\w_{1:k} \in \mathbb{R}^{k}$. Different choices for $g$ include the squared function $g(\w) = \w^2$ and the soft-plus function $g(\w) = \log(1 + \exp(\w))$. The operator $\mathcal{M}_k: f \mapsto\mathcal{M}_k(f)$ in~\eqref{eq:monotone_comp_parameterization} maps the unconstrained function $f\colon \mathbb{R}^{k} \rightarrow \mathbb{R}$ into a monotone function.
Thus, using this representation for $S_k$, we can focus on approximating $f$.
Similarly to the numerical experiments in~\cite{Baptista2020}, we consider linear expansions for $f$ given by $f(\w_{1:k}) = \sum_{j} \alpha_j \psi_{j}(\w_{1:k})$, where $\psi_j$ are prescribed features and $\alpha_j \in \mathbb{R}$ are unknown coefficients. In this work, we consider tensor products of normalized Hermite functions with respect to each variable, i.e., $\psi_{j}(\w_{1:k}) = \prod_{i=1}^{k} \psi_{j,i}(\w_i)$, but we note that $f$ can also be parameterized using other features or non-linear expansions (e.g., neural networks). 

Given a set of summary statistics, the coefficients $\boldsymbol{\alpha} = (\alpha_j)$ can be found using a gradient-based optimization technique, such as the BFGS or Newton--CG algorithms~\cite{nocedal2006numerical}. To identify the set of features in $f$, we use the Adaptive Transport Map algorithm introduced in~\cite{Baptista2020}. This procedure begins with an empty set of features $\Lambda_{0}$, i.e., the zero function for $f$, and incrementally adds feature to at a time that most reduces the value of the objective function.  To determine the stopping point, i.e., the maximum number of features, we use a validation set. In our numerical experiments, we partition the data by using 80\% of the samples to learn the map and the remaining 20\% to evaluate the negative log-likelihood of the approximate conditional density. If the minimum negative log-likelihood doesn't improve for at least 20 iterations, we stop the adaptive procedure and return the approximation for the basis size with the lowest validation error. In practice, we also limit the number of features to $p=100$ features for 2-dimensional maps and $p=250$ features for higher-dimensional maps. During the iteration, we evaluated the integrals in~\eqref{eq:monotone_comp_parameterization} numerically using 16 Gaussian quadrature points for each sample. We observed that a larger number of points did not affect the features selected by the adaptive procedure.
\end{appendices}

\end{document}